\documentclass[12pt]{article}
\usepackage{psfrag,graphicx,subfigure,amssymb}

\makeatletter
  
  \@addtoreset{equation}{section}
\makeatother

\begin{document}

%
%
\begin{titlepage}

\renewcommand{\thefootnote}{\fnsymbol{footnote}}

\begin{flushright}
MIFP-02-09\\
OU-HET 418\\
hep-th/0211232 \\
November 2002
\end{flushright}

\bigskip

\begin{center}
{\Large \bf Open Wilson Lines as States of Closed String}\\

\bigskip
\bigskip
Koichi MURAKAMI\footnote{E-mail: 
   kmurakami@physics.tamu.edu}\\
\bigskip
{\small \it George P. and
Cynthia W. Mitchell Institute for Fundamental Physics,\\
Texas A\&M University, College Station,
TX 77843-4242, USA
}

\bigskip
Toshio NAKATSU\footnote{E-mail:
    nakatsu@het.phys.sci.osaka-u.ac.jp}\\
\bigskip
{\small \it Department of Physics,
        Graduate School of Science, Osaka University,\\
        Toyonaka, Osaka 560-0043, Japan
}
\end{center}

\bigskip
\bigskip

\begin{abstract}

System of a D-brane in bosonic string theory
on a constant $B$ field background 
is studied in order to obtain further insight 
into the bulk-boundary duality. 
Boundary states which describe arbitrary numbers of 
open-string tachyons and gluons are given. 
UV behaviors of field theories on the non-commutative 
world-volume are investigated by using these states. 
We take zero-slope limits of 
generating functions of one-loop amplitudes 
of gluons (and open-string tachyons)
in which the region of the small open-string proper time
is magnified. 
Existence of $B$ field
allows the limits to be slightly different from
the standard field theory limits of closed-string.
They enable us
to capture world-volume theories at a trans-string scale. 
In this limit the generating functions 
are shown to be factorized by two curved open Wilson lines 
(and their analogues) and become 
integrals on the space of paths  
with a Gaussian distribution around straight lines.  
These indicate a possibility that field theories 
on the non-commutative world-volume   
are topological at such a trans-string scale.
We also give a proof of 
the Dhar-Kitazawa conjecture 
by making an explicit correspondence 
between the closed-string states and the paths. 
Momentum eigenstates of closed-string or 
momentum loops also play an important role 
in these analyses.  

\end{abstract}

\setcounter{footnote}{0}
\renewcommand{\thefootnote}{\arabic{footnote}}

\end{titlepage}
%
%

\newtheorem{formula}{Formula}[section]

\newtheorem{definition}{Definition}[section]
\newtheorem{proposition}{Proposition}[section]
\newtheorem{theorem}{Theorem}[section]
\newtheorem{lemma}{Lemma}[section]
\newtheorem{remark}{Remark}[section]
\newcommand{\Bll}{\Bigl\langle}
\newcommand{\Brr}{\Bigr\rangle}
\newcommand{\bll}{\bigl\langle}
\newcommand{\brr}{\bigr\rangle}
\newcommand{\Bl}{\Bigl |}
\newcommand{\Br}{\Bigr |}
\newcommand{\bl}{\bigl |}
\newcommand{\br}{\bigr |}

\newcommand{\Vg}{V_{A}}
\newcommand{\hVg}{\hat{V}_{A}}

\section{Introduction}

Systems of interacting open- and closed-strings 
play important roles in several aspects of string theory. 
It has been found out that
theories even formulated as pure closed-strings
have open-string sectors when they are accompanied by
D-branes \cite{Polchinski}. 
One of the most important features of 
such open-closed mixed systems is a duality 
between open- and closed-strings. 
This duality becomes manifest, for instance, 
by seeing one-loop diagrams of open-string 
as tree propagations of closed-string
through modular transformations 
on the string world-sheets.
In the systems of D-branes, 
this duality should be a rationale for 
correspondence between gauge theory on the world-volumes
(open-string sector) and gravity theory in the bulk space-time
(closed-string sector). 
AdS/CFT correspondence \cite{Maldacena}\cite{GKP}\cite{Witten}
may be regarded as one of the most remarkable examples
of such bulk-boundary correspondence. 
This viewpoint was emphasized also 
in \cite{JKY}.

Nevertheless it becomes very difficult to establish the 
correspondence between the two. 
This is essentially because light particles (IR effects)
in the one sector are realized by summing up whole
massive towers (UV effects) in the other sector.
When a constant $B$ field background of closed-string 
is turned on, however, 
we  expect that the situation could be
drastically changed.
On this background,
the world-volumes of D-branes become
non-commutative~\cite{C-D-S}\cite{DH}\cite{Chu-Ho}\cite{AAS}
\cite{Seiberg-Witten}.
Low-energy effective theories of the open-string sector,
therefore, become field theories on the non-commutative 
world-volumes.
It was pointed out \cite{UV-IR mixing} that
in such theories there happens a mixing of the UV and the IR.
This UV/IR mixing gives 
\cite{UV-IR mixing} 
us a chance to capture some effects
of light particles in the bulk gravity theories,
e.g.\ gravitons, by investigating
non-commutative field theories on the world-volumes.
To pursue such possibility, 
non-planar one-loop amplitudes of open-string were
studied on this background
\cite{Andreev-Dorn}\cite{Kiem-Lee}%
\cite{BCR}\cite{Gomis et al}\cite{Liu-Michelson}.
Higher loops were investigated as well
in \cite{2-loop}.

Coupling of non-commutative D-branes to closed-string
in the bulk has attracted much interest
particularly from the viewpoint of bulk-boundary correspondence
\cite{H-I}\cite{M-R}.
It was pointed out in \cite{HKLL}\cite{Garousi}
that the generalized star-products arise in
disk amplitudes consisting of a closed-string vertex operator
and open-string vertex ones
on a constant $B$ field background.
The generalized star products are also 
found to appear in straight open Wilson lines 
by the expansions in powers of non-commutative gauge fields
\cite{Mehen-Wise}.
These observations were combined in \cite{Okawa-Ooguri}
and it was shown there that in the zero-slope limit of
\cite{Seiberg-Witten}
disk amplitudes of
a closed-string tachyon and arbitrary numbers of
gauge fields on this background
give rise to a straight open Wilson line.
Thus it has been clarified that
open Wilson lines play a role in the correspondence
between the bulk gravity and the non-commutative gauge
theories.

Open Wilson lines found in \cite{IIKK}
are remarkable gauge invariant objects
in non-commutative gauge theories.
Gauge invariant operators in non-commutative gauge theories
can be constructed in the forms of local operators
smeared along straight Wilson
lines~\cite{IIKK}\cite{GHI}\cite{Das-Rey}. 
It was also shown in \cite{Okawa-Ooguri} that
coupling of closed-string graviton to non-commutative
gauge theories actually is an operator of this form.

Open Wilson lines need not be straight
in order to be gauge invariant.
Taking account of the above role
played by straight open Wilson lines in the bulk-boundary duality,
Dhar and Kitazawa \cite{Dhar-Kitazawa} 
conjectured that 
curved open Wilson lines fluctuating around a straight path 
should be brought about by higher level (stringy) states of 
closed-string. 
They showed that the coupling of graviton state 
is in fact the gauge invariant operator which appears
in the leading coefficients in harmonic expansions of curved
open Wilson lines around a straight path.

~

\noindent
It is well-known that the so-called boundary states 
\cite{CLNY1}\cite{ghost boundary state}\cite{CLNY3} 
provide a description of D-branes in closed-string theory. 
Open- and closed-strings interact on the branes 
in space-time. 
Therefore it is further expected that 
boundary states admit to describe these interactions. 
In this article, 
receiving the above development and understanding, 
we study the system of a D-brane  
in bosonic string theory
on a constant $B$ field background. 
In order to obtain further insight 
into the bulk-boundary duality, 
we first exploit 
the boundary state formalism 
to include states which describe 
arbitrary numbers of open-string tachyons and gluons. 
Open-string legs of these boundary states, 
that is, 
tachyons and gluons, 
need not to be on-shell. 
Thereby it becomes possible to study the duality.

On a constant $B$ field background,   
boundary state without open-string legs 
has been constructed 
in \cite{CLNY1}\cite{ghost boundary state}\cite{CLNY3}. 
It is briefly reviewed in the next section. 
Let us denote the state 
by $|B\rangle$. 
This state does not lead any couplings 
of a closed-string to open-string excitations 
on the world-volume. 
We may regard $|B\rangle$ as 
a perturbative vacuum of closed-string 
in the presence of the brane. 
It is also possible to 
interpret the state as a Bogolubov transform of 
the $SL_{2}(\mathbb{C})$-invariant vacuum of 
closed-string :
$\left|B\right \rangle =g |\mathbf{0}\rangle$, 
where $g$ is a suitable generator of the transformation.

In Section \ref{sec:tachyon boundary state} 
we construct boundary states of open-string tachyons. 
Basic idea of the construction may be explained as follows.  
We first examine an insertion of 
a closed-string tachyon   
into the brane.  
Let us suppose that the closed-string world-sheet  
is an infinite semi-cylinder $(\sigma,\tau)$ 
with $\tau \geq 0$ and that 
the brane or the state $|B\rangle$ is located at $\tau=0$. 
We may describe the insertion by 
$\lim_{\tau \rightarrow 0+} V_T(\sigma,\tau)|B \rangle$, 
where $V_T$ is a closed-string tachyon vertex operator. 
However, there occurs a singularity within $V_T$ at 
the world-sheet boundary. This originates in correlations 
between the chiral and the anti-chiral sectors 
which are caused by the brane. 
We then regularize such a singularity and define 
a renormalized tachyon vertex operator $V^{ren}_T$. 
It turns out surprisingly that 
the renormalized operator becomes 
an open-string tachyon vertex at the boundary. 
In our prescription of the renormalization,  
disk Green's function and the transform $g^{-1}V_Tg$ 
play essential roles. 
Boundary state of a single open-string tachyon is given by 
$\lim_{\tau \rightarrow 0+} V^{ren}_T(\sigma,\tau)|B \rangle$.  
Boundary states of arbitrary numbers of open-string tachyons 
turn out to be realized by the successive insertions of 
the above renormalized operators. 
As for boundary states of gluons, 
they can be obtained  by taking the same steps 
as illustrated above,  
except that we need to introduce a suitable local operator 
of closed-string in place of $V_T$ and renormalize 
it in a suitable fashion. 
These are presented 
in Section \ref{sec:gluon boundary state}. 
Consistency of the constructions is examined from 
several aspects at some length.  
In particular, we compute 
closed-string tree propagations between these boundary states 
and make sure that they reproduce the corresponding open-string 
one-loop amplitudes.
These are presented 
separately in Section \ref{sec:tachyon boundary state}
for the tachyons and 
in Section \ref{sec:gluon boundary state}
for the gluons. 
It is also worth noting that the present constructions 
are relevant on a vanishing $B$ field background.

Low-energy description of world-volume theories 
of D-branes is obtained by taking 
a zero-slope limit $\alpha' \rightarrow 0$ so that 
it makes all perturbative stringy states 
($\mbox{mass}^2 \sim 1/\alpha'$) of open-string infinitely 
heavy and decouple from the light states.  
As regards one-loop amplitudes of open-string,  
this can be achieved 
by taking the limit with $\alpha' \tau^{(o)}$ fixed.  
Here $\tau^{(o)}$ is the proper time on the world-sheet 
of open-string.  
The fixed parameter $s^{(o)}\equiv \alpha' \tau^{(o)}$ 
becomes the Schwinger parameter of one-loop amplitudes of 
low-energy effective world-volume theories.  
It is shown 
\cite{Andreev-Dorn}\cite{Kiem-Lee}%
\cite{BCR}\cite{Gomis et al}\cite{Liu-Michelson}\cite{S.J.Rey et al} 
that one-loop amplitudes of open-string tachyons on a constant 
$B$ field background reduce to those of 
a non-commutative (tachyonic) scalar field theory on the world-volume. 
The above zero-slope limit can be interpreted 
as magnification of the string amplitudes in the vicinity 
of $\tau^{(o)}=+\infty$. 
In Section \ref{sec:UV NC scalar}
we investigate a possible 
UV behavior of this non-commutative field theory. 
In order to know the UV behavior one needs to focus on the 
region $s^{(o)}\approx 0$.  
For the description we take the following route.
We introduce a parameter $s^{(c)}\equiv \alpha' |\tau^{(c)}|$,
where $\tau^{(c)}$ is the closed-string proper time.
$s^{(o)}$ and $s^{(c)}$ are related to each other
through the modular transformation of the world-sheet 
by $s^{(c)}=2\alpha'^{2}/s^{(o)}$.
Then we take a zero-slope limit with $s^{(c)}$ fixed. 
Existence of $B$ field
allows us to make the limit slightly different from 
the standard field theory limit of closed-string 
(gravity limit).  
This enables us to capture the world-volume theories.
This zero-slope limit is magnification of 
the string amplitudes 
in the vicinity of $\tau^{(o)}=0$  and 
hence the region $s^{(o)}\approx 0$. 
Strictly speaking, this is  
a trans-string scale of the world-volume theories.
Generating function of 
one-loop amplitudes of open-string tachyons 
is found out to be factorized at the limit 
into two (analogues of) straight open Wilson lines
exchanging closed-string tachyons.
In general, one-loop amplitudes of open-string are 
factorized by a tower of closed-string states. 
In gravity limit 
the propagations of closed-string tachyons become dominant.
Although the present limit is slightly different from
gravity limit, 
these zero-slope limits share the common property. 
UV behavior of the non-commutative gauge theory 
is explored in Section \ref{sec:UV NC gauge} 
by following the same route. 
Generating function of one-loop amplitudes of gluons 
is also found out to be factorized at this limit  
into two straight open Wilson lines
exchanging closed-string tachyons. 
These analyses indicate a possibility 
that field theories on the non-commutative world-volume   
become topological at such a trans-string scale 
of the world-volume.

The conjecture \cite{Dhar-Kitazawa} 
leads us to expect that 
curved open Wilson lines somehow factorize 
generating functions of one-loop amplitudes 
of gluons or open-string tachyons. 
The standard factorization made by particle 
states of closed-string seems to be little use. 
Instead, 
momentum eigenstates of closed-string are 
used for the factorization. 
Closed-string momentum $P(\sigma)$ is 
a momentum loop (a loop in the momentum space) 
while closed-string coordinate $X(\sigma)$ is 
a space-time loop. 
Section \ref{sec:comments on eigenstates}
is devoted to introduction of coordinate and 
momentum eigenstates of closed-string.
We also provide 
some observations on their relations 
with boundary states. 
{}Factorizations made by the momentum eigenstates 
are examined 
in Section~\ref{sec:open Wilson lines (I)} for the tachyons 
and 
in Section \ref{sec:open Wilson lines (II)} for the gluons. 
It turns out that 
in the zero-slope limit
(the previous UV limit of the world-volume theories) 
momentum loops become curves in the non-commutative world-volume 
and that 
open Wilson lines along these curves appear 
in the factorizations.
The generating functions 
are factorized by two curved open Wilson lines. 
More precisely, 
they become integrals 
on the space of curves with a Gaussian distribution
around straight lines.  
As regards fluctuations from the straight lines, 
width of the distribution becomes so sharp 
that the integrals reduce to the previous 
factorizations by straight open Wilson lines.

We start Section \ref{sec:open Wilson lines (II)}
by giving a proof of the conjecture 
made by Dhar and Kitazawa. 
We introduce the closed-string state 
$
|\Omega (P)\rangle
=
:\exp 
\left( 
i\int_0^{2\pi} 
   d\sigma 
   P_{\mu}(\sigma)\hat{X}^{\mu}(\sigma)
\right)
:
|\mathbf{0} \rangle
$, where $P(\sigma)$ is a momentum loop. 
This state is not 
an eigenstate of closed-string momentum 
but serves as a generating function of 
(generally off-shell) closed-string states. 
Overlap between 
$|\Omega(P)\rangle$ and 
boundary states with open-string legs 
is a generating function of 
couplings of all the closed-string states 
to the non-commutative D-brane. 
The overlap with the boundary states of gluons 
is shown to become 
a curved open Wilson line 
in the zero-slope limit. 
We make an explicit correspondence 
between the closed-string states and the coefficients
of harmonic expansions of the curve.  
These provide the proof. 
We also show that 
in the zero-slope limit
the momentum eigenstate is identified with the state
$| \Omega (P) \rangle$ after some manipulation. 
This accounts for the previous factorizations
by curved open Wilson lines.

Wilson line is invariant under reparametrizations of the path. 
The corresponding transformations on the string world-sheet 
are reparametrizations of the boundary. 
The reparametrization invariance of boundary states 
is fulfilled by imposing the Ishibashi condition 
\cite{Ishibashi condition} 
or equivalently 
the BRST invariance on these states. 
As regards the boundary states of open-string tachyon 
and gluon it is shown 
in Sections \ref{sec:tachyon boundary state} 
and \ref{sec:gluon boundary state}
that 
their reparametrization invariance 
is equivalent to the on-shell conditions.   
On the other hand we do not require any condition 
on gluons to obtain open Wilson lines. 
This puzzle is solved by seeing that 
the Ishibashi condition or the BRST invariance 
becomes null in the present zero-slope limit.

It is observed in Section \ref{sec:discussions} that 
all the boundary states constructed so far in this article 
are eigenstates of the closed-string momentum operators.
The eigenvalues are essentially delta functions  
on the world-sheet boundary. 
Their boundary actions are computed by following 
the prescription given in \cite{CLNY3}.
It turns out that they are the standard boundary 
actions used in the path-integral formalism 
of the world-sheet theory. 
After a speculation based on these observations 
we finally make a conjecture on the duality 
between open- and closed- strings.

In Appendix \ref{sec:open-string tensors}
world-volume and space-time tensors used in the text 
are summarized including their relations. 
In Appendix \ref{sec:formulae}
some formulae of creation and annihilation modes 
are described. These are necessary for our 
computations of several string amplitudes 
in the text. 
Oscillator realizations of coordinate and momentum 
eigenstates are presented 
in Appendix \ref{sec:eigen}.


\section{A Short Course on Boundary States}

Let us consider the system of D$p$-brane 
in bosonic string theory. 
We take the following closed-string background :
\begin{eqnarray}
&&ds^{2}=g_{MN}dx^{M}dx^{N}
    =g_{\mu\nu}dx^{\mu}dx^{\nu}+g_{ij}dx^{i}dx^{j}~,
    \nonumber\\
&& B=\frac{1}{2}B_{\mu\nu}dx^{\mu}\wedge dx^{\nu}~. 
\label{eq:bg}
\end{eqnarray}
Here $g_{MN}$ is a flat space-time metric,  
which we refer to as closed-string metric. 
Two-form gauge field $B_{\mu\nu}$ is constant. 
We divide the space-time directions into two pieces,
$x^{M}=(x^{\mu},x^{i})$, 
where $\mu=0,1,\cdots,p$ and $i=p+1,\cdots,D-1$. 
The directions $x^{\mu}$ are supposed to be parallel 
to the D$p$-brane. 
The directions $x^{i}$ are perpendicular 
to the brane.

Closed-string may capture the brane. 
Relevant action of a closed-string takes the form : 
\begin{equation}
S[X]=\frac{1}{4\pi\alpha'}\int_{\Sigma} d\tau d\sigma 
  \Bigl \{
  \partial_{a}X^{M} \partial^{a}X^{N}g_{MN}
  -i2\pi\alpha' \epsilon^{ab}\partial_{a}X^{\mu}
   \partial_{b}X^{\nu}B_{\mu\nu}
  \Bigr \}~,
 \label{eq:action1}
\end{equation}
where $\epsilon^{ab}$
is the antisymmetric tensor 
on the world-sheet with 
$\epsilon^{\tau\sigma}\equiv 1$.
The world-sheet $\Sigma$ is a disk or an infinite 
semi-cylinder, and we use the cylinder coordinates 
$(\tau,\sigma)$ 
($\tau \geq 0$ and $0\leq \sigma < 2\pi$). 
Closed-string interacts with the brane at the 
boundary of the world-sheet, that is, at $\tau=0$. 
The second term of the action is an integration of 
the two-form (its pull-back) on $\Sigma$. 
Since it is an exact two-form, 
by applying the Stokes theorem we can recast 
it into a boundary integral :
\begin{equation}
S[X]
=\frac{1}{4\pi\alpha'}\int_{\Sigma} d\tau d\sigma 
  \partial_{a}X^{M} \partial^{a}X^{N}g_{MN}
  -\frac{i}{2}\int_{\partial\Sigma} d\sigma
    B_{\mu\nu}X^{\mu} \partial_{\sigma}X^{\nu}~.
    \label{eq:action2}
\end{equation}
Energy-momentum tensor $T_{ab}(\sigma,\tau)$ 
is obtained from the action. 
Since the integration of the two-form  
is independent of the world-sheet metric,
$T_{ab}$ acquires 
the standard form without the $B$-field.

Closed-string coordinates $X^{M}(\sigma,\tau)$
have mode expansions of the form, 
\begin{equation}
X^{M}(\sigma,\tau)=\hat{x}_{0}^{M}
-i\alpha'\hat{p}_{0}^{M}\tau
+i\sqrt{\frac{\alpha'}{2}}
  \sum_{n\neq 0}\left(\frac{\alpha^{M}_{n}}{n}e^{-n(\tau+i\sigma)}+
      \frac{\tilde{\alpha}^{M}_{n}}{n}e^{-n(\tau-i\sigma)}\right)~,
 \label{eq:modeX}
\end{equation}
and the standard first quantization requires   
the following commutation relations :
\begin{eqnarray}
&&[\hat{x}^{M}_{0}, \hat{p}^{N}_{0}]=ig^{MN}~,
\quad
[\hat{x}_{0}^{M},\hat{x}^{N}_{0}]
=[\hat{p}^{M}_{0},\hat{p}^{N}_{0}]=0~,\nonumber\\
&&[\alpha^{M}_{m},\alpha^{N}_{n}]=mg^{MN}\delta_{m+n}~,
\quad
[\tilde{\alpha}^{M}_{m}, \tilde{\alpha}^{N}_{n}]
=mg^{MN}\delta_{m+n}~,
\quad
[\alpha^{M}_{m}, \tilde{\alpha}^{N}_{n}]=0~.
\label{eq:CCR}
\end{eqnarray}
The energy-momentum tensor generates reparametrizations 
of the world-sheet. 
Their infinitesimal forms turn out to be the Virasoro 
algebras with the central charges equal to $D$. 
The Virasoro generators are given by the expansions, 
$
T_{zz}(z)
$
$
=
\sum_{m\in\mathbb{Z}}L_{m}z^{-m-2}
$
and 
$
T_{\overline{z}\overline{z}}(\overline{z}) 
$
$
=
\sum_{m\in\mathbb{Z}}\tilde{L}_{m}\bar{z}^{-m-2}
$. 
Here we use the complex coordinates $(z,\bar{z})$ 
instead of the cylinder coordinates. 
They are related by 
$(z,\bar{z})=(e^{\tau+i\sigma},e^{\tau-i\sigma})$.
$L_{m}$ and $\tilde{L}_{m}$ 
are generators of the chiral and the anti-chiral sectors
respectively. 
These have the following representations in 
terms of the oscillator modes : 
\begin{equation}
L_{m}= \frac{1}{2}:\sum_{n\in \mathbb{Z}}g_{MN}\alpha^{M}_{n}
        \alpha^{N}_{m-n}:~, \quad
\tilde{L}_{m} 
     =\frac{1}{2} : \sum_{n\in \mathbb{Z}}g_{MN}
  \tilde{\alpha}^{M}_{n}   \tilde{\alpha}^{N}_{m-n}:~,
\label{closed Virasoro generators}
\end{equation}
where
$
\alpha^{M}_{0}
=\tilde{\alpha}^{M}_{0}
=\sqrt{\frac{\alpha'}{2}}p^{M}_{0}
$.
$: \ :$ denotes the standard normal ordering
with respect to the $SL_{2}(\mathbb{C})$-invariant vacuum of
closed-string $|\mathbf{0}\rangle$:
\begin{equation}
:\hat{x}^{M}_{0} \hat{p}^{N}_{0}:~=~
 :\hat{p}^{N}_{0}\hat{x}^{M}_{0}:~=~
 \hat{x}^{M}_{0} \hat{p}^{N}_{0}~,
\qquad
:\alpha^{M}_{-n}\alpha^{N}_{n}:~=~
:\alpha^{N}_{n}\alpha^{M}_{-n}:~=~
\alpha^{M}_{-n}\alpha^{N}_{n}~,
\label{closed-string normal ordering}
\end{equation}
for $n\geq 1$ and the similar prescription 
for the anti-chiral modes.

Closed-string can find the brane through 
the boundary conditions imposed at $\tau=0$. 
We take a variation of the action : 
\begin{eqnarray}
\delta S[X]
&=& 
 -\frac{1}{2\pi\alpha'}
    \int_{\Sigma} d\tau d\sigma
         \delta X^{M} g_{MN}
         \Bigl(
               \partial_{\tau}\partial_{\tau}
                +
               \partial_{\sigma}\partial_{\sigma}
          \Bigr)
   X^{N} 
\nonumber\\
&& 
+\frac{1}{2\pi\alpha'}
     \int_{\partial \Sigma}d\sigma
           \left[ 
              \delta X^{\mu}
                 \Bigl(
                    g_{\mu\nu}\partial_{\tau}X^{\nu}
                    -
                    2\pi i\alpha'B_{\mu\nu}
                        \partial_{\sigma}X^{\nu}
                 \Bigr)
              + 
               \delta X^{i}g_{ij}
                   \partial_{\tau}X^{j}
\right]~.
\end{eqnarray}
Vanishing of the first term (bulk term) 
for an arbitrary 
$\delta X^{M}(\sigma,\tau)$ gives  
the equation of motions of the string-coordinates. 
Vanishing of the second term (boundary term) 
needs to be considered separately for 
$\delta X^{\mu}(\sigma,0)$ and 
$\delta X^{i}(\sigma,0)$. 
The interaction with the brane is possible 
everywhere on its world-volume. 
Thus $\delta X^{\mu}(\sigma,\tau)$ are arbitrary 
at the boundary. We need to impose 
the Neumann boundary condition on
$X^{\mu}$  
\footnote{Precisely speaking, this is a mixed
boundary condition. We refer to this boundary
condition as the Neumann boundary condition because it
reduces to the Neumann boundary condition in the absence
of $B$ field. 
In this paper we use this terminology otherwise stated.
}. 
On the other hand $\delta X^i(\sigma,\tau)$ need 
to vanish at the boundary. 
For the vanishing we impose the Dirichlet boundary 
condition on $X^i$ :
\begin{eqnarray}
\Bigl[
     g_{\mu\nu}\partial_{\tau}X^{\nu}
          -2\pi i\alpha'
     B_{\mu\nu}\partial_{\sigma}X^{\nu}
\Bigr]
(\sigma,0)=0~,
~~~~~~
X^{i}(\sigma,0)
=x^{i}_{0}~.
\label{eq:bc2}
\end{eqnarray}
Here 
$x^{i}_{0}$ 
is a position of the brane in the space-time.
Owing to these boundary conditions we refer 
the directions $x^{\mu}$ and $x^i$ respectively 
as the Neumann and the Dirichlet directions.

The above conditions are used to introduce 
a boundary state 
in the first quantized picture of closed-string. 
Let us denote it by $| B \rangle$. 
It is a state which satisfies the following conditions :
\begin{eqnarray}
\Bigl[
      g_{\mu\nu}\partial_{\tau}X^{\nu}
           -2\pi i\alpha
      B_{\mu\nu}\partial_{\sigma}X^{\nu}
\Bigr](\sigma,0)
|B\rangle 
=
0~,~~~~~
X^{i}(\sigma,0)
|B\rangle 
= 
x^{i}_{0}
|B\rangle~.
\label{eq:bstate1}
\end{eqnarray}
These are linear constraints on $|B\rangle$ 
and determine the state modulo its normalization. 
It turns out that the state satisfies the Ishibashi 
condition \cite{Ishibashi condition} as follows :
\begin{eqnarray}
\left( L_n-\tilde{L}_{-n} \right)
| B \rangle =0
\qquad 
\mbox{for $\forall n$}~.
\label{Ishibashi condition} 
\end{eqnarray}

The above state can be interpreted as a 
perturbative vacuum of closed-string in the presence 
of the brane. 
Since there are no correlations between string-coordinates 
of the Neumann and the Dirichlet directions, we can factorize  
the state into a product :
\begin{eqnarray}
\bl B \brr 
= 
\bl B_{N} \brr 
\otimes 
\bl B_{D} \brr,
\end{eqnarray}
where the subscripts $N$ and $D$ denote 
the corresponding boundary conditions. 
We may refer to these two states respectively 
as the Neumann and the Dirichlet boundary states 
for short. 
They satisfy the following constraints :
\begin{equation}
\Bigl[
    g_{\mu\nu}\partial_{\tau}X^{\nu}
        -2\pi i\alpha'
    B_{\mu\nu}\partial_{\sigma}X^{\nu}
\Bigr]
(\sigma,0)
\bl B_{N} \brr=0~,
\quad 
X^{i}(\sigma,0) \bl B_{D} \brr
=
x^{i}_{0}\bl B_{D} \brr~.
\label{eq:bstateND}
\end{equation}
These constraints may be handled in the 
oscillator representations. 
For the Neumann boundary state   
it can be read as follows :
\begin{eqnarray}
\hat{p}_{0\mu}
|B_{N}\rangle
=0~,
~~~~
\Bigl( 
    E_{\mu\nu}\alpha^{\nu}_{n}
    +
    E^{T}_{\mu\nu}\tilde{\alpha}^{\nu}_{-n}
\Bigr)
   |B_{N}\rangle
=0\qquad 
\mbox{for $\forall n\neq 0$},
\label{eq:bstate2} 
\end{eqnarray}
where $E_{\mu \nu}$ and its transpose $E^{T}_{\mu\nu}$
are tensors
defined as 
\begin{eqnarray}
E_{\mu \nu}
=
g_{\mu \nu}+
2\pi \alpha' 
B_{\mu \nu}~,~~
E_{\mu \nu}^T
=
g_{\mu \nu}-
2\pi \alpha' 
B_{\mu \nu}~. 
\label{def of E}
\end{eqnarray}
As regards the Dirichlet boundary state 
the constraint can be read as follows :
\begin{eqnarray}
\hat{x}_{0}^{i}
|B_{D}\rangle
=
x_{0}^{i} 
|B_{D}\rangle~,
~~~~
\left(  
    \alpha^{i}_{n}-\tilde{\alpha}^{i}_{-n} 
\right)
|B_{D}\rangle
=
0 
\qquad \mbox{for}~~\forall n\neq 0~.
\label{eq:bstate3}
\end{eqnarray}

\subsubsection*{Boundary state $|B_D \rangle$}
Let us describe these boundary states 
including their normalizations. 
The above constraints can be solved 
without any difficulty 
in the oscillator representations. 
Determinations of their normalization factors 
need to be cared. 
We start with the Dirichlet boundary state.

The Dirichlet boundary state turns out 
to be given by 
\begin{eqnarray}
\bl B_D \brr 
\equiv 
 \left( 
   \frac{(2\pi^2 \alpha')^d}{\mbox{det} g_{ij}}
 \right)^{\frac{1}{4}} 
 \prod_{n=1}^{\infty}\exp 
 \left\{
   \frac{1}{n}g_{ij}\alpha_{-n}^i\tilde{\alpha}_{-n}^{j} 
 \right\}
 |x_{0D}\rangle, 
\label{state Bd}
\end{eqnarray}
where we put $d=D-p-1$. 
The state
$
|x_{0D} \rangle
=
|x_{0D}=(x^{i}_{0})\rangle
$ 
is an eigenstate of 
the zero modes $\hat{x}_{0}^{i}$ 
with eigenvalues $x_{0}^{i}$.  
Momentum representation of $|x_{0D}\rangle$ is 
given by 
\begin{eqnarray}
|x_{0D}\rangle = \frac{1}{(2\pi)^{d/2}}\int d^{d}k \,
|k_{D}\rangle e^{-ik_ix_0^i}~,
\label{eigenstate of x0}
\end{eqnarray}
where $|k_{D}\rangle=|k_{D} = (k_{i})\rangle$
denotes an eigenstate 
of the momentum zero modes $\hat{p}_{0i}$ with eigenvalues $k_{i}$
defined as
\begin{eqnarray}
|k_{D} \rangle = e^{i k_i\hat{x}_0^i}| {\bf 0} \rangle~. 
\label{eigenstate of p0}
\end{eqnarray} 
The dual state can be obtained by taking the BPZ conjugation : 
\begin{eqnarray}
\bll B_D \br 
=
\left( 
\frac{(2\pi^2 \alpha')^d}{\mbox{det} g_{ij}}
\right)^{\frac{1}{4}} 
\langle x_{0D}|
\prod_{n=1}^{\infty}\exp 
\left\{
\frac{1}{n}g_{ij}\alpha_{-n}^i\tilde{\alpha}_{-n}^{j} 
\right\}.
\label{dual state Bd}
\end{eqnarray}

It is easy to see that the state (\ref{state Bd}) 
satisfies the conditions (\ref{eq:bstate3}). 
Since they are linear conditions,  
normalization factor of the state should be 
determined by other means.  
We fixed it to be 
$\left( 
\frac{(2\pi^2 \alpha')^d}{\mbox{det} g_{ij}}
\right)^{\frac{1}{4}}$  
in the above. 
It is determined by claiming that  
closed-string propagations 
along the Dirichlet directions reproduce
the vacuum one-loop amplitude of 
open-string which satisfies 
the Dirichlet-Dirichlet (D-D) boundary conditions.

\begin{figure}
\psfrag{pitauc}{$\pi |\tau^{(c)}|$}
\psfrag{0}{$0$}
\begin{center}
\includegraphics[height=7cm]{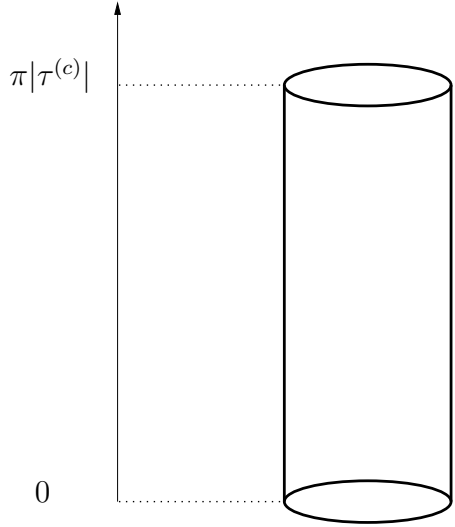}
\caption{
{\small 
Evolution of closed-string by an imaginary time 
$\pi |\tau^{(c)}|$.}}
\label{cylinder}
\end{center}
\end{figure}

\begin{figure}
\psfrag{tauo}{$\tau^{(o)}$}
\psfrag{0}{$0$}
\begin{center}
\includegraphics[height=7cm]{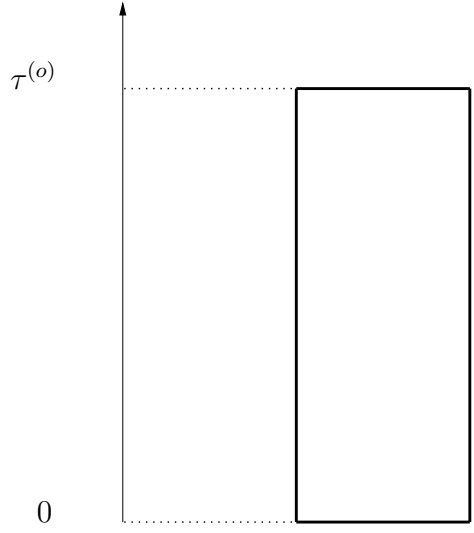}
\caption{
{\small 
Evolution of open-string by an imaginary time $\tau^{(o)}$.
Two bold horizontal lines are identified.}}
\label{strip}
\end{center}
\end{figure}

We parametrize propagations of closed-string by 
$\tau^{(c)} \in i \mathbb{R}_{\geq 0}$. 
Closed-string evolves  
by an imaginary time $\pi |\tau^{(c)}|$
with the Hamiltonian $L_0+\tilde{L_0}$. 
See Figure \ref{cylinder}.
Closed-string propagations along the Dirichlet directions 
are measured by 
$\bll B_D |q_c^{\frac{L_0+\tilde{L}_0}{2}}| B_D \brr$, 
where we put $q_c=e^{2\pi i \tau^{(c)}}$. 
This amplitude can be easily calculated by using the 
formula in Appendix \ref{sec:formulae}. 
It turns out to be
\begin{eqnarray}
\bll B_D |
q_c^{\frac{L_0+\tilde{L}_0}{2}}
| B_D \brr
&=& 
\left(
      \frac{\alpha'}{2}
\right)^{\frac{d}{2}}
\int 
\frac{dk}{\sqrt{\mbox{det}g_{ij}}}
q_c^{\frac{\alpha'}{4}k_ig^{ij}k_j} 
\prod_{n=1}^{\infty}(1-q_c^n)^{-d}, 
\nonumber \\
&=& 
e^{\frac{\pi i \tau^{(c)}d}{12}}
\left( -i\tau^{(c)} \right)^{-\frac{d}{2}}
\eta 
\left( 
   \tau^{(c)}
\right)^{-d},
\label{Bd-Bd}
\end{eqnarray}
where $\eta (\tau)\equiv q^{1/24}\prod_{n=1}^{\infty}(1-q^n)$ 
with $q=e^{2\pi i \tau}$. 
Contribution of the world-sheet reparametrization ghosts 
is excluded in Eq.(\ref{Bd-Bd}). 
It will be included in string propagations 
along the Neumann directions. 
The corresponding amplitude of open-string 
is given by $\mbox{Tr}_{D-D}e^{-\tau^{(o)}L_0}$. 
Open-string propagates by an imaginary time $\tau^{(o)}$
$(\in \mathbb{R}_{\geq 0})$ 
with the Hamiltonian $L_0$. 
See Figure \ref{strip}. 
The trace is taken over the sector of open-string 
which satisfies the D-D boundary conditions. 
The amplitude becomes as follows : 
\begin{eqnarray}
\mbox{Tr}_{D-D}e^{-\tau^{(o)}L_0}
&=& 
e^{-\frac{\tau^{(o)}d}{24}}
\eta
\left(
   -\frac{\tau^{(o)}}{2\pi i}
\right)^{-d}~.
\label{D-D}
\end{eqnarray}

The standard argument allows us to interpret 
these open-string amplitudes  
as the free propagations of closed-string. 
Two imaginary times are related 
with each other by $\tau^{(o)}=2\pi i/\tau^{(c)}$. 
This leads us to write Eq.(\ref{D-D}) as 
\begin{eqnarray}
\mbox{Tr}_{D-D}e^{-\tau^{(o)}L_0}
=
\left.
e^{-\frac{\pi i d}{12 \tau^{(c)}}}
\left( -i \tau^{(c)} \right)^{-\frac{d}{2}}
\eta
\left(
   \tau^{(c)}
\right)^{-d}
\rule{0em}{1ex}
\right|_{\tau^{(c)}=\frac{2 \pi i}{\tau^{(o)}}},
\label{D-D2}
\end{eqnarray}
where the modular transformation
$
\eta 
\left(
   -\frac{1}{\tau^{(c)}}
\right)^3
=
(-i\tau^{(c)})^{3/2}
\eta
\left(
   \tau^{(c)}
\right)^3$ 
is used.  A comparison between Eqs.(\ref{Bd-Bd}) and (\ref{D-D2}) 
gives the identity :
\begin{eqnarray}
\bll B_D |q_c^{\frac{L_0+\tilde{L}_0}{2}}| B_D \brr
=
e^{\frac{\pi i d}{12}\left(\tau^{(c)}+\frac{1}{\tau^{(c)}}\right)}
\left. 
\mbox{Tr}_{D-D}e^{-\tau^{(o)}L_0} 
\right|_{\tau^{(o)}= \frac{2\pi i}{\tau^{(c)}}}. 
\label{open-closed for D}
\end{eqnarray}
Therefore the propagations of closed-string 
reproduce correctly 
the corresponding one-loop amplitude of open-string 
except the factor 
$e^{\frac{\pi i d}{12}(\tau^{(c)}+1/\tau^{(c)})}$.  
This factor turns out to be 
canceled by a similar one appearing in closed-string 
propagations along the Neumann directions.

\subsubsection*{Boundary state $|B_N\rangle$}
The Neumann boundary state is given \cite{CLNY1} by 
\begin{eqnarray}
\bl B_N \brr 
\equiv 
\left( 
     \frac{\mbox{det}^2 E_{\mu \nu}}
           {(2\alpha')^{p+1}(-\mbox{det}g_{\mu \nu})}
\right)^{\frac{1}{4}}
\prod_{n=1}^{\infty} \exp 
\left\{ 
-\frac{1}{n}
N_{\mu \nu}\alpha_{-n}^{\mu}\tilde{\alpha}_{-n}^{\nu}
\right\} 
|{\bf 0} \rangle~,
\label{state Bn}
\end{eqnarray}
where $N_{\mu\nu}$ is a tensor defined by 
\begin{equation}
N_{\mu\nu}=\left(g\frac{1}{E}E^{T}\right)_{\mu\nu}~.
\label{def of N}
\end{equation}
The dual state is obtained by the BPZ conjugation. 
Its explicit form is as follows :
\begin{eqnarray}
\bll B_N \br =
\left( 
\frac{\mbox{det}^2 E_{\mu \nu}}
{(2\alpha')^{p+1}(-\mbox{det}g_{\mu \nu})}
\right)^{\frac{1}{4}}
\langle \mathbf{0} |
\prod_{n=1}^{\infty}
\exp \left\{ 
-\frac{1}{n}
N_{\mu \nu}\alpha_{n}^{\mu}\tilde{\alpha}_{n}^{\nu}
\right\}~.
\label{dual state Bn}
\end{eqnarray}
The above normalization factor of the state 
is chosen to reproduce the related 
one-loop amplitudes of open-string. 
This will be shown in the next section.

\section{Open-String Tachyons in Closed-String Theory}
\label{sec:tachyon boundary state}
As can be seen in Eqs.(\ref{state Bd}) and (\ref{state Bn})
these boundary states are the Bogolubov transforms of 
perturbative vacua. 
Generators of the transformations are   
$\prod_{n=1}^{\infty}
\exp
\left\{
\frac{1}{n}g_{ij}
\alpha_{-n}^i\tilde{\alpha}_{-n}^j
\right\}$ for the Dirichlet boundary 
and 
$\prod_{n=1}^{\infty}
\exp 
\left\{-\frac{1}{n}N_{\mu \nu}
\alpha_{-n}^{\mu}\tilde{\alpha}_{-n}^{\nu}
\right\}$ for the Neumann boundary. 
We can expect that information on the boundary 
conditions are all encoded in these generators.

Let ${\cal O}(\sigma,\tau)$ be a local 
operator of closed-string. 
Action of this operator on a boundary state 
$g | {\bf 0} \rangle$, where $g$ is a generator, 
can be written as follows :
\begin{eqnarray}
{\cal O}(\sigma,\tau) g 
| {\bf 0} \rangle 
=
g \times 
g^{-1}{\cal O}(\sigma,\tau) g 
| {\bf 0} \rangle.
\end{eqnarray} 
One may expect that 
$g^{-1}{\cal O}(\sigma,\tau) g$ 
describes the Bogolubov transform 
of this local operator and ask 
the physical implication particularly 
from the viewpoint of boundary conformal 
field theories. 
However, story is not so simple. 
In general,  
$g^{-1}{\cal O}(\sigma,\tau) g$ 
turns out to be singular. 
More precisely it becomes singular at 
$\tau=0$, where the boundary state resides. 
This reflects the fact that 
the system under consideration is actually 
a system of closed- and open-strings. 
We wish to make an idea of the Bogolubov 
transformations of these local operators 
rigorous. 
For this purpose, we have to perform a regularization
by which the above singularity becomes tractable. 
This leads us to define
$\mbox{ad}_{g^{-1}}{\cal O}(\sigma,\tau)$, 
which becomes regular at the world-sheet boundary. 
It is a local operator and interpreted as 
the adjoint transform of ${\cal O}(\sigma,\tau)$ 
by $g^{-1}$. We will find out their physical 
interpretation.   
In this section we concentrate on 
tachyon vertex operators. 
Since we are mainly  
interested in the world-volume theory of $p$-brane,  
we restrict ourselves to the Bogolubov transform associated 
with the Neumann boundary state. We denote the generator 
by $g_N$, 
\begin{eqnarray}
g_N \equiv 
\prod_{n=1}^{\infty}
\exp 
\left\{-\frac{1}{n}N_{\mu \nu}
\alpha_{-n}^{\mu}\tilde{\alpha}_{-n}^{\nu}
\right\}.
\label{gN}
\end{eqnarray}

\subsection{Bogolubov transformation and renormalization}

Let $V_T(\sigma,\tau;k)$ be closed-string tachyon vertex 
operator of momentum $k$. 
Momentum $k$ is supposed to have only components 
along the Neumann directions. 
An explicit form is given by 
\begin{eqnarray}
V_T(\sigma,\tau;k)
&=&
:e^{ik_{\mu}X^{\mu}(\sigma,\tau)}:,
\nonumber \\
&=& 
\exp(ik_{\mu}\hat{x}^{\mu}_0)
\times 
|z|^{\alpha'k_{\mu}\hat{p}_0^{\mu}}
\times 
\prod_{n=1}^{\infty}
\exp \left[ 
\sqrt{\frac{\alpha'}{2}}
\frac{1}{n}k_{\mu}
    (\alpha_{-n}^{\mu}z^n
            +\tilde{\alpha}_{-n}^{\mu}\bar{z}^n) \right] 
\nonumber \\
&& ~~
\times 
\prod_{n=1}^{\infty}
\exp \left[ 
-\sqrt{\frac{\alpha'}{2}}
\frac{1}{n}k_{\mu}
    (\alpha_{n}^{\mu}z^{-n}
            +\tilde{\alpha}_{n}^{\mu}\bar{z}^{-n}) \right]. 
\label{closed-string tachyon vertex}
\end{eqnarray}

To discuss the Bogolubov transform  
it is convenient to write down the transforms 
of oscillator modes of the string coordinates. 
These can be read as :
\begin{eqnarray}
g_N^{-1}\alpha_{n}^{\mu}g_N
=\alpha_{n}^{\mu}
-{(g^{-1}N)^{\mu}}_{\nu}\tilde{\alpha}_{-n}^{\nu},~~~~~
g_N^{-1}\tilde{\alpha}_{n}^{\mu}g_N
=\tilde{\alpha}_{n}^{\mu}
-{(g^{-1}N^{T})^{\mu}}_{\nu}\alpha_{-n}^{\nu}, 
\label{transforms of massive modes}
\end{eqnarray}
for $n \geq 1$. 
The modes $\alpha_n^{\mu}$ 
and $\tilde{\alpha}_n^{\mu}$ 
for $n \leq 0$  
are kept intact.   
(We put $\alpha_0^{\mu}=\tilde{\alpha}_0^{\mu}\equiv 
\sqrt{\alpha'/2}\hat{p}_0^{\mu}$.) 
The above mixture of the creation- and annihilation-modes 
makes the transform $g_N^{-1}V_T(\sigma,\tau ; k)g_N$ 
singular at $\tau=0$ 
or equivalently, since we put $z=e^{\tau+i\sigma}$,  
at $|z|=1$. 
It can be written in the following form for $|z|>1$ :  
\begin{eqnarray}
g_N^{-1}V_T(\sigma,\tau;k)g_N
&=&
\left(
\frac{|z|^2}{|z|^2-1}
\right)^{-\frac{\alpha'}{2}
          k_{\mu}(g^{-1}Ng^{-1})^{\mu \nu}k_{\nu}}
\times
\mbox{ad}_{g_N^{-1}}V_T(\sigma,\tau;k).  
\label{transform of VT}
\end{eqnarray}
Here we introduce a local operator
$\mbox{ad}_{g_N^{-1}}V_T$ 
which we interpret as 
the Bogolubov transform of $V_T$. 
It takes the form of
\begin{eqnarray}
\mbox{ad}_{g_N^{-1}}V_T(\sigma,\tau;k)=
{\cal N}(\bar{z};k)\tilde{{\cal N}}(z;k)
V_T(\sigma,\tau;k), 
\label{normal ordered transformed VT}
\end{eqnarray}
where ${\cal N}(\bar{z};k)$ and $\tilde{{\cal N}}(z;k)$ 
are operators consisting only of the creation modes :
\begin{eqnarray}
{\cal N}(\bar{z};k)&=&
\prod_{n=1}^{\infty}
\exp \left\{ 
\sqrt{\frac{\alpha'}{2}}\frac{1}{n}
k_{\mu}{(g^{-1}N^T)^{\mu}}_{\nu}\alpha_{-n}^{\nu}
\bar{z}^{-n}
\right\},
\nonumber \\
\tilde{{\cal N}}(z;k)&=&
\prod_{n=1}^{\infty}
\exp \left\{ 
\sqrt{\frac{\alpha'}{2}}\frac{1}{n}
k_{\mu}{(g^{-1}N)^{\mu}}_{\nu}\tilde{\alpha}_{-n}^{\nu}
z^{-n}
\right\}.
\label{N and tilded N}
\end{eqnarray}

Singularity of $g^{-1}_{N}V_T(\sigma,\tau ;k)g_{N}$ comes from 
the factor 
$
\left(
\frac{|z|^2}{|z|^2-1}
\right)^{-\frac{\alpha'}{2}
          k_{\mu}(g^{-1}Ng^{-1})^{\mu \nu}k_{\nu}}
$
in Eq.(\ref{transform of VT}).
Because of this factor, the transform
(\ref{transform of VT}) becomes singular
at $\tau=0$ where the boundary state $\bl B_{N} \brr$
is located.
As will be seen soon, this factor should be subtracted
in our construction of open-string tachyon vertex operator
from a closed-string local operator.
This factor can therefore be regarded 
as a renormalization factor 
of open-string tachyon vertex operator 
under its interpretation in terms of closed-string. 
Putting this factor aside for a while,  
let us examine the operators  $\mbox{ad}_{g_N^{-1}}V_T$.  
We first discuss their operator product expansion (OPE). 
It is convenient to recall the OPE between 
closed-string tachyon vertex operators. 
It can be read from the expansion 
(\ref{closed-string tachyon vertex}) as follows :  
\begin{equation}
V_T(\sigma_1,\tau_1;k^{(1)})
V_T(\sigma_2,\tau_2;k^{(2)})
= 
|z_1-z_2|^{\alpha'k_{\mu}^{(1)}g^{\mu \nu}k_{\nu}^{(2)}}
\times  
:
V_T(\sigma_1,\tau_1;k^{(1)})
V_T(\sigma_2,\tau_2;k^{(2)})
:~, 
\label{OPE VT}
\end{equation}
for $|z_1|\geq |z_2|$ and $z_1 \neq z_2$. 
The OPE under consideration changes from Eq.(\ref{OPE VT}). 
The modification comes from operator products between 
${\cal N}$, $\tilde{{\cal N}}$ and $V_T$. 
It can be calculated by using the expansions 
(\ref{N and tilded N}). We finally obtain : 
\begin{eqnarray}
&&
\mbox{ad}_{g_N^{-1}}V_T(\sigma_1,\tau_1;k^{(1)}) 
\mbox{ad}_{g_N^{-1}}V_T(\sigma_2,\tau_2;k^{(2)}) 
\nonumber \\
&&
~~
=
\exp 
\Bigl\{
     -k_{\mu}^{(1)}
          \mathbb{G}^{\mu \nu}
             (z_1,\bar{z}_1|z_2,\bar{z}_2) 
      k^{(2)}_{\nu}
\Bigr \}
\times  
:
\mbox{ad}_{g_N^{-1}}V_T(\sigma_1,\tau_1;k^{(1)})
\mbox{ad}_{g_N^{-1}}V_T(\sigma_2,\tau_2;k^{(2)})
:~, 
\nonumber \\
\label{OPE renormalized tachyon vertex}
\end{eqnarray}
for 
$
|z_1|\geq |z_2|, |z_1z_2|\geq 1
$ 
and 
$
z_1 \neq z_2
$. 
Here 
$
\mathbb{G}^{\mu \nu}
(z_1,\bar{z}_1|z_2,\bar{z}_2)
$ 
is Green's function on the unit disk $|z|\geq 1$  
in the presence of a constant $B$ field. 
It is defined by 
\begin{eqnarray}
\mathbb{G}^{\mu \nu}
        (z_1,\bar{z}_1|z_2,\bar{z}_2)
\equiv 
\frac{ \langle {\bf 0} |
         X^{\mu}(\sigma_1,\tau_1)
         X^{\nu}(\sigma_2,\tau_2)
       |B_{N}\rangle }
     { \langle {\bf 0}|B_N\rangle }
-
\frac{ \langle {\bf 0}|
           \hat{x}_0^{\mu}\hat{x}_0^{\nu}
       |{\bf 0}\rangle}
     {\langle {\bf 0}|{\bf 0} \rangle}.
\label{def of disk Green function}
\end{eqnarray}
The RHS can be evaluated by 
using the Bogolubov transforms 
(\ref{transforms of massive modes}) 
and written down explicitly as follows : 
\begin{eqnarray}
\mathbb{G}^{\mu \nu}
     (z_1,\bar{z}_1|z_2,\bar{z}_2)
&=&
-\frac{\alpha'}{2}
   g^{\mu \nu} \ln |z_1-z_2|^2 
-\frac{\alpha'}{2}
  (g^{-1}Ng^{-1})^{\mu \nu}
  \ln \left( 1-\frac{1}{z_1\bar{z}_2}\right)
\nonumber \\
&&
-\frac{\alpha'}{2}
   (g^{-1}N^Tg^{-1})^{\mu \nu}
   \ln \left( 1-\frac{1}{\bar{z}_1z_2}\right). 
\label{disk Green function}
\end{eqnarray}

Let us recall that 
a system only of closed-strings admits
a holomorphic factorization. 
Particularly there is no correlation between 
chiral and anti-chiral pieces,  
$X^{\mu}(z)$ and $\tilde{X}^{\mu}(\bar{z})$, of 
string coordinates 
$X^{\mu}(\sigma,\tau)=X^{\mu}(z)+\tilde{X}^{\mu}(\bar{z})$. 
The factorized term of 
$\mathbb{G}^{\mu \nu}
(z_1,\bar{z}_1|z_2,\bar{z}_2)$ in Eq.(\ref{disk Green function}) 
is a sum of the correlations 
$\langle X^{\mu}(z_1)X^{\nu}(z_2)\rangle$ 
and 
$\langle \tilde{X}^{\mu}(\bar{z}_1)
\tilde{X}^{\nu}(\bar{z}_2)\rangle$. 
The second and third terms 
are not factorized and they are respectively 
the correlations  
$\langle X^{\mu}(z_1)\tilde{X}^{\nu}(\bar{z}_2)\rangle$
and 
$\langle \tilde{X}^{\mu}(\bar{z}_1)X^{\nu}(z_2)\rangle$. 
These correlations are characteristic of open-string theory. 
Green's function  
provides a nice description of the singular factor 
in Eq.(\ref{transform of VT}).
Using this terminology we can write it as 
follows:
\begin{equation}
\left(
      \frac{|z|^2}{|z|^2-1}
\right)^{-\frac{\alpha'}{2}k_{\mu}
          (g^{-1}Ng^{-1})^{\mu \nu}k_{\nu}}
=\frac{\left\langle k \right| V_{T}(\sigma,\tau;k)
       \left| B_{N}\right\rangle}
       {\left\langle \mathbf{0}\left|B_{N}\right\rangle\right.}
=
\exp \left\{ 
     ik_{\mu}ik_{\nu}
     \langle 
         X^{\mu}(z)\tilde{X}^{\nu}(\bar{z}) 
     \rangle
    \label{eq:singular factor as self-contraction of VT}
\right\}~,
\end{equation}
where the first equality follows from
Eq.(\ref{transform of VT}) and
$\langle k| \mbox{ad}_{g_{N}^{-1}} V_{T} (\sigma,\tau;k)
  |\mathbf{0}\rangle
  / \left\langle \mathbf{0}\left|B_{N}\right\rangle\right.=1$.
Taking account of this expression, 
the transform (\ref{transform of VT}) indicates 
that boundary states give rise to extra correlations, 
i.e.\ correlations between 
the chiral and the anti-chiral sectors, 
by the amount of 
$\langle X^{\mu}(z_{1}) \tilde{X}^{\nu}(\bar{z}_{2})\rangle$.
In the presence of the world-sheet boundary, 
or equivalently in the open-closed mixed system,
correlations between the chiral and anti-chiral sectors 
exist even in the closed string sector, and we need to take 
care of them. As pointed out in \cite{ghost boundary state}, 
this is a direct result of the fact that 
boundary states on the unit circle reflect a vertex operator at $z$
making its mirror image at $1/\bar{z}$ :
  $e^{ik_{\mu}X^{\mu}(z)}|B_{N}\rangle
    =e^{ik_{\mu}{(g^{-1}N)^{\mu}}_{\nu}
         \tilde{X}^{\nu}\left(\frac{1}{z}\right)}
         |B\rangle$ (see also \cite{BCR}).
This brings about a short distance singularity at the boundary 
$z=1/\bar{z}$. 
Since we intend to construct open-string vertex operators
in terms of closed-string,  
we need to carry out a renormalization to manage this type of
singularity.


The OPE (\ref{OPE renormalized tachyon vertex}) itself  
strongly suggests an interpretation of   
$\mbox{ad}_{g_N^{-1}}V_T$ from the open-string viewpoint.  
To pursue such a possibility 
we introduce a renormalized tachyon vertex operator 
$V_T^{ren}$ by subtracting the above singular factor :  
\begin{eqnarray}
V_T^{ren}(\sigma,\tau;k)
\equiv 
\left(
   \frac{|z|^2}{|z|^2-1}
\right)^{\frac{\alpha'}{2}
         k_{\mu}(g^{-1}Ng^{-1})^{\mu \nu}k_{\nu}}
\times 
V_T(\sigma,\tau;k). 
\label{def of ren VT}
\end{eqnarray}
In other words we have 
\begin{eqnarray}
g_N^{-1}V_T^{ren}(\sigma,\tau;k)g_N
=
\mbox{ad}_{g_N^{-1}}V_T(\sigma,\tau;k). 
\label{def2 of ren VT}
\end{eqnarray}
We call $V_T^{ren}(\sigma,\tau;k)$ 
renormalized {\it open-string} tachyon vertex operator  
with momentum $k$.

We need to explain why we call 
it open-string tachyon vertex operator. 
Let us consider 
an action of $\mathrm{diff} S^1$, the group of 
diffeomorphisms of a circle, on $V_T^{ren}$. 
We take 
$L_{n}-\tilde{L}_{-n}$ ($n \in \mathbb{Z}$) 
as the generators.  
It is convenient to recall that  
$L_{n}$ act on $V_T$ in the following manner :  
\begin{eqnarray}
\Bigl[ 
   L_n, V_T(\sigma,\tau;k) 
\Bigr]=
\left\{ 
     z^{n+1}\partial_z
     +\frac{\alpha'}{4}
        k_{\mu}g^{\mu \nu}k_{\nu}
          (n+1)z^n 
\right\} 
V_T(\sigma,\tau;k).   
\label{Ln on VT}
\end{eqnarray}
As regards $\tilde{L}_{n}$ 
their actions are the same as above 
except replacing $z$ by $\bar{z}$.  
The action of $\mathrm{diff} S^1$ on $V_T^{ren}$ 
becomes slightly different from 
a sum of the above Virasoro actions. 
The modification comes from 
an existence of the renormalization factor 
in Eq.(\ref{def of ren VT}). 
It can be read as follows : 
\begin{eqnarray}
&&
\left[ 
       L_n-\tilde{L}_{-n},
       V^{ren}_T(\sigma,\tau;k) 
\right]
\nonumber \\
&&~~~~~~~~
=
\left\{ 
  z^{n+1}\partial_z
  -\bar{z}^{-(n-1)}\partial_{\bar{z}}
  +\frac{\alpha'}{4}k_{\mu}g^{\mu \nu}k_{\nu}
        \left((n+1)z^n+(n-1)\bar{z}^{-n}\right) 
\right. 
\nonumber \\
&&~~~~~~~~~~~~
\left.
  +\frac{\alpha'}{2}
    k_{\mu}(g^{-1}Ng^{-1})^{\mu \nu}k_{\nu}
      \frac{z^n-\bar{z}^{-n}}{|z|^2-1}
\right\}
V^{ren}_{T}(\sigma,\tau;k).
\label{Kn on ren VT}
\end{eqnarray}
An implication may be seen 
by taking the limit $\tau \rightarrow 0+$ in the equation. 
By using the cylinder coordinates instead of 
$(z, \bar{z})$ the limit turns out to be as follows :
\begin{eqnarray}
&&
\left[ 
    L_n-\tilde{L}_{-n},
    \lim_{\tau \rightarrow 0+}
          V^{ren}_T(\sigma,\tau;k)
\right]
\nonumber \\
&&~~~~~~~~~
=
\left\{ 
    -ie^{i n \sigma}\frac{d}{d\sigma}
    +n\alpha'k_{\mu}G^{\mu \nu}k_{\nu} 
           e^{in \sigma}
\right\}
\lim_{\tau \rightarrow 0+}
V^{ren}_T(\sigma,\tau;k), 
\label{Kn on lim ren VT}
\end{eqnarray}
where we translate the closed-string background tensors 
$g_{\mu \nu}$ and $B_{\mu \nu}$
into the open-string background tensors 
$G_{\mu \nu}$ and $\theta^{\mu \nu}$. 
We may call these two kinds of background tensors 
respectively closed- and open-string tensors for short. 
They are related \cite{Seiberg-Witten} with each other by 
\begin{eqnarray}
\left( E^{-1} \right)^{\mu \nu}
=
G^{\mu \nu}
+
\frac{1}{2\pi \alpha'}\theta^{\mu \nu}~.
\label{def of G and theta}
\end{eqnarray} 
In an actual derivation of Eq.(\ref{Kn on lim ren VT}) 
we use the following identities : 
\begin{eqnarray}
g^{\mu \nu}
&=&
G^{\mu \nu}-\frac{1}{(2\pi \alpha')^2}
\theta^{\mu \rho} G_{\rho \rho'} \theta^{\rho' \nu},
\nonumber \\
g^{\mu \rho}N_{\rho \rho'}g^{\rho' \nu}
&=&
\left( G^{\mu \nu}+\frac{1}{(2\pi \alpha')^2}
\theta^{\mu \rho} G_{\rho \rho'} \theta^{\rho' \nu} 
\right)
+\frac{1}{\pi \alpha'}\theta^{\mu \nu}. 
\label{equalities2 and 3}
\end{eqnarray}
These can be obtained from Eq.(\ref{def of G and theta}). 
Related formulas are summarized in
Appendix \ref{sec:open-string tensors}.

The action (\ref{Kn on lim ren VT}) is 
completely the same as the action of the Virasoro algebra on 
open-string tachyon vertex operator with the same 
momentum. In addition to the OPE 
(\ref{OPE renormalized tachyon vertex}) 
this is the reason why we identify $V_{T}^{ren}$ 
with renormalized open-string tachyon vertex operators.

We can also make the above consideration 
in terms of boundary states. 
Let us consider the state  
$V^{ren}_T(\sigma,\tau;k) \bl B_N \brr$.
By using the relation (\ref{def2 of ren VT}) 
we can write it in a form, 
$
g_{N}
$ 
$
\mbox{ad}_{g_N^{-1}}V_{T}(\sigma,\tau;k)
|{\bf 0} \rangle
$. 
An explicit form of 
$\mbox{ad}_{g_N^{-1}}V_{T}$ has been given 
in Eq.(\ref{normal ordered transformed VT}). 
Thereby we can express the state  
in the oscillator representation and then take  
the limit $\tau \rightarrow 0+$ without ambiguity. 
We call thus obtained state  
$\bl B_N ;(\sigma,k) \brr$.   
It turns out to have the following form : 
\begin{eqnarray}
\lefteqn{
\bl 
B_N ;(\sigma,k) 
\brr 
\equiv 
\lim_{\tau \rightarrow 0+} 
V_T^{ren}(\sigma,\tau;k)
\bl 
B_N 
\brr }
\nonumber \\
&&
=
\left( 
   \frac{\mbox{det}^2E_{\mu \nu}}
        {(2\alpha')^{(p+1)}(-\mbox{det}g_{\mu \nu})}
\right)^{1/4}
g_N 
\times 
\lim_{\tau \rightarrow 0+}
\mbox{ad}_{g_N^{-1}}V_T^{ren}(\sigma,\tau;k) 
|{\bf 0} \rangle
\nonumber \\
&&
=
\left( 
   \frac{-\mbox{det} G_{\mu \nu}}
        {(2\alpha')^{(p+1)}}
\right)^{1/4}
\prod_{n=1}^{\infty}
\exp 
\left[ 
  \sqrt{2\alpha'}\frac{1}{n}k_{\mu}
  \left\{
     \left(
           \frac{1}{E^{T}}g
     \right)^{\mu}_{\nu}
       \alpha_{-n}^{\nu}
          e^{in\sigma}
      +
     \left(
           \frac{1}{E}g
     \right)^{\mu}_{\nu}
       \tilde{\alpha}_{-n}^{\nu}
          e^{-in\sigma}
   \right\} 
\right]
g_{N}|k \rangle.  
\nonumber \\ 
\label{one tachyon BN}
\end{eqnarray}
Here we express the normalization factor 
of $\bl B_N \brr$ in terms of 
the open-string tensors. 
This translation is done by using the identity,
$G_{\mu \nu}=E^T_{\mu \rho}g^{\rho \rho'}E_{\rho' \nu}$, 
which also follows from Eq.(\ref{def of G and theta}).

Action of $\mathrm{diff} S^1$ on this state can be obtained from 
Eq.(\ref{Kn on lim ren VT}). It can be read as  
\begin{eqnarray}
\left(
L_n-\tilde{L}_{-n}
\right) \bl B_N ; (\sigma,k)\brr 
=
(-i)\left\{ 
e^{in\sigma}\frac{d}{d\sigma}+in\alpha'
k_{\mu}G^{\mu \nu}k_{\nu}e^{in\sigma}
\right\} 
\bl B_N ; (\sigma,k) \brr~,  
\label{Kn on one tachyon BN}
\end{eqnarray}
for $n \in \mathbb{Z}$. 
The Ishibashi condition imposed on 
$\bl B_N ; (\sigma,k)\brr$,
i.e.\ vanishing of the RHS of the above equation 
for an arbitrary $n$,
(strictly speaking, modulo a total derivative with 
respect to $\sigma$),
requires 
$\alpha'k_{\mu}G^{\mu \nu}k_{\nu}=1$. 
It is the on-shell condition 
of open-string tachyon.

\subsection{String field theory viewpoint}

The action of $\mathrm{diff} S^1$ on the boundary state has  
an interpretation in terms of string field theory.

Fundamental ingredient in string field theory 
is a BRST charge $Q$. 
It is a grassmann-odd operator obeying the usual relations, 
$Q^2=0$ and 
$T_{ab}(\sigma,\tau)=
\left\{ Q, b_{ab}(\sigma,\tau)\right\}$, 
where $T_{ab}$ and $b_{ab}$ 
are respectively total energy-momentum tensor 
and anti-ghost field of a world-sheet theory.
In the case of bosonic closed-string field theory 
\cite{Zwiebach} world-sheet theory consists 
of closed-string coordinates $X^{I}$ (a matter system) 
and the world-sheet reparametrization ghosts.
The ghost system is described by $(b,c)$ for the chiral part 
and $(\tilde{b},\tilde{c})$ for the anti-chiral part. 
Closed-string BRST charge $Q_c$  is decomposed into 
$Q_c=q+\tilde{q}$. 
The chiral part $q$ has the following form : 
\begin{eqnarray}
q=
\sum_{n}c_nL_{-n}^{X}+c_0L_0^{(b,c)}
+\sum_{m,n \neq 0}nc_nc_mb_{-n-m},
\label{q}
\end{eqnarray}
where 
$L_n^{X,(b,c)}$ are the Virasoro generators  
of the matter and the ghost systems.  
$c_n$ and $b_n$ are the Fourier modes of 
ghost and anti-ghost fields : 
$c(\sigma,\tau)=\sum_{n}c_ne^{-n(\tau+i\sigma)}$ 
and 
$b(\sigma,\tau)=\sum_{n}b_ne^{-n(\tau+i\sigma)}$. 
The anti-chiral part $\tilde{q}$ has the same form as the above 
except replacing the chiral quantities with the anti-chiral ones. 
These $q$ and $\tilde{q}$ are nilpotent independently  
and satisfy the relations, $\left\{q,b_n\right\}=L_n$ 
and 
$\left\{\tilde{q},\tilde{b}_n\right\}=\tilde{L}_n$. 
Here 
$L_{n}=L_{n}^X+L_{n}^{(b,c)}$ 
and 
$\tilde{L}_{n}=\tilde{L}_{n}^X+\tilde{L}_{n}^{(b,c)}$ 
are the Virasoro generators of the total system.

Let ${\cal H}_c^X$ and ${\cal H}_c^{ghost}$ be respectively 
the Hilbert spaces of matter and ghost systems. 
We put ${\cal H}^{aux}={\cal H}_c^X \otimes {\cal H}_c^{ghost}$. 
Closed-string Hilbert space ${\cal H}_c$ consists of vectors $\psi$ 
of ${\cal H}^{aux}$ which satisfy the conditions, 
\begin{eqnarray}
(b_0-\tilde{b}_0)\psi=0,~~~~~(L_0-\tilde{L}_0)\psi=0. 
\label{state condition of closed-string}
\end{eqnarray}
Necessity of the two conditions is explained 
in \cite{Zwiebach} from the perspective of two-dimensional 
conformal field theories.

Now we want to interpret the state $\bl B_N;(\sigma,k)\brr$ 
as a state of the closed-string Hilbert space. 
Since it is a boundary state of the Neumann directions 
we first extend it to a vector of ${\cal H}_c^X$ 
by tensoring the boundary state (\ref{state Bd}) 
of the Dirichlet directions. 
As for the ghost sector an appropriate state 
is known. It is given \cite{ghost boundary state} by 
\begin{eqnarray}
\bl B \brr_{ghost}=
\prod_{n=1}^{\infty}
\exp \left\{ 
c_{-n}\tilde{b}_{-n}+\tilde{c}_{-n}b_{-n}
\right\} 
(c_0+\tilde{c}_0)c_1\tilde{c}_1 
|{\bf 0} \rangle, 
\label{ghost boundary state}
\end{eqnarray}
and satisfies the following boundary conditions :
\begin{eqnarray}
(c_n+\tilde{c}_{-n})\bl B \brr_{ghost}=0,~~~~ 
(b_n-\tilde{b}_{-n})\bl B \brr_{ghost}=0, 
\label{boundary condition Bghost}
\end{eqnarray}
for $n \in \mathbb{Z}$. 
It can be seen from these conditions that 
$\bl B \brr_{ghost}$ satisfies the Ishibashi condition.   
Further tensoring the ghost boundary state we obtain  
$\left(\bl B_N;(\sigma,k)\brr  
\otimes \bl B_D \brr \right)
     \otimes    \bl B \brr_{ghost}$
of  ${\cal H}^{aux}$. 
We call it $\bl B;(\sigma,k) \brr_{tot}$.  
This state satisfies the first condition in 
Eq.(\ref{state condition of closed-string}) because of
Eq.(\ref{boundary condition Bghost}). 
As for the second, 
the $n=0$ case of Eq.(\ref{Kn on one tachyon BN}) 
gives $(L_0-\tilde{L}_0)\bl B;(\sigma,k) \brr_{tot}$
$=(-i)\partial_{\sigma}\bl B;(\sigma,k) \brr_{tot}$. 
This means that we need to integrate over $\sigma$ to regard 
the state as a vector of ${\cal H}_c$.

Physical states of closed-string are identified with 
the BRST invariant states of ${\cal H}_c$. 
In order to describe how $Q_c$ acts on the boundary state 
we note a formula related with the ghost boundary state : 
Let $|{\cal O}\rangle_X$ be a state of ${\cal H}_c^X$. 
Action of $Q_c$ on 
$|{\cal O}\rangle_X \otimes \bl B \brr_{ghost} $ 
is expressed as
\begin{eqnarray}
Q_c
|{\cal O}\rangle_X \otimes \bl B \brr_{ghost}
=
\sum_{n}c_{-n}
(L_n^X-\tilde{L}^X_{-n})|{\cal O}\rangle_{X}\otimes 
\bl B \brr_{ghost}
\end{eqnarray}
This can be derived by the standard calculation. 
The action on the boundary state can be read as 
follows :
\begin{eqnarray}
&&
Q_c 
\left[ 
\int_0^{2\pi}
d \sigma
\bl B;(\sigma,k) \brr_{tot}
\right]
\nonumber \\
&&
~~~~
=
\int_0^{2\pi}
d \sigma
\sum_{n}c_{-n}(L_n^X-\tilde{L}_{-n}^X)
\bl B;(\sigma,k) \brr_{tot}
\nonumber \\
&&
~~~~
= 
(-i)\int_0^{2\pi}
d \sigma
\sum_{n}e^{in \sigma}
\left(
\frac{d}{d \sigma}+in\alpha'k_{mu}G^{\mu \nu}k_{\nu}
\right)c_{-n}
\bl B;(\sigma,k) \brr_{tot}, 
\label{Qc on one tachyon B}
\end{eqnarray}
where Eq.(\ref{Kn on one tachyon BN}) is used in the 
second equality. 
The BRST invariance requires, as we stated previously, 
the on-shell condition of open-string tachyon.

The above action can be identified with the BRST transformation 
of open-string tachyon vertex operator. 
$c_{n}$ in Eq.(\ref{Qc on one tachyon B}) correspond to  
the same modes of ghost field 
appearing in open-string field theory. 
Actually Eq.(\ref{Qc on one tachyon B}) is an example of 
the formula given in \cite{Nakatsu}. It states that the 
closed-string BRST charge $Q_c$ acts on boundary states 
as the generator of the BRST transformation
of open-string field. 
It is used there to show a macroscopic description of 
the cubic open-string field theory is given by a boundary 
open-string field theory.

\subsection{Boundary states with $M$ open-string tachyons}

The previous construction can be generalized to those 
of boundary states relevant to the description of
$M$ open-string tachyons.
Let $V^{ren}_{T}(\sigma_r,\tau_r;k^{(r)})$ 
($1 \leq r \leq M$)  
be the renormalized open-string tachyon vertex operators. 
Also let $z_r \equiv e^{\tau_r+i\sigma_r}$ be 
distinct points on the infinite semi-cylinder satisfying 
the condition, $|z_1|\geq |z_2| \geq \cdots |z_M| \geq 1$. 
We start with the following state :
\begin{eqnarray} 
V_{T}^{ren}(\sigma_1,\tau_1;k^{(1)})
\cdots 
V_{T}^{ren}(\sigma_M,\tau_M;k^{(M)})
\bl B_N \brr .
\end{eqnarray} 
We translate the state into a form, 
$g_N$ 
$\prod_{r}$
$\mbox{ad}_{g_N^{-1}}V_T^{ren}(\sigma_r,\tau_r;k^{(r)})$  
$| {\bf 0} \rangle$ 
by the relation (\ref{def2 of ren VT}). 
We use the OPE (\ref{OPE renormalized tachyon vertex}) 
for a further evaluation. Finally we arrive at 
the following representation : 
\begin{eqnarray}
&&
V_T^{ren}(\sigma_1,\tau_1;k^{(1)}) 
V_T^{ren}(\sigma_2,\tau_2;k^{(2)})
\cdots 
V_T^{ren}(\sigma_M,\tau_M;k^{(M)})
\bl B_N \brr 
\nonumber \\
&&
~~
=
\left( 
\frac{\mbox{det}^2E_{\mu \nu}}
     {(2\alpha')^{p+1}(-\mbox{det}g_{\mu \nu})}
\right)^{1/4}
\prod_{r < s}^M
\exp 
\Bigl\{ 
-
k_{\mu}^{(r)}
\mathbb{G}^{\mu \nu}
(z_r,\bar{z}_r|z_s,\bar{z}_s)
k_{\nu}^{(s)} 
\Bigr\} 
\nonumber \\
&&
~~~~~~~~~~~
\times 
g_N 
:
\mbox{ad}_{g_N^{-1}}V_T(\sigma_1,\tau_1;k^{(1)})
\mbox{ad}_{g_N^{-1}}V_T(\sigma_2,\tau_2;k^{(2)})
\cdots 
\mbox{ad}_{g_N^{-1}}V_T(\sigma_M,\tau_M;k^{(M)})
: 
| {\bf 0} \rangle 
\nonumber \\
&&
~~
=
\left( 
 \frac{\mbox{det}^2E_{\mu \nu}}
 {(2\alpha')^{p+1}(-\mbox{det}g_{\mu \nu})}
\right)^{1/4}
\prod_{r < s}^M
\exp 
\Bigl\{ 
  -k_{\mu}^{(r)}
  \mathbb{G}^{\mu \nu}
  (z_r,\bar{z}_r|z_s,\bar{z}_s)
  k_{\nu}^{(s)} 
\Bigr\} 
\nonumber \\
&&
~~~~~~~~~~~
\times 
\prod_{n=1}^{\infty}
\exp 
\left[ 
  \sqrt{\frac{\alpha'}{2}}\frac{1}{n}
  \sum_{r=1}^M
    k_\mu^{(r)}
     \left( 
        z_r^n\delta^{\mu}_{\nu}+\bar{z}_r^{-n}
        {(g^{-1}N^T)^{\mu}}_{\nu}
     \right)
     \alpha_{-n}^{\nu} 
\right]
\nonumber \\
&&
~~~~~~~~~~~
\times 
\prod_{n=1}^{\infty}
\exp 
\left[ 
   \sqrt{\frac{\alpha'}{2}}\frac{1}{n}
   \sum_{r=1}^M
     k_\mu^{(r)}
       \left( 
          \bar{z}_r^n\delta^{\mu}_{\nu}+z_r^{-n}
           {(g^{-1}N)^{\mu}}_{\nu}
       \right)
     \tilde{\alpha}_{-n}^{\nu} 
\right]
g_N 
\left| \sum_{r=1}^Mk^{(r)} \right\rangle~~~.
\label{pre M open-string tachyon state} 
\end{eqnarray}

Boundary state relevant to the description of $M$ off-shell 
open-string tachyons may be obtained from the above 
state by taking the same steps as for $M=1$.  
This state will be called 
$
\bl 
B_N ; 
(\sigma_1,k^{(1)}),
$ 
$
\cdots, 
$
$
(\sigma_M,k^{(M)})
\brr
$. 
It is defined 
by the limit $\forall \tau_r \rightarrow 0+$ 
of Eq.(\ref{pre M open-string tachyon state}). 
In the course of taking the limit 
we do not meet any difficulty 
for the oscillator part. 
What we need to take care are 
the products
$
\prod_{r < s}^M
\exp 
$
$
\left\{ 
-k_{\mu}^{(r)}
\mathbb{G}^{\mu \nu}
(z_r,\bar{z}_r|z_s,\bar{z}_s)
k_{\nu}^{(s)} 
\right\}
$.  
Behavior of Green's function 
at the boundary turns out to be as follows : 
\begin{eqnarray}
&&
\lim _{\tau_{r,s}\rightarrow 0+}
\mathbb{G}^{\mu \nu}
      (z_r,\bar{z}_r 
          | z_s, \bar{z}_s) 
\nonumber \\
&&
~~~~~~~~~
=
-2\alpha'G^{\mu \nu}
\ln |e^{i\sigma_r}-e^{i\sigma_s}|
+
\frac{i}{2\pi}\theta^{\mu \nu}
\left\{ 
   \sigma_r-\sigma_s
    -\pi 
   \epsilon(\sigma_r-\sigma_s)
\right\}, 
\label{boundary limit of disk Green function}
\end{eqnarray}
where $\epsilon (x)$ is the sign function defined as
$\epsilon(x)=1$ for $x >0$ and 
$=-1$ for $x<0$. 
Taking account of this behavior of Green's function,
the limit $\forall \tau_{r}\rightarrow 0+$ 
becomes as follows :
\begin{eqnarray}
&&
\left| B_N ; 
(\sigma_1,k^{(1)}),(\sigma_2,k^{(2)}), 
\cdots, (\sigma_M,k^{(M)})\right\rangle  
\nonumber \\
&& 
~~~~
\equiv 
\lim_{\forall \tau_r \rightarrow 0+}
V^{ren}_T(\sigma_1,\tau_1;k^{(1)})
V^{ren}_T(\sigma_2,\tau_1;k^{(2)})
\cdots 
V^{ren}_T(\sigma_M,\tau_M;k^{(M)})
\bl B_N \brr 
\nonumber \\
&&
~~~~
= 
\left( 
   \frac{-\mbox{det} G_{\mu \nu}}
        {(2\alpha')^{p+1}}
\right)^{1/4}
\prod_{r < s}^M
e^{\frac{i}{2}
     k_{\mu}^{(r)}
        \theta^{\mu \nu}
     k_{\nu}^{(s)}
   \epsilon(\sigma_r-\sigma_s)}
\nonumber \\
&&
~~~~~~~
\times 
\prod_{r < s}^M
e^{-\frac{i}{2\pi}
      k_{\mu}^{(r)}
        \theta^{\mu \nu}
      k_{\nu}^{(s)}
      (\sigma_r-\sigma_s)}
\times
\prod_{r < s}^M 
|e^{i\sigma_r}-e^{i\sigma_s}|
^{2\alpha'k_{\mu}^{(r)}G^{\mu \nu}k_{\nu}^{(s)}}
\nonumber \\
&&
~~~~~~~
\times 
\prod_{n=1}^{\infty}
\exp 
\left[
  \frac{\sqrt{2\alpha'}}{n}
  \sum_{r=1}^M
    k_{\mu}^{(r)}
      \left \{
          {\left(
              \frac{1}{E^T}g
          \right)^{\mu}}_{\nu}
             \alpha_{-n}^{\nu}
             e^{i n \sigma_r}
           +
           {\left(
              \frac{1}{E}g
           \right)^{\mu}}_{\nu}
              \tilde{\alpha}_{-n}^{\nu}
              e^{-i n \sigma_r}
      \right \}
\right] \nonumber\\
&&~~~~~~~
\times 
g_N 
\left| 
\sum_{r=1}^Mk^{(r)} 
\right\rangle.
\label{M open-string tachyon boundary state}
\end{eqnarray}

We wish to interpret these states  
as boundary states with off-shell 
open-string tachyons. 
This is verified by comparing 
closed-string tree propagations between these states 
with the corresponding open-string one-loop amplitudes 
and testing their coincidence. 
These will be done in the next subsection. 

It is worth mentioning that 
the present construction is also applicable in a vanishing 
$B$ field background and in particular 
the boundary states
(\ref{M open-string tachyon boundary state}) become 
available just by putting $B_{\mu\nu}=0$.

\subsubsection{Construction of dual boundary state}

To describe the closed-string propagations 
we need the dual boundary states. 
Although they are given by the BPZ conjugation, 
we apply the previous formalism to their construction
as well
in order to gain some insight into them. 
Instead of $g_N$ we use the BPZ dual : 
\begin{eqnarray}
g_{N}^{\dagger}
=\prod_{n=1}^{\infty}
\exp 
\left\{
     -\frac{1}{n}N_{\mu \nu}
        \alpha_{n}^{\mu}
        \tilde{\alpha}_{n}^{\nu}
\right\}.
\label{gN dagger}
\end{eqnarray}

The transform of tachyon vertex operator is 
given by 
$
g_{N}^{\dagger} 
V_{T}(\sigma,\tau;k)
\left( g_{N}^{\dagger} \right)^{-1}
$. 
It can be written in the following form 
for $|z|<1$ :
\begin{eqnarray}
g_{N}^{\dagger} 
V_{T}(\sigma,\tau;k)
\left( g_{N}^{\dagger} \right)^{-1}
=
\left( 
   1-|z|^2 
\right)^{  \frac{\alpha'}{2}
           k_{\mu}
            (g^{-1}Ng^{-1})^{\mu \nu}
           k_{\nu}} 
\times 
\mbox{ad}_{g_{N}^{\dagger}} V_{T}(\sigma,\tau;k). 
\label{dual transform of VT}
\end{eqnarray}
Here we introduce a local operator 
$
\mbox{ad}_{g_{N}^{\dagger}} V_{T}
$, 
which we interpret 
as the (dual) Bogolubov transform of 
$
V_{T}(\sigma,\tau;k)
$.
It takes the form of
\begin{eqnarray}
\mbox{ad}_{g_{N}^{\dagger}} V_{T}(\sigma,\tau;k)
=
V_{T}(\sigma,\tau;k)
{\cal N}_{\infty}(\bar{z};k)
\tilde{{\cal N}}_{\infty}(z;k),
\end{eqnarray}
where 
$
{\cal N}_{\infty}(\bar{z};k)
$
and 
$
\tilde{{\cal N}}_{\infty}(z;k)
$
are operators consisting only of the annihilation modes :
\begin{eqnarray}
{\cal N}_{\infty}(\bar{z};k)
&=& 
\prod_{n=1}^{\infty}
\exp 
\left\{
         -\sqrt{\frac{\alpha'}{2}}
         \frac{1}{n}
          k_{\mu}
              {(g^{-1}N^T)^{\mu}}_{\nu}
          \alpha_n^{\nu}
          \bar{z}^{n}
\right\},
\nonumber \\
\tilde{{\cal N}}_{\infty}(z;k)
&=& 
\prod_{n=1}^{\infty}
\exp 
\left\{
         -\sqrt{\frac{\alpha'}{2}}
         \frac{1}{n}
          k_{\mu}
              {(g^{-1}N)^{\mu}}_{\nu}
          \tilde{\alpha}_n^{\nu}
          z^{n}
\right\}.
\end{eqnarray}
Singularity of 
$
g_{N}^{\dagger} 
V_{T}(\sigma,\tau;k)
\left( g_{N}^{\dagger} \right)^{-1}
$ 
comes from  
$
\left( 
   1-|z|^2 
\right)^{  \frac{\alpha'}{2}
           k_{\mu}
            (g^{-1}Ng^{-1})^{\mu \nu}
           k_{\nu}} 
$ 
in Eq.( \ref{dual transform of VT}). 
It becomes the same as the singular factor appearing 
in Eq.(\ref{transform of VT}) 
after changing the coordinate $z$ to $1/z$.

Similar calculation to what is made to obtain 
Eq.(\ref{OPE renormalized tachyon vertex}) leads 
to the OPE : 
\begin{eqnarray}
&& 
\mbox{ad}_{g_{N}^{\dagger}} 
V_{T}(\sigma_1,\tau_1;k^{(1)})
\mbox{ad}_{g_{N}^{\dagger}} 
V_{T}(\sigma_2,\tau_2;k^{(2)})
\nonumber \\
&&
~~
=
\exp 
\Bigl\{
   -k_{\mu}^{(1)}
         \mathbb{G}^{\mu \nu}_{\infty}
            (z_1,\bar{z}_1|z_2,\bar{z}_2)
    k_{\nu}^{(2)}
\Bigr\}
\times 
:
\mbox{ad}_{g_{N}^{\dagger}} 
V_{T}(\sigma_1,\tau_1;k^{(1)})
\mbox{ad}_{g_N^{\dagger}}
V_{T}(\sigma_2,\tau_2;k^{(2)})
:~~~ , 
\nonumber \\ 
\label{dual OPE renormalized tachyon vertex}
\end{eqnarray}
for 
$
|z_1| \geq |z_2|, |z_1z_2| \leq 1
$ 
and 
$
z_1 \neq z_2
$. 
Here 
$
\mathbb{G}^{\mu \nu}_{\infty}
  (z_1,\bar{z}_1|z_2,\bar{z}_2)
$ 
is Green's function on the unit disk 
$
|z| \leq 1
$. 
It is defined by 
\begin{eqnarray}
\mathbb{G}^{\mu \nu}_{\infty}
         (z_1,\bar{z}_1|z_2,\bar{z}_2)
\equiv 
\frac{
       \langle B_N |
            X^{\mu}(\sigma_1,\tau_1)
            X^{\nu}(\sigma_2,\tau_2)
       |{\bf 0}\rangle}
     {  
       \langle B_N| {\bf 0}\rangle}
-
\frac{
        \langle {\bf 0}|
            \hat{x}_0^{\mu}\hat{x}_0^{\nu}
        |{\bf 0}\rangle}
     {  
        \langle {\bf 0}|{\bf 0} \rangle},  
\label{def of dual disk Green function}
\end{eqnarray}
and written down explicitly as follows :  
\begin{eqnarray}
\mathbb{G}^{\mu \nu}_{\infty}
    (z_1,\bar{z}_1|z_2,\bar{z}_2)
&=&
-\frac{\alpha'}{2}
   g^{\mu \nu} 
      \ln |z_1-z_2|^2 
-\frac{\alpha'}{2}
   (g^{-1}Ng^{-1})^{\mu \nu}
      \ln 
        \left( 1-z_1\bar{z}_2 \right)
\nonumber \\
&&
-\frac{\alpha'}{2}
    (g^{-1}N^Tg^{-1})^{\mu \nu}
       \ln 
          \left( 1-\bar{z}_1z_2 \right). 
\label{dual disk Green function}
\end{eqnarray}

Similarly to
Eq.(\ref{eq:singular factor as self-contraction of VT}), 
the singular factor appearing in Eq.(\ref{dual transform of VT})
is expressed as the self-contraction of $V_{T}(\sigma,\tau;k)$
between its chiral and anti-chiral parts.
This may be seen as follows.
Let $\langle X^{\mu}(z_{1})\tilde{X}^{\nu}(\bar{z}_{2})
     \rangle_{\infty}$
be the chiral--anti-chiral correlation of Green's function
$\mathbb{G}_{\infty}^{\mu\nu}(z_{1},\bar{z}_{1}|z_{2},\bar{z}_{2})$~:
$\langle X^{\mu}(z_{1})\tilde{X}^{\nu}(\bar{z}_{2})\rangle_{\infty}
     \equiv -\frac{\alpha'}{2}
     \left(g^{-1}Ng^{-1}\right)^{\mu\nu} \ln
     \left(1-z_{1}\bar{z}_{2}\right)$.
Using this quantity, we can write the singular factor as
\begin{equation}
\left(1-|z|^{2}\right)^{\frac{\alpha'}{2}
           k_{\mu}\left(g^{-1}Ng^{-1}\right)^{\mu\nu}k_{\nu}}
 = \frac{ \left\langle B_{N}\right| V_{T}(\sigma,\tau;k) 
            \left|-k\right\rangle}
          {\left\langle B_{N}\left|\mathbf{0}\right\rangle\right.}
   =\exp\left\{ ik_{\mu}ik_{\nu}
                \left\langle X^{\mu}(z)\bar{X}^{\nu}(\bar{z})
                \right\rangle_{\infty} \right\}~.
\end{equation}

Dual of the renormalized open-string tachyon operator, 
which we call $V_{T \infty}^{ren}$, is given 
by subtracting the singular factor 
appearing in Eq.(\ref{dual transform of VT}) as follows:
\begin{eqnarray}
V_{T \infty}^{ren}(\sigma,\tau;k)
= 
\left( 1-|z|^2 \right)^{-\frac{\alpha'}{2}
k_{\mu}(g^{-1}Ng^{-1})^{\mu \nu}k_{\nu}} 
\times 
V_T(\sigma,\tau ; k).
\label{def of dual ren VT}
\end{eqnarray}
We remark that the $\mathrm{diff} S^1$ action on the dual
vertex operator
reduces to Eq.(\ref{Kn on lim ren VT}) after we take the 
limit $\tau \rightarrow 0-$.

Let us describe the dual boundary state. 
It is obtained by following the same route  
as taken for Eq.(\ref{M open-string tachyon boundary state}). 
Let $V_{T \infty}^{ren}(\sigma_r,\tau_r;k^{(r)})$ 
($1 \leq r \leq M$) be the dual vertex operators. 
Let $z_r=e^{\tau_r+i\sigma_r}$ be distinct points on 
the infinite semi-cylinder satisfying the condition, 
$1 \geq |z_1| \geq |z_2| \geq \cdots \geq |z_M|$. 
We consider the state 
$\bll B_N \br 
V_{T \infty}^{ren}(\sigma_1,\tau_1;k^{(1)})
\cdots V_{T \infty}^{ren}(\sigma_M,\tau_M;k^{(M)})$ 
and take the limit $ \forall \tau_r \rightarrow 0-$. 
Thus obtained state, which we call 
$\left \langle  B_N ; 
(\sigma_1,k^{(1)}),
\cdots, (\sigma_M,k^{(M)})\right| $, 
is the dual. Explicitly it is given by 
\begin{eqnarray}
&&
\left \langle  B_N ; 
(\sigma_1,k^{(1)}),(\sigma_2,k^{(2)}), 
\cdots, (\sigma_M,k^{(M)})\right|  
\nonumber \\
&&
~~
\equiv 
\lim_{\forall \tau_r \rightarrow 0-}
\bll B_N \br 
V_{T \infty}^{ren}
\left( \sigma_1,\tau_1;k^{(1)} \right)
V_{T \infty}^{ren}
\left( \sigma_2,\tau_2;k^{(2)} \right)
\cdots V_{T \infty}^{ren}(\sigma_M,\tau_M;k^{(M)}) 
\nonumber \\
&&
~~
=
\left( 
\frac{-\mbox{det} G_{\mu \nu}}{(2\alpha')^{p+1}}
\right)^{1/4}
\prod_{r < s}^M
e^{-\frac{i}{2}k_{\mu}^{(r)}
\theta^{\mu \nu}k_{\nu}^{(s)}
\epsilon(\sigma_r-\sigma_s)}
\nonumber \\
&&
~~~~~~~
\times 
\prod_{r < s}^M
e^{\frac{i}{2\pi}k_{\mu}^{(r)}
\theta^{\mu \nu}k_{\nu}^{(s)}(\sigma_r-\sigma_s)}
\times 
\prod_{r < s}^M 
|e^{i\sigma_r}-e^{i\sigma_s}|
^{2\alpha'k_{\mu}^{(r)}G^{\mu \nu}k_{\nu}^{(s)}}
\nonumber \\
&&
~~~~~~~
\times 
\left
\langle 
-\sum_{r=1}^Mk^{(r)} 
\right|
g_N^{\dagger}
\prod_{n=1}^{\infty}
\exp 
\left[
-\frac{\sqrt{2\alpha'}}{n}
\sum_{r=1}^M
k_{\mu}^{(r)}
\left \{
{\left(\frac{1}{E^T}g
\right)^{\mu}}_{\nu}
\alpha_{n}^{\nu}
e^{-i n \sigma_r}
+
{\left(\frac{1}{E}g
\right)^{\mu}}_{\nu}
\tilde{\alpha}_{n}^{\nu}
e^{i n \sigma_r}
\right \}
\right]. 
\nonumber \\ 
\label{M open-string tachyon dual boundary state} 
\end{eqnarray}

\subsubsection{Closed-string propagation}

The hamiltonian operator of closed-string is 
$L_0+\tilde{L}_0-2$. 
We can conveniently parametrize 
propagations of closed-string 
by $\tau^{(c)} \in i \mathbb{R}_{\geq 0}$. 
The evolution by an imaginary time 
$\pi |\tau^{(c)}|$ is given by the operator 
$q_c^{\frac{1}{2}(L_0+\tilde{L}_0-2)}$, 
where we put $q_c=e^{2\pi i \tau^{(c)}}$. 
In this subsection we only consider 
the propagations along the Neumann directions. 
Propagations of closed-string between the boundary 
states (\ref{M open-string tachyon boundary state}) 
are measured by the following amplitudes :
\begin{eqnarray}
\left \langle B_N ; 
(\sigma_{M+1},k^{(M+1)}),\cdots, (\sigma_{M+N},k^{(M+N)})
\right| 
q_c^{\frac{1}{2}(L_0+\tilde{L}_0-2)} 
\left| 
B_N;
(\sigma_1,k^{(1)}),\cdots,(\sigma_{M},k^{(M)})
\right \rangle. 
\label{def of tachyon amplitude by boundary state}
\end{eqnarray}

We calculate these amplitudes by using the oscillator 
representations. It is useful to recall the following 
equalities :
\begin{eqnarray}
q_c^{-\frac{1}{2}(L_0+\tilde{L}_0-2)}
\alpha_n^{\mu}
q_c^{\frac{1}{2}(L_0+\tilde{L}_0-2)}
=
q_c^{\frac{n}{2}}\alpha_n^{\mu},
~~~~~
q_c^{-\frac{1}{2}(L_0+\tilde{L}_0-2)}
\tilde{\alpha}_n^{\mu}
q_c^{\frac{1}{2}(L_0+\tilde{L}_0-2)}
=
q_c^{\frac{n}{2}}\tilde{\alpha}_n^{\mu}, 
\end{eqnarray}
and 
\begin{eqnarray}
\langle -k |
q_c^{\frac{1}{2}(L_0+\tilde{L}_0)}
| k' \rangle 
=
q_c^{\frac{\alpha'}{4}g_{\mu \nu}k_{\mu}k_{\nu}}
\delta^{(p+1)}(k+k'),
\end{eqnarray}
where we put $\delta^{(p+1)}(k)\equiv \prod_{\mu}\delta(k_{\mu})$. 
First we plug the representations 
(\ref{M open-string tachyon boundary state}) 
and 
(\ref{M open-string tachyon dual boundary state}) 
into Eq.(\ref{def of tachyon amplitude by boundary state}). 
We then evaluate the amplitudes by taking account of  
the above equalities.  
They turn out to have the following factorized form : 
\begin{eqnarray}
&&
\left \langle B_N ; 
(\sigma_{M+1},k^{(M+1)}),\cdots, (\sigma_{M+N},k^{(M+N)})
\right| 
q_c^{\frac{1}{2}(L_0+\tilde{L}_0-2)} 
\left| 
B_N;
(\sigma_1,k^{(1)}),\cdots,(\sigma_{M},k^{(M)})
\right \rangle 
\nonumber \\
&&
=
\left( 
\frac{-\mbox{det}G_{\mu \nu}}{(2\alpha')^{(p+1)}}
\right)^{1/2}
\delta^{(p+1)}
\left( \sum_{r=1}^{M+N}k^{(r)}
\right) 
\times 
q_c^{-1-\frac{\alpha'}{4}\sum_{r=1}^M\sum_{s=M+1}^{M+N}
k_{\mu}^{(r)}g^{\mu \nu}k_{\nu}^{(s)}}
\nonumber \\
&&
~~~~
\times
\prod_{1\leq r<s \leq M}
e^{\frac{i}{2}k_{\mu}^{(r)}\theta^{\mu \nu}k_{\nu}^{(s)}
\epsilon(\sigma_r-\sigma_s)}
\times 
\prod_{M+1 \leq r<s \leq M+N}
e^{-\frac{i}{2}k_{\mu}^{(r)}\theta^{\mu \nu}k_{\nu}^{(s)}
\epsilon(\sigma_r-\sigma_s)}
\nonumber \\
&&
~~~~
\times 
\prod_{1 \leq r < s \leq M}
e^{-\frac{i}{2\pi}k_{\mu}^{(r)}
\theta^{\mu \nu}k_{\nu}^{(s)}(\sigma_r-\sigma_s)}
\times 
\prod_{M+1 \leq r < s \leq M+N} 
e^{\frac{i}{2\pi}k_{\mu}^{(r)}
\theta^{\mu \nu}k_{\nu}^{(s)}(\sigma_r-\sigma_s)}
\nonumber \\
&& 
~~~~
\times
\prod_{1 \leq r < s \leq M} 
|e^{i\sigma_r}-e^{i\sigma_s}|
^{2\alpha'k_{\mu}^{(r)}G^{\mu \nu}k_{\nu}^{(s)}}
\times 
\prod_{M+1 \leq r < s \leq M+N} 
|e^{i\sigma_r}-e^{i\sigma_s}|
^{2\alpha'k_{\mu}^{(r)}G^{\mu \nu}k_{\nu}^{(s)}}
\nonumber \\
&&
~~~~
\times 
F
\left(
q_c,\left\{\sigma_r \right\},\left\{k^{(r)}\right\}
\right)~~, 
\label{pre tachyon amplitude by boundary state}
\end{eqnarray}
where $F$ denotes contributions from the massive modes 
of closed-string. It is given by the infinite products : 
\begin{eqnarray}
&&
F
\left(
q_c,\left\{\sigma_r \right\},\left\{k^{(r)}\right\}
\right)
\nonumber \\
&&
\equiv 
\prod_{n=1}^{\infty}
\langle 0 |
\exp 
\left[
-\frac{q_c^{n}}{n}
\alpha_{n}^{\mu}N_{\mu \nu}\tilde{\alpha}_{n}^{\nu}
\right. 
\nonumber \\
&&
~~~~~~~~~~~~~~~~~~~~~
\left. 
-\frac{\sqrt{2\alpha'}q_c^{n/2}}{n}
\sum_{r=M+1}^{M+N}
k_{\mu}^{(r)}
\left\{
{\left(\frac{1}{E^T}g
\right)^{\mu}}_{\nu}
\alpha_{n}^{\nu}
e^{-i n \sigma_r}
+
{\left(\frac{1}{E}g
\right)^{\mu}}_{\nu}
\tilde{\alpha}_{n}^{\nu}
e^{i n \sigma_r}
\right \}
\right] 
\nonumber \\
&&
~~~~~~~~~
\times 
\exp 
\left[
-\frac{1}{n}\alpha_{-n}^{\mu}
N_{\mu \nu}\tilde{\alpha}_{-n}^{\nu}
\right.
\nonumber \\
&&
~~~~~~~~~~~~~~~~~~~~~
\left.
+\frac{\sqrt{2\alpha'}}{n}
\sum_{r=1}^M
k_{\mu}^{(r)}
\left \{
{\left(\frac{1}{E^T}g
\right)^{\mu}}_{\nu}
\alpha_{-n}^{\nu}
e^{i n \sigma_r}
+
{\left(\frac{1}{E}g
\right)^{\mu}}_{\nu}
\tilde{\alpha}_{-n}^{\nu}
e^{-i n \sigma_r}
\right \}
\right] 
| 0 \rangle~~. 
\nonumber \\
\label{def of F} 
\end{eqnarray}

Further evaluations of the infinite products  
are carried out in Appendix \ref{sec:formulae}. 
We just quote the result obtained there. 
The contributions from the massive modes turn out to be 
as follows:
\begin{eqnarray}
&&
F
\left(
q_c,\left\{\sigma_r \right\},\left\{k^{(r)}\right\}
\right)
\nonumber \\
&&
=
\prod_{n=1}^{\infty}
\left( 1-q_c^n \right)^{-p-1}
\nonumber \\
&&
~~
\times 
\prod_{1 \leq r < s \leq M}
\left[
\frac{
      \prod_{n=1}^{\infty}
        \left( 1-e^{i(\sigma_r-\sigma_s)}q_c^n \right)
        \left( 1-e^{-i(\sigma_r-\sigma_s)}q_c^n \right)}
     {\prod_{n=1}^{\infty}
        \left( 1-q_c^n \right)^2}
\right]^{2\alpha'k_{\mu}^{(r)}G_{\mu \nu}k_{\nu}^{(s)}}
\nonumber \\
&&
~~
\times 
\prod_{M+1 \leq r < s \leq M+N}
\left[
     \frac{
          \prod_{n=1}^{\infty}
             \left( 1-e^{i(\sigma_r-\sigma_s)}q_c^n \right)
             \left( 1-e^{-i(\sigma_r-\sigma_s)}q_c^n \right)}
          {\prod_{n=1}^{\infty}
                      \left( 1-q_c^n \right)^2}
\right]^{2\alpha'k_{\mu}^{(r)}G_{\mu \nu}k_{\nu}^{(s)}}
\nonumber \\
&&
~~
\times 
\prod_{r=1}^M
\prod_{s=M+1}^{M+N}
\left[
    \frac{
      \prod_{n=0}^{\infty}
         \left( 1-e^{i(\sigma_r-\sigma_s)}q_c^{n+1/2}\right)
         \left( 1-e^{-i(\sigma_r-\sigma_s)}q_c^{n+1/2}\right)}
         {\prod_{n=1}^{\infty}
                   \left( 1-q_c^n\right)^2}
\right]^{2\alpha'k_{\mu}^{(r)}G_{\mu \nu}k_{\nu}^{(s)}}. 
\nonumber \\
\label{result on F}
\end{eqnarray}

For the comparison with open-string one-loop calculation 
it is convenient to write down 
the amplitudes (\ref{pre tachyon amplitude by boundary state}) 
by using the elliptic $\theta$-functions. 
Appropriate $\theta$-functions are $\theta_{1,4}(\nu|\tau)$.  
They have the following representations :
\begin{eqnarray}
\theta_1(\nu|\tau)&=& 
2 q^{\frac{1}{8}}\sin \pi \nu 
\prod_{n=1}^{\infty}
(1-q^n)(1-e^{2\pi i \nu}q^n)(1-e^{-2\pi i \nu}q^n),
\nonumber \\
\theta_4(\nu|\tau)&=& 
\prod_{n=1}^{\infty}(1-q^n)
\prod_{n=0}^{\infty}
(1-e^{2\pi i \nu}q^{n+\frac{1}{2}})
(1-e^{-2\pi i \nu}q^{n+\frac{1}{2}}),  
\label{elliptic theta functions}
\end{eqnarray}
where we put $q=e^{2 \pi i \tau}$.

We first pay attention to  
$
|e^{i\sigma_r}-e^{i\sigma_s}|
^{2\alpha'k_{\mu}^{(r)}G^{\mu \nu}k_{\nu}^{(s)}}
$ 
in Eq.(\ref{pre tachyon amplitude by boundary state}). 
It can be combined with infinite products   
in Eq.(\ref{result on F}) and gives rise to 
$
\left[ 
    \eta(\tau^{(c)})^{-3}
    \theta_1 
      \left(  \frac{ |\sigma_r-\sigma_s| }{2\pi} 
            \left|  \tau^{(c)} 
      \right. \right)
\right]
^{2\alpha'k_{\mu}^{(r)}G^{\mu \nu}k_{\nu}^{(s)}}
$.
Here we have taken into account that  
$\sigma_{r,s}$ run only from $0$ to $2\pi$. 
We next consider 
$
q_c^{-\frac{\alpha'}{4}\sum_{r=1}^M\sum_{s=M+1}^{M+N}
k_{\mu}^{(r)}g^{\mu \nu}k_{\nu}^{(s)}}
$ 
in Eq.(\ref{pre tachyon amplitude by boundary state}). 
Using the first relation of (\ref{equalities2 and 3})
it can be written in terms of the open-string tensors 
as follows :
\begin{eqnarray}
&&
q_c^{-\frac{\alpha'}{4}\sum_{r=1}^M\sum_{s=M+1}^{M+N}
k_{\mu}^{(r)}g^{\mu \nu}k_{\nu}^{(s)}}
\nonumber \\
&&
~~~~
=
q_c^{
\frac{1}{16 \pi^2 \alpha'}
\sum_{r=1}^M\sum_{s=M+1}^{M+N}
k_{\mu}^{(r)}(\theta G \theta)^{\mu \nu}k_{\nu}^{(s)}
}
\times 
q_c^{
-\frac{\alpha'}{4}\sum_{r=1}^M\sum_{s=M+1}^{M+N}
k_{\mu}^{(r)}G^{\mu \nu}k_{\nu}^{(s)}}. 
\end{eqnarray} 
Let us introduce $K_{\mu}$ 
as the total momentum of the $in$-state : 
$K \equiv \sum_{r=1}^{M}k^{(r)}$. 
The first factor can be written as 
$
q_{c}^{-\frac{1}{16 \pi^2 \alpha'}
         K_{\mu}(\theta G \theta)^{\mu \nu}K_{\nu}}
$ 
due to the momentum conservation.
We then manage  
$
q_c^{-\frac{\alpha'}{4}
k_{\mu}^{(r)}G^{\mu \nu}k_{\nu}^{(s)}}
$
in the second factor. 
It can be combined with 
another infinite products  
in Eq.(\ref{result on F}) 
and gives rise to  
$
\left[ 
    \eta(\tau^{(c)})^{-3}
    \theta_4 
      \left( \frac{(\sigma_r-\sigma_s)}{2\pi}
            \left| \tau^{(c)} 
      \right. \right)
\right]
^{2\alpha'k_{\mu}^{(r)}G^{\mu \nu}k_{\nu}^{(s)}}
$.  
Gathering these expressions 
we finally find out : 
\begin{eqnarray}
&&
\left \langle B_N ; 
(\sigma_{M+1},k^{(M+1)}),\cdots, (\sigma_{M+N},k^{(M+N)})
\right| 
q_c^{\frac{1}{2}(L_0+\tilde{L}_0-2)} 
\left| 
B_N;
(\sigma_1,k^{(1)}),\cdots,(\sigma_{M},k^{(M)})
\right \rangle 
\nonumber \\
&&
=
\left( 
\frac{-\mbox{det}G_{\mu \nu}}{(2\alpha')^{(p+1)}}
\right)^{1/2}
\delta^{p+1}
\left(\sum_{r=1}^{M+N}k^{(r)}
\right) 
\nonumber \\
&&
~~~~
\times
\prod_{1\leq r<s \leq M}
e^{\frac{i}{2}k_{\mu}^{(r)}\theta^{\mu \nu}k_{\nu}^{(s)}
\epsilon(\sigma_r-\sigma_s)}
\times 
\prod_{M+1 \leq r<s \leq M+N}
e^{-\frac{i}{2}k_{\mu}^{(r)}\theta^{\mu \nu}k_{\nu}^{(s)}
\epsilon(\sigma_r-\sigma_s)}
\nonumber \\
&&
~~~~
\times 
q_c^{\frac{p-25}{24}} 
\eta(\tau^{(c)})^{-p+1}
\nonumber \\
&&
~~~~
\times 
q_{c}^{-\frac{1}{16 \pi^2 \alpha'}
         K_{\mu}(\theta G \theta)^{\mu \nu}K_{\nu}}
\times 
\prod_{r=1}^M\prod_{s=M+1}^{M+N}
\left[ 
    \eta(\tau^{(c)})^{-3}
    \theta_4 
       \left.\left( \frac{\sigma_r-\sigma_s}{2\pi} 
             \right| \tau^{(c)} 
       \right)
\right]^{2\alpha'k_{\mu}^{(r)}G^{\mu \nu}k_{\nu}^{(s)}}
\nonumber \\
&& 
~~~~
\times 
\prod_{r=1}^{M}
e^{\frac{\sigma_r}
        {2\pi i}k_{\mu}^{(r)}\theta^{\mu \nu}K_{\nu}}
\times 
\prod_{1 \leq r < s \leq M} 
\left[ 
    \eta(\tau^{(c)})^{-3}
    \theta_1 
       \left.\left( \frac{|\sigma_r-\sigma_s|}{2\pi} 
             \right| \tau^{(c)} 
       \right)
\right]^{2\alpha'k_{\mu}^{(r)}G^{\mu \nu}k_{\nu}^{(s)}}
\nonumber \\
&&
~~~~
\times 
\prod_{r=M+1}^{M+N}
e^{\frac{\sigma_r}
        {2\pi i}k_{\mu}^{(r)}\theta^{\mu \nu}K_{\nu}}
\times 
\prod_{M+1 \leq r < s \leq M+N} 
\left[ 
    \eta(\tau^{(c)})^{-3}
    \theta_1 
       \left.\left( \frac{|\sigma_r-\sigma_s|}{2\pi} 
             \right| \tau^{(c)} 
       \right)
\right]^{2\alpha'k_{\mu}^{(r)}G^{\mu \nu}k_{\nu}^{(s)}}, 
\nonumber \\
\label{tachyon amplitude by boundary state}
\end{eqnarray}
where the products of 
$
e^{\frac{\sigma_r}
        {2\pi i}k_{\mu}^{(r)}\theta^{\mu \nu}K_{\nu}}
$ 
come from the following translations 
of the corresponding terms in  
Eq.(\ref{pre tachyon amplitude by boundary state}) :
\begin{eqnarray} 
\prod_{1 \leq r < s \leq M}
e^{-\frac{i}{2\pi}k_{\mu}^{(r)}
\theta^{\mu \nu}k_{\nu}^{(s)}(\sigma_r-\sigma_s)}
&=&
\prod_{r=1}^{M}
e^{\frac{\sigma_r}
        {2\pi i}k_{\mu}^{(r)}\theta^{\mu \nu}
          \left(\sum_{s=1}^Mk_{\nu}^{(s)}\right)}, 
\nonumber \\
\prod_{M+1 \leq r < s \leq M+N} 
e^{\frac{i}{2\pi}k_{\mu}^{(r)}
\theta^{\mu \nu}k_{\nu}^{(s)}(\sigma_r-\sigma_s)}
&=&
\prod_{r=M+1}^{M+N}
e^{\frac{\sigma_r}
        {2\pi i}k_{\mu}^{(r)}\theta^{\mu \nu}
         \left(-\sum_{s=M+1}^{M+N}k_{\nu}^{(s)}\right)}. 
\end{eqnarray}
We also note that 
contribution of the world-sheet reparametrization ghosts, 
which turns out to be 
$
\prod_{n=1}^{\infty}
(1-q_c^n)^2
$, 
has been included in 
Eq.(\ref{tachyon amplitude by boundary state}).  
Together with the ghost contribution, 
$q_c^{-1}$ 
in Eq.(\ref{pre tachyon amplitude by boundary state}) 
and 
$
\prod_{n=1}^{\infty}(1-q_c^n)^{-p-1}
$ in $F$ 
give rise to 
$
q_c^{\frac{p-25}{24}}\eta(\tau^{(c)})^{-p+1}
$
in Eq.(\ref{tachyon amplitude by boundary state}).

\subsection{Comparison with open-string one-loop calculation}
\label{sec:open-string tachyon one-loop}

The amplitudes 
(\ref{tachyon amplitude by boundary state}), 
combined with closed-string propagations along the Dirichlet 
directions, 
should be compared with open-string tachyon one-loop amplitudes. 
We begin this subsection 
with a brief exposition on the open-string calculation 
in a constant $B$ field background of closed-string.

\subsubsection{Open-string one-loop calculation}

Calculation of open-string one-loop amplitudes in a constant 
$B$ field background is a direct extension of
that in the case of $B=0$ (see e.g.\ \cite{Green-Schwarz-Witten}) 
except a quantization of open-string. 
Let us first describe the quantization. 
Action of a bosonic open-string in this background is given by 
\begin{eqnarray}
S=
\frac{1}{\pi \alpha'}
\int\frac{d\bar{\rho} \wedge d\rho}{2i}
\Bigl\{
g_{MN}
\bar{\partial}_{\rho}X^{M}(\rho,\bar{\rho})
\partial_{\rho}X^{N}(\rho,\bar{\rho})
+
2\pi\alpha'B_{\mu \nu}
\bar{\partial}_{\rho}X^{\mu}(\rho,\bar{\rho})
\partial_{\rho}X^{\nu}(\rho,\bar{\rho})
\Bigr\},
\label{open-string action}
\end{eqnarray}
where the upper half-plane, $\mbox{Im} \rho \geq 0$, 
is taken as the world-sheet. 
An infinite strip 
$(\tau,\sigma)$ ($0 \leq \sigma \leq  \pi$) 
can be mapped to the upper half-plane 
by $\rho = e^{\tau+i\sigma}$. 
Taking variation of the action,
one can find that 
the Neumann and the Dirichlet boundary conditions
are imposed  
on open-string coordinates $X^{\mu}$ and $X^{i}$ respectively: 
\begin{eqnarray}
\left. 
g_{\mu \nu}
(\bar{\partial}_{\rho}-\partial_{\rho})X^{\nu} 
+2\pi \alpha'B_{\mu \nu}
(\bar{\partial}_{\rho}+\partial_{\rho})X^{\nu}
\rule{0em}{2.5ex}
\right|_{\rho=\bar{\rho}}
=0 ,
~~~~~~~
\left.
X^i \right|_{\rho=\bar{\rho}}
=x_0^i.
\label{open-string boundary condition}
\end{eqnarray}
These boundary conditions can be managed
in the mode expansions of $X^M$.  
They turn out to be as follows :
\begin{eqnarray}
X^{\mu}(\rho,\bar{\rho})
&=&
\hat{x}_0^{\mu}
-i\alpha'\hat{p}_0^{\mu}\ln |\rho|^2
+i\sqrt{\frac{\alpha'}{2}}
\sum_{n \neq 0}\frac{\alpha^{\mu}_{n}}{n}
(\rho^{-n}+\bar{\rho}^{-n})
\nonumber \\
&&
-\frac{i}{2 \pi}{(\theta G)^{\mu}}_{\nu}
\left\{ \hat{p}_0^{\nu} \ln \left(\frac{\rho}{\bar{\rho}}\right)
-\frac{1}{\sqrt{2\alpha'}}
\sum_{n \neq 0}
\frac{\alpha^{\nu}_n}{n}(\rho^{-n}-\bar{\rho}^{-n})
\right\}, 
\nonumber \\
X^i(\rho,\bar{\rho})
&=&
x_0^i+
i\sqrt{\frac{\alpha'}{2}}\sum_{n \neq 0}
\frac{\alpha^i_{n}}{n}(\rho^{-n}-\bar{\rho}^{-n}), 
\label{open-string mode expansion}
\end{eqnarray}
where we translate the closed-string tensors into 
the open-string ones.  
The quantization is prescribed in \cite{Chu-Ho}\cite{AAS}. 
It turns out to be realized by the following 
commutation relations\footnote{
Notations used for open-string are 
similar to those for closed-string 
as far as they do not cause any confusion.}
:
\begin{eqnarray}
&&
[\hat{x}_0^{\mu},\hat{x}_0^{\nu}]=i \theta^{\mu \nu},~~~
[\hat{p}_0^{\mu},\hat{p}_0^{\nu}]=0,~~~
[\hat{x}_0^{\mu}, \hat{p}_0^{\nu}]=i G^{\mu \nu},
\nonumber \\
&&
[\alpha_m^{\mu},\alpha_n^{\nu}]=mG^{\mu \nu}\delta_{m+n},~~~
[\alpha_m^{i},\alpha_n^{j}]=mg^{i j}\delta_{m+n}. 
\label{CR of open-string modes }
\end{eqnarray}
The commutation relation of $\hat{x}_0^{\mu}$ indicates 
the non-commutativity of the world-volume 
\cite{C-D-S}\cite{DH}.

Related with a prescription of normal orderings  
relative to the $SL_2(\mathbb{R})$-invariant vacuum of 
open-string, 
it is convenient to introduce \cite{S-K-T}
commutative zero-modes 
$
\tilde{x}_0^{\mu} 
\equiv 
\hat{x}_0^{\mu}
+\frac{1}{2}{(\theta G)^{\mu}}_{\nu}
\hat{p}_0^{\nu}$ 
instead 
of the non-commutative ones. 
They satisfy the standard canonical relations ; 
$
[\tilde{x}_0^{\mu},\tilde{x}_0^{\nu}]=
[\hat{p}_0^{\mu},\hat{p}_0^{\nu}]=0
$ 
and 
$
[\tilde{x}_0^{\mu}, \hat{p}_0^{\nu}]=i G^{\mu \nu}
$.
In what follows we adopt the standard normal ordering 
of $(\tilde{x}_0^{\mu},\hat{p}_0^{\mu}, 
\alpha_n^{\mu},\alpha_n^i)$.   
It is also denoted by $:~~:~$.

The Virasoro generators $L_n$ can be obtained 
from the energy-momentum tensor  
$
T(\rho)
=
-\frac{1}{\alpha'}g_{MN}
\partial_{\rho}X^{M}\partial_{\rho}X^{N}
$ 
by 
$
T(\rho)
=\sum_{m}
L_{m}\rho^{-m-2}
$.
They turn out to have the following forms :
\begin{eqnarray}
L_{0}
&=&
\alpha'G^{\mu \nu}\hat{p}_0^{\mu}\hat{p}_0^{\nu}
+
\left(
\sum_{m=1}^{\infty}
G_{\mu \nu}\alpha_{-m}^{\mu}\alpha_{m}^{\nu}
+
g_{ij}\alpha_{-m}^i\alpha_{m}^j
\right),
\nonumber \\
L_{n}
&=&
\frac{1}{2}
\sum_{m \in \mathbb{Z}}
\left(
G_{\mu \nu}\alpha_{n-m}^{\mu}\alpha_{m}^{\nu}
+
g_{ij}\alpha_{n-m}^i\alpha_{m}^j 
\right)
~~~~~~~
\mbox{for}~~n \neq 0, 
\label{open-string Virasoro generator}
\end{eqnarray}
where we put 
$\alpha_0^{\mu}\equiv \sqrt{2\alpha'}\hat{p}_0^{\mu}$. 
Among these generators 
open-string propagator is given by 
$\Delta \equiv \frac{1}{L_0-1}$. 
Evolution of open-string on the upper half-plane 
is described by $\xi^{L_0-2}$ 
due to an integral representation  
$\Delta = \int_0^{1}d \xi \,\xi^{L_0-2}$.

It is useful to observe the open-string coordinates, 
particularly $X^{\mu}(\rho,\bar{\rho})$, 
at the boundary of the world-sheet. 
We may use the radial coordinates 
$(\xi,\sigma)$, where $\xi \equiv |\rho|$, 
as a parametrization of the upper half-plane. 
They can be read as follows :
\begin{eqnarray}
\left. 
X^{\mu}(\rho,\bar{\rho})
\right|_{\sigma=0}
&=&
\tilde{x}_0^{\mu}
-\frac{1}{2}{(\theta G)^{\mu}}_{\nu}
\hat{p}_0^{\nu}
-2i\alpha'\hat{p}_0^{\mu}\ln \xi 
+i\sqrt{2\alpha'}
\sum_{n \neq 0}\frac{\alpha_n^{\mu}}{n}\xi^{-n},
\nonumber \\
\left.
X^{\mu}(\rho,\bar{\rho})
\right|_{\sigma=\pi}
&=&
\tilde{x}_0^{\mu}
+\frac{1}{2}{(\theta G)^{\mu}}_{\nu}
\hat{p}_0^{\nu}
-2i\alpha'\hat{p}_0^{\mu}\ln \xi 
+i\sqrt{2\alpha'}
\sum_{n \neq 0}\frac{\alpha_n^{\mu}}{n}(-\xi)^{-n}.
\label{open-string at boundary}
\end{eqnarray}
At both boundaries $\sigma =0$ and $\sigma =\pi$,  
effect of the non-commutativity of the world-volume 
(the $\theta$-dependence) is encoded only 
in the zero modes while the massive part acquires 
the standard mode expansion. 
Open-string amplitudes are correlation functions 
between local operators inserted at the boundaries.  
It follows from Eq.(\ref{open-string at boundary}) 
that $\theta$-dependence of local operators 
such as tachyon- and gluon-vertices are encoded 
only in their zero modes. 
One-loop amplitudes can be obtained 
in the operator formalism simply 
by taking a trace of these operators   
with the propagators inserting among them. 
The integral representation of the propagator 
may be used. 
The zero mode dependent part decouples from the others in 
$\xi^{L_0-2}$.  
Therefore, for the particular cases of 
tachyons and gluons,  
only the zero-modes of open-string 
are influenced by the non-commutativity of the world-volume.

\begin{figure}
\psfrag{rho}{$\rho$}
\psfrag{0}{$0$}
\psfrag{1}{$1$}
\psfrag{-1}{$-1$}
\psfrag{k1}{$k^{(1)}$}
\psfrag{kM}{$k^{(M)}$}
\psfrag{kM+1}{$k^{(M+1)}$}
\psfrag{kM+N}{$k^{(M+N)}$}
\psfrag{zeta1}{$\xi_1$}
\psfrag{zetaM}{$\xi_M$}
\psfrag{-zetaM+1}{$-\xi_{M+1}$}
\psfrag{-zetaM+N}{$-\xi_{M+N}$}
\begin{center}
\includegraphics[height=14cm]{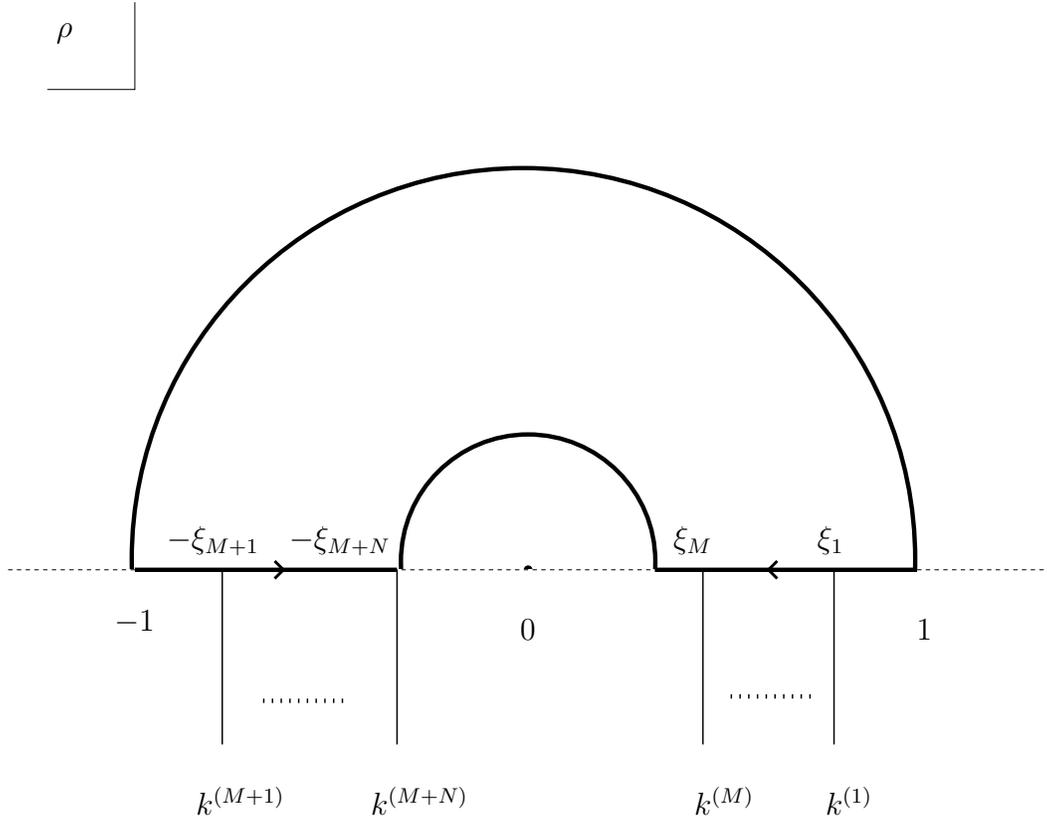}
\caption{
{\small 
Open-string one-loop diagram drawn on the upper half-plane.
Two semi-circles are identified with each other.}}
\label{upper-half-plane}
\end{center}
\end{figure}

\begin{figure}
\psfrag{u}{$u$}
\psfrag{0}{$0$}
\psfrag{pii}{$i \pi$}
\psfrag{k1}{$k^{(1)}$}
\psfrag{kM}{$k^{(M)}$}
\psfrag{kM+1}{$k^{(M+1)}$}
\psfrag{kM+N}{$k^{(M+N)}$}
\psfrag{nu1}{$\nu_1$}
\psfrag{nuM}{$\nu_M$}
\psfrag{zetaM+1+pii}{$\nu_{M+1}+i\pi$}
\psfrag{nuM+N+pii}{$\nu_{M+N}+i\pi$}
\psfrag{tauo}{$\tau^{(o)}$}
\begin{center}
\includegraphics[height=14cm]{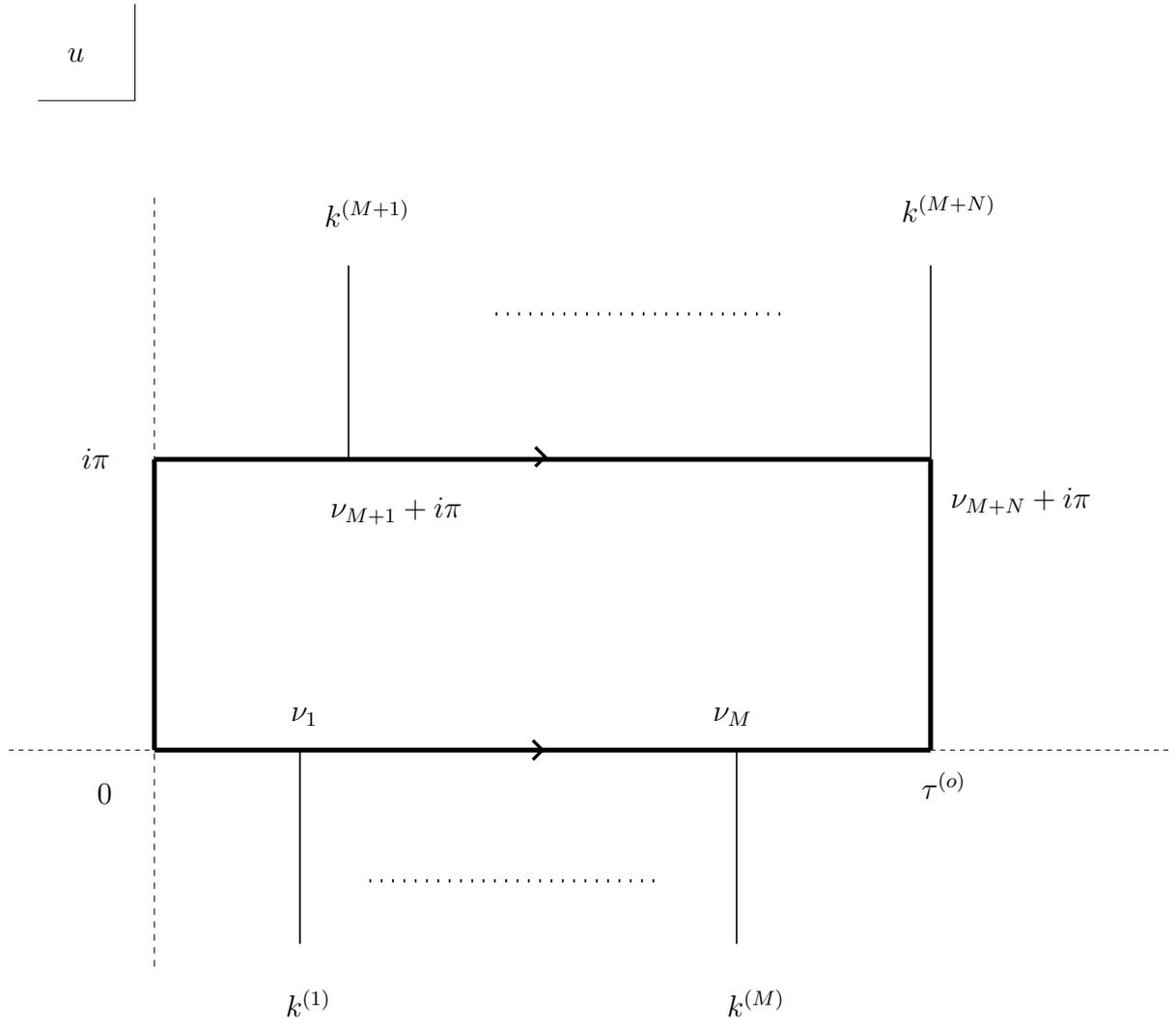}
\caption{
{\small 
Open-string one-loop diagram drawn on the $u$-plane.
Two bold vertical lines are identified with each other.}}
\label{open-string diagram}
\end{center}
\end{figure}

Let us consider one-loop amplitudes of tachyon vertex operators. 
We introduce open-string tachyon vertex operator of momentum $k$ 
by 
\begin{eqnarray}
V_T^{open}(\xi,\sigma;k)
\equiv 
\left. 
:e^{ik_{\mu} X^{\mu}(\rho,\bar{\rho})}: 
\right|_{\rho=\xi e^{i\sigma}}. 
\label{open-string tachyon vertex}
\end{eqnarray}
We consider the scattering process of $M+N$ tachyons 
with momenta $k^{(r)}$. 
It is worth noting that
the momenta $k^{(r)}$ have only components 
along the Neumann directions.
It is
because the propagation of open-string is restricted
along the D-brane world-volume. 
Diagram describing 
the one-loop scattering process can be drawn 
on the upper half-plane as depicted in  
Figure \ref{upper-half-plane}.
An open-string evolves along the radial direction 
from the outer semi-circle to the inner one 
and interacts with the tachyon vertices at its ends.  
Two semi-circles are identified 
with each other and thereby the diagrams is  
interpreted as an open-string one-loop.
The corresponding tachyon amplitude, which we call  
$
I_{T}
\left( 
   (k^{(1)},\cdots,k^{(M)})
    ;
   (k^{(M+1)},\cdots,k^{(M+N)})
\right)
$, 
is given by  a sum of traces of 
their products arranged 
in cyclically distinct orders
with keeping their partial cyclic orderings 
at the both ends : 
\begin{eqnarray}
&&
I_{T}
\left( \left( k^{(1)},\cdots,k^{(M)} \right) ;
       \left( k^{(M+1)},\cdots,k^{(M+N)} \right)
\right)
\nonumber \\
&&
~~
\equiv
\mbox{Tr} 
\left\{ 
\Delta 
V_T^{open}\left(1,0;k^{(1)}\right)
\cdots 
\Delta 
V_T^{open}\left(1,0;k^{(M)}\right)
\right.
\nonumber \\
&&
~~~~~~~~~~~~~~~~~~~~~
\times
\left.
\Delta 
V_T^{open}\left(1,\pi;k^{(M+1)}\right)
\cdots 
\Delta 
V_T^{open}\left(1,\pi;k^{(M+N)}\right)
\right\}
\nonumber \\
&&
~~~~~~~~
+~~\cdots\cdots. 
\label{def of open-string tachyon amplitude}
\end{eqnarray}
We can evaluate the amplitude in the standard manner. 
Let us recall that 
the tachyon vertex operator enjoys the properties :
\begin{eqnarray}
\xi^{L_0}V_T^{open}(1,0;k)\xi^{-L_0}
&=&
\xi^{\alpha'k_{\mu}G^{\mu \nu}k_{\nu}}
V_T^{open}(\xi,0;k),
\nonumber \\
\xi^{L_0}V_T^{open}(1,\pi;k)\xi^{-L_0}
&=&
\xi^{\alpha'k_{\mu}G^{\mu \nu}k_{\nu}}
V_T^{open}(\xi,\pi;k).
\label{Evolution of open-string tachyon vertex}
\end{eqnarray}
Combined with the integral form of $\Delta$,
this enables us to write  
the amplitude as follows :
\begin{eqnarray}
&&
I_{T}
\left( \left( k^{(1)},\cdots,k^{(M)} \right) ; 
       \left( k^{(M+1)},\cdots,k^{(M+N)} \right)
\right)
\nonumber \\
&&
~~
=
\int_{0 \leq \xi_{M+N} \leq \xi_{M+N-1} 
        \leq \cdots \leq \xi_1 \leq 1}
\prod_{r=1}^{M+N}
\frac{d \xi_r}{\xi_r} 
\times 
\prod_{r=1}^{M+N}
\xi_r^{\alpha'k_{\mu}^{(r)}G_{\mu \nu}k_{\nu}^{(r)}}
\nonumber \\
&&
~~~~~~~~~~
\times 
\mbox{Tr}
\left\{
V_T^{open}\left(\xi_1,0;k^{(1)}\right)\cdots
V_T^{open}\left(\xi_M,0;k^{(M)}\right) 
\right.
\nonumber \\
&&
~~~~~~~~~~~~~~~
\left.
\times 
V_T^{open}\left(\xi_{M+1},\pi;k^{(M+1)}\right)\cdots
V_T^{open}\left(\xi_{M+N},\pi;k^{(M+N)}\right) 
\xi_{M+N}^{L_0-1}
\right\} 
\nonumber \\
&&
~~~~
+
~~
\cdots\cdots. 
\label{integral open-string tachyon amplitude}
\end{eqnarray}
The coordinates $\xi_r$, 
which appear in the RHS as insertion points 
of the tachyon vertices, 
can be thought to provide a parametrization of 
the diagram. 
Another parametrization may be obtained 
by mapping the diagram to a cylinder 
with width $\pi$.  
See Figure \ref{open-string diagram}.
Correspondingly $\xi_r$ are mapped to $\nu_r$ 
by $\nu_r=-\ln \xi_r$. 
We put $\tau^{(o)}\equiv -\ln \xi_{M+N}$. 
We may use $\tau^{(o)}$ instead of $\nu_{M+N}$. 
The evolution of open-string becomes manifest 
in this parameterization. 
In the figure  
an open-string with width $\pi$ 
evolves along the real axis from the origin 
to $\tau^{(0)}$, interacting with the tachyon 
vertices inserted at $\nu_r$ or $\nu_r +i\pi$.

After a little calculation of the traces 
in the RHS of 
Eq.(\ref{integral open-string tachyon amplitude}) 
we can obtain the amplitude in an integral form. 
In terms of $\tau^{(o)}$ and $\nu_r$ 
this integral can be written 
as follows : 
\begin{eqnarray}
&&
I_{T}
\left( \left( k^{(1)},\cdots,k^{(M)} \right) ; 
       \left( k^{(M+1)},\cdots,k^{(M+N)} \right)
\right)
\nonumber \\
&&
=
\sqrt{-\mbox{det}G_{\mu \nu}}
\delta^{(p+1)}
     \left(\sum_{r=1}^{M+N}k^{(r)}
     \right)
\nonumber \\
&&
~~~~
\times 
\prod_{1 \leq r < s \leq M}
e^{-\frac{i}{2}
          k_{\mu}^{(r)}\theta^{\mu \nu}k_{\nu}^{(s)}}
\times 
\prod_{M+1 \leq r < s \leq M+N}
e^{\frac{i}{2} 
         k_{\mu}^{(r)}\theta^{\mu \nu}k_{\nu}^{(s)}}
\nonumber \\
&&
~~~~
\times 
\int 
   d\tau^{(o)} 
\prod_{r=1}^{M+N-1}
   d\nu_r
\left( 
    \frac{\pi}{\alpha' \tau^{(o)}} 
\right)^{\frac{p+1}{2}} 
e^{\frac{26-D}{24}\tau^{(o)}}
\eta 
\left(
     -\frac{\tau^{(o)}}{2\pi i} 
\right)^{-D+2} 
\nonumber \\
&&
~~~~~~~
\times 
e^{\frac{1}{4\alpha' \tau^{(o)}}
     K_{\mu}(\theta G \theta)^{\mu \nu}K_{\nu}} 
\times 
\prod_{r=1}^{M+N}
e^{-i\frac{\nu_r}{\tau^{(o)}}
    k_{\mu}^{(r)}\theta^{\mu \nu}K_{\nu}}
\nonumber \\
&&
~~~~~~~
\times 
\prod_{r=1}^M
\prod_{s=M+1}^{M+N}
\left[ 
       e^{-\frac{(\nu_s-\nu_r)^2}{2\tau^{(0)}}}
       \eta 
           \left( \frac{-\tau^{(o)}}{2\pi i} \right)^{-3}
       \theta_1 
           \left( \frac{(\nu_s+i\pi) -\nu_r}{2\pi i} \right| 
           \left. \frac{-\tau^{(o)}}{2\pi i} \right) 
\right]^{2\alpha'k_{\mu}^{(r)}G^{\mu \nu}k_{\nu}^{(s)}}
\nonumber \\
&&
~~~~~~~
\times 
\prod_{1 \leq r < s \leq M}
\left[ 
       ie^{-\frac{(\nu_s-\nu_r)^2}{2\tau^{(0)}}}
       \eta 
           \left( \frac{-\tau^{(o)}}{2\pi i} \right)^{-3}
       \theta_1 
           \left( \frac{\nu_s-\nu_r}{2\pi i} \right| 
           \left. \frac{-\tau^{(o)}}{2\pi i} \right) 
\right]^{2\alpha'k_{\mu}^{(r)}G^{\mu \nu}k_{\nu}^{(s)}} 
\nonumber \\
&&
~~~~~~~
\times  
\prod_{M+1 \leq r < s \leq M+N}
\left[ 
       ie^{-\frac{(\nu_s-\nu_r)^2}{2\tau^{(0)}}}
       \eta 
           \left( \frac{-\tau^{(o)}}{2\pi i} \right)^{-3}
       \theta_1 
           \left( \frac{\nu_s-\nu_r}{2\pi i} \right| 
           \left. \frac{-\tau^{(o)}}{2\pi i} \right) 
\right]^{2\alpha'k_{\mu}^{(r)}G^{\mu \nu}k_{\nu}^{(s)}},  
\nonumber \\
\label{tachyon one-loop by open-string parameters}  
\end{eqnarray}
where the integral is performed over the region :
\begin{eqnarray}
&&
\tau^{(o)} \in \mathbb{R}_{\geq 0}, 
\nonumber \\
&&
0 \leq \nu_1 \leq \cdots 
\leq \nu_M \leq \tau^{(0)},
\nonumber \\
&&
0 \leq \nu_{M+1} \leq \cdots 
\leq \nu_{M+N-1} \leq \tau^{(0)}.  
\label{moduli by open-string parameters}
\end{eqnarray} 
Contribution of the world-sheet reparametrization ghosts, 
which is equal to 
$
\prod_{n=1}^{\infty}(1-e^{-n\tau^{(o)}})^2
$, 
has been included. We  note that $K_{\mu}$ 
used in the above is given by  
$K_{\mu}=\sum_{r=1}^{M}k_{\mu}^{(r)}$.

\subsubsection{Comparison with one-loop amplitudes of open-string}

\begin{figure}
\psfrag{v}{$v$}
\psfrag{0}{$0$}
\psfrag{2pii}{$2\pi$}
\psfrag{k1}{$k^{(1)}$}
\psfrag{kM}{$k^{(M)}$}
\psfrag{kM+1}{$k^{(M+1)}$}
\psfrag{kM+N}{$k^{(M+N)}$}
\psfrag{sigma1}{$\sigma_1$}
\psfrag{sigmaM}{$\sigma_M$}
\psfrag{sigmaM+1+pitauc}{$\sigma_{M+1}+\pi\tau^{(c)}$}
\psfrag{sigmaM+N+pitauc}{$\sigma_{M+N}+\pi\tau^{(c)}$}
\psfrag{pitauc}{$\pi \tau^{(c)}$}
\begin{center}
\includegraphics[height=14cm]{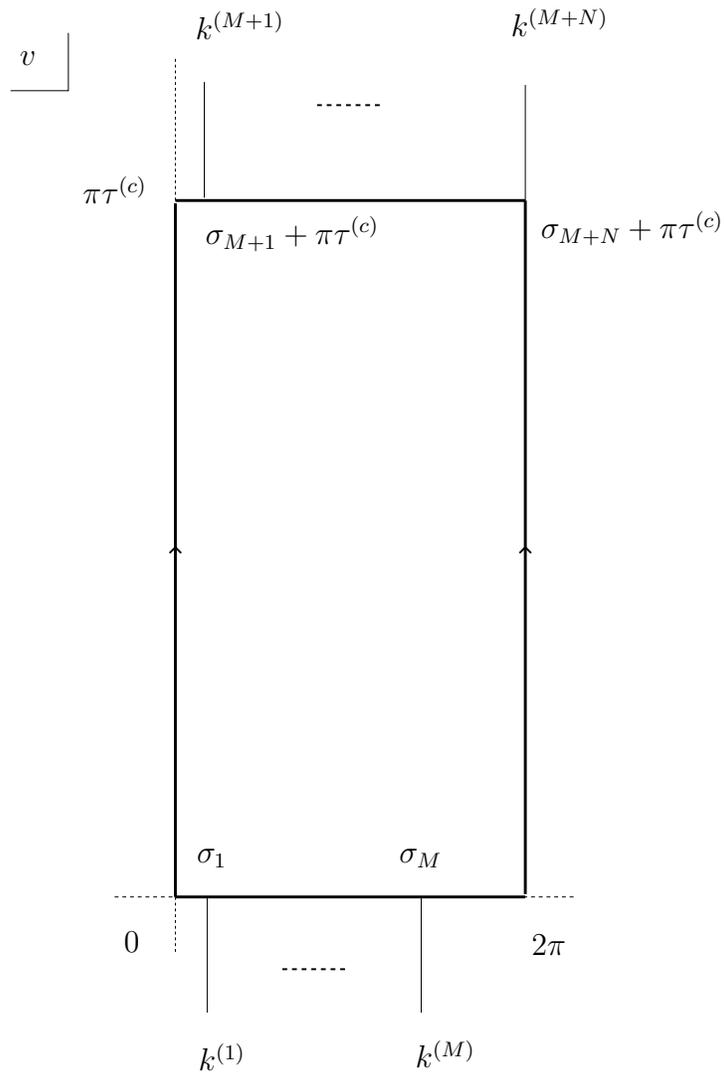}
\caption{
{\small 
Open-string one-loop diagram drawn on the $v$-plane.
Two bold vertical lines are identified with each other.}}
\label{closed-string diagram}
\end{center}
\end{figure}

To compare the open-string one-loop amplitude
(\ref{tachyon one-loop by open-string parameters})
with 
Eq.(\ref{tachyon amplitude by boundary state}) 
we map the cylinder 
drawn on the $u$-plane (Figure \ref{open-string diagram}) 
to a cylinder with width $2\pi^2/\tau^{(o)}$ 
on the $v$-plane by the conformal transformation 
$v=2\pi u/\tau^{(o)}$. 
See Figure \ref{closed-string diagram}.
$\nu_r$ are mapped to $\sigma_r$ by 
$\sigma_r=2\pi \nu_r/\tau^{(o)}$. We put 
$\tau^{(c)}\equiv 2\pi i/\tau^{(o)}$. 
These provide a new parametrization of the diagram, 
which makes an evolution of closed-string manifest. 
A closed-string with circumference $2 \pi$ evolves 
along the imaginary axis, starting from
the real axis where it interacts with
 $M$ open-string tachyons 
and ending on $\mathrm{Im} v=\pi |\tau^{(c)}|$
where it interacts with
$N$ open-string tachyons.
We note that the insertion point of
the $(M+N)$-th open-string tachyon 
is fixed at $\sigma_{M+N}=2\pi$.

Let us describe the one-loop scattering amplitude 
of $M+N$ tachyons as an integral 
with respect to $\tau^{(c)}$ and $\sigma_r$.
By making use of the modular transform,  
$
\theta_1 
( \nu /\tau | -1/\tau ) 
= 
-i \sqrt{-i\tau}e^{\pi i \nu^2/\tau}
\theta_1 ( \nu | \tau )
$, 
we can express several correlations appearing  
in Eq.(\ref{tachyon one-loop by open-string parameters}) 
by means  of $\tau^{(c)}$ and $\sigma_r$ :  
\begin{eqnarray}
&&
e^{-\frac{(\nu_s-\nu_r)^2}{2\tau^{(0)}}}
\eta 
   \left( 
      \frac{-\tau^{(o)}}{2\pi i}
   \right)^{-3}
\theta_1 
   \left( 
       \frac{(\nu_s+i\pi)-\nu_r}{2\pi i} 
   \right| 
   \left. 
        \frac{-\tau^{(o)}}{2\pi i} 
   \right) 
= 
(-i\tau^{(c)})^{-1}
\eta(\tau^{(c)})^{-3}
\theta_4  
   \left.\left( 
       \frac{\sigma_s-\sigma_r}{2\pi} 
   \right |
      \tau^{(c)} 
   \right ), 
\nonumber \\
&&
ie^{-\frac{(\nu_s-\nu_r)^2}{2\tau^{(0)}}}
\eta 
   \left( 
       \frac{-\tau^{(o)}}{2\pi i} 
   \right)^{-3}
\theta_1 
   \left( 
       \frac{\nu_s-\nu_r}{2\pi i} 
   \right| 
   \left. 
      \frac{-\tau^{(o)}}{2\pi i} 
   \right) 
= 
(-i\tau^{(c)})^{-1}
\eta(\tau^{(c)})^{-3}
\theta_1 
    \left.\left( 
        \frac{\sigma_s-\sigma_r}{2\pi}
    \right |
        \tau^{(c)} 
    \right ). 
\nonumber \\
\label{modular trans of tachyon correlations}
\end{eqnarray}
Other ingredients can be also expressed  
in terms of these parameters.  
The integral turns out to be as follows : 
\begin{eqnarray}
&&
I_{T}
\left( \left( k^{(1)},\cdots,k^{(M)} \right) ; 
       \left( k^{(M+1)},\cdots,k^{(M+N)} \right)
\right)
\nonumber \\
&&
=
\left(
\frac{-\mbox{det}G_{\mu \nu}}{(2\alpha')^{p+1}}
\right)^{1/2}
\delta^{(p+1)}
     \left(\sum_{r=1}^{M+N}k^{(r)}
     \right)
\nonumber \\
&&
~~~~
\times 
\prod_{1 \leq r < s \leq M}
e^{-\frac{i}{2}
          k_{\mu}^{(r)}\theta^{\mu \nu}k_{\nu}^{(s)}}
\times 
\prod_{M+1 \leq r < s \leq M+N}
e^{\frac{i}{2} 
         k_{\mu}^{(r)}\theta^{\mu \nu}k_{\nu}^{(s)}}
\nonumber \\
&&
~~~~
\times 
(-2\pi i)
\int_{i0}^{i \infty} 
d\tau^{(c)} 
\int 
\prod_{r=1}^{M+N-1}
d\sigma_r~
  e^{\frac{26-D}{24}\frac{2\pi i}{\tau^{(c)}}}
  \left(-i\tau^{(c)}\right)^{-\frac{d}{2}}
  \eta (\tau^{(c)})^{-D+2} 
\nonumber \\
&&
~~~~~~~~~~~
\times  
\prod_{r=1}^{M+N}
 \left(
     -i\tau^{(c)}
  \right)^{\alpha'k_{\mu}^{(r)}G^{\mu \nu}k_{\nu}^{(r)}-1}
\nonumber \\
&&
~~~~~~~~~~~
\times 
e^{\frac{\tau^{(c)}}{8\pi i\alpha'}
     K_{\mu}(\theta G \theta)^{\mu \nu}K_{\nu}} 
\times
\prod_{r=1}^M
\prod_{s=M+1}^{M+N}
\left[ 
  \eta(\tau^{(c)})^{-3}
  \theta_4  
        \left.\left( 
          \frac{\sigma_s-\sigma_r}{2\pi} 
        \right |
          \tau^{(c)} 
        \right )
\right]^{2\alpha'k_{\mu}^{(r)}G^{\mu \nu}k_{\nu}^{(s)}}
\nonumber \\
&&
~~~~~~~~~~~
\times 
\prod_{r=1}^{M}
e^{\frac{\sigma_r}{2\pi i}k_{\mu}^{(r)}
\theta^{\mu \nu}K_{\nu}}
\times 
\prod_{1 \leq r < s \leq M}
\left[ 
\eta(\tau^{(c)})^{-3}
\theta_1 
\left.\left( 
\frac{\sigma_s-\sigma_r}{2\pi} 
\right |
\tau^{(c)} 
\right )       
\right]^{2\alpha'k_{\mu}^{(r)}G^{\mu \nu}k_{\nu}^{(s)}} 
\nonumber \\
&&
~~~~~~~~~~~
\times  
\prod_{r=M+1}^{M+N}
e^{\frac{\sigma_r}{2\pi i}k_{\mu}^{(r)}
\theta^{\mu \nu}K_{\nu}}
\times 
\prod_{M+1 \leq r < s \leq M+N}
\left[ 
 \eta(\tau^{(c)})^{-3}
  \theta_4  
        \left.\left( 
          \frac{\sigma_s-\sigma_r}{2\pi} 
        \right |
          \tau^{(c)} 
        \right )
\right]^{2\alpha'k_{\mu}^{(r)}G^{\mu \nu}k_{\nu}^{(s)}}, 
\nonumber \\
\label{tachyon one-loop by closed-string parameters}  
\end{eqnarray}
where the integration region becomes as follows : 
\begin{eqnarray}
&&
\tau^{(c)} \in i \mathbb{R}_{\geq 0}, 
\nonumber \\
&&
0 \leq \sigma_1 \leq \cdots 
\leq \sigma_M \leq 2 \pi,
\nonumber \\
&&
0 \leq \sigma_{M+1} \leq \cdots 
\leq \sigma_{M+N-1} \leq  2\pi.  
\label{moduli by closed-string parameters}
\end{eqnarray}

The above integral form allows us to express  
the one-loop amplitude of $M+N$ open-string tachyons    
by using the closed-string amplitude given in   
Eq.(\ref{tachyon amplitude by boundary state}). 
We need to include the contribution of closed-string propagations 
along the Dirichlet directions. 
This is carried out by  multiplying
Eq.(\ref{tachyon amplitude by boundary state}) by  
$
\bll B_D \br 
q_c^{\frac{L_0+\tilde{L}_{0}}{2}}
\bl B_D \brr
$ 
$
=
q_c^{d/24}(-i\tau^{(c)})^{-d/2}
\eta(\tau^{(c)})^{-d}
$.
Thus we finally obtain the following equality : 
\begin{eqnarray}
&&
I_{T}
\left( \left( k^{(1)},\cdots,k^{(M)} \right) ; 
       \left( k^{(M+1)},\cdots,k^{(M+N)} \right)
\right)
\nonumber \\
&&
=
(-2\pi i)
\int_{i0}^{i \infty} 
d\tau^{(c)}  
~e^{
     2\pi i
        \left( 
         \tau^{(c)}+\frac{1}{\tau^{(c)}}
        \right) 
        \frac{26-D}{24}
    }
\prod_{r=1}^{M+N}
\left(-i \tau^{(c)}\right)
^{\alpha'k_{\mu}^{(r)}G^{\mu \nu}k_{\nu}^{(r)}-1}
\nonumber \\
&&
~~~~~~~~~~~~~~~
\left[ 
\int \prod_{r=M+1}^{M+N-1}d\sigma_r
\left \langle 
   B_D 
\right | 
\otimes 
\left \langle  
   B_N ; (\sigma_{M+1},k^{(M+1)}),\cdots, 
           (\sigma_{M+N},k^{(M+N)}) 
\right |
\right] 
\nonumber \\
&&
~~~~~~~~~~~~~~~~~~~~~~
\times q_c^{\frac{L_0+\tilde{L}_0-2}{2}} 
\left[ 
\int \prod_{r=1}^{M} d\sigma_r  
\left | 
         B_N ; (\sigma_{1},k^{(1)}),\cdots, 
           (\sigma_{M},k^{(M)}) 
\right \rangle  
\otimes 
\left |
     B_D 
\right \rangle   
\right], 
\nonumber \\
\label{equivalence with boundary state formalism}
\end{eqnarray}
where the integration is performed on 
the region
(\ref{moduli by closed-string parameters}).  
Additional two factors appearing in the RHS 
vanish when we impose 
the on-shell conditions on $k^{(r)}$ at the critical 
dimensions ($D=26$).
Thus we showed that our boundary states
(\ref{M open-string tachyon boundary state})
correctly reproduce the corresponding open-string
one-loop amplitudes.
We note that this computation
implies as well that normalization factor used 
in Eq.(\ref{state Bn}) is properly chosen to
reproduce the corresponding amplitudes.

\section{UV limit of Non-Commutative Scalar Field Theory}
\label{sec:UV NC scalar}

Low-energy description of world-volume theories 
of $p$-brane can be obtained 
by taking a zero-slope limit 
$\alpha' \rightarrow 0$ 
so that it makes all perturbative 
stringy states ($\mbox{mass}^2 \sim 1/\alpha'$) 
of open-string infinitely heavy and decouple 
from the light states.   
Low-energy effective theory relevant to open-string tachyons 
in the presence of a constant $B$ field is 
a scalar field theory on the non-commutative world-volume.  
In general, quantum field theories 
on a non-commutative space  
suffer the UV-IR mixing \cite{UV-IR mixing} 
which originates in the non-commutativity. 
It causes 
\cite{renormalization of non-commutative field theory} 
a serious problem 
on the renormalization prescription of these theories.  
In this section, 
based on the results obtained so far, 
we investigate the UV behavior of 
the non-commutative scalar field theory. 
For this sake  
it is convenient to start with a brief description 
of the above zero-slope limit of the one-loop amplitudes  
of $M+N$ tachyons. 
Our study in this section is restricted to 
the case of $25$-brane in the critical dimensions.

The amplitude is described in 
(\ref{tachyon one-loop by open-string parameters})
as an integral, where $\tau^{(o)}$ and $\nu_r$ 
are used as the integral variables.   
Since these parameters make an evolution of open-string 
manifest in the scattering process, 
we call them open-string parameters. 
For the same reason we call 
$\tau^{(c)}$ and $\sigma_r$ 
closed-string parameters. 
The zero-slope limit of the amplitude
(\ref{tachyon one-loop by open-string parameters})
will be 
a field theory one-loop amplitude, 
particularly written in the Schwinger representation
\cite{Andreev-Dorn}\cite{Kiem-Lee}
\cite{BCR}\cite{Gomis et al}\cite{Liu-Michelson}\cite{S.J.Rey et al}.  
Open-string parameters are translated to 
the Schwinger parameters 
in the low-energy world-volume theory.  
{}For the dimensional reasons we put 
\begin{eqnarray}
s^{(o)}\equiv \alpha' \tau^{(o)}, ~~~~~
T_r  \equiv \alpha' \nu_r.  
\label{Schwinger parameters of scalar field theory}
\end{eqnarray}
In order to obtain a proper field theory limit of 
open-string the field theory parameters   
$s^{(o)}$ and $T_r$
need to be fixed under the zero-slope limit. 
Simultaneously we also need to fix open-string tensors 
$G_{\mu \nu}$ and $\theta^{\mu \nu}$ since they  
describe a classical geometry of the world-volume.  
In an actual derivation of the zero-slope limit 
we first rewrite the integral in terms of these field theory 
parameters and then pick up the dominant contribution  
of the $\alpha'$-expansion. 
Let us take a look at correlations appearing 
in Eq.(\ref{tachyon one-loop by open-string parameters}). 
Terms which become dominant after the above translations  
can be read as follows :
\begin{eqnarray}   
&&
\left[
e^{-\frac{(\nu_s-\nu_r)^2}{2\tau^{(0)}}}
   \eta 
     \left( 
          \frac{-\tau^{(o)}}{2\pi i} 
     \right)^{-3}
   \theta_1 
      \left( 
          \frac{(\nu_s+i\pi)-\nu_r}{2\pi i}
      \right| 
      \left. 
         \frac{-\tau^{(o)}}{2\pi i} 
      \right) 
\right]^{2\alpha'k_{\mu}^{(r)}G^{\mu \nu}k_{\nu}^{(s)}}
\nonumber \\
&&
~~~~~~~~~~~~~~~~~~~~~~~~~~~~
\approx 
\left[
e^{-\frac{(T_s-T_r)^2}{s^{(o)}}+|T_s-T_r|}
\right]^{k_{\mu}^{(r)}G^{\mu \nu}k_{\nu}^{(s)}}, 
\nonumber \\  
&&
\left[
ie^{-\frac{(\nu_s-\nu_r)^2}{2\tau^{(0)}}}
    \eta 
       \left( 
           \frac{-\tau^{(o)}}{2\pi i}
       \right)^{-3}
    \theta_1 
       \left( 
           \frac{\nu_s-\nu_r}{2\pi i} 
       \right| 
       \left. 
           \frac{-\tau^{(o)}}{2\pi i} 
       \right) 
\right]^{2\alpha'k_{\mu}^{(r)}G^{\mu \nu}k_{\nu}^{(s)}}
\nonumber \\
&&
~~~~~~~~~~~~~~~~~~~~~~~~~~~~
\approx
\left[
e^{-\frac{(T_s-T_r)^2}{s^{(o)}}+|T_s-T_r|}
\right]^{k_{\mu}^{(r)}G^{\mu \nu}k_{\nu}^{(s)}}. 
\end{eqnarray}
We also note that the dominant contribution of 
$\eta (-\tau^{(o)}/2\pi i)^{-24}$ in 
Eq.(\ref{tachyon one-loop by open-string parameters}) 
is $e^{s^{(o)}/\alpha'}$. Gathering these estimations 
the following integral 
\cite{Andreev-Dorn}\cite{Kiem-Lee}
\cite{BCR}\cite{Gomis et al}\cite{Liu-Michelson}\cite{S.J.Rey et al} 
turns out to be the zero-slope limit :
\begin{eqnarray}
&&
I_{T}
\left( \left( k^{(1)},\cdots,k^{(M)} \right) ; 
       \left( k^{(M+1)},\cdots,k^{(M+N)} \right)
\right)
\nonumber \\
&&
\approx 
\frac{\pi^{13}}{\alpha'^{M+N}}
\sqrt{-\mbox{det} G_{\mu \nu}}
\delta^{26} \left( \sum_{r=1}^{M+N}k^{(r)} \right)
\prod_{1 \leq r < s \leq M}
e^{-\frac{i}{2}
          k_{\mu}^{(r)}\theta^{\mu \nu}k_{\nu}^{(s)}}
\times
\prod_{M+1 \leq r < s \leq M+N}
e^{\frac{i}{2} 
         k_{\mu}^{(r)}\theta^{\mu \nu}k_{\nu}^{(s)}}
\nonumber \\
&&
~~~~
\times 
\int ds^{(o)} 
(s^{(o)})^{-13}
\exp 
\left \{ 
     \frac{s^{(o)}}{\alpha'}
      +
     \frac{K_{\mu}(\theta G \theta)^{\mu \nu} 
             K_{\nu}}{4 s^{(o)}}
\right \}   
\int 
\prod_{r=1}^{M+N-1} dT_r 
\nonumber \\
&&
~~~~~~~~~~~~~~~
\times 
\prod_{r=1}^{M+N}
e^{  -i\frac{T_r}{s^{(0)}}
        k_{\mu}^{(r)}\theta^{\mu \nu}K_{\nu}}
\times 
\prod_{r < s}^{M+N}
\left[ 
e^{-\frac{(T_r-T_s)^2}{s^{(o)}}+|T_r-T_s|}
\right]^{k_{\mu}^{(r)}G^{\mu \nu}k_{\nu}^{(s)}}, 
\label{open-string FTL of one-loop amplitude}
\end{eqnarray}
where the integral is performed over the region :
\begin{eqnarray}
&&
s^{(o)} \in \mathbb{R}_{\geq 0},
\nonumber \\
&&
0 \leq T_1 \leq \cdots \leq T_M \leq s^{(o)}, 
\nonumber \\
&& 
0 \leq T_{M+1} \leq \cdots T_{M+N-1} \leq s^{(o)}.
\label{moduli in open-string FTL} 
\end{eqnarray}

Eq.(\ref{open-string FTL of one-loop amplitude}) can be 
identified with one-loop amplitude of open-string 
tachyon $\phi$ (a scalar particle of $\mbox{mass}^2=-1/\alpha'$) 
living on the non-commutative world-volume. 
It is the amplitude obtained from 
the corresponding one-loop Feynman diagram 
consisting of $M+N$ trivalent vertices. 
Each of the vertices represents the following 
cubic interaction : 
\begin{eqnarray}
{\cal L}_{int}=
\frac{1}{3}\phi \star \phi \star \phi (x).  
\end{eqnarray}
Here we introduce the Moyal product 
(an associative non-commutative product) by 
\begin{eqnarray}
f \star g(x) \equiv 
\lim_{y \rightarrow x} 
\exp \left\{ \frac{i}{2}\theta^{\mu \nu}
             \frac{\partial}{\partial x^{\mu}}
             \frac{\partial}{\partial y^{\nu}} 
     \right\}
f(x)g(y). 
\label{def of Moyal product}
\end{eqnarray}
The Feynman rule becomes a little complicated due to the 
above Moyal products. 
But the Feynman integral can be evaluated by using the standard 
technique \cite{Itzkson-Zuber} and translated to 
Eq.(\ref{open-string FTL of one-loop amplitude}).

We can find 
$
\exp 
\left \{
    \frac{s^{(o)}}{\alpha'}
       +
    \frac{ K_{\mu}(\theta G \theta)^{\mu \nu}K_{\nu} }
             {4s^{(o)}}            
\right \}
=
\exp 
\left \{
     \frac{s^{(o)}}{\alpha'}
\right \}
\times 
\exp 
\left \{ 
        \frac{ K_{\mu}(\theta G \theta)^{\mu \nu}K_{\nu} }
             {4s^{(o)}}
\right \} 
$
in the integral of 
Eq.(\ref{open-string FTL of one-loop amplitude}).
The first term comes from the Schwinger representation 
of the tachyon propagator, 
$
\exp 
\left \{
  -s^{(0)}(p^2-\frac{1}{\alpha'})
\right \}
$.
This is the standard term which appears 
in ordinary field theory one-loop amplitudes. 
The second term can be understood \cite{UV-IR mixing} 
as a UV regularization 
(a regularization of the integration near $s^{(o)}=0$). 
It is a curious regularization since 
it depends on the external momentum, 
$K=\sum_{r=1}^{M}k^{(r)}$. 
This feature can be thought of as a characteristic of 
quantum field theories on a non-commutative space and 
makes field theoretical description of physics at high energy 
scale difficult. It should be also noted that 
integrations in Eq.(\ref{open-string FTL of one-loop amplitude}) 
have potential singularities at $T_r=T_s$ ($r \neq s$). 
These singularities can be already seen in the integral 
(\ref{tachyon one-loop by open-string parameters}). 
Although we do not describe their regularization here,  
prescription used \cite{Nakatsu} for the cubic open-string 
field theory will be effective.

\subsection{UV limit of non-commutative scalar field theory}

Both parameters 
$(\tau^{(o)},\nu_r)$ and $(\tau^{(c)},\sigma_r)$ 
are regarded as coordinates of the moduli 
space of conformal classes of cylinder with 
$M+N$ punctures at the boundaries.  
For each values of the closed-string parameters 
we obtain a graph as depicted in Figure 
\ref{closed-string diagram}. 
Set of these graphs is the above moduli space. 
Strictly speaking, these graphs are 
representatives of the conformal classes
and thus it is possible to make different choices.
For each values of the open-string parameters 
we obtain a graph as depicted in 
Figure \ref{open-string diagram}. 
It can be identified 
in the standard manner with 
a metrized trivalent one-loop ribbon graph. 
Metric of the graph is given by the open-string 
parameters. These metrized ribbon graphs 
can be chosen as the representatives. 
Therefore the open-string parameters give 
another set of coordinates of the moduli space.

The above moduli space has several ends. 
Typically we have two ends at 
$\tau^{(o)}=0$ and $\tau^{(o)}=+\infty$. 
The previous zero-slope limit of open-string 
is related with the end at the infinity  
as can be seen from 
Eq.(\ref{Schwinger parameters of scalar field theory}). 
Actual procedure to obtain the zero-slope limit  
(\ref{open-string FTL of one-loop amplitude}) 
shows that the field theory amplitude is 
a suitable magnification of the integral 
(\ref{tachyon one-loop by open-string parameters}) 
on an infinitesimal neighbourhood around this end. 
We want to know the UV behavior of the world-volume 
theory. 
For such a purpose we need to focus on 
the region $s^{(o)} \approx 0$.  
One possible resolution may be obtained by taking 
a zero-slope limit such that it magnifies an 
infinitesimal neighbourhood around another end 
located at $\tau^{(o)}=0$. 
Near this end use of the closed-string parameters 
will be effective as the open-string 
parameters near the infinity.  
Let us examine a zero-slope limit which will be 
taken by fixing the following parameters :
\begin{eqnarray}
s^{(c)}\equiv \alpha' |\tau^{(c)}|,~~~~\sigma_r.  
\label{field theory parameters of closed-string}
\end{eqnarray}
In the course of taking the limit 
we also fix open-string tensors 
$G_{\mu \nu}$ and $\theta^{\mu \nu}$ 
in order to capture the world-volume theory.    
It follows from Eq.(\ref{equalities2 and 3}) that 
this makes the limit different from the 
standard field theory limit of closed-string.  
Transformation between the open- and the closed-string
parameters leads  
$s^{(c)}=2\pi \alpha'\cdot \frac{\alpha'}{s^{(o)}}$ 
and 
$\sigma_r=2\pi \frac{T_r}{s^{(o)}}$. 
For $s^{(c)}$ and $\sigma_r$ to remain finite 
the original field theory parameters need to satisfy 
$s^{(o)}\ll \alpha'$ and $T_r \approx s^{(o)}$. 
We cannot neglect effects of 
all the perturbative stringy states of open-string 
at such a trans-string scale. Hence the limit in question 
includes their effects.

Relevant integral form of one-loop amplitude of 
$M+N$ open-string tachyons is given by 
Eq.(\ref{tachyon one-loop by closed-string parameters}) 
using the closed-string parameters or equivalently 
by Eq.(\ref{equivalence with boundary state formalism}) 
using the boundary states. 
We generalize the amplitude by dropping out the factor 
$
\prod_{r=1}^{M+N}
(-i\tau^{(c)})
^{\alpha'k_{\mu}^{(r)}G^{\mu \nu}k_{\nu}^{(r)}-1} 
$
in the integral and introduce the following 
quantity
\footnote{We put $p=25$ and $D=26$.} :
\begin{eqnarray}
&&
J_{T}
\left( \left( k^{(1)},\cdots,k^{(M)} \right) ; 
       \left( k^{(M+1)},\cdots,k^{(M+N)} \right)
\right)
\nonumber \\
&&
\equiv 
(-2\pi i)
\int_{i0}^{i \infty} 
d\tau^{(c)}  
\left[ 
\int \prod_{r=M+1}^{M+N-1}d\sigma_r
\left \langle  
   B_N ; (\sigma_{M+1},k^{(M+1)}),\cdots, 
           (\sigma_{M+N},k^{(M+N)}) 
\right |
\right] 
\nonumber \\
&&
~~~~~~~~~~~~~~~~~~~~~~
\times q_c^{\frac{L_0+\tilde{L}_0-2}{2}} 
\left[ 
\int \prod_{r=1}^{M} d\sigma_r  
\left | 
         B_N ; (\sigma_{1},k^{(1)}),\cdots, 
           (\sigma_{M},k^{(M)}) 
\right \rangle     
\right], 
\label{def of JT}
\end{eqnarray}
where the integral is performed over the region 
(\ref{moduli by closed-string parameters}). 
$\sigma_{M+N}$ is fixed to $2\pi$.  
When $k^{(r)}$ satisfy the on-shell conditions  
the above $J_T$ coincides with the original amplitude.  
It is an off-shell generalization slightly 
different from the original one.

Let us take the zero-slope limit. 
We first look at correlations which appear  
in Eq.(\ref{def of JT}). They can be read from 
Eq.(\ref{tachyon one-loop by closed-string parameters}). 
Terms which become dominant after the translations  
(\ref{field theory parameters of closed-string}) 
are as follows :
\begin{eqnarray}   
&&
\left[
   \eta 
     \left( 
         \tau^{(c)}
     \right)^{-3}
   \theta_4
      \left.
      \left( 
         \frac{\sigma_s-\sigma_r}{2\pi} 
      \right|
         \tau^{(c)} 
      \right) 
\right]^{2\alpha'k_{\mu}^{(r)}G^{\mu \nu}k_{\nu}^{(s)}}
\approx 
e^{\frac{\pi}{2}s^{(c)}k_{\mu}^{(r)}G^{\mu \nu}k_{\nu}^{(s)}}, 
\nonumber \\  
&&
\left[
    \eta 
       \left( 
          \tau^{(c)}
       \right)^{-3}
    \theta_1
       \left. 
       \left( 
           \frac{\sigma_s-\sigma_r}{2\pi } 
       \right|  
          \tau^{(c)}
       \right) 
\right]^{2\alpha'k_{\mu}^{(r)}G^{\mu \nu}k_{\nu}^{(s)}}
\approx
1. 
\end{eqnarray}
We also note that the dominant contribution of 
$\eta(\tau^{(c)})^{-24}$ in 
Eq.(\ref{tachyon one-loop by closed-string parameters}) 
is $e^{\frac{2\pi}{\alpha'}s^{(c)}}$. 
Gathering these estimations the following integral 
turns out to be the zero-slope limit :
\begin{eqnarray}
&&
J_{T}
\left( \left( k^{(1)},\cdots,k^{(M)} \right) ; 
       \left( k^{(M+1)},\cdots,k^{(M+N)} \right)
\right)
\nonumber \\
&&
\approx 
\frac{2\pi}{\alpha'}
\delta^{26} 
  \left(
       \sum_{r=1}^{M+N}k^{(r)}
  \right)
\left[ 
\int ds^{(c)}
\exp \left \{ -\frac{\pi s^{(c)}}{2}
        \left( 
        K_{\mu}g^{\mu \nu}K_{\nu}-\frac{4}{\alpha'}
        \right) 
     \right \}
\right]
\nonumber \\
&&
~~~~~
\times 
    \left( 
       \frac{-\mbox{det}G_{\mu \nu}}{(2\alpha')^{26}}
    \right)^{1/4}
    \int \prod_{r=1}^{M} d\sigma_r
        \prod_{1 \leq r < s \leq M}
            e^{-\frac{i}{2}k_{\mu}^{(r)}
                 \theta^{\mu \nu}k_{\nu}^{(s)}}
        \prod_{r=1}^M
            e^{-ik_{\mu}^{(r)}\theta^{\mu \nu}K_{\nu}
                \frac{\sigma_r}{2\pi}}
\nonumber \\
&&
~~~~~
\times 
    \left( 
       \frac{-\mbox{det}G_{\mu \nu}}{(2\alpha')^{26}}
    \right)^{1/4}
    \int \prod_{r=M+1}^{M+N-1} d\sigma_r
          \prod_{M+1 \leq r < s \leq M+N}
              e^{\frac{i}{2}k_{\mu}^{(r)}
                    \theta^{\mu \nu}k_{\nu}^{(s)}}
          \prod_{r=M+1}^{M+N}
              e^{-ik_{\mu}^{(r)}\theta^{\mu \nu}K_{\nu}
                    \frac{\sigma_r}{2\pi}}~~, 
\nonumber \\
\label{closed-string FTL of one-loop amplitude}
\end{eqnarray}
where the integral is performed over the region :
($\sigma_{M+N}$ is fixed to $2\pi$.)
\begin{eqnarray}
&&
s^{(c)} \in \mathbb{R}_{\geq 0}, 
\nonumber \\
&&
0 \leq \sigma_1 \leq \cdots \leq \sigma_M \leq 2\pi, 
\nonumber \\
&&
0 \leq \sigma_{M+1} \leq \cdots \leq \sigma_{M+N-1} \leq 2\pi. 
\label{moduli in closed-string FTL}
\end{eqnarray}
Here the closed-string metric $g^{\mu \nu}$ should be 
understood as 
$g^{\mu \nu}
\approx 
-\frac{(\theta G \theta)^{\mu \nu}}{(2\pi \alpha')^2}$ 
by using Eq.(\ref{equalities2 and 3}).

It is important to compare the above zero-slope limit 
with the previous one. 
In Eq.(\ref{closed-string FTL of one-loop amplitude}) 
the Schwinger propagator  
$
\exp \left \{ -\frac{\pi s^{(c)}}{2}
        \left( 
        K_{\mu}g^{\mu \nu}K_{\nu}-\frac{4}{\alpha'}
        \right) 
     \right \}
$
is integrated and gives rise to the propagator of 
{\it closed-string tachyon} of momentum $K$, 
while the counterpart in 
Eq.(\ref{open-string FTL of one-loop amplitude})
consists of two parts 
one of which is standard in field theories 
and the other is a curious regularization 
factor originated in the non-commutativity.  
The factors 
$
\left[ 
   \exp \left \{
     -\frac{(T_r-T_s)^2}{s^{(o)}}+|T_r-T_s|
        \right \}
\right]^{k_{\mu}^{(r)}G^{\mu \nu}k_{\nu}^{(s)}} 
$ 
found in Eq.(\ref{open-string FTL of one-loop amplitude})
describe correlations 
between two open-string tachyons inserted at 
$T_r$ and $T_s$ in the Schwinger time. 
These correlations are made by propagations of 
open-string tachyon between them and particularly 
have their origin in kinetic energy 
$p^2$ of the propagating open-string tachyon.
These correlations are lost in 
Eq.(\ref{closed-string FTL of one-loop amplitude}). 
This indicates that 
open-string tachyon becomes topological at the 
trans-string scale 
and that it is described by a one-dimensional 
topological field theory.  
Vanishing of the correlations 
originates in the modular transforms
(\ref{modular trans of tachyon correlations}). 
So this is a stringy effect.

\subsection{Factorization by straight open Wilson lines}

We now consider generating function 
of amplitudes (\ref{def of JT}) and examine the zero-slope 
limit. Let $\varphi(k)$ be the Fourier modes of 
open-string tachyon field :
$
\phi(x)=
\int \frac{dk}{(2\pi)^{13}}
\varphi(k)e^{ik_{\mu}x^{\mu}}. 
$
Since it is a real scalar field, the Fourier modes satisfy 
$\overline{\varphi(k)}=\varphi(-k)$, where 
the overline denotes complex conjugation. 
We examine the zero-slope limit of the following generating 
function :
\begin{eqnarray}
&&
\sum_{M=0}^{\infty}\sum_{N=1}^{\infty}
\int \prod_{r=1}^{M+N}
\frac{dk^{(r)}}{(2\pi)^{13}}
J_T
\left(
\left( k^{(1)},\cdots,k^{(M)}
\right)
;
\left(-k^{(M+1)},\cdots,-k^{(M+N)}
\right)
\right)
\nonumber \\
&&
~~~~~~~~~~~~~~~~~~~~~~~~~
\times 
\prod_{r=1}^{M} 
\varphi(k^{(r)})
\prod_{r=M+1}^{M+N}
\varphi(-k^{(r)}).  
\label{def of generating function of tachyon amplitudes}
\end{eqnarray}

Let us start with taking the zero-slope limit term by term 
in the sum. The total momentum is conserved in each term.  
It can be translated to the following integral :
\begin{eqnarray}
\lefteqn{
\delta^{26} 
  \left(
     \sum_{r=1}^Mk^{(r)}-\sum_{r=M+1}^{M+N}k^{(r)}
  \right)
=
\int dp\, 
\delta^{26} 
  \left(
     p+\sum_{r=1}^Mk^{(r)}
  \right)
\delta^{26} 
  \left(
     p+\sum_{r=M+1}^{M+N}k^{(r)}
  \right)
}
\nonumber \\
&&
=
\int dp \,
\delta^{26} \;
  \left(
     p+\sum_{r=1}^Mk^{(r)}
  \right)
  e^{-\frac{i}{2}p_{\mu}\theta^{\mu\nu}\sum_{r=1}^{M}k^{(r)}_{\nu}}
  \;
  \delta^{26} 
  \left(
     p+\sum_{r=M+1}^{M+N}k^{(r)}
  \right)
  e^{\frac{i}{2}p_{\mu}\theta^{\mu\nu}\sum_{r=M+1}^{M+N}k^{(r)}_{\nu}}
\nonumber \\
&&
= 
\int dp 
\left(
\int 
\frac{dx}{(2\pi)^{26}}
e^{ip_{\mu}x^{\mu}}
  e^{ik_{\mu}^{(r)}x^{\mu}}
\right)
e^{-\frac{i}{2}p_{\mu}\theta^{\mu\nu}\sum_{r=1}^{M}k^{(r)}_{\nu}}
\nonumber \\
&&
\hspace{5em}
\times 
\left(
\int 
\frac{dy}{(2\pi)^{26}}
e^{-ip_{\mu}y^{\mu}}
e^{-ik_{\mu}^{(r)}y^{\mu}}
\right)
e^{\frac{i}{2}p_{\mu}\theta^{\mu\nu}\sum_{r=M+1}^{M+N}k^{(r)}_{\nu}}. 
\label{manipulation of delta}
\end{eqnarray}
The zero-slope limit can be obtained by using 
Eq.(\ref{closed-string FTL of one-loop amplitude}). 
It turns out to be  expressed in terms of the Moyal products. 
The above rearrangement of the momentum conservation is used
in this course. 
It yields the following products :
\begin{eqnarray}
&&
\int 
\prod_{r=1}^M
\frac{dk^{(r)}}{(2\pi)^{13}}
e^{ip_{\mu}x^{\mu}}
e^{-\frac{i}{2}p_{\mu}\theta^{\mu\nu}\sum_{r=1}^{M}k^{(r)}_{\nu}}
\prod_{1 \leq r < s \leq M}
e^{-i\frac{i}{2}k_{\mu}^{(r)}
         \theta^{\mu \nu}k_{\nu}^{(s)}}
\prod_{r=1}^M
\varphi(k^{(r)})
e^{ik_{\mu}^{(r)}
   \left(
      x^{\mu}
      +\theta^{\mu \nu}p_{\nu}\frac{\sigma_r}{2\pi}
   \right)}
\nonumber \\
&&
~~~~
=
e^{ip_{\mu}x^{\mu}}
\star
\phi(x_0(\sigma_1))\star \phi(x_0(\sigma_2))
\star \cdots \star 
\phi(x_0(\sigma_M)).
\end{eqnarray}
Here we introduce straight line $x_0(\sigma)$ 
in the world-volume by 
$
x^{\mu}_0(\sigma)\equiv 
x^{\mu}+\theta^{\mu \nu}p_{\nu}\frac{\sigma}{2\pi}
$
with $0\leq \sigma \leq 2\pi$,
and the Moyal products are taken with respect to $x$.
After these replacements 
the zero-slope limit turns out to be :
\begin{eqnarray} 
&&
\int \prod_{r=1}^{M+N}
\frac{dk^{(r)}}{(2\pi)^{13}}
J_T
\left(
\left( k^{(1)},\cdots,k^{(M)}
\right)
;
\left(-k^{(M+1)},\cdots,-k^{(M+N)}
\right)
\right)
\nonumber \\
&&
~~~~~~~~~~~~~~~~~~~~~~
\times
\prod_{r=1}^{M} 
\varphi(k^{(r)})
\prod_{r=M+1}^{M+N}
\varphi(-k^{(r)})
\nonumber \\
&&
\approx 
\frac{2\pi}{\alpha'}
\int dp 
\left[
 \int_{0}^{\infty}ds^{(c)}
    \exp 
     \left\{ -\frac{\pi s^{(c)}}{2}
        \left( 
           p_{\mu}g^{\mu \nu}p_{\nu}-\frac{4}{\alpha'}
        \right)
     \right\}
\right]
\nonumber \\
&&
~~~~~~~
\times 
  \left( 
     \frac{-\mbox{det} G_{\mu \nu}}
          {(2\alpha')^{26}}
  \right)^{\frac{1}{4}}
\int 
\frac{dx}{(2\pi)^{26}}
    e^{ip_{\mu}x^{\mu}}  \star
\left[
  \int 
    \prod_{r=1}^{M}d\sigma_r 
        \phi(x_0(\sigma_1))
             \star \cdots \star 
        \phi(x_0(\sigma_M))
\right]
\nonumber \\
&&
~~~~~~~
\times 
  \left( 
     \frac{-\mbox{det} G_{\mu \nu}}
          {(2\alpha')^{26}}
  \right)^{\frac{1}{4}}
\overline{
\int 
\frac{dx}{(2\pi)^{26}}
    e^{ip_{\mu}x^{\mu}}  \star
\left[
  \int 
      \prod_{r=M+1}^{M+N-1}d\sigma_r 
           \phi(x_0(\sigma_{M+1}))
                  \star  \cdots \star 
           \phi(x_0(\sigma_{M+N}))
\right]
}~~, 
\nonumber\\
\label{closed-string FTL by Moyal products}
\end{eqnarray}
where the integrations on $\sigma_r$ are performed over the 
region (\ref{moduli in closed-string FTL}).

The zero-slope limit of the generating function 
(\ref{def of generating function of tachyon amplitudes})
can be  obtained by summing up  
Eq.(\ref{closed-string FTL by Moyal products}) 
with respect to $M$ and $N$. 
Let us introduce path-ordered exponential 
along a line $x^{\mu}(\sigma)$. 
It is defined by 
\begin{eqnarray}
{\cal P}_{\star}
\left[
  \exp 
     \left\{ 
        \int_{0}^{2\pi}d\sigma \phi(x(\sigma))
     \right\}  
\right]
\equiv 
\sum_{M=0}^{\infty}
\int_{0 \leq \sigma_1 \leq \cdots \leq \sigma_M \leq 2\pi}
\prod_{r=1}^M d\sigma_r 
 \phi(x(\sigma_1))
             \star \cdots \star 
        \phi(x(\sigma_M))~~. 
\end{eqnarray}
This is an analogue of the Wilson line in gauge theories. 
Each summation of the Moyal products of 
Eq.(\ref{closed-string FTL by Moyal products}) gives 
the path-ordered exponential along the straight line. 
The zero-slope limit can be written as follows :
\begin{eqnarray}
&&
\sum_{M=0}^{\infty}\sum_{N=1}^{\infty}
\int \prod_{r=1}^{M+N}
\frac{dk^{(r)}}{(2\pi)^{13}}
J_T
\left(
\left( k^{(1)},\cdots,k^{(M)}
\right)
;
\left(-k^{(M+1)},\cdots,-k^{(M+N)}
\right)
\right)
\nonumber \\
&&
~~~~~~~~~~~~~~~~~~~~~~~~~
\times 
\prod_{r=1}^{M} 
\varphi(k^{(r)})
\prod_{r=M+1}^{M+N}
\varphi(-k^{(r)})
\nonumber \\
&&
\approx 
\frac{2\pi}{\alpha'}
\int dp 
\left[
 \int_{0}^{\infty}ds^{(c)}
    \exp 
     \left\{ -\frac{\pi s^{(c)}}{2}
        \left( 
           p_{\mu}g^{\mu \nu}p_{\nu}-\frac{4}{\alpha'}
        \right)
     \right\}
\right]
\nonumber \\
&&
~~~~~~~
\times 
  \left( 
     \frac{-\mbox{det} G_{\mu \nu}}
          {(2\alpha')^{26}}
  \right)^{\frac{1}{4}}
\int 
\frac{dx}{(2\pi)^{26}}
    e^{ip_{\mu}x^{\mu}}  \star
{\cal P}_{\star}
\left[
  \exp 
    \left\{
       \int_{0}^{2\pi}d\sigma \phi(x_0(\sigma))  
    \right\}
\right]   
\nonumber \\
&&
~~~~~~~
\times 
  \left( 
     \frac{-\mbox{det} G_{\mu \nu}}
          {(2\alpha')^{26}}
  \right)^{\frac{1}{4}}
\overline{
\int 
\frac{dx}{(2\pi)^{26}}
    e^{ip_{\mu}x^{\mu}}  \star
{\cal P}_{\star}
\left[
  \exp 
    \left\{
       \int_{0}^{2\pi}d\sigma \phi(x_0(\sigma))  
    \right\}
\right]
\star \phi(x_0(2\pi))  
}~~.  
\nonumber\\
\label{closed-string FTL by straight open Wilson lines}
\end{eqnarray}
Insertion of $\phi(x_0(2\pi))$ makes the above equation 
asymmetric. This term has its origin in one-loop 
calculation of open-string or tachyon field on the world-volume. 
The one-loop diagram has a $U(1)$ symmetry which generates 
a constant simultaneous shift 
of the Schwinger parameters. This $U(1)$ 
symmetry is fixed in the previous discussion by putting 
$\tau^{(o)}=-\ln \xi_{M+N}$.  
It becomes equivalent to $\sigma_{M+N}=2\pi$.
Modulo fixing the $U(1)$ symmetry, 
Eq.(\ref{closed-string FTL by straight open Wilson lines}) 
shows that the generating function is factorized 
at the zero-slope limit  
into a sum of products of 
$
{\cal P}_{\star}
\left[ 
e^{\int_0^{2\pi}
     d\sigma \phi(x_0(\sigma))}
\right]
$ 
and that open Wilson lines with the same velocity 
$
\frac{dx^{\mu}_0(\sigma)}{d \sigma}
=
\frac{\theta^{\mu \nu}p_{\nu}}{2\pi}
$ 
interact with each other by 
exchanging closed-string tachyon 
with momentum $p_{\mu}$.

A factorization of one-loop amplitudes 
of non-commutative scalar field theory is considered   
in 
\cite{S.J.Rey et al}\cite{one-loop of non-commutative scalar field}
and claimed there 
to be realized  
at the low-energy of this field theory 
by using 
the path-ordered exponentials. 
The procedure of the zero-slope limit
adopted in those papers
are quite different form that  of the present paper.
Our limiting procedure is not the conventional one
to obtain the field theory limit of open-string.
In fact the present limit would reduce to the field theory
limit of closed-string if the $B$ field were vanishing.
We would like to emphasize that the existence of
the $B$ field makes the limit different from the
conventional field theory limit of closed-string.

\section{Coordinate and Momentum Eigenstates of 
Closed-String}\label{sec:comments on eigenstates}  

We make a digression from the previous discussion 
on the world-volume theory.  
In this section we introduce coordinate and momentum 
eigenstates of closed-string and make some observation 
on their relation with the boundary states. 
These eigenstates will play important roles in 
the subsequent discussions.

We begin with a simple remark on closed-string 
momentum currents. These currents are introduced 
by taking functional derivatives of the action :
$
P_{M}(\sigma,\tau)
=
i\frac{\delta S[X]}
      {\delta \partial_{\tau}
             X^{M}(\sigma,\tau)}
$. 
We have two expressions of the action, 
(\ref{eq:action1})
and 
(\ref{eq:action2}). 
They provide two momentum currents, 
which coincide for the Dirichlet directions but become 
slightly different from each other for the Neumann 
directions.  
When the action (\ref{eq:action2}) is used 
one obtains
\begin{equation}
P_{\mu}(\sigma,\tau)
\equiv 
\frac{i}{2\pi\alpha'}
  g_{\mu\nu}
    \partial_{\tau}
       X^{\nu}(\sigma,\tau)~,
\label{eq:P}
\end{equation}
while the action (\ref{eq:action1}) leads to 
\begin{equation}
P^{(B)}_{\mu}(\sigma,\tau)
\equiv 
\frac{i}{2\pi\alpha'}
\Bigl(         
       g_{\mu\nu}
        \partial_{\tau}X^{\nu}(\sigma,\tau)
         -
       i2\pi\alpha'B_{\mu\nu}
         \partial_{\sigma}X^{\nu}(\sigma,\tau)
\Bigr)~.
\label{eq:PB}
\end{equation}
Their conserved charges become the same, 
that is, momentum $p_{\mu}$, since  
these are integrals of the currents 
over the circle. 
Whichever momentum current one adopts as the conjugate 
variable of $X^{\mu}(\sigma,\tau)$, 
canonical quantization of closed-string leads to 
the same commutation relations (\ref{eq:CCR}).

Let $\hat{X}^{M}(\sigma)$ be closed-string coordinate 
operators. They are simply given by 
$
\hat{X}^{M}(\sigma)
\equiv 
X^{M}(\sigma,0)
$. 
The conjugate momentum operators 
$\hat{P}_M(\sigma)$ are those operators satisfying 
$
\left[
\hat{X}^{M}(\sigma),
\right.
$
$
\left.
\hat{P}_{N}(\sigma')
\right]
$
$
=i\delta^{M}_{N}
\delta(\sigma-\sigma')
$
\footnote{
The delta function on the circle is 
given by 
\begin{eqnarray}
\delta(\sigma)=
\frac{1}{2\pi}\sum_{n\in\mathbb{Z}}e^{in\sigma}
~~~~~~(0\leq \sigma <2\pi)~.
\nonumber
\end{eqnarray}
}.
These are given by the above momentum currents. 
Due to the existence of 
two different currents we have two operators,  
$
\hat{P}_{M}(\sigma)
\equiv 
P_{M}(\sigma,0)
$ 
and 
$
\hat{P}^{(B)}_{M}(\sigma)
\equiv 
P^{(B)}_{M}(\sigma,0)
$. 
All these operators are periodic with respect to $\sigma$. 
We may provide their mode expansions similar to 
Eq.(\ref{eq:modeX}). For the later discussion 
these turn out to be inconvenient. Instead we use 
the following mode expansions :
\begin{eqnarray}
\hat{X}^{M}(\sigma)
&=&
\hat{x}^{M}_{0}
+
\frac{1}{\sqrt{2}} 
\sum_{n=1}^{\infty}
     \left(
        \hat{\chi}_{n}^{M} e^{-in\sigma}
         +
        \hat{\chi}^{\dagger M}_{n}e^{in\sigma}
 \right)~,
\nonumber\\
\hat{P}_{M}(\sigma)
&=&
\frac{1}{2\pi}
 \left[
    \hat{p}_{0M}
     +
    \frac{1}{\sqrt{2}}
        \sum_{n=1}^{\infty}
           \left(
             \hat{\psi}_{n M}e^{-in\sigma}
             +
             \hat{\psi}^{\dagger}_{n M}e^{in\sigma}
           \right)
\right]~,
\nonumber\\
\hat{P}^{(B)}_{M}(\sigma)
&=&
\frac{1}{2\pi}
\left[
     \hat{p}_{0M}
     +
     \frac{1}{\sqrt{2}}
        \sum_{n=1}^{\infty}
          \left(
            \hat{\varrho}_{nM}e^{-in\sigma}
            +
            \hat{\varrho}^{\dagger}_{nM}e^{in\sigma}
          \right)
\right]~,
\label{modes of coordinates and momntum operators}
\end{eqnarray}
where 
$\hat{\chi}^{\dagger M}_{n}$, 
$\hat{\psi}^{\dagger}_{nM}$
and 
$\hat{\varrho}^{\dagger}_{nM}$ 
are the hermitian conjugates  
of the corresponding ones. 
Their commutation relations can be read as follows :
\begin{eqnarray}
&&[\hat{x}_{0}^{M},\hat{p}_{0N}]=i\delta^{M}_{N}~,
\quad
[\hat{\chi}^{M}_{m},\hat{\psi}^{\dagger}_{nN}]
=[\hat{\chi}^{\dagger M}_{m},\hat{\psi}_{n N}]
 =2i\delta^{M}_{N}\delta_{m,n}~,
\nonumber\\
&&
 [\hat{\chi}^{M}_{m},\hat{\varrho}^{\dagger}_{n N}]
=[\hat{\chi}^{\dagger M}_{m},\hat{\varrho}_{n N}]
 =2i\delta^{M}_{N}\delta_{m,n}~,
\quad \mbox{otherwise$=0$}~.
\end{eqnarray}

These oscillator modes are linear sums of the 
standard oscillator modes of $X^M(\sigma,\tau)$. 
For the reader's convenience we attach below the dictionary :
\begin{eqnarray}
\hat{\chi}^{M}_{n}
=
i\frac{\sqrt{\alpha'}}{n}
   \left(
        \alpha^{M}_{n}
         -
        \tilde{\alpha}^{M}_{-n}
   \right)
&,&\quad
\hat{\chi}^{\dagger M}_{n}
=
-i\frac{\sqrt{\alpha'}}{n}
   \left(
       \alpha^{M}_{-n}
        -
        \tilde{\alpha}^{M}_{n} 
    \right)~,
\nonumber\\
\hat{\psi}_{nM} 
= 
\frac{1}{\sqrt{\alpha'}}g_{MN}
    \left(
       \alpha^{N}_{n}
        +
       \tilde{\alpha}^{N}_{-n}
     \right)
&,&\quad
\hat{\psi}^{\dagger}_{nM}
=
\frac{1}{\sqrt{\alpha'}}g_{MN}
    \left(
      \alpha_{-n}^{N}
       +
      \tilde{\alpha}_{n}^{N}
\right)~,
\nonumber\\
\hat{\varrho}_{nM} 
= 
\frac{1}{\sqrt{\alpha'}}
   \left(
     E_{MN}\alpha^{N}_{n}
       +
     E^{T}_{MN}\tilde{\alpha}_{-n}^{N}
   \right)
&,& \quad
\hat{\varrho}^{\dagger}_{nM}
=
\frac{1}{\sqrt{\alpha'}}
   \left(
        E_{MN}\alpha^{N}_{-n}
        +
        E^{T}_{MN}\tilde{\alpha}^{N}_{n}
    \right)~, 
  \label{eq:identification2}
\end{eqnarray}
where the $(i,j)$-components of $E$ are understood 
as $g_{ij}$. 
Therefore we have 
$
\hat{\psi}_{ni}=\hat{\varrho}_{ni}
$ 
and 
$
\hat{\psi}^{\dagger}_{ni}=\hat{\varrho}^{\dagger}_{ni}
$.

\subsection{Coordinate and momentum eigenstates}
Coordinate and momentum eigenstates are respectively 
eigenstates of $\hat{X}(\sigma)$ 
and $\hat{P}(\sigma)$ or $\hat{P}^{(B)}(\sigma)$. 
We describe them in some detail.  
These are closed-string extensions of 
those obtained \cite{GJ} for open-string.

\subsubsection*{Coordinate eigenstates} 
We first describe the coordinate eigenstates. 
They are characterized by two conditions ;  
(i) they are eigenstates of the string coordinate 
operators, 
\begin{eqnarray}
\hat{X}^M(\sigma)
|X \rangle 
=
X^M(\sigma) 
|X \rangle,
\end{eqnarray}
and (ii) they satisfy the orthonormality 
condition, 
\begin{eqnarray}
\langle X | X' \rangle 
=
\prod_{0 \leq \sigma < 2\pi}
\delta(X(\sigma)-X'(\sigma)),
\label{eq:orthonormality of X bs} 
\end{eqnarray}
where we abbreviate the superscript $M$ in the RHS. 
Dual states $\langle X |$ are defined by the hermitian 
conjugation. It is done by simultaneous operations of 
the BPZ and complex conjugations.

The above conditions determine the eigenstates. 
We factorize them into products of 
the Neumann and the Dirichlet sectors :
\begin{eqnarray}
|X \rangle 
=
|X_N \rangle 
\otimes 
|X_D \rangle, 
\end{eqnarray}
where 
$|X_{N,D}\rangle$ are the coordinate eigenstates 
of the corresponding directions. 
Let us provide oscillator representations 
of these states.  
Taking account of the mode expansions 
(\ref{modes of coordinates and momntum operators}) 
we parametrize the eigenvalues 
in terms of complex variables 
$\chi_n^M$ ($n \in \mathbb{Z}_{\geq 1}$) besides 
the zero-modes $x_0^M$. We put 
\begin{eqnarray}
X^{M}(\sigma)
=
x^{M}_{0}
+
\frac{1}{\sqrt{2}} 
\sum_{n=1}^{\infty}
\left(
        \chi_{n}^{M} e^{-in\sigma}
         +
        \overline{\chi}^{M}_{n}e^{in\sigma}
\right)~.
\end{eqnarray}
The eigenstates are given by the following infinite products :
\begin{eqnarray}
|X_{N}\rangle 
=
|x_{0N}\rangle 
\otimes 
\prod_{n=1}^{\infty}
\left|
  \chi_{n},\bar{\chi}_{n}
\right\rangle_N~, 
~~~
|X_{D}\rangle 
=
|x_{0D}\rangle 
\otimes 
\prod_{n=1}^{\infty}
\left|
     \chi_{n},\bar{\chi}_{n}
\right\rangle_{D}. 
\label{eq:es-X}
\end{eqnarray}
Here 
$
|x_{0N}\rangle
=
|x_{0N}=(x^{\mu}_{0})\rangle
$
(
resp. 
$
|x_{0D}\rangle
$
)
are the normalized eigenstates of the zero modes
$\hat{x}^{\mu}_{0}$ 
(resp. $\hat{x}^{i}_{0}$)
with eigenvalues 
$x^{\mu}_{0} \in \mathbb{R}^{p+1}$
(resp. $x^{i}_0$), 
and their normalization conditions are 
\begin{equation}
\langle x'_{0N}|x_{0N}\rangle 
= 
\prod_{\mu=0}^{p}
\delta 
\left(
      x^{\prime\mu}_{0}-x^{\mu}_{0}
\right)~,
\quad
\langle x'_{0D}|x_{0D}\rangle 
= 
\prod_{i=p+1}^{D-1}
\delta 
\left(
      x^{\prime i}_{0}-x^{i}_{0}
\right)~.
\label{eq:norm-x0}
\end{equation}
$
\left|\chi_{n},\bar{\chi}_{n}\right\rangle_{N}
$
(resp. 
$
\left|\chi_{n},\bar{\chi}_{n}\right\rangle_{D}
$
)
are the normalized eigenstates of 
$
(\hat{\chi}^{\mu}_{n},\hat{\chi}^{\dagger\mu}_{n})
$ 
(resp. 
$
(\hat{\chi}^{i}_{n},\hat{\chi}^{\dagger i}_{n})
$)
with eigenvalues
$
\left(\chi^{\mu}_{n},
\bar{\chi}^{\mu}_{n}\right)
$
(resp. 
$
(\chi^{i}_{n},\bar{\chi}^{i}_{n})
$). 
Their normalization conditions are chosen as  
\begin{eqnarray}
{}_{N}\!\left\langle \chi^{\prime}_{n}, \bar{\chi}^{\prime}_{n}\right|
\left.\chi_{n},\bar{\chi}_{n}\right\rangle_{N}
 &=& \prod_{\mu=0}^{p} 
  \delta \left(\mathrm{Re}\chi^{\prime\mu}_{n}-\mathrm{Re}\chi^{\mu}_{n}
         \right)
  \delta \left( \mathrm{Im}\chi^{\prime\mu}_{n}-\mathrm{Im}\chi^{\mu}_{n}
         \right)~, \nonumber\\
{}_{D}\!\left\langle \chi^{\prime}_{n}, \bar{\chi}^{\prime}_{n}\right|
\left.\chi_{n},\bar{\chi}_{n}\right\rangle_{D}
 &=& \prod_{i=p+1}^{D-1} 
  \delta \left(\mathrm{Re}\chi^{\prime i}_{n}-\mathrm{Re}\chi^{i}_{n}
         \right)
  \delta \left( \mathrm{Im}\chi^{\prime i}_{n}-\mathrm{Im}\chi^{i}_{n}
         \right)~.
 \label{eq:norm-chi}
\end{eqnarray}
The orthonormality (\ref{eq:orthonormality of X bs})
follows from the above conditions 
imposed on each eigenstates :
\begin{eqnarray}
\langle X| X' \rangle &=& 
\langle x_0 | x_0' \rangle 
\prod_{n=1}^{\infty}
\langle \chi_{n},\bar{\chi}_n|\chi'_{n},\bar{\chi}'_{n}\rangle
\nonumber \\
&=& 
\delta(x_0-x_0')
\prod_{n=1}^{\infty}
\delta(\mbox{Re}\chi_n-\mbox{Re}\chi'_n)
\delta(\mbox{Im}\chi_n-\mbox{Im}\chi'_n)
\nonumber \\
&=& 
\prod_{0 \leq \sigma < 2\pi}
\delta(X(\sigma)-X'(\sigma)),
\end{eqnarray}
where we put 
$
|x_{0}\rangle \equiv 
$
$
|x_{0N}\rangle 
\otimes 
|x_{0D}\rangle
$
and 
$
|\chi_{n},\bar{\chi}_{n}\rangle
\equiv 
$
$
|\chi_{n},\bar{\chi}_{n}\rangle_{N}
\otimes 
|\chi_{n},\bar{\chi}_{n}\rangle_{D}
$.

The steps to obtain oscillator realizations of eigenstates 
$
|\chi_{n},\bar{\chi}_{n}\rangle_{N}
$
and 
$ 
|\chi_{n},\bar{\chi}_{n}\rangle_{D}
$ 
are presented in Appendix \ref{sec:eigen}
\footnote{
The following dictionary might be useful 
for the correspondence : 
\begin{eqnarray}
&&\hat{\chi}^{M}_{n}
  =\hat{\phi}^{(\mathrm{I})M}_{n}+i\hat{\phi}^{(\mathrm{II})M}_{n}~,
\quad
\hat{\psi}_{nM}
  =\hat{\pi}^{(\mathrm{I})}_{nM}+i\hat{\pi}^{(\mathrm{II})}_{nM}~,
\quad
\hat{\varrho}_{nM}
 =\hat{\varpi}^{(\mathrm{I})}_{nM}+i \hat{\varpi}^{(\mathrm{II})}_{nM}~,
\nonumber\\
&& \hat{\chi}^{\dagger M}_{n}
  =\hat{\phi}^{(\mathrm{I})M}_{n}-i\hat{\phi}^{(\mathrm{II})M}_{n}~,
\quad
 \hat{\psi}^{\dagger}_{nM}
  =\hat{\pi}^{(\mathrm{I})}_{nM}-i\hat{\pi}^{(\mathrm{II})}_{nM}~,
\quad
\hat{\varrho}^{\dagger}_{nM}
 =\hat{\varpi}^{(\mathrm{I})}_{nM}-i \hat{\varpi}^{(\mathrm{II})}_{nM}~.
\nonumber 
\end{eqnarray}
}. 
They turn out to be as follows : 
\begin{eqnarray}
\lefteqn{
\left|\chi_{n},\bar{\chi}_{n}\right\rangle_{N} =
 \left( \frac{n}{\pi\alpha'}\right)^{\frac{p+1}{2}}
  \sqrt{-\det g_{\mu\nu}} }\nonumber\\
  && \times \exp \left[ 
     \frac{1}{n}\alpha^{\mu}_{-n}g_{\mu\nu}\tilde{\alpha}^{\nu}_{-n}
     - \frac{i}{\sqrt{\alpha'}}\left(
            \chi^{\mu}_{n}g_{\mu\nu}\alpha^{\nu}_{-n}
            +\bar{\chi}^{\mu}_{n}g_{\mu\nu}\tilde{\alpha}^{\nu}_{-n}
          \right)
     -\frac{n}{2\alpha'}\bar{\chi}^{\mu}_{n}g_{\mu\nu}\chi^{\nu}_{n}
     \right] |0 \rangle~,
 \label{eq:chi-chi N}\\
\lefteqn{
  \left|\chi_{n},\bar{\chi}_{n}\right\rangle_{D} = 
    \left(\frac{n}{\pi\alpha'}\right)^{\frac{d}{2}}
    \sqrt{\det g_{ij}} } \nonumber\\
  && \times \exp\left[\frac{1}{n}\alpha^{i}_{-n}g_{ij}
                           \tilde{\alpha}^{\nu}_{j}
                -\frac{i}{\sqrt{\alpha'}}
            \left(\chi^{i}_{n}g_{ij}\alpha^{i}_{-n}
                  +\bar{\chi}^{i}_{n} g_{ij} \tilde{\alpha}^{j}_{-n}
            \right)
               -\frac{n}{2\alpha'} \bar{\chi}^{i}_{n}g_{ij}
                                   \chi^{j}_{n} \right]
          |0\rangle~.
 \label{eq:chi-chi D}
\end{eqnarray}

The coordinate eigenstates 
provide a complete basis of the Hilbert space
$
\mathcal{H}^{X}_{c}
=
\mathcal{H}^{X}_{cN} 
\otimes 
\mathcal{H}^{X}_{cD}
$,
where 
$\mathcal{H}^{X}_{cN}$ 
and 
$\mathcal{H}^{X}_{cD}$ 
stand for the Neumann and the Dirichlet sectors. 
We have the following completeness relation : 
\begin{eqnarray}
1&=& \int \left[dX\right] \;|X\rangle \langle X|
 \nonumber\\
 &=& \int d^{D}x_{0} \prod_{n=1}^{\infty}
  \left(\prod_{M=0}^{D-1}
     \frac{d\bar{\chi}^{M}_{n}d\chi^{M}_{n}}{2i}\right)
     \left|\chi_{n},\bar{\chi}_{n}\right\rangle
      |x_{0}\rangle \langle x_{0}|
     \left\langle \chi_{n},\bar{\chi}_{n}\right|~.
\end{eqnarray}

\subsubsection*{Momentum eigenstates}

We start with eigenstates of $\hat{P}(\sigma)$. 
They are characterized by two conditions ;
(i) they are eigenstates of $\hat{P}(\sigma)$, 
\begin{eqnarray}
\hat{P}_M(\sigma)|P \rangle 
=
P_M(\sigma)| P \rangle, 
\end{eqnarray}
and (ii) they satisfy the orthonormality condition, 
\begin{eqnarray}
\langle P | P' \rangle 
=
\prod_{0 \leq \sigma < 2\pi}
\delta 
(P(\sigma)-P'(\sigma)). 
\end{eqnarray}
These conditions determine the eigenstates. 
We factorize them into products of the Neumann 
and the Dirichlet sectors. 
\begin{eqnarray}
|P \rangle 
=
|P_N \rangle \otimes
|P_D \rangle.
\end{eqnarray}

Let us concentrate on the Neumann directions. 
Construction of eigenstates of the Dirichlet sector 
is parallel. 
We parametrize the eigenvalues 
in terms of complex variables 
$\psi_{n \mu}$ ($n \in \mathbb{Z}_{\geq 1}$) besides 
the zero-modes $p_{0 \mu}$. We put 
\begin{eqnarray}
P_{\mu}(\sigma)
&=&
\frac{1}{2\pi}
 \left[
    p_{0 \mu}
     +
    \frac{1}{\sqrt{2}}
        \sum_{n=1}^{\infty}
           \left(
             \psi_{n \mu}e^{-in\sigma}
             +
             \overline{\psi}_{n \mu}e^{in\sigma}
           \right)
\right]~.
\end{eqnarray}
The eigenstates are given by the following infinite products :
\begin{eqnarray}
|P_N \rangle 
=
|p_{0N} \rangle 
\otimes 
\prod_{n=1}^{\infty}
| \psi_n,\overline{\psi}_n \rangle_N, 
\label{state PN}
\end{eqnarray}
where 
$
|\psi_n,\overline{\psi}_n \rangle_N 
$
denote the normalized  eigenstates of 
$(\hat{\psi}_{n\mu},\hat{\psi}^{\dagger}_{n\mu})$ 
with  eigenvalues
$(\psi_{n\mu},\bar{\psi}_{n\mu})$, 
and the normalization conditions are taken as  
\begin{equation}
{}_N\!
\langle \psi'_{n},\bar{\psi}'_{n}|
\psi_{n},\bar{\psi}_{n}\rangle_N
=\prod_{\mu=0}^{p}
 \delta\left(\mathrm{Re}\psi'_{n\mu}-\mathrm{Re}\psi_{n\mu}\right)
 \delta\left(\mathrm{Im}\psi'_{n\mu}-\mathrm{Im}\psi_{n\mu}\right)~.
\end{equation}
These conditions ensure the orthonormality of $|P_N \rangle$. 
Oscillator realizations of  
$|\psi_{n},\bar{\psi}_{n}\rangle_N$ 
are given in appendix \ref{sec:eigen}. 
They turn out to be as follows : 
\begin{eqnarray}
\lefteqn{\left|\psi_{n},\bar{\psi}_{n}\right\rangle_N 
=
  \left(\frac{\alpha'}{\pi n}\right)^{\frac{p+1}{2}}
   \frac{1}{\sqrt{-\det g_{\mu\nu}}}}
\nonumber\\
&& \times \exp \left[-\frac{1}{n} \alpha^{\mu}_{-n}g_{\mu\nu}
   \tilde{\alpha}^{\nu}_{-n}
   +\frac{\sqrt{\alpha'}}{n}
    \left(\psi_{n\mu}\alpha^{\mu}_{-n}
          +\bar{\psi}_{n\mu}\tilde{\alpha}^{\mu}_{-n}\right)
    -\frac{\alpha'}{2n}\bar{\psi}_{n\mu}g^{\mu\nu} \psi_{n\nu}
    \right]
    |0 \rangle~.
 \label{eq:psi-psi}
\end{eqnarray}

Similarly to $|X_N \rangle$,  
momentum eigenstates $|P_N \rangle$ 
provide another complete basis of $\mathcal{H}^{X}_{cN}$.
The completeness relation is
\begin{eqnarray}
1&=& \int \left[ dP_{N} \right]\; |P_{N}\rangle\langle P_{N}|
\nonumber\\
 &=& \int d^{p+1}p_{0} \prod_{n=1}^{\infty}
          \left( \prod_{\mu=0}^{p}
                 \frac{d\bar{\psi}_{n\mu} d\psi_{n\mu}}{2i}\right)
     \left|\psi_{n}, \bar{\psi}_{n}\right\rangle_N 
          |p_{0N}\rangle
     \langle p_{0N}| 
     {}_N\!\left\langle \psi_{n},\bar{\psi}_{n}\right|~.
\label{partition unity by PN}
\end{eqnarray}

Next we describe eigenstates 
of $\hat{P}^{(B)}(\sigma)$. 
They are determined by similar conditions 
to those imposed on $\hat{P}(\sigma)$ and are 
factorized into products of the Neumann and 
the Dirichlet sectors :
\begin{eqnarray}
|P^{(B)}\rangle 
=
|P^{(B)}_N \rangle 
\otimes 
|P_D \rangle. 
\end{eqnarray}
We also concentrate on the Neumann directions. 
Let us parametrize the eigenvalues $P^{(B)}_{\mu}(\sigma)$ 
by complex variables $\varrho_{n \mu}$ ($n \in \mathbb{Z}_{\geq 1}$) 
and the zero-modes $p_{0 \mu}$ as 
\begin{eqnarray}
P^{(B)}_{\mu}(\sigma)
=
\frac{1}{2\pi}
\left[
     p_{0\mu}
     +
     \frac{1}{\sqrt{2}}
        \sum_{n=1}^{\infty}
          \left(
            \varrho_{n\mu}e^{-in\sigma}
            +
            \overline{\varrho}_{n\mu}e^{in\sigma}
          \right)
\right].
\label{eq:PBN-eigenvalue}
\end{eqnarray}
The eigenstates are given by the following infinite products :
\begin{eqnarray}
|P^{(B)}_N \rangle 
=
|p_{0N}\rangle 
\otimes 
\prod_{n=1}^{\infty}
|\varrho_{n},\bar{\varrho}_{n}\rangle_N, 
\label{state PBN}
\end{eqnarray}
where 
$|\varrho_{n},\bar{\varrho}_{n}\rangle_N$ 
are the normalized eigenstates of 
$(\hat{\varrho}_{n\mu}, \hat{\varrho}^{\dagger}_{n\mu})$
with eigenvalues 
$(\varrho_{n\mu}, \bar{\varrho}_{n\mu})$. 
We normalize them by 
\begin{equation}
{}_N\!\langle \varrho'_{n},\bar{\varrho}'_{n}|
        \varrho_{n},\bar{\varrho}_{n}\rangle_N
 = \prod_{\mu=0}^{p}
   \delta(\mathrm{Re}\varrho'_{n\mu}-\mathrm{Re}\varrho_{n\mu})
   \delta(\mathrm{Im}\varrho'_{n\mu}-\mathrm{Im}\varrho_{n\mu})~.
\end{equation}
The orthonormality of $|P_N^{(B)}\rangle$ follows from these 
conditions. 
Oscillator realizations of these states are given in the appendix. 
They turn out to be as follows :
\begin{eqnarray}
\lefteqn{
\left|\varrho_{n}, \bar{\varrho}_{n}\right\rangle_N
  =
\left(\frac{\alpha'}{n\pi} \right)^{\frac{p+1}{2}}
   \sqrt{-\det
   \left(\frac{1}{E^{T}}g\frac{1}{E}\right)^{\mu\nu}} }
\nonumber\\
&& \times \exp \left[
  -\frac{1}{n}\alpha^{\mu}_{-n}N_{\mu\nu}\tilde{\alpha}^{\nu}_{-n}
  +\frac{\sqrt{\alpha'}}{n}\left\{ 
    \varrho_{n\mu} {\left(\frac{1}{E^{T}}g\right)^{\mu}}_{\nu}
    \alpha^{\nu}_{-n}
    +\bar{\varrho}_{n\mu} {\left(\frac{1}{E}g\right)^{\mu}}_{\nu}
      \tilde{\alpha}^{\nu}_{-n}\right\}
\right. \nonumber\\
&& \hspace{3.5em} \left.
 -\frac{\alpha'}{2n}\bar{\varrho}_{n\mu}
    \left(\frac{1}{E^{T}}g\frac{1}{E}\right)^{\mu\nu}
    \varrho_{n\nu}\right] |0\rangle~.
 \label{eq:varrho-varrho}
\end{eqnarray}

Momentum eigenstates $|P^{(B)}_N \rangle $  
provide a complete basis of $\mathcal{H}^{X}_{cN}$ as well.
The completeness relation reads as follows : 
\begin{eqnarray}
1&=& \int \left[dP_N^{(B)}\right]\; \left|P_N^{(B)}\right\rangle
     \left\langle P_N^{(B)} \right| \nonumber\\
 &=& \int d^{p+1}p_{0} \prod_{n=1}^{\infty} \left(
     \prod_{\mu=0}^{p}
     \frac{d\bar{\varrho}_{n\mu} d\varrho_{n\mu}}{2i}\right)
     \left|\varrho_{n},\bar{\varrho}_{n}\right\rangle_N
     |p_{0N}\rangle \langle p_{0N}|
     {}_N\!\left\langle \varrho_{n},\bar{\varrho}_{n}\right|~.
\end{eqnarray}

\subsection{Some Observations}

The previous constructions give all the eigenstates in terms of 
infinite products of the correctly normalized eigenstates of each 
massive modes. For instance 
the eigenstates 
$|X_N \rangle$ and $|X_D \rangle$ in 
(\ref{eq:es-X}) are given by the infinite products 
of 
$|\chi_n,\overline{\chi}_n \rangle_N$ 
and 
$|\chi_n,\overline{\chi}_n \rangle_D$.  
One can find in Eqs.(\ref{eq:chi-chi N}) 
and (\ref{eq:chi-chi D})
that the normalization factors are  respectively 
$
\left(
   \frac{n}{\pi \alpha'}
\right)^{\frac{p+1}{2}}
\sqrt{-\mbox{det}g_{\mu \nu}}
$
and 
$
\left(
   \frac{n}{\pi \alpha'}
\right)^{\frac{d}{2}}
\sqrt{\mbox{det}g_{i j}}
$. 
Infinite products of these constants become  
the normalization factors of 
$|X_N \rangle$ and $|X_D \rangle$. 
Let us denote them by  
$
\mathcal{C}^{(N)}_{X}\equiv 
$
$\prod_{n=1}^{\infty}
\left[ n^{\frac{p+1}{2}}
\left(
   \frac{-\det g_{\mu\nu}}
        {\left(\pi\alpha'\right)^{p+1}}
\right)^{\frac{1}{2}}
\right]
$
and 
$
\mathcal{C}^{(D)}_{X}\equiv 
$
$\prod_{n=1}^{\infty}
\left[ n^{\frac{d}{2}}
\left(
   \frac{\det g_{ij}}
        {\left(\pi\alpha'\right)^{d}}
\right)^{\frac{1}{2}}
\right]
$. 
The eigenstates can be written as follows :
\begin{eqnarray}
&&
|X_{N}\rangle 
\nonumber \\
&&
= \mathcal{C}^{(N)}_{X}
    \prod_{n=1}^{\infty}
    \exp \left[ 
     \frac{1}{n}\alpha^{\mu}_{-n}g_{\mu\nu}\tilde{\alpha}^{\nu}_{-n}
     -\frac{i}{\sqrt{\alpha'}}\left(
            \chi^{\mu}_{n}g_{\mu\nu}\alpha^{\nu}_{-n}
            +\bar{\chi}^{\mu}_{n}g_{\mu\nu}\tilde{\alpha}^{\nu}_{-n}
          \right)
     -\frac{n}{2\alpha'}\bar{\chi}^{\mu}_{n}g_{\mu\nu}\chi^{\nu}_{n}
     \right] 
     |x_{0N}\rangle~, \nonumber\\
&&
|X_{D}\rangle 
\nonumber\\
&&
= 
\mathcal{C}^{(D)}_{X}
\prod_{n=1}^{\infty}
\exp 
  \left[ 
       \frac{1}{n}\alpha^{i}_{-n}
        g_{ij}\tilde{\alpha}^{j}_{-n}
        -\frac{i}{\sqrt{\alpha'}}
          \left(
              \chi^{i}_{n}g_{ij}
                        \alpha^{j}_{-n}
                +
               \bar{\chi}^{i}_{n}g_{ij}
                     \tilde{\alpha}^{j}_{-n}
          \right)
        -\frac{n}{2\alpha'}
            \bar{\chi}^{i}_{n}g_{ij}\chi^{j}_{n}
     \right] 
|x_{0D}\rangle~,
\nonumber \\  \label{coordinates 2}
\end{eqnarray}
As is performed in \cite{CLNY3}, one may evaluate 
$\mathcal{C}^{(N,D)}_{X}$ by 
using the zeta-function regularization.
This regularization scheme leads the following 
identities 
\footnote{
$\zeta(s)=\sum_{n=1}^{\infty}\frac{1}{n^{s}}$;
$\sum_{n=1}^{\infty}1=\zeta(0)=-\frac{1}{2}$,
$\sum_{n=1}^{\infty}\left(-\ln n\right)=\zeta^{\prime}(0)
 =-\frac{1}{2}\ln(2\pi)$ etc.}: 
\begin{equation}
\prod_{n=1}^{\infty} a 
 =a^{\zeta (0)}  =a^{-\frac{1}{2}}~,
\quad
\prod_{n=1}^{\infty} n^{\alpha}
  =\exp \left(-\alpha\zeta^{\prime}(0)\right)
  =\left(2\pi\right)^{\frac{\alpha}{2}}~,
\end{equation}
and thus we obtain 
\begin{equation}
\mathcal{C}_{X}^{(N)}
 =\left( \frac{\left(2\pi^{2}\alpha'\right)^{p+1}}
              {-\det g_{\mu\nu}} \right)^{\frac{1}{4}}~,
\quad
\mathcal{C}_{X}^{(D)}=\left(\frac{(2\pi^{2}\alpha')^{d}}{\det g_{ij}}
                      \right)^{\frac{1}{4}}~.
   \label{CX}
\end{equation}

It is important to recall that the boundary state 
$\bl B_{D} \brr$ is an eigenstate of $\hat{X}^{i}(\sigma)$
with eigenvalues ${X}^{i}(\sigma)=x^{i}_{0}$.
We have
$\bl B_{D} \brr \propto \bl X_{D}=x_{0D}\brr$. 
If one uses the zeta-function regularization and adopts  
the above $\mathcal{C}_X^{(D)}$ as the normalization 
constant, these two states become precisely identical :
\begin{equation}
\bl B_{D}\brr =\bl X_{D}=x_{0D} \brr~.
\label{Bd as Xd}
\end{equation}

Let us also recall 
that in quantum mechanics of a single particle,
coordinate eigenstates $|x\rangle$ 
are described by
$|x\rangle =e^{-ix\hat{p}}|x=0\rangle$,
where $\hat{p}$ is the momentum operator.
One may find an analogous realization for the 
string coordinate eigenstates. 
Taking account of Eq.(\ref{Bd as Xd}) 
one can infer the following one :
\begin{equation}
|X_{D}\rangle = :\exp\left(-i\int^{2\pi}_{0} d\sigma
     X^{i}(\sigma)\hat{P}_{i}(\sigma) \right): 
    \bl B_{D};x_{0D}=0\brr~,
  \label{eq:x-shift}
\end{equation}
where 
$\bl B_{D};x_{0D}=0\brr$ 
is the state (\ref{Bd as Xd})
with $x_{0}^{i}=0$.
This turns out to be the case. 
Using the previous parametrization of the eigenvalues 
one can find 
\begin{equation}
\int^{2\pi}_{0} d\sigma
 X^{i}(\sigma)\hat{P}_{i}(\sigma)
  =x^{i}_{0}\hat{p}^{i}_{0}
   +\frac{g_{ij}}{2\sqrt{\alpha'}}
    \sum_{n=1}^{\infty}\left\{ \left(
      \chi_{n}^{i}\alpha^{j}_{-n}
         +\bar{\chi}^{i}_{n}\tilde{\alpha}^{j}_{-n} \right)
      +\left(\bar{\chi}_{n}^{i}\alpha^{j}_{n}
           +\chi_{n}^{i} \tilde{\alpha}^{j}_{n}\right) \right\}~.
\end{equation}
It is easy to see that the normal ordered exponential 
in the RHS of Eq.(\ref{eq:x-shift}) reproduces 
$|X_D \rangle$. 
The formula (\ref{eq:ee}) may be useful.

The above discussion is also applicable to the string 
momentum eigenstates. We can write these states as follows :
\begin{eqnarray}
&&|P_{N}\rangle 
\nonumber\\
&&
=\mathcal{C}_{P_N} \prod_{n=1}^{\infty}
\exp \left[-\frac{1}{n} \alpha^{\mu}_{-n}g_{\mu\nu}
   \tilde{\alpha}^{\nu}_{-n}
   +\frac{\sqrt{\alpha'}}{n}
    \left(\psi_{n\mu}\alpha^{\mu}_{-n}
          +\bar{\psi}_{n\mu}\tilde{\alpha}^{\mu}_{-n}\right)
    -\frac{\alpha'}{2n}\bar{\psi}_{n\mu}g^{\mu\nu} \psi_{n\nu}
    \right] |p_{0N}\rangle~,~~
\nonumber \\
&&
\left| P_N^{(B)}\right\rangle 
\nonumber\\
&&
=\mathcal{C}_{P_N^{(B)}}
 \prod_{n=1}^{\infty}
 \exp \left[
  -\frac{1}{n}\alpha^{\mu}_{-n}N_{\mu\nu}\tilde{\alpha}^{\nu}_{-n}
  +\frac{\sqrt{\alpha'}}{n}\left\{ 
    \varrho_{n\mu} {\left(\frac{1}{E^{T}}g\right)^{\mu}}_{\nu}
    \alpha^{\nu}_{-n}
    +\bar{\varrho}_{n\mu} {\left(\frac{1}{E}g\right)^{\mu}}_{\nu}
      \tilde{\alpha}^{\nu}_{-n}\right\}
\right. \nonumber\\
&& \hspace{8em} \left.
 -\frac{\alpha'}{2n}\bar{\varrho}_{n\mu}
    \left(\frac{1}{E^{T}}g\frac{1}{E}\right)^{\mu\nu}
    \varrho_{n\nu}\right]
 |p_{0N}\rangle~,
 \label{eq:PN-state}
\end{eqnarray}
where the normalization factors 
$\mathcal{C}_{P_N}$
and 
$
\mathcal{C}_{P_N^{(B)}} 
$
are originally given by the infinite products 
as follow from 
Eqs.(\ref{eq:psi-psi}) and (\ref{eq:varrho-varrho}). 
They are regularized to  
\begin{eqnarray}
\mathcal{C}_{P_N}
=
\left(
   \frac{-\det g_{\mu\nu}}
        {(2\alpha')^{p+1}}
\right)^{\frac{1}{4}}~,~~~~
\mathcal{C}_{P^{(B)}_N}
=
\left(
   \frac{\det^{2}E_{\mu\nu}}
        {(2\alpha')^{p+1}(-\det g_{\mu\nu})}
\right)^{\frac{1}{4}}~.
\label{CPB}
\end{eqnarray}

Boundary state $\bl B_{N} \brr$ is an eigenstate of  
$\hat{P}^{(B)}_{\mu}(\sigma)$ with vanishing eigenvalue. 
We have $\bl B_{N} \brr \propto |P^{(B)}=0\rangle$. 
If we use the above $\mathcal{C}_{P^{(B)}_N}$ 
as the normalization constant, these two states become 
identical including their normalizations :
\begin{equation}
\bl B_{N} \brr = \left|P_N^{(B)}=0\right\rangle~.
\end{equation}

Momentum eigenstates can be realized 
in the quantum mechanics by 
$
|p \rangle
=
e^{ip \hat{x}}|p=0\rangle
$, 
where $\hat{x}$ is the position operator. 
We can find an analogous realization of  
the string momentum eigenstates.  
It turns out to be as follows : 
\begin{eqnarray}
\left|P_N^{(B)}\right\rangle
&=& \prod_{n=1}^{\infty}\exp \left[
   -\frac{\alpha'}{4n}\bar{\varrho}_{n\mu}
     \left\{g^{-1}\left(g-\frac{1}{2}\left(N-N^{T}\right)\right)g^{-1}
     \right\}^{\mu\nu}\varrho_{n\nu}\right]\nonumber\\
&& \times
    :\exp\left(i\int^{2\pi}_{0}d\sigma P^{(B)}_{\mu}(\sigma)
               \hat{X}^{\mu}(\sigma)\right):
    \bl B_{N}\brr~.
    \label{eq:PNB-shift}
\end{eqnarray}

For the later convenience we provide a similar observation 
for the eigenstates $\left| P_{N} \right\rangle$ as well.
Let $\left| B_{N} \right\rangle_{B=0}$ be the state
which is obtained  from the Neumann boundary state
 $\left| B_{N} \right\rangle$ by putting $B_{\mu\nu}=0$:
\begin{equation}
 \left| B_{N} \right\rangle_{B=0}
   = \left( \frac{- \det g_{\mu\nu}}{(2\alpha')^{p+1}}\right)^{\frac{1}{4}}
     \prod_{n=1}^{\infty}
     \exp \left(-\frac{1}{n} \alpha^{\mu}_{-n} g_{\mu\nu}
                             \tilde{\alpha}^{\nu}_{-n}\right)
     \left| \mathbf{0} \right\rangle~.
\label{eq:BN with zero B}
\end{equation}
This is the Neumann boundary state 
in a vanishing $B$ field background.
One can readily find that this is an eigenstate of
$\hat{P}_{\mu}(\sigma)$ with zero eigenvalue:
$\hat{P}_{\mu}(\sigma) \left|B_{N}\right\rangle_{B=0} = 0$
and hence 
$\left|B_{N}\right\rangle_{B=0}\propto \left|P_{N}=0\right\rangle$.
Comparing definitions (\ref{eq:PN-state}) and
(\ref{eq:BN with zero B}), we find that these states
coincide with each other including their normalizations,
if we adopt $C_{P_{N}}$ in Eq.(\ref{CPB})
as the normalization of the momentum eigenstates.

Concerning the state $\left| P_{N} \right\rangle$,
a similar formula to Eq.(\ref{eq:PNB-shift}) becomes
\begin{equation}
\left| P_{N} \right\rangle
 = \prod_{n=1}^{\infty}
      \exp \left[ -\frac{\alpha'}{4n}
                   \bar{\psi}_{n\mu} g^{\mu\nu} \psi_{n\nu} \right]
  \, : \exp \left(i \int^{2\pi}_{0} d\sigma
                         P_{\mu}(\sigma) \hat{X}^{\mu}(\sigma)
                    \right) :
     \left| B_{N} \right\rangle_{B=0}~.
     \label{eq:PN-shift}
\end{equation}
This has the same form as Eq.(\ref{eq:PNB-shift})
with putting $B_{\mu\nu}=0$.

\section{Open Wilson Lines in Closed-String Theory (I)}
\label{sec:open Wilson lines (I)}

At the zero-slope limit, which is introduced 
in Eq.(\ref{field theory parameters of closed-string}) 
and its below in order to capture the UV behavior 
of the world-volume theory, 
the generating function of one-loop amplitudes 
of open-string tachyons is shown to exhibit  
the factorized form 
(\ref{closed-string FTL by straight open Wilson lines}). 
It is expressed as a sum of products of 
two Wilson lines 
(strictly speaking, their analogues) 
along the same straight lines $x_0(\sigma)$, 
multiplying the propagators of closed-string tachyon. 
This shows that closed-string tachyon $T$ has a tadpole 
interaction with the open Wilson line.  
It can be written in a form,
$
\int dx T(x)
{\cal P}_{\star}
\left[ 
e^{\int_0^{2\pi}
     d\sigma \phi(x_0(\sigma))}
\right]
$.   
Its origin in string theory can be found 
in Eq.(\ref{def of JT}). 
The interaction simply comes from the closed-string tachyon 
modes of boundary states, 
$
\left \langle K 
\left | 
         B_N ; (\sigma_{1},k^{(1)}),
\right. \right.
$
$
\cdots,
$
$
\left. 
     (\sigma_{M},k^{(M)}) 
\right \rangle     
$.

Actually these boundary states have 
all the components of perturbative closed-string states. 
At the level of string amplitudes 
all of them  
propagate between the boundary states and 
contribute to the amplitudes. 
It is shown in \cite{Okawa-Ooguri} that 
the straight open Wilson line can couple 
with the on-shell graviton and,  
as will be seen in the later section, 
it can be generalized to the off-shell. 
This indicates that 
all the perturbative closed-string states 
have tadpole interactions 
with the open Wilson line.
Therefore we may 
unfasten the zero-slope limit 
(\ref{closed-string FTL by straight open Wilson lines})
so that propagations of all these states 
are made manifest. 
In this section we pursue such a possibility. 
We also restrict to the case of $25$-brane in the critical 
dimensions.

Let us provide a general perspective on this issue 
before we start calculations. 
First of all, 
it can be expected \cite{Dhar-Kitazawa} that 
closed-string propagations 
including gravitons  
fluctuate the straight line  
appearing in Eq.(\ref{closed-string FTL by Moyal products}) 
and transform it into curved ones.  
In other words we can expect that 
there are correlations between 
their deviations from the straight line 
and 
the propagations of closed-string states. 
These curves will appear as the corresponding Wilson lines, 
and 
we may factorize the generating function 
into a sum of these products  
at the zero-slope limit. 
Analogously to the standard factorization 
of closed-string amplitudes 
the sum must be taken originally 
over the perturbative closed-string states.
The above correspondence between curves and states 
will enable us to translate the sum  
as an integral over the space of curves. 
This integral may be suppressed by a suitable weight as 
the straight line is suppressed in 
Eq.(\ref{closed-string FTL by straight open Wilson lines}) 
by the closed-string tachyon propagator.
It is amazing that one can interpret the tachyon propagator 
as a propagator of the straight open Wilson line 
by notifying 
$
p_{\mu}g^{\mu \nu}p_{\nu}
\approx 
\alpha'^{-2}
\dot{x}^{\mu}_0G_{\mu \nu}\dot{x}^{\nu}_0
$.  
The factorization 
(\ref{closed-string FTL by straight open Wilson lines})
may be obtained from the aforementioned integral by 
integrating out the fluctuations.   
We may say that the straight open Wilson line is 
the average.

\subsection{Factorization by closed-string momentum eigenstates}

In order to justify the above perspective 
let us first factorize the string amplitudes 
by an insertion of a partition of unity. 
Use of that constructed from 
closed-string momentum eigenstates turns out to be relevant. 
It is given in Eq.(\ref{partition unity by PN}) as 
$
1=
\int [dP_N]~
| P_N \rangle 
\langle P_N |,
$ 
where the eigenvalue $P_{\mu}(\sigma)$ is 
parametrized by 
$
P_{\mu}(\sigma)=
\frac{1}{2\pi}
\left[
p_{0 \mu}+
\frac{1}{\sqrt{2}}
\sum_{n=1}^{\infty}
\left( 
\psi_{n \mu}e^{-in\sigma}
+
\overline{\psi}_{n \mu}
e^{in \sigma}
\right)
\right]
$.

We factorize the amplitudes 
(\ref{def of tachyon amplitude by boundary state}) 
by an insertion of 
Eq.(\ref{partition unity by PN}) :
\begin{eqnarray}
&&
\left \langle B_N ; 
(\sigma_{M+1},k^{(M+1)}),\cdots, (\sigma_{M+N},k^{(M+N)})
\right| 
q_c^{\frac{1}{2}(L_0+\tilde{L}_0-2)} 
\left| 
B_N;
(\sigma_1,k^{(1)}),\cdots,(\sigma_{M},k^{(M)})
\right \rangle 
\nonumber \\
&&
=
\int [dP_N]
\left \langle 
   B_N ; 
     (\sigma_{M+1},k^{(M+1)}),
          \cdots, (\sigma_{M+N},k^{(M+N)})
\right|
   q_c^{\frac{1}{4}(L_0+\tilde{L}_0-2)} 
\Bl
   P_N 
\Brr
\nonumber \\
&&
~~~~~~~~~~~~~~~~~~~
\times  
\Bll 
   P_N 
\Br 
   q_c^{\frac{1}{4}(L_0+\tilde{L}_0-2)} 
\left| 
    B_N ;
       (\sigma_1,k^{(1)}),
             \cdots,(\sigma_{M},k^{(M)})
\right \rangle. 
\label{def of momentum factorization of tachyon amplitudes}
\end{eqnarray}
Two factors appearing in the above can be calculated 
by using the oscillator representations. 
These are given 
in Eq.(\ref{M open-string tachyon boundary state}) 
(and Eq.(\ref{M open-string tachyon dual boundary state})) 
for the boundary states,
and Eq.(\ref{eq:PN-state}) 
(and its hermitian conjugate)  
for the momentum eigenstates. 
Technical difficulties we may meet in the calculations 
are evaluations of matrix elements of the following type :
\begin{eqnarray}
&&
\langle 0 |
\exp 
\left[
  -
  \frac{1}{n}q_c^{n/2}
  \alpha_{n}^{\mu}g_{\mu \nu}\tilde{\alpha}_{n}^{\nu}
  + 
  \frac{\sqrt{\alpha'}}{n}
  q_c^{n/4}
  \left(
  \overline{\psi}_{n \mu}\alpha^{\mu}_{n}
  +
  \psi_{n \mu}\tilde{\alpha}^{\mu}_{n}
  \right)
\right]
\nonumber \\
&&
~~~
\times 
\exp 
\left[
   -
   \frac{1}{n}
   \alpha_{-n}^{\mu}N_{\mu \nu}\tilde{\alpha}_{-n}^{\nu}
   +
   \frac{\sqrt{2\alpha'}}{n}
   \sum_{r=1}^M
   k_{\mu}^{(r)}
   \left \{
     \left(\frac{1}{E^T}g
        \right)^{\mu}_{\nu}
     \alpha_{-n}^{\nu}
     e^{i n \sigma_r}
     +
     \left(\frac{1}{E}g
         \right)^{\mu}_{\nu}
     \tilde{\alpha}_{-n}^{\nu}
     e^{-i n \sigma_r}
   \right \}
\right] 
| 0 \rangle.
\nonumber \\
\label{difficult matrix element}
\end{eqnarray}
These can be computed by using Eq.(\ref{eq:vevee}) 
in Appendix \ref{sec:formulae}. 
To describe the result it turns out useful to rescale complex 
variables $\psi_n$ to $2q_c^{n/4}\psi_n$. 
After these rescalings 
the factorization can be written as follows :
\begin{eqnarray}
&&
\int [dP_N]
\left \langle 
   B_N ; 
     (\sigma_{M+1},k^{(M+1)}),
          \cdots, (\sigma_{M+N},k^{(M+N)})
\right|
   q_c^{\frac{1}{4}(L_0+\tilde{L}_0-2)} 
\Bl
   P_N 
\Brr
\nonumber \\
&&
~~~~~~~~~~~~~~~~~~~
\times  
\Bll 
   P_N 
\Br 
   q_c^{\frac{1}{4}(L_0+\tilde{L}_0-2)} 
\left| 
    B_N ;
       (\sigma_1,k^{(1)}),
             \cdots,(\sigma_{M},k^{(M)})
\right \rangle. 
\nonumber \\
&&
=
\left( 
\frac{-\mbox{det}G_{\mu \nu}}{(2\alpha')^{26}}
\right)^{\frac{1}{2}}
\int dp_0 
\delta^{26}
\left( p_0+\sum_{r=M+1}^{M+N}k^{(r)}
\right) 
\delta^{26}
\left( p_0-\sum_{r=1}^{M}k^{(r)}
\right) 
\times 
q_c^{\frac{\alpha'}{4}
     p_{0 \mu}g^{\mu \nu}p_{0 \nu}-1}
\nonumber \\
&&
~~
\times
\prod_{1\leq r<s \leq M}
e^{\frac{i}{2}k_{\mu}^{(r)}\theta^{\mu \nu}k_{\nu}^{(s)}
\epsilon(\sigma_r-\sigma_s)}
\times 
\prod_{M+1 \leq r<s \leq M+N}
e^{-\frac{i}{2}k_{\mu}^{(r)}\theta^{\mu \nu}k_{\nu}^{(s)}
\epsilon(\sigma_r-\sigma_s)}
\nonumber \\
&&
~~
\times 
\prod_{1 \leq r < s \leq M}
e^{-\frac{i}{2\pi}k_{\mu}^{(r)}
\theta^{\mu \nu}k_{\nu}^{(s)}(\sigma_r-\sigma_s)}
\times 
\prod_{M+1 \leq r < s \leq M+N} 
e^{\frac{i}{2\pi}k_{\mu}^{(r)}
\theta^{\mu \nu}k_{\nu}^{(s)}(\sigma_r-\sigma_s)}
\nonumber \\
&& 
~~
\times
\prod_{1 \leq r < s \leq M} 
|e^{i\sigma_r}-e^{i\sigma_s}|
^{2\alpha'k_{\mu}^{(r)}G^{\mu \nu}k_{\nu}^{(s)}}
\times 
\prod_{M+1 \leq r < s \leq M+N} 
|e^{i\sigma_r}-e^{i\sigma_s}|
^{2\alpha'k_{\mu}^{(r)}G^{\mu \nu}k_{\nu}^{(s)}}
\nonumber \\
&&
~~
\times 
\prod_{n=1}^{\infty}
\exp 
\left[
    -\frac{2\alpha'q_c^{\frac{n}{2}}}{n}
    \left(
       \sum_{r=1}^M e^{-in\sigma_r}k_{\mu}^{(r)}
    \right)     
    \left( 
         \frac{1}{E}
          g
         \frac{1}{E-q_c^{\frac{n}{2}}E^T} 
    \right)^{\mu \nu}
    \left( 
       \sum_{r=1}^M e^{in\sigma_r}k_{\nu}^{(r)}
    \right)
\right.
\nonumber \\
&&
~~~~~~~~~~~~~~~~~~~~~~~
\left. 
    -\frac{2\alpha'q_c^{\frac{n}{2}}}{n}
    \left(
       \sum_{r=M+1}^{M+N} e^{in\sigma_r}k_{\mu}^{(r)}
    \right)     
    \left( 
         \frac{1}{E}
           g
         \frac{1}{E-q_c^{\frac{n}{2}}E^T} 
    \right)^{\mu \nu}
    \left( 
       \sum_{r=M+1}^{M+N} e^{-in\sigma_r}k_{\nu}^{(r)}
    \right)
\right]
\nonumber \\
&&
~~
\times 
\prod_{n=1}^{\infty}
\int 
\frac{d\overline{\psi}_n d\psi_n}{(2i)^{26}}
\left(
      \frac{\alpha'}{4n\pi q_c^{\frac{n}{2}}}
\right)^{26}
     \frac{-\mbox{det} g_{\mu \nu}}
     {\mbox{det}^2(g-q_c^{\frac{n}{2}}N)_{\mu \nu}}
\nonumber \\
&&
~~~~~~
\times 
\exp 
\left[
   -
   \frac{\alpha'}{4n}
   \overline{\psi}_{n \mu}
   \left(
      \frac{1}{q_c^{\frac{n}{2}}g}
      +
      \frac{1}{g-q_c^{\frac{n}{2}}N^T}N^T\frac{1}{g}
      +
      \frac{1}{g}N\frac{1}{g-q_c^{\frac{n}{2}}N}
   \right)^{\mu \nu}
   \psi_{n \nu}
\right]
\nonumber \\
&&
~~~~~~
\times
\exp 
\left[ 
   \frac{\alpha'}{n\sqrt{2}}
   \psi_{n \mu}
   \left( 
       \frac{1}{E^T-q_c^{\frac{n}{2}}E}
   \right)^{\mu \nu}
   \left( 
       \sum_{r=1}^M e^{-in\sigma_r}k_{\nu}^{(r)}
   \right)
\right. 
\nonumber \\
&&
~~~~~~~~~~~~~~~~~~~~~~~~~~~~~~~~~
\left.
   +
   \frac{\alpha'}{n \sqrt{2}}
   \overline{\psi}_{n \mu}
   \left( 
       \frac{1}{E-q_c^{\frac{n}{2}}E^T}
   \right)^{\mu \nu}
   \left( 
       \sum_{r=1}^M e^{in\sigma_r}k_{\nu}^{(r)}
   \right)
\right]
\nonumber \\    
&&
~~~~~~
\times
\exp 
\left[ 
   -\frac{\alpha'}{n\sqrt{2}}
   \overline{\psi}_{n \mu}
   \left( 
       \frac{1}{E^T-q_c^{\frac{n}{2}}E}
   \right)^{\mu \nu}
   \left( 
       \sum_{r=M+1}^{M+N} e^{in\sigma_r}k_{\nu}^{(r)}
   \right)
\right.
\nonumber \\
&&
~~~~~~~~~~~~~~~~~~~~~~~~~~~~~~~~~
\left.
   -
   \frac{\alpha'}{n\sqrt{2}}
   \psi_{n \mu}
   \left( 
       \frac{1}{E-q_c^{\frac{n}{2}}E^T}
   \right)^{\mu \nu}
   \left( 
       \sum_{r=M+1}^{M+N} e^{-in\sigma_r}k_{\nu}^{(r)}
   \right)
\right]. 
\label{momentum factorization of tachyon amplitude}
\end{eqnarray}

The above factorization may be compared with 
the previous expression of the amplitudes. 
Terms in the first four lines can be found exactly 
in Eq.(\ref{pre tachyon amplitude by boundary state}). 
The other terms describe a factorization of $F$ 
in the same equation.
If we integrate out 
$\psi_{n \mu}$ and $\overline{\psi}_{n \mu}$ 
they provide Eq.(\ref{result on F}).  
These variables describe fluctuations 
of $P_{\mu}(\sigma)$. 
Corresponding degrees of freedom of 
closed-string are given by 
the massive modes. 
The factorization of $F$ in terms of those complex 
variables are plausible 
since $F$ is the sum of contributions of 
these massive modes.

\subsection{Factorization by open Wilson lines}

We examine the zero-slope limit of the factorization 
(\ref{momentum factorization of tachyon amplitude}). 
The limit we discuss is same  
as that investigated previously to capture the UV behavior 
of the world-volume theory. 
It is taken by fixing parameters 
$s^{(c)}=\alpha'|\tau^{(c)}|$ and $\sigma_r$ besides 
the open-string tensors. We also keep $\psi_n$ and 
$\overline{\psi}_n$ intact. 
They do not scale under the limit. 
It should be noticed that the complex variables 
used originally to parametrize $P_{\mu}(\sigma)$ 
do scale under the limit. 
This is because $\psi_n$ and $\overline{\psi}_n$ 
in (\ref{momentum factorization of tachyon amplitude}) 
are the rescaled ones introduced  
by multiplying the original variables by 
$2q_c^{n/4}$.

We first focus on the integral over $\psi_n$ 
and $\overline{\psi}_n$ in 
Eq.(\ref{momentum factorization of tachyon amplitude}). 
Let us start by considering the first exponential
in the integral.
Scaling part in the exponent is expressed  
by means of the closed-string tensors and 
$q_c=e^{2\pi i\tau^{(c)}}$. 
It has the form :  
\begin{eqnarray}
\alpha'
\left(\frac{1}{q_c^{\frac{n}{2}}g}
+
\frac{1}{g-q_c^{\frac{n}{2}}N^T}N^T\frac{1}{g}
+
\frac{1}{g}N\frac{1}{g-q_c^{\frac{n}{2}}N}
\right). 
\end{eqnarray}
Dominant contribution at the limit clearly 
comes from the first term :
\begin{eqnarray}
\alpha'
\left(\frac{1}{q_c^{\frac{n}{2}}g}
+
\frac{1}{g-q_c^{\frac{n}{2}}N^T}N^T\frac{1}{g}
+
\frac{1}{g}N\frac{1}{g-q_c^{\frac{n}{2}}N}
\right)
\approx
q_c^{-\frac{n}{2}}\alpha'g^{-1}
\approx
-\frac{\theta G \theta}{4\pi^2q_c^{\frac{n}{2}}\alpha'},  
\end{eqnarray}
where we use equalities (\ref{equalities2 and 3}).
We turn to the determinant factors,
which is also described by the closed-string tensors. 
Their contributions can be read as follows :
\begin{eqnarray} 
\left(
      \frac{\alpha'}{4n\pi q_c^{\frac{n}{2}}}
\right)^{26}
     \frac{\mbox{det} g}
     {\mbox{det}^2(g-q_c^{\frac{n}{2}}N)}
\approx
\left( 
  \frac{1}{4n\pi}
\right)^{26}
\mbox{det}
\left[
q_c^{-\frac{n}{2}}\alpha'
g^{-1}
\right]
\approx
\left( 
  \frac{1}{2\pi}
\right)^{26}
\mbox{det}
\left[- 
\frac{\theta G \theta}
     {8\pi^2nq_c^{\frac{n}{2}}\alpha'}
\right].
\end{eqnarray}
The other two exponentials in the integral 
have similar forms. It is enough to know 
the behavior of 
$\alpha'(E^T-q_c^{\frac{n}{2}}E)^{-1}$ 
in their exponents. 
It can be read by using Eq.(\ref{def of G and theta}) 
as follows :
\begin{eqnarray}
\frac{\alpha'}{E^T-q_c^{\frac{n}{2}}E}
\approx 
\frac{\alpha'}{E^T}
\approx 
-\frac{\theta}{2\pi}.
\end{eqnarray}
As we studied previously,
contributions of the first four lines in the factorization 
(\ref{momentum factorization of tachyon amplitude})
are to give rise to 
the straight open Wilson lines.
The last pieces we need to estimate are 
the exponentials whose exponents are bilinear 
of $e^{in\sigma_r}k^{(r)}$ with the weights, 
$
\alpha'
q_c^{\frac{n}{2}}
E^{-1}g
(E-q_c^{\frac{n}{2}}E^T)^{-1}
$. 
It is enough to know the behavior of these weights. 
Again using Eqs. 
(\ref{def of G and theta}) and  (\ref{equalities2 and 3}) 
it can be read as 
$ 
\alpha'
q_c^{\frac{n}{2}}
E^{-1}g
(E-q_c^{\frac{n}{2}}E^T)^{-1}
$
$
\approx 
-\alpha'q_c^{\frac{n}{2}}
G^{-1}
\approx 
0
$.
Thus we can neglect these pieces in the zero-slope limit.

Collecting all these estimations we can obtain the 
zero-slope limit of the factorization.
It turns out to be as follows :
\begin{eqnarray}
&&
\int [dP_N]
\left \langle 
   B_N ; 
     (\sigma_{M+1},k^{(M+1)}),
          \cdots, (\sigma_{M+N},k^{(M+N)})
\right|
   q_c^{\frac{1}{4}(L_0+\tilde{L}_0-2)} 
\Bl
   P_N 
\Brr
\nonumber \\
&&
~~~~~~~~~~~~~~~~~~~
\times  
\Bll 
   P_N 
\Br 
   q_c^{\frac{1}{4}(L_0+\tilde{L}_0-2)} 
\left| 
    B_N ;
       (\sigma_1,k^{(1)}),
             \cdots,(\sigma_{M},k^{(M)})
\right \rangle
\nonumber \\
&&
\approx 
\int dp_0 
\exp 
  \left\{
       -\frac{\pi s^{(c)}}{2}
         \left( 
             p_{0 \mu}g^{\mu \nu}p_{0 \nu}
             -
             \frac{4}{\alpha'}
         \right)
  \right\}
\nonumber \\
&&
~~
\times
\prod_{n=1}^{\infty}
\int 
\frac{d\overline{\psi}_nd\psi_n}{(4 \pi i)^{26}}
\left\{-
\mbox{det}
  \left(- 
     \frac{\theta G \theta}
          {8\pi^2nq_c^{\frac{n}{2}}\alpha'}
  \right)  \right\} \;
\exp 
  \left\{
     \frac{\overline{\psi}_{n \mu}
            \left( \theta G \theta \right)^{\mu \nu}
           \psi_{n \nu}}
          {16\pi^2n q_c^{\frac{n}{2}}\alpha'}
  \right\}
\nonumber \\
&&
~~~~~~~~~~
\times 
\left( 
     \frac{-\mbox{det}G_{\mu \nu}}{(2\alpha')^{26}}
\right)^{\frac{1}{4}}
\delta^{26}
     \left(
       p_0- \sum_{r=1}^{M}k^{(r)}
     \right) 
\prod_{1\leq r<s \leq M}
   e^{\frac{i}{2}k_{\mu}^{(r)}\theta^{\mu \nu}k_{\nu}^{(s)}
       \epsilon(\sigma_r-\sigma_s)}
\nonumber \\
&&
~~~~~~~~~~~~~~~~
\times
\prod_{r=1}^{M}
\exp 
\left[
  -\frac{i}{2\pi}k_{\mu}^{(r)}\theta^{\mu \nu}
  \left\{ 
    p_{0 \nu}\sigma_r 
    +
    \frac{1}{\sqrt{2}}
    \sum_{n=1}^{\infty}\frac{i}{n}
       \left( 
          \psi_{n \nu}e^{-in\sigma_r}
           -
          \overline{\psi}_{n \nu}e^{in\sigma_r} 
       \right)
  \right\}
\right]
\nonumber \\
&&
~~~~~~~~~~
\times 
\left( 
     \frac{-\mbox{det}G_{\mu \nu}}{(2\alpha')^{26}}
\right)^{\frac{1}{4}}
\delta^{26}
     \left(
       p_0+ \sum_{r=M+1}^{M+N}k^{(r)}
     \right) 
\prod_{M+1\leq r<s \leq M+N}
   e^{-\frac{i}{2}k_{\mu}^{(r)}\theta^{\mu \nu}k_{\nu}^{(s)}
       \epsilon(\sigma_r-\sigma_s)}
\nonumber \\
&&
~~~~~~~~~~~~~~~~
\times
\prod_{r=M+1}^{M+N}
\exp 
\left[
  -\frac{i}{2\pi}k_{\mu}^{(r)}\theta^{\mu \nu}
  \left\{ 
    p_{0 \nu}\sigma_r 
    +
    \frac{1}{\sqrt{2}}
    \sum_{n=1}^{\infty}\frac{i}{n}
       \left( 
          \psi_{n \nu}e^{-in\sigma_r}
           -
          \overline{\psi}_{n \nu}e^{in\sigma_r}
       \right)
  \right\}
\right].
\nonumber \\
\label{FTL of momentum factorization of tachyon amplitude}
\end{eqnarray}  
In the above  
$q_c=e^{-\frac{2\pi s^{(c)}}{\alpha'}}$
and the closed-string metric is understood as 
$
g^{\mu \nu} 
\approx 
-\frac{(\theta G \theta)^{\mu \nu}}
      {(2\pi \alpha')^2}
$. 
Both of the integrals with respect to
$p_0$ and $\psi_n$ $(\bar{\psi_n})$ 
become Gaussian integrals. 
The Gaussian weight of $p_0$ is 
$
\pi s^{(c)}g^{-1} 
\approx 
-s^{(c)}
\frac{\theta G \theta}
     {4\pi \alpha'^2}
$ 
while the weight of $\psi_n$ is 
$
-
\frac{\theta G \theta}
     {8\pi^2 n q_c^{\frac{n}{2}}\alpha'}
$, 
which is equal to 
$
-s^{(c)}
\frac{\theta G \theta}
     {8\pi \alpha'^2}
\times 
\left(
   \frac{n \pi s^{(c)}}{\alpha'}
\right)^{-1}
e^{\frac{n \pi s^{(c)}}{\alpha'}}
$.
Therefore the fluctuation of $\psi_n$ is suppressed 
relative to that of $p_0$ exponentially by 
$
\sqrt{
\left(
   \frac{n \pi s^{(c)}}{\alpha'}
\right)
e^{-\frac{n \pi s^{(c)}}{\alpha'}}
}
$. 
The mean value of $\psi_n$ $(\overline{\psi}_n)$ is 
$O(q_c^{\frac{n}{2}}\alpha')$. 
Since it is negligible 
the integration of $\psi_n$ and $\overline{\psi}_n$ 
can be accomplished by just replacing it with
the unity\footnote{
The Gaussian integral
gives suitable products of 
$
e^{\frac{2\alpha'q_c^{\frac{n}{2}}}{n}
   (e^{\pm in \sigma_r}k_{\mu}^{(r)})
    G^{\mu \nu}
   (e^{\mp in \sigma_r}k_{\nu}^{(r)})}
$.
It can be neglected at the zero-slope limit. 
Actually speaking,  these integrals cancel 
with the subdominant terms 
which we abandon at the preceding paragraph.}.
After these replacements 
the above factorization leads to 
Eq.(\ref{closed-string FTL of one-loop amplitude}).

Let us describe the zero-slope limit of 
the generating function by using the factorization 
(\ref{FTL of momentum factorization of tachyon amplitude}). 
We start with $J_T$ defined by (\ref{def of JT}). 
The LHS of
Eq.(\ref{FTL of momentum factorization of tachyon amplitude}) 
is the momentum eigenstate factorization of the amplitude 
(\ref{def of tachyon amplitude by boundary state}).
Hence 
(\ref{FTL of momentum factorization of tachyon amplitude}) 
leads to a factorization of $J_T$ 
at the zero-slope limit. 
As regards the delta functions we translate them 
into integrals as given in (\ref{manipulation of delta}). 
By a straightforward calculation we obtain 
the following factorization of $J_T$ at the zero-slope 
limit :
\begin{eqnarray}
&&
J_T
\left( 
       (k^{(1)},\cdots,k^{(M)});
       (k^{(M+1)},\cdots,k^{(M+N)})
\right)
\nonumber \\
&&
\approx 
\int 
dp_0
\prod_{n=1}^{\infty}
\frac{d\overline{\psi}_nd\psi_n}
     {(4\pi i)^{26}}
\nonumber \\
&&
~~~
\times 
\left[
  \frac{2\pi}{\alpha'}
  \int_{0}^{\infty} ds^{(c)}
     \exp 
       \left\{
          -\frac{\pi s^{(c)}}{2}
          \left( 
              p_{0 \mu}g^{\mu \nu}p_{0 \nu} 
               -
              \frac{4}{\alpha'}
          \right)
       \right\}
\right.
\nonumber \\
&&
~~~~~~~~~~~~~~~~~~~
\left.    
\times 
    \prod_{n=1}^{\infty}
    \left\{
       -\mbox{det}
         \left( 
                \frac{-\theta G \theta}
                {8 \pi^2 n q_c^{\frac{n}{2}}\alpha' }
         \right)
       \exp 
          \left\{
                \frac{\overline{\psi}_{n \mu}
                      (\theta G \theta)^{\mu \nu}\psi_{n \nu}}
                 {16 \pi^2 n q_c^{\frac{n}{2}}\alpha'}
          \right\}
      \right\}
\right]
\nonumber \\
&&
~~~
\times 
\left( 
  \frac{-\mbox{det} G_{\mu \nu}}
       {(2\alpha')^{26}}
\right)^{\frac{1}{4}}
\int 
\frac{dx}{(2\pi)^{26}}
e^{ip_{0 \mu}x^{\mu}}
e^{-\frac{i}{2}p_{0\mu}\theta^{\mu\nu}\sum_{r=1}^{M}k_{\nu}^{(r)}}
  \nonumber\\
 && \hspace{10em} \times
\int 
\prod_{r=1}^{M}
d\sigma_r 
\prod_{1 \leq r < s \leq M}
e^{-\frac{i}{2}k_{\mu}^{(r)}\theta^{\mu \nu}k_{\nu}^{(s)}}
\prod_{r=1}^M
e^{ik_{\mu}^{(r)}x^{\mu}(\sigma_r)}
\nonumber \\
&&
~~~
\times 
\left( 
  \frac{-\mbox{det} G_{\mu \nu}}
       {(2\alpha')^{26}}
\right)^{\frac{1}{4}}
\int 
\frac{dx}{(2\pi)^{26}}
e^{-ip_{0 \mu}x^{\mu}}
e^{-\frac{i}{2}p_{0\mu}\theta^{\mu\nu}\sum_{r=M+1}^{M+N}k_{\nu}^{(r)}}
  \nonumber\\
  && \hspace{10em} \times
\int 
\prod_{r=M+1}^{M+N-1}
d\sigma_r 
\prod_{M+1 \leq r < s \leq M+N}
e^{\frac{i}{2}k_{\mu}^{(r)}\theta^{\mu \nu}k_{\nu}^{(s)}}
\prod_{r=M+1}^{M+N}
e^{ik_{\mu}^{(r)}x^{\mu}(\sigma_r)}.
\label{FTL of momentum factorization of JT}
\end{eqnarray}  
Here we introduce a curve 
$x(\sigma)$ ($0 \leq \sigma \leq 2\pi$)
parametrized by $p_0$ and $(\psi_n,\overline{\psi}_n)$ 
as follows :
\begin{eqnarray}
x^{\mu}(\sigma)
\equiv 
x^{\mu}
+
\frac{\theta^{\mu \nu}}{2 \pi}
\left\{ 
p_{0 \nu}\sigma 
+
\frac{1}{\sqrt{2}}
\sum_{n=1}^{\infty}
\frac{i}{n}
\left( 
     \psi_{n \nu}e^{-in\sigma}
       -
     \overline{\psi}_{n \nu}e^{in \sigma}
\right)
\right\}. 
\label{curve}
\end{eqnarray}
The integrations of $\sigma_r$ 
are performed over the region :
\begin{eqnarray}
&&
0 \leq \sigma_1 \leq \cdots 
\leq \sigma_M \leq 2\pi, 
\nonumber \\
&&
0 \leq \sigma_{M+1} \leq \cdots 
\leq \sigma_{M+N-1} \leq 2\pi.
\end{eqnarray}

The generating function is defined by 
(\ref{def of generating function of tachyon amplitudes}). 
The zero-slope limit can be obtained from 
(\ref{FTL of momentum factorization of JT}) by 
multiplying 
$
\int \prod_r \frac{dk^{(r)}}{(2\pi)^{13}}
\prod_r \varphi(k^{(r)})
$
and summing up with respect to $M$ and $N$. 
These are the same steps as taken previously 
to obtain the factorization 
by the straight open Wilson lines. 
At the present factorization they give us 
the following zero-slope limit :
\begin{eqnarray}
&&
\sum_{M=0}^{\infty}\sum_{N=1}^{\infty}
\int \prod_{r=1}^{M+N}
\frac{dk^{(r)}}{(2\pi)^{13}}
J_T
\left(
\left( k^{(1)},\cdots,k^{(M)}
\right)
;
\left(-k^{(M+1)},\cdots,-k^{(M+N)}
\right)
\right)
\nonumber \\
&&
~~~~~~~~~~~~~~~~~~~~~~~~~
\times 
\prod_{r=1}^{M} 
\varphi(k^{(r)})
\prod_{r=M+1}^{M+N}
\varphi(-k^{(r)})
\nonumber \\
&&
\approx 
\int 
dp_0
\prod_{n=1}^{\infty}
\frac{d\overline{\psi}_nd\psi_n}
     {(4\pi i)^{26}}
\nonumber \\
&&
~~~
\times 
\left[
  \frac{2\pi}{\alpha'}
  \int_{0}^{\infty} ds^{(c)}
     \exp 
       \left\{
          -\frac{\pi s^{(c)}}{2}
          \left( 
              p_{0 \mu}g^{\mu \nu}p_{0 \nu} 
               -
              \frac{4}{\alpha'}
          \right)
       \right\}
\right.
\nonumber \\
&&
~~~~~~~~~~~~~~~~~~~
\left.    
\times 
    \prod_{n=1}^{\infty}
    \left\{
       -\mbox{det}
         \left( 
                \frac{-\theta G \theta}
                {8 \pi^2 n q_c^{\frac{n}{2}}\alpha' }
         \right)
       \exp 
          \left\{
                \frac{\overline{\psi}_{n \mu}
                      (\theta G \theta)^{\mu \nu}\psi_{n \nu}}
                 {16 \pi^2 n q_c^{\frac{n}{2}}\alpha'}
          \right\}
      \right\}
\right]
\nonumber \\
&&
~~~
\times 
\left( 
  \frac{-\mbox{det} G_{\mu \nu}}
       {(2\alpha')^{26}}
\right)^{\frac{1}{4}}
\int 
\frac{dx}{(2\pi)^{26}}
e^{ip_{0 \mu}x^{\mu}}  \star
{\cal P}_{\star}
\left[
  \exp 
    \left\{
       \int_{0}^{2\pi}d\sigma \phi(x(\sigma))  
    \right\}
\right] 
\nonumber \\
&&
~~~
\times 
\left( 
  \frac{-\mbox{det} G_{\mu \nu}}
       {(2\alpha')^{26}}
\right)^{\frac{1}{4}}
\overline{
\int 
\frac{dx}{(2\pi)^{26}}
e^{ip_{0 \mu}x^{\mu}} \star
{\cal P}_{\star}
\left[
  \exp 
    \left\{
       \int_{0}^{2\pi}d\sigma \phi(x(\sigma))  
    \right\}
\right]
\star
\phi(x(2\pi))
}.
\label{closed-string FTL by open Wilson lines}
\end{eqnarray}
Modulo fixing the $U(1)$ symmetry 
the generating function is factorized into a sum or 
an integral of products of two open Wilson lines 
taken along the curves. 
The first term in the above integral 
can be thought of as a measure for the curves. 
Actually we can decompose  $x(\sigma)$ into 
$
x_0(\sigma)+\delta x(\sigma) 
$.
Here $x_0(\sigma)$ is a straight line 
which connects two ends of the curve. 
$\delta x(\sigma)$ describes a fluctuation 
from the straight line and becomes $2\pi$-periodic. 
$x$ and $p_0$ in Eq.(\ref{curve}) are determined 
by $x_0(\sigma)$ while $\psi_n$ are determined by 
$\delta x(\sigma)$.  
It may be also interesting to describe the curves 
in terms of loops in the momentum space. Let $p(\sigma)$ 
be a loop in the momentum space. This gives a curve 
$x(\sigma)$ by 
$
\dot{x}^{\mu}(\sigma)=\theta^{\mu \nu}p_{\nu}(\sigma)
$. 
$p_0$ and $\psi_n$ in Eq.(\ref{curve}) are determined by 
$p=\int_0^{2\pi}d\sigma p(\sigma)$ 
and 
$\psi_n 
= 
\sqrt{2}\int_0^{2\pi}d \sigma e^{in \sigma}p(\sigma)$.

The integration of $\delta x(\sigma)$ 
are suppressed exponentially compared with 
$x_0(\sigma)$. It can be accomplished by replacing 
the integration variables with their mean values
and becomes
unity at the zero-slope limit. 
Then the above factorization reduces to  
that by the straight open Wilson lines.


\section{Gluons in Closed-String Theory}
\label{sec:gluon boundary state}

Our study of gauge theory starts from this section. 
An analogue of gluon vertex operator of open-string 
is introduced in closed-string theory. 
Investigation of its Bogolubov transformation leads 
a renormalization of this operator. 
It will be shown that the renormalized operator enjoys 
the standard properties of (open-string) gluon vertex operator, 
including the action of the Virasoro algebra. 
These operators, acting on the Neumann boundary state, 
give rise to boundary states which turn out to be identified
with the boundary states 
of (open-string) off-shell gluons. 
In particular we will show that closed-string tree amplitudes 
between these states coincide with the corresponding gluon one-loop 
amplitudes of open-string. 
Our discussion in this section goes almost parallel 
to Section \ref{sec:tachyon boundary state}
where the boundary states of off-shell 
open-string tachyons are constructed.

\subsection{Bogolubov transformation and renormalization}

In closed-string theory an analogue of gluon-vertex 
operator may be taken as 
\begin{eqnarray}
\Vg \left(\sigma,\tau ;k\right)
&\equiv&:A_{\mu}(k) \partial_{\sigma}
  X^{\mu}(\sigma,\tau)
  \cdot e^{ik_{\nu}X^{\nu}(\sigma,\tau)}: \nonumber\\
&=&
  i:A_{\mu}(k) \left(z\partial-\bar{z}\bar{\partial}\right)
  X^{\mu}(z,\bar{z})\cdot e^{ik_{\nu}X^{\nu}(z,\bar{z})}:~,
\end{eqnarray}
where $(p+1)$-vectors $k_{\mu}$ and $A_{\mu}(k)$ are 
the momentum and the polarization vectors. 
The polarization vector is the Fourier transform of 
$U(1)$ gauge field $\mathcal{A}_{\mu}(x)$ :
\begin{equation}
\mathcal{A}_{\mu}(x)=\int \frac{d^{p+1}k}{(2\pi)^{\frac{p+1}{2}}}
   A_{\mu}(k)e^{ik_{\nu}x^{\nu}}~.
  \label{eq:F.T.-gluon}
\end{equation} 
The gauge field takes value in $\mathbb{R}$. 
This yields $\overline{A_{\mu}(k)}=A_{\mu}(-k)$.
While we concentrate on $U(1)$ gauge group in this paper,
it can be straightforwardly generalized to $U(N)$ 
by assigning the Chan-Paton indices.

Let us express the above operator in an auxiliary form. 
This often makes subsequent calculation facile. 
Let $a \in \mathbb{R}$ be an auxiliary parameter. 
We write the operator in an exponential form :
\begin{equation}
  \Vg (\sigma,\tau;k) =
        \left. i \frac{\partial}{\partial a}
        \hVg (\sigma,\tau;k;a) \right|_{a=0}~,
      \label{VG and hVG}
\end{equation}
with
\begin{eqnarray}
\lefteqn{\hVg (\sigma,\tau;k;a) \equiv 
   \; : e^{
      i\big\{ k_{\mu}-a A_{\mu}(k)\partial_{\sigma}\big\}
      X^{\mu}(\sigma,\tau)} :}\nonumber\\
&&= \exp\left(ik_{\mu}\hat{x}_{0}^{\mu}\right)\,
    |z|^{\alpha'k_{\mu}\hat{p}^{\mu}_{0}} \nonumber\\
&& \quad
   \times \prod_{n=1}^{\infty}
   \exp \left[ \sqrt{\frac{\alpha'}{2}}\frac{1}{n}
            \left\{\bigg(k_{\mu}-ina A_{\mu}(k)\bigg)
                       \alpha_{-n}^{\mu}z^{n}
                   +\bigg(k_{\mu}+ina A_{\mu}(k)\bigg)
                        \tilde{\alpha}_{-n}^{\mu} \bar{z}^{n}
            \right\}\right] \nonumber\\
&& \quad
   \times \prod_{n=1}^{\infty}
   \exp \left[-\sqrt{\frac{\alpha'}{2}}\frac{1}{n}
            \left\{ \bigg(k_{\mu}+ina A_{\mu}(k)\bigg)
                        \alpha_{n}^{\mu}z^{-n}
                    +\bigg(k_{\mu}-ina A_{\mu}(k)\bigg)
                        \tilde{\alpha}_{n}^{\mu}\bar{z}^{-n}
            \right\}\right]~.
\end{eqnarray}
The relation (\ref{VG and hVG}) implies that 
the terms proportional to higher powers of $a$ 
become irrelevant to the amplitudes.

We consider the Bogolubov transformation for
$\hat{V}_{A}$ generated by $g_{N}$ given in Eq.(\ref{gN}).
Using (\ref{transforms of massive modes}) we can 
write down the transform as follows : 
\begin{equation}
g_{N}^{-1} \hVg \left(\sigma,\tau;k;a\right) g_{N}
= R(\tau;a)\;
  \mbox{ad}_{g_{N}^{-1}}\hVg (\sigma,\tau;k;a)~,
 \label{transform of hVG}
\end{equation}
where
\begin{eqnarray}
R(\tau;a)
&\equiv&\left(\frac{|z|^{2}}{|z|^{2}-1}\right)^{-\frac{\alpha'}{2}
      k_{\mu}\left(g^{-1}Ng^{-1}\right)^{\mu\nu}k_{\nu}}\nonumber\\
 && \times
 \exp\left[
    -ia\frac{\alpha'}{2} A_{\mu}(k)
   \left(g^{-1}\left(N-N^{T}\right)g^{-1}\right)^{\mu\nu} k_{\nu}
   \frac{1}{|z|^{2}-1} 
      \right.
   \nonumber\\
&& \hspace{4em}\left. 
     -a^{2} \frac{\alpha'}{2} A_{\mu}(k)
            \left(g^{-1}Ng^{-1}\right)^{\mu\nu} A_{\nu}(k)
            \left(\frac{|z|}{|z|^{2}-1}\right)^{2} 
        \right]~. 
  \label{R}
\end{eqnarray}
The operator $\mbox{ad}_{g_{N}^{-1}} \hVg (\sigma,\tau;k;a)$ 
is defined to be 
\begin{equation}
\mbox{ad}_{g_{N}^{-1}} \hVg (\sigma,\tau;k;a)
 = \mathcal{M}_{A}(\bar{z};k;a)\tilde{\mathcal{M}}_{A}(z;k;a)
 \hVg(\sigma,\tau;k;a)~,
 \label{normal ordered transformed hVG}
\end{equation}
where $\mathcal{M}_{A}(\bar{z};k;a)$ and $\tilde{\mathcal{M}}_{A}(z;k;a)$
consist of the creation modes alone and take the forms of
\begin{eqnarray}
\mathcal{M}_{A}(\bar{z};k;a) &=& \prod_{n=1}^{\infty}
   \exp\left\{\sqrt{\frac{\alpha'}{2}}\frac{1}{n}
          \bigg(k_{\mu}-in a A_{\mu}(k)\bigg)
          {\left(g^{-1}N^{T}\right)^{\mu}}_{\nu}\alpha^{\nu}_{-n}
          \bar{z}^{-n}\right\}~, \nonumber\\
\tilde{\mathcal{M}}_{A}(z;k;a) &=& \prod_{n=1}^{\infty}
    \exp \left\{ \sqrt{\frac{\alpha'}{2}}\frac{1}{n}
          \bigg(k_{\mu}+in a A_{\mu}(k)\bigg)
          {\left(g^{-1}N\right)^{\mu}}_{\nu}\tilde{\alpha}^{\nu}_{-n}
          z^{-n} \right\}~.
  \label{M and tilded M}
\end{eqnarray}

In the transform (\ref{transform of hVG}) 
we can find the same singularity structure 
as that of the tachyon.
It is due to the factor $R(\tau;a)$, 
which is responsible for
the singularity of the transform (\ref{transform of hVG})
at the world-sheet boundary,
$\tau=0 (\Leftrightarrow |z|=1)$.
This factor contains an exponential in addition to 
the same factor which appeared in the transform 
$g_N^{-1}V_Tg_N$. 
Let us make a few comments about this extra factor. 
The first term of the exponent is proportional to $a$.
The closed-string tensor $g^{-1}(N-N^T)g^{-1}$ 
used there is
translated into $\frac{2}{\pi\alpha'} \theta^{\mu\nu}$. 
The second term becomes irrelevant since it is proportional to $a^2$. 
Therefore the net effect of this extra factor is to give a term 
proportional to $\theta$. 
Particularly in the absence of a constant $B$ field, 
$R(\tau ; a)$ reduces to that of the tachyon.

The singular factor which appears in the transform of the tachyon 
vertex operator has been expressed 
in Eq.(\ref{eq:singular factor as self-contraction of VT}) as the 
self-contraction between the chiral and the anti-chiral parts 
of $V_T$. 
In the present case as well,
we can think of the singular factor as 
such a self-contraction of $\hat{V}_{A}(\sigma,\tau;k;a)$. 
In fact, we can express $R(\tau;a)$
by using the chiral--anti-chiral correlation 
$\langle X^{\mu}(z_{1})\tilde{X}^{\nu}(\bar{z}_{2})\rangle
 =-\frac{\alpha'}{2}\left(g^{-1}Ng^{-1}\right)^{\mu\nu}
   \ln \left(1-\frac{1}{z_{1}\bar{z}_{2}}\right)$
of Green's function
$\mathbb{G}^{\mu\nu} (z_{1},\bar{z}_{1}|z_{2},\bar{z}_{2})$
as follows :
\begin{eqnarray}
R(\tau;a)
=
\exp \left[ -\left(k_{\mu}-ia A_{\mu}(k)z\partial\right)
                 \left(k_{\nu}+ia A_{\nu}(k)
                                     \bar{z}\bar{\partial}\right)
          \left\langle X^{\mu}(z)\tilde{X}^{\nu}(\bar{z}) \right\rangle
         \right]~.
   \label{R in terms of Green function}
\end{eqnarray}
As mentioned in the analysis of tachyon, 
boundary states induce correlations between 
the chiral and the anti-chiral sectors, and 
this feature is characteristic of systems of 
interacting closed- and open-strings.

We wish to find out a relation between 
$\mbox{ad}_{g_N^{-1}} \hat{V}_A$  and  
open-string gluon vertex operators. 
Previous discussions to establish the relation 
between $\mbox{ad}_{g_N^{-1}} V_T$ 
and open-string tachyon vertex operators  
consist of : 
i) Description of the OPE between 
$\mbox{ad}_{g_N^{-1}} V_T$, which turns out to be given 
in (\ref{OPE renormalized tachyon vertex}) by using 
Green's function  
$\mathbb{G}^{\mu\nu} (z_{1},\bar{z}_{1}|z_{2},\bar{z}_{2})$ 
and allows us to introduce the renormalized open-string 
tachyon vertex operators $V_T^{ren}$, 
ii) description of the diff$S^1$ action on $V_T^{ren}$, 
which turns out to reduce at the boundary 
to the standard Virasoro action 
on open-string tachyon vertex operators, 
and iii) reproduction of open-string one-loop tachyon amplitudes 
by the boundary state formalism. 
We repeat these analyses as for gluons in the following.

Let us examine the OPE between $\mbox{ad}_{g_N^{-1}}\hat{V}_A$. 
It is convenient to start with the OPE 
between the auxiliary operators. 
The standard calculation leads to 
\begin{eqnarray}
\lefteqn{
 \hVg \left(\sigma_{1},\tau_{1};k^{(1)};a_{1}\right)
 \hVg \left(\sigma_{2},\tau_{2};k^{(2)};a_{2}\right)}
 \nonumber\\
&&= \exp \left[i\left\{ k_{(1)}-ia_{1} A_{\mu}(k^{(1)})
    \left(z_{1}\partial_{z_{1}}
        -\bar{z}_{1}\bar{\partial}_{\bar{z}_{1}}\right)
    \right\} \right. \nonumber\\
&& \hspace{3em} \left.\times
         i\left\{ k_{\mu}^{(2)}-ia_{2} A_{\mu}(k^{(2)})
    \left(z_{2}\partial_{z_{2}}
      -\bar{z}_{2}\bar{\partial}_{\bar{z}_{2}}\right)
    \right\} 
    \left(-\frac{\alpha'}{2}g^{\mu\nu}
          \ln \left|z_{1}-z_{2}\right|^{2}
    \right)\right] \nonumber\\
&& \quad \;
   \times :\hVg \left(\sigma_{1},\tau_{1};k^{(1)};a_{1}\right)
 \hVg \left(\sigma_{2},\tau_{2};k^{(2)};a_{2}\right):~,
\end{eqnarray}
for $|z_{1}|\geq |z_{2}|$ and $z_{1}\neq z_{2}$. 
Since $\mbox{ad}_{g_N^{-1}}\hat{V}_A$ have the form given 
in (\ref{normal ordered transformed hVG}), 
we can obtain
their OPE from the above 
by taking account of the OPEs between 
${\cal M}_A,\tilde{{\cal M}}_A$ and $\hat{V}_A$. 
We find that  
\begin{eqnarray}
\lefteqn{
\mbox{ad}_{g_{N}^{-1}}
   \hVg \left(\sigma_{1},\tau_{1};k^{(1)};a_{1}
                 \right)\;
\mbox{ad}_{g_{N}^{-1}}
     \hVg \left(\sigma_{2},\tau_{2};k^{(2)};a_{2}
     \right)}\nonumber\\
&&=\exp
   \left[ i \bigg\{k_{\mu}^{(1)}
      -ia_{1} A_{\mu}(k^{(1)})
          \left(z_{1}\partial_{z_{1}}
                -\bar{z}_{1}\bar{\partial}_{\bar{z}_{1}} \right)
           \bigg\} \right.\nonumber\\
&& \hspace{3.5em}\left. \times   i \bigg\{k_{\mu}^{(2)}
      -ia_{2} A_{\mu}(k^{(2)})
          \left(z_{2}\partial_{z_{2}}
                -\bar{z}_{2}\bar{\partial}_{\bar{z}_{2}} \right)
           \bigg\}
           \mathbb{G}^{\mu\nu}(z_{1},\bar{z}_{1}|z_{2},\bar{z}_{2})
           \right] \nonumber\\
&& \quad \; \times 
  : \mbox{ad}_{g_{N}^{-1}}
   \hVg \left(\sigma_{1},\tau_{1};k^{(1)};a_{1}
                 \right)\;
\mbox{ad}_{g_{N}^{-1}}
     \hVg \left(\sigma_{2},\tau_{2};k^{(2)};a_{2}
     \right) :~,
  \label{OPE renormalized gluon}
\end{eqnarray}
for $|z_{1}|\geq |z_{2}|$, $|z_{1}z_{2}|\geq 1$ and $z_{1}\neq z_{2}$.
This is a direct extension of the tachyon case 
(\ref{OPE renormalized tachyon vertex}) and suggests an interpretation 
of $\mbox{ad}_{g_N^{-1}}\hat{V}_A$ in terms of the gluon vertex 
operator.

Let us introduce the renormalized operator $\hat{V}_A^{ren}$ 
by the following subtraction :
\begin{equation}
\hVg^{ren}\left(\sigma,\tau;k;a\right)
\equiv 
e^{\frac{i}{2\pi}aA_{\mu}(k)\theta^{\mu\nu}k_{\nu}}
  R(\tau;a)^{-1}
  \hVg \left(\sigma,\tau;k;a\right)~.
 \label{def of renormalized hVG}
\end{equation}
In other words, 
\begin{equation}
g^{-1}_{N} \hVg^{ren} \left(\sigma,\tau;k;a\right) g_{N}
=e^{\frac{i}{2\pi}aA_{\mu}(k)\theta^{\mu\nu}k_{\nu}}
  \;\mbox{ad}_{g^{-1}_{N}}\hVg (\sigma,\tau;k;a)~.
\label{renormalized operators}
\end{equation}
In addition to the subtraction of the singular factor 
$R(\tau;a)$, a finite subtraction is made in the above by 
the multiplication of 
$e^{\frac{i}{2\pi}a A_{\mu}(k)\theta^{\mu\nu}k_{\nu}}$.
Although it vanishes in the absence of $B$ field,  
this factor becomes surely necessary in general. 
Without this subtraction we cannot reproduce  
even the standard Virasoro action which we present below. 
The renormalized operator (\ref{def of renormalized hVG})
 gives rise to $V_A^{ren}$  
by the relation (\ref{VG and hVG}). 
We refer to $\Vg^{ren}(\sigma,\tau;k)$ as 
renormalized gluon vertex operator with momentum $k_{\mu}$
and polarization $A_{\mu}(k)$.
It can be written as follows : 
\begin{eqnarray}
\lefteqn{
  \Vg^{ren} \left(\sigma,\tau;k\right)
   \equiv \left. i \frac{\partial}{\partial a}
           \hVg^{ren} \left(\sigma,\tau;k;a\right) \right|_{a=0}}
       \nonumber\\
&& = \left( \frac{|z|^{2}}{|z|^{2}-1}
    \right)^{\frac{\alpha'}{2} k_{\mu}
             \left(g^{-1}Ng^{-1}\right)^{\mu\nu} k_{\nu}} 
    \nonumber\\
&& \hspace{2em}\times
       \left\{ \Vg (\sigma,\tau;k)
         -\frac{1}{\pi} A_{\mu}(k)\theta^{\mu\nu} k_{\nu}
        \left(\frac{1}{|z|^{2}-1}+\frac{1}{2}\right)
        V_{T}(\sigma,\tau;k) \right\}~,
  \label{def of renormalized VG}
\end{eqnarray}
where $V_{T}(\sigma,\tau;k)$ 
$\left(=\hVg(\sigma,\tau;k;a=0)\right)$ 
is the closed-string tachyon vertex operator.

Next we will examine the action of diff$S^1$ on the above 
renormalized operators. As a shortcut we use the description  
in terms of boundary states. 
Relevant states are introduced as 
$\Bigl| \hat{B}_{N} [A] ;a;(\sigma,k)\Bigr\rangle$
and $\Bigl| B_{N}[A];  (\sigma,k) \Bigr\rangle$. 
The first states are defined by using $\hat{V}^{ren}_A$ 
as follows :
\begin{eqnarray}
\lefteqn{
\Bigl| \hat{B}_{N}[A];a; (\sigma,k) \Bigr\rangle
\equiv \lim_{\tau\rightarrow 0+} \hVg^{ren} (\sigma,\tau;k;a)
           \left| B_{N} \right\rangle} 
\nonumber \\ 
&&= \left( \frac{\left(\det E_{\mu\nu}\right)^{2}}
                {\left(2\alpha'\right)^{p+1}(-\det g_{\mu\nu})}
    \right)^{\frac{1}{4}} \lim_{\tau\rightarrow 0+}
    g_{N} \left(g_{N}^{-1}\hVg^{ren}(\sigma,\tau;k;a) g_{N}\right)
    |\mathbf{0}\rangle 
\nonumber\\
&&=\left( \frac{-\det G_{\mu\nu}}
                {\left(2\alpha'\right)^{p+1}}
    \right)^{\frac{1}{4}}
    e^{\frac{i}{2\pi} aA_{\mu}(k)\theta^{\mu\nu}k_{\nu}}\,g_{N}
    \lim_{\tau\rightarrow 0+}
    \mbox{ad}_{g^{-1}_{N}}\hVg (\sigma,\tau;k;a)
    |\mathbf{0}\rangle 
\nonumber\\
&&=\left( \frac{-\det G_{\mu\nu}}
                {\left(2\alpha'\right)^{p+1}}
    \right)^{\frac{1}{4}}
    e^{\frac{i}{2\pi}aA_{\mu}(k)\theta^{\mu\nu}k_{\nu}}
\nonumber\\
&& \hspace{1.5em} \times
    \prod_{n=1}^{\infty} \exp
      \left[ \frac{\sqrt{2\alpha'}}{n} \left\{
         \left(k_{\mu}-ina A_{\mu}(k)\right)
         {\left(\frac{1}{E^{T}} g\right)^{\mu}}_{\nu}
         \alpha^{\nu}_{-n} e^{in\sigma}
     \right. \right. 
\nonumber\\
&& \hspace{10em}\left.\left.
     +\left(k_{\mu}+ina  A_{\mu}(k)\right)
      {\left(\frac{1}{E}g\right)^{\mu}}_{\nu}
      \tilde{\alpha}^{\nu}_{-n} e^{-in\sigma}
      \right\}\right]g_{N} |k_{N}\rangle~,
\label{aux 1 gluon bs}
\end{eqnarray}
where Eqs.(\ref{normal ordered transformed hVG})
and (\ref{renormalized operators}) are utilized 
to obtain the oscillator representation. 
The second states are defined similarly by using
$V_A^{ren}$ as  
\begin{eqnarray}
\Bigl| B_{N} [A];  (\sigma,k) \Bigr\rangle
 &\equiv& \lim_{\tau\rightarrow 0+} \Vg^{ren} (\sigma,\tau;k)
          \left| B_{N} \right\rangle 
 = \left. i \frac{\partial}{\partial a}
     \Bigl| \hat{B}_{N}[A];a; (\sigma,k) \Bigr\rangle
     \right|_{a=0}~.
\end{eqnarray}
The oscillator representation can be obtained from 
(\ref{aux 1 gluon bs}). This yields  
\begin{eqnarray}
\lefteqn{\Bigl|B_{N}[A]; (\sigma,k)\Bigr\rangle
         =\left. i \frac{\partial}{\partial a}
         \Bigl| \hat{B}_{N}[A]; a; (\sigma,k)\Bigr\rangle
         \right|_{a=0}
}\nonumber\\
&& = \left[-\frac{1}{2\pi}A_{\mu}(k)\theta^{\mu\nu}k_{\nu}
  +\sum_{n=1}^{\infty}\sqrt{2\alpha'}A_{\mu}(k)
      \left\{ {\left(\frac{1}{E^{T}}g\right)^{\mu}}_{\nu}
              \alpha^{\nu}_{-n} e^{in\sigma}
             - {\left(\frac{1}{E}g\right)^{\mu}}_{\nu}
             \tilde{\alpha}^{\nu}_{-n} e^{-in\sigma}
       \right\}\right]\nonumber\\
&& \hspace{1.5em}\times
  \bl B_{N}; (\sigma,k)\brr~,
\label{1 gluon bs}
\end{eqnarray}
where $\bl B_{N}; (\sigma,k) \brr$ is the boundary state with
a single open-string tachyon  given in (\ref{one tachyon BN}).

Let us recall that generators of diff$S^1$ are 
$L_{m}-\tilde{L}_{-m}$ $(m\in \mathbb{Z})$. 
Their actions on the states
$\Bigl| \hat{B}_{N}[A];a; (\sigma,k)\Bigr\rangle$ 
turn out to be as follows :
\begin{eqnarray}
\lefteqn{
\left( L_{m}-\tilde{L}_{-m} \right)
  \Bigl|\hat{B}_{N} [A];a; (\sigma,k) \Bigr\rangle}
  \nonumber\\
&&= e^{im\sigma} \left[
   \sqrt{2\alpha'}\sum_{n=1}^{\infty}\left\{
     k_{\mu}-i(n+m)a A_{\mu}(k) \right\}
   {\left(\frac{1}{E^{T}}g\right)^{\mu}}_{\nu} \alpha^{\nu}_{-n}
   e^{in\sigma} \right. \nonumber\\
&&  \hspace{4em} -\sqrt{2\alpha'} \sum_{n=1}
    \left\{k_{\mu}+i(n-m)a A_{\mu}(k) \right\}
    {\left(\frac{1}{E}g\right)^{\mu}}_{\nu}
    \tilde{\alpha}^{\nu}_{-n}e^{-in\sigma}
 \nonumber\\
&& \hspace{4em}
   +\alpha' mk_{\mu}G^{\mu\nu}k_{\nu}
   +iaA_{\mu}(k)
     \left(\frac{m}{2\pi}\theta^{\mu\nu}
           -\alpha'  m^{2} G^{\mu\nu}\right)
     k_{\nu} \nonumber\\
&& \hspace{4em}
     -a^{2} \alpha' \frac{1}{6}m(m-1)(m+1)
       A_{\mu}(k)\,G^{\mu\nu}A_{\nu}(k) \Bigg] 
   \Bigl|\hat{B}_{N}[A];a; (\sigma,k)\Bigr\rangle~,
\end{eqnarray}
for $\forall m \in \mathbb{Z}$.
As regards the states 
$\Bigl| B_{N} [A]; (\sigma,k)\Bigr\rangle$, 
the actions of $L_{n}-\tilde{L}_{-n}$ can be
read from the above 
by differentiating it with respect to $a$
and then setting $a=0$. 
We obtain
\begin{eqnarray}
\left(L_{m}-\tilde{L}_{-m}\right)
  \Bigl| B_{N} [A]; (\sigma,k)\Bigr\rangle
&=& e^{im\sigma} \left\{
       -i\frac{\partial}{\partial \sigma}
       +m\left(\alpha'k_{\mu}G^{\mu\nu}k_{\nu}+1\right)
       \right\} \Bigl| B_{N}[A]; (\sigma,k)\Bigr\rangle
    \nonumber\\
&& 
      +e^{im\sigma}\alpha' m^{2}k_{\mu}G^{\mu\nu}A_{\nu}(k)\;
       \bl B_{N};(\sigma,k) \brr~.
   \label{eq:Ishibashi-Virasoro}
\end{eqnarray}
The RHS can be compared
with the standard Virasoro action on (open-string) gluon 
vertex operators of the same momenta and polarizations. 
We find that they are identical with each other.

The Ishibashi condition imposed on 
$\Bigl | B_{N}[A]; (\sigma,k)\Bigr \rangle$, 
namely the vanishing of the RHS 
of Eq.(\ref{eq:Ishibashi-Virasoro}),
modulo total derivative with respect to $\sigma$,
requires 
the well-known physical state condition of 
the gauge field : 
\begin{equation}
\alpha' k_{\mu}G^{\mu\nu} k_{\nu}=0~,
\quad k_{\mu} G^{\mu\nu} A_{\nu}(k) =0~. 
 \label{physical condition of gluon}
\end{equation} 
By the same argument presented for the tachyon the above action 
is translated in closed-string field theory 
to the action of the BRST charge $Q_c$. 
The BRST invariance of these boundary states becomes 
precisely the on-shell condition of the gauge field.

\subsection{Boundary states with $M$ gluons}
\subsubsection{Construction of boundary states with $M$ gluons}

Previous constructions 
(\ref{aux 1 gluon bs}) and (\ref{1 gluon bs}) are 
generalized to the cases of $M$ off-shell gluons. 
Let us start with the states 
\begin{eqnarray}
\lefteqn{
\hVg^{ren}\left(\sigma_{1},\tau_{1};k^{(1)};a_{1}\right)
 \cdots
 \hVg^{ren}\left(\sigma_{M},\tau_{M};k^{(M)};a_{M}\right)
 \left| B_{N}\right\rangle
 }\nonumber\\
&&=\left(\frac{-\det G_{\mu\nu}}{(2\alpha')^{p+1}}\right)^{\frac{1}{4}}
g_{N} \left( g^{-1}_{N} 
     \hVg^{ren}(\sigma_{1},\tau_{1};k^{(1)};a_{1}) 
     g_{N}\right) \cdots \nonumber\\
&& \hspace{10.5em} \cdots
     \left( g^{-1}_{N} 
     \hVg^{ren}(\sigma_{M},\tau_{M};k^{(M)};a_{M}) 
     g_{N}\right) 
|\mathbf{0}\rangle~, 
\end{eqnarray}
where the insertion points 
$z_{r}=e^{\tau_{r}+i\sigma_{r}}$ $(r=1,2,\ldots,M)$
of the auxiliary renormalized operators 
$\hat{V}_A^{ren}$ 
are distinct with each other, satisfying 
the condition 
$|z_{1}|\geq |z_{2}|\geq \cdots |z_{M}| \geq 1$.
The oscillator representation can be obtained by taking account of 
the relation (\ref{renormalized operators}) and then using the OPE 
(\ref{OPE renormalized gluon}). 
It turns out to be
\begin{eqnarray}
\lefteqn{
\hVg^{ren}\left(\sigma_{1},\tau_{1};k^{(1)};a_{1}\right)
 \cdots
 \hVg^{ren}\left(\sigma_{M},\tau_{M};k^{(M)};a_{M}\right)
 \left| B_{N}\right\rangle
 }\nonumber\\
&& =\left(\frac{-\det G_{\mu\nu}}{(2\alpha')^{p+1}}
    \right)^{\frac{1}{4}}
     \prod_{r=1}^{M}
    e^{\frac{i}{2\pi} a_{r}A_{\mu}(k^{(r)})\theta^{\mu\nu}k^{(r)} }
      \nonumber\\
&& \hspace{2em} \times
    \prod_{r<s}^{M} \exp \left[
    i \bigg\{ k_{\mu}^{(r)}-ia_{r} A_{\mu}(k^{(r)})
        \left(z_{r}\partial_{z_{r}}
            -\bar{z}_{r}\bar{\partial}_{\bar{z}_{r}} \right)\bigg\}
     \right.
   \nonumber\\
&& \hspace{6.5em} \left.
   \times 
    i \bigg\{ k_{\mu}^{(s)}-ia_{s} A_{\mu}(k^{(s)})
        \left(z_{s}\partial_{z_{s}}
            -\bar{z}_{s}\bar{\partial}_{\bar{z}_{s}} \right)\bigg\}
     \mathbb{G}^{\mu\nu} (z_{r},\bar{z}_{r}|z_{s},\bar{z}_{s})
     \right] \nonumber\\
&& \hspace{2em} \times  
    \prod_{n=1}^{\infty}
    \exp \Bigg[ \sqrt{\frac{\alpha'}{2}}\frac{1}{n}
         \sum_{r=1}^{M} \bigg\{
        \left(k^{(r)}_{\mu}-ina_{r}  A_{\mu}(k^{(r)})\right)
        \left(z_{r}^{n}\delta^{\mu}_{\nu}+\bar{z}_{r}^{-n}
              {\left(g^{-1}N^{T}\right)^{\mu}}_{\nu}
        \right) \alpha^{\nu}_{-n}  \nonumber\\
&& \hspace{11.5em} 
   + \left(k^{(r)}_{\mu}+ina_{r} A_{\mu}(k^{(r)})\right)
     \left( \bar{z}_{r}\delta^{\mu}_{\nu}
          +z_{r}^{-n}{\left(g^{-1}N\right)^{\mu}}_{\nu}
     \right) \tilde{\alpha}^{\nu}_{-n}
     \bigg\} \Bigg]\nonumber\\
&& \hspace{2em}\times
  g_{N} \left|\sum_{r=1}^{M}k_{N}^{(r)}\right\rangle~.
\label{M gluons + BN}
\end{eqnarray}

Limits $\forall\tau_{r}\rightarrow 0+$ of the above 
states may be interpreted as auxiliary boundary states of $M$ 
off-shell gluons. They are auxiliary since the parameters $a_{r}$ 
are still included. This limiting process 
corresponds to sending the operators onto the world-sheet boundary. 
We then obtain the following states :
\begin{eqnarray}
\lefteqn{
\Bigl| \hat{B}_{N}[A];\{a_{r}\};
   (\sigma_{1},k^{(1)}),(\sigma_{2},k^{(2)}),
   \cdots,(\sigma_{M},k^{(M)})\Bigr\rangle
}\nonumber\\
&& \equiv \lim_{\forall \tau_{r}\rightarrow 0+}
\hVg^{ren}\left(\sigma_{1},\tau_{1};k^{(1)};a_{1}\right)
 \cdots
 \hVg^{ren}\left(\sigma_{M},\tau_{M};k^{(M)};a_{M}\right)
 \left| B_{N}\right\rangle \nonumber\\
&& = \left(\frac{-\det G_{\mu\nu}}{\left(2\alpha'\right)^{p+1}}
     \right)^{\frac{1}{4}}     
    e^{ \frac{i}{2\pi} \sum_{r=1}^{M}
    \left( a_{r}A_{\mu}(k^{(r)})
             -\sigma_{r} k_{\mu}^{(r)} \right)
      \theta^{\mu\nu} \sum_{s=1}^{M}k_{\nu}^{(s)}}
  \prod_{r<s}^{M} e^{\frac{i}{2}k_{\mu}^{(r)}\theta^{\mu\nu}k_{\nu}^{(s)}
                     \epsilon (\sigma_{r}-\sigma_{s})}
   \nonumber\\
&& \hspace{2em} \times \prod_{r<s}^{M}
    \exp \left[2\alpha' G^{\mu\nu} \left(
        k^{(r)}_{\mu}-a_{r} A_{\mu}(k^{(r)})
        \partial_{\sigma_{r}}\right) 
        \left(
        k^{(s)}_{\nu}-a_{s} A_{\nu}(k^{(s)})
        \partial_{\sigma_{s}}\right)
        \ln \left|e^{i\sigma_{r}} -e^{i\sigma_{s}} \right| 
        \right]  \nonumber\\
&& \hspace{2em} \times \prod_{n=1}^{\infty} \exp\left[
      \frac{\sqrt{2\alpha'}}{n} \sum_{r=1}^{M} \left\{
      \left(k_{\mu}^{(r)}-ina_{r} A_{\mu}(k^{(r)})\right)
      {\left(\frac{1}{E^{T}}g\right)^{\mu}}_{\nu}
      \alpha^{\nu}_{-n} e^{in\sigma_{r}} \right. \right. \nonumber\\
&& \hspace{12em} \left.\left.
      +\left(k^{(r)}_{\nu}+ina_{r} A_{\mu}(k^{(r)})\right)
      {\left(\frac{1}{E}g\right)^{\mu}}_{\nu} \tilde{\alpha}^{\nu}_{-n}
      e^{-in\sigma_{r}}\right\}\right]\,
      g_{N} \left|\sum_{r=1}^{M}k^{(r)}_{N}\right\rangle~.
      \nonumber\\
\label{aux M gluon bs}
\end{eqnarray}

We can repeat the above argument with $\hat{V}_A^{ren}$ replaced
by $V_A^{ren}$. This gives rise to the states  
\begin{eqnarray}
\Bigl| B_{N}[A]; (\sigma_{1},k^{(1)}),\cdots, (\sigma_{M},k^{(M)})
                    \Bigr\rangle
\equiv
 \lim_{\forall \tau_{r}\rightarrow 0+}
  \Vg^{ren} (\sigma_{1},\tau_{1};k^{(1)})
  \cdots
  \Vg^{ren} (\sigma_{M},\tau_{M};k^{(M)})
  \bl B_{N}\brr. 
\label{def M gluon bs}
\end{eqnarray}
These states are expected to be identified with 
boundary states of $M$ off-shell gluons.
In order to justify this identification, we should
evaluate closed-string free propagations between  
these boundary states
and compare them with the corresponding open-string one-loop amplitudes. 
Nevertheless,  
the oscillator representation of the states
(\ref{def M gluon bs}) becomes too complicated to evaluate  
the amplitudes. Taking account of the relation 
(\ref{def of renormalized VG}) it becomes convenient 
to use instead the following equivalent realization : 
\begin{eqnarray}
&&
\Bigl| B_{N}[A]; (\sigma_{1},k^{(1)}),\cdots, (\sigma_{M},k^{(M)})
                    \Bigr\rangle
\nonumber \\
&&~~~~~~~~~~~~~~
= i^{M} 
     \left. \prod_{r=1}^M 
       \frac{\partial}{\partial a_{r}}
     \right|_{a_{r}=0}
     \Bigl| \hat{B}_{N}[A];\{a_{r}\};(\sigma_{1},k^{(1)}),
                 \cdots,(\sigma_{M},k^{(M)}) \Bigr\rangle~, 
\label{M gluon bs}
\end{eqnarray}
where the operation 
$ 
\left. \prod_{r=1}^M 
       \frac{\partial}{\partial a_{r}}
\right|_{a_{r}=0}
$
means putting $a_{r}=0$ $(\forall r)$ after the differentiations.

We note that 
the boundary states (\ref{M gluon bs}) become 
available in the absence of $B$ field 
by putting $B_{\mu\nu}=0$.

\subsubsection{Construction of dual boundary states}

The dual boundary states are required in order to 
obtain the closed-string tree amplitudes. 
We construct these states by taking the same route 
as the tachyon case.

We begin by considering the Bogolubov transform of $\hVg$
generated by $g^{\dagger}_{N}$ which is
the BPZ dual of $g_{N}$ given in Eq.(\ref{gN dagger}). 
We have
\begin{equation}
g^{\dagger}_{N} \hVg (\sigma,\tau;k;a) 
          \left(g^{\dagger}_{N}\right)^{-1}
 = R_{\infty} (\tau,a)
   \; \,\mbox{ad}_{g_{N}^{\dagger}} \hVg (\sigma,\tau;k;a)~.
\end{equation}
Here we introduce a regular operator 
$\mbox{ad}_{g^{\dagger}_{N}}\hVg (\sigma,\tau;k;a)$ defined as 
\begin{eqnarray}
\mbox{ad}_{g^{\dagger}_{N}}\hVg (\sigma,\tau;k;a)
= \hVg (\sigma,\tau;k;a)
  \mathcal{M}_{A \infty}(\bar{z};k;a)
  \tilde{\mathcal{M}}_{A \infty} (z;k;a)~,
\end{eqnarray}
with
\begin{eqnarray}
\mathcal{M}_{A \infty}(\bar{z};k;a) &=&
   \prod_{n=1}^{\infty} \exp \left[ -\sqrt{\frac{\alpha'}{2}}
      \frac{1}{n} \left( k_{\mu}+ina  A_{\mu}(k) \right)
        {\left(g^{-1}N^{T}\right)^{\mu}}_{\nu} \alpha^{\nu}_{n}
        \bar{z}^{n} \right]~,\nonumber\\
\tilde{\mathcal{M}}_{A \infty}(z;k;a) &=&
    \prod_{n=1}^{\infty} 
    \exp \left[-\sqrt{\frac{\alpha'}{2}}
         \frac{1}{n} \left(k_{\mu}-ina A_{\mu}(k)\right)
             {\left(g^{-1}N\right)^{\mu}}_{\nu}
             \tilde{\alpha}^{\nu}_{n}z^{n}
          \right]~.
\end{eqnarray}
$R_{\infty}(\tau;a)$ in the above denotes the singular factor. 
It takes the same form as $R(\tau;a)$ with $z$ and $\theta^{\mu\nu}$
replaced by $1/z$ and $-\theta^{\mu\nu}$ respectively :
\begin{eqnarray}
R_{\infty}(\tau,a)
    &=&\left(1-|z|^{2}\right)^{\frac{\alpha'}{2}
             k_{\mu}(g^{-1}Ng^{-1})^{\mu\nu}k_{\nu}}
    \nonumber\\
&&\times
        \exp \left[ -a^{2}\frac{\alpha'}{2} A_{\mu}(k)
                \left(g^{-1}Ng^{-1}\right)^{\mu\nu} A_{\nu}(k)
                \left( \frac{|z|}{1-|z|^{2}}\right)^{2} \right.
      \nonumber\\
&& \hspace{4em} \left. +i a \frac{\alpha'}{2} A_{\mu}(k)
     \left(g^{-1}\left(N-N^{T}\right)g^{-1}\right)^{\mu\nu}
     k_{\nu}
     \frac{|z|^{2}}{1-|z|^{2}} \right]~. 
\end{eqnarray}
This factor represents 
the self-contraction between the chiral and the anti-chiral
pieces of $\hVg $. 
It can be written in terms of 
the chiral--anti-chiral correlation
$\langle X^{\mu}(z_{1}) \tilde{X}^{\nu}(\bar{z}_{2})
 \rangle_{\infty}$
of Green's function
$\mathbb{G}_{\infty}^{\mu\nu}(z_{1},\bar{z}_{1}|z_{2},\bar{z}_{2})$ :
\begin{eqnarray}
R_{\infty}(\tau;a)
 &=& \exp \left[ -\left(k_{\mu}-ia  A_{\mu}(k)z\partial\right)
                  \left(k_{\nu}+ia A_{\nu}(k)\bar{z}\bar{\partial}    
             \right) \left\langle
              X^{\mu}(z)\tilde{X}^{\nu}(\bar{z})\right\rangle_{\infty}
           \right]~.
\end{eqnarray}

Dual of the auxiliary renormalized gluon vertex operator, 
$\hat{V}_{A\infty}^{ren}(\sigma,\tau;k;a)$ is introduced 
by the following subtractions : 
\begin{equation}
\hat{V}^{ren}_{A\infty}\left(\sigma,\tau;k;a\right)
 = e^{-\frac{i}{2\pi}aA_{\mu}(k) \theta^{\mu\nu}k_{\nu}}
  R_{\infty}(\tau;a)^{-1}
   \hVg (\sigma,\tau;k;a)~, 
\label{def of dual aux gluon operator}
\end{equation}
where the finite subtraction has been made 
as required previously. 
Dual of the renormalized gluon operator, 
which we call $V_{A\infty}^{ren}(\sigma,\tau;k)$, 
is obtained from the above 
by using the relation (\ref{VG and hVG}) :
\begin{eqnarray}
\lefteqn{
  V_{A\infty}^{ren} (\sigma,\tau;k;a)
 \equiv \left. i \frac{\partial}{\partial a}
  \hat{V}^{ren}_{A \infty}(\sigma,\tau;k;a)
  \right|_{a=0} \nonumber
 }\\
&&= \left(1-|z|^{2}\right)^{-\frac{\alpha'}{2} k_{\mu}
                        \left(g^{-1}Ng^{-1}\right)^{\mu\nu}
                        k_{\nu}}
    \nonumber\\
&& \hspace{1.5em} \times
     \left\{ V_{A}(\sigma,\tau;k)
             +\frac{1}{\pi} \left(A_{\mu}(k)\theta^{\mu\nu}k_{\nu}\right)
              \left(\frac{|z|^{2}}{1-|z|^{2}}+\frac{1}{2}\right)
              V_{T}(\sigma,\tau;k)
     \right\}~.
\end{eqnarray}

Let us describe the dual boundary states. 
We first consider the states 
$$\bll B_{N} \br 
\hat{V}^{ren}_{A\infty}(\sigma_{1},\tau_{1};k^{(1)};a_{1})
\cdots
\hat{V}^{ren}_{A\infty}(\sigma_{M},\tau_{M};k^{(M)};a_{M})$$
where the insertion points are all distinct satisfying 
the condition, 
$1\geq |z_{1}| \geq |z_{2}| \geq \cdots \geq |z_{M}|$. 
We then take the limit $\forall \tau_{r}\rightarrow 0-$ of these 
states. Thus we obtain the duals. 
They take the forms of
\begin{eqnarray}
\lefteqn{
\Bigl\langle \hat{B}_{N}[A];\{a_{r}\};
      (\sigma_{1},k^{(1)}),(\sigma_{2},k^{(2)}),\cdots,
      (\sigma_{M},k^{(M)}) \Bigr|
}\nonumber\\
&&\equiv\lim_{\forall \tau_{r}\rightarrow 0-}
  \bll B_{N} \br 
\hat{V}^{ren}_{A\infty}(\sigma_{1},\tau_{1};k^{(1)};a_{1})
\cdots
\hat{V}^{ren}_{A\infty}(\sigma_{M},\tau_{M};k^{(M)};a_{M})
\nonumber\\
&&= \left(\frac{-\det G_{\mu\nu}}{(2\alpha')^{p+1}}\right)^{\frac{1}{4}}
  e^{-\frac{i}{2\pi}\sum_{r,s=1}^{M}\left(a_{r}A_{\mu}(k^{(r)})
                        -\sigma_{r} k^{(r)}_{\mu}\right)
     \theta^{\mu\nu}k^{(s)}_{\nu}}
 \prod_{r<s}^{M} e^{-\frac{i}{2}k_{\mu}^{(r)}\theta^{\mu\nu}
     k_{\nu}^{(s)} \epsilon (\sigma_{r}-\sigma_{s})}
     \nonumber\\
&&
~~~~
\times \prod_{r<s}^{M} \exp \Bigl[2 \alpha' G^{\mu\nu}
     \left(k_{\mu}^{(r)}
        -a_{r}A_{\mu}(k^{(r)})\partial_{\sigma_{r}}\right)
           \left(k_{\mu}^{(s)}
        -a_{s} A_{\mu}(k^{(s)})\partial_{\sigma_{s}}\right)
       \ln \left| e^{i\sigma_{r}}-e^{i\sigma_{s}}\right| \Bigr]
   \nonumber\\
&&  
~~~~ 
\times
  \left\langle -\sum_{r=1}^{M}k_{N}^{(r)}\right| g^{\dagger}_{N}
 \nonumber\\
&& \hspace{4em} \times
   \prod_{n=1}^{\infty} \exp \left[
     -\frac{\sqrt{2\alpha'}}{n} \sum_{r=1}^{M}
      \left\{ \left(k^{(r)}_{\mu}+ina_{r} A_{\mu}(k^{(r)})\right)
          {\left(\frac{1}{E^{T}} g\right)^{\mu}}_{\nu}
          \alpha^{\nu}_{n} e^{-in\sigma_{r}} \right.\right.
    \nonumber\\
&& \hspace{14em} \left.\left.
     +\left(k^{(r)}_{\mu}-ina_{r} A_{\mu}(k^{(r)})\right)
          {\left(\frac{1}{E} g\right)^{\mu}}_{\nu}
          \tilde{\alpha}^{\nu}_{n} e^{in\sigma_{r}}
    \right\}\right]~. 
\label{dual aux M gluon bs}
\end{eqnarray}
The oscillator representation in the above 
is derived from Eq.(\ref{def of dual aux gluon operator})
and the following OPE between 
$\mbox{ad}_{g^{\dagger}_{N}}\hVg$ :
\begin{eqnarray}
 \lefteqn{
 \mbox{ad}_{g^{\dagger}_{N}}\hVg (\sigma_{1}, \tau_{1};
    k^{(1)};a_{1})\;\;
 \mbox{ad}_{g^{\dagger}_{N}} \hVg (\sigma_{2}, \tau_{2};
    k^{(2)};a_{2})
 }\nonumber\\
 &&= \exp \left[ -\bigg\{k_{\mu}^{(1)}
     -ia_{1} A_{\mu}(k^{(1)})
     \left(z_{1}\partial_{1}-\bar{z}_{1}\bar{\partial}_{1}\right)
     \bigg\} \right. \nonumber\\
 && \hspace{4em}\left. \times \bigg\{
    k_{\nu}^{(2)}-ia_{2} A_{\nu}(k^{(2)})
       \left(z_{2}\partial_{2}-\bar{z}_{2}\bar{\partial}_{2}\right)
     \bigg\}
     \mathbb{G}^{\mu\nu}_{\infty} (z_{1},\bar{z}_{1}|z_{2},\bar{z}_{2})
     \right]\nonumber\\
 && \hspace{2em} \times
   : \mbox{ad}_{g^{\dagger}_{N}}\hVg (\sigma_{1}, \tau_{1};
    k^{(1)};a_{1})\;\;
 \mbox{ad}_{g^{\dagger}_{N}} \hVg (\sigma_{2}, \tau_{2};
    k^{(2)};a_{2}) :~,
\end{eqnarray}
for $|z_{1}|\geq |z_{2}|$, $|z_{1}z_{2}|\leq 1$ and
$z_{1}\neq z_{2}$. 
Duals to the states (\ref{def M gluon bs}) are obtained from 
(\ref{dual aux M gluon bs}) as follows :
\begin{eqnarray}
&&
\Bigl \langle 
B_{N}[A]; (\sigma_{1},k^{(1)}),\cdots, (\sigma_{M},k^{(M)})
\Bigr |
\nonumber \\
&&~~~~~~~~~~~~~~
= i^{M} 
     \left. \prod_{r=1}^M 
       \frac{\partial}{\partial a_{r}}
     \right|_{a_{r}=0}
     \Bigl \langle 
          \hat{B}_{N}[A];\{a_{r}\};(\sigma_{1},k^{(1)}),
                 \cdots,(\sigma_{M},k^{(M)}) \Bigr|~. 
\label{dual M gluon bs}
\end{eqnarray}

\subsubsection{Closed-string propagation}

We now come to computations of closed-string tree amplitudes
between the boundary states (\ref{def M gluon bs}). 
It is a straightforward generalization of 
what we did for the boundary states of off-shell 
open-string tachyons but becomes much complicated.

To avoid an unnecessary complication 
and  make the result transparent  
let us calculate the following amplitudes :
\begin{eqnarray} 
&&
  \Bigl\langle  \hat{B}_{N}[A];\{a_{r}\};
    (\sigma_{M+1},k^{(M+1)}),\cdots, (\sigma_{M+N},k^{(M+N)}) 
  \Bigr|
  \nonumber\\
&& 
~~~~~~~~~~~~~~~~~~~~~~~
 \times
 q_{c}^{\frac{1}{2} \left(L_{0}+\tilde{L}_{0}-2 \right)}
 \Bigl| \hat{B}_{N}[A];\{a_{r}\};(\sigma_{1},k^{(1)}), \cdots,
                 (\sigma_{M},k^{(M)}) 
 \Bigr \rangle,  
\label{def of aux gluon amplitude by bs}
\end{eqnarray}
where $q_{c}=e^{2\pi i \tau^{(c)}}$. 
These amplitudes reduce to the amplitudes in question 
by using the relations 
(\ref{M gluon bs}) and (\ref{dual M gluon bs}).

We can factorize the amplitudes into products of 
two kinds of contributions 
from the zero-modes and the massive modes 
of closed-string. 
By using the oscillator representations 
(\ref{aux M gluon bs}) and (\ref{dual aux M gluon bs}) 
these become as follows :  
\begin{eqnarray}
&&
  \left\langle  \hat{B}_{N}[A];\{a_{r}\};
    (\sigma_{M+1},k^{(M+1)}),\cdots, (\sigma_{M+N},k^{(M+N)}) \right|
  \nonumber\\
&& \hspace{5em} \times
 q_{c}^{\frac{1}{2} \left(L_{0}+\tilde{L}_{0}-2 \right)}
 \left| \hat{B}_{N}[A];\{a_{r}\};(\sigma_{1},k^{(1)}), \cdots,
                 (\sigma_{M},k^{(M)}) \right \rangle
\nonumber\\
&&=\left(\frac{-\det G_{\mu\nu}}{(2\alpha')^{p+1}}
   \right)^{\frac{1}{2}}\;
   \delta^{(p+1)}
       \left(\sum_{r=1}^{M+N} k^{(r)} \right)\;\;
   q_{c}^{-1-\frac{\alpha'}{4}\sum_{r=1}^{M}\sum_{s=M+1}^{M+N}
          k_{\mu}^{(r)} g^{\mu\nu} k_{\nu}^{(s)}}
  \nonumber\\
&& 
~~
\times
   \prod_{1\leq r <s \leq M}
   e^{\frac{i}{2}k_{\mu}^{(r)}\theta^{\mu\nu} k_{\nu}^{(s)}
      \epsilon( \sigma_{r} - \sigma_{s}) }
   \prod_{M+1 \leq r<s \leq M+N}
   e^{-\frac{i}{2}k_{\mu}^{(r)}\theta^{\mu\nu} k_{\nu}^{(s)}
      \epsilon( \sigma_{r} - \sigma_{s}) }
   \nonumber\\
&& 
~~\times
    e^{\frac{i}{2\pi} \sum_{r,s=1}^{M}\left(a_{r}A_{\mu}(k^{(r)})
                         -\sigma_{r} k_{\mu}^{(r)}\right)
         \theta^{\mu\nu}k_{\nu}^{(s)}}
    \;
    e^{- \frac{i}{2\pi}
           \sum_{r,s=M+1}^{M+N}\left(a_{r}A_{\mu}(k^{(r)})
                         -\sigma_{r} k_{\mu}^{(r)}\right)
         \theta^{\mu\nu}k_{\nu}^{(s)}}
   \nonumber\\
&& 
~~\times
  \prod_{1\leq r<s \leq M} \exp \left[
   2 \alpha' G^{\mu\nu}
      \left(k_{\mu}^{(r)}-a_{r} A_{\mu}(k^{(r)})
            \partial_{\sigma_{r}}   \right) 
       \left(k_{\mu}^{(s)}-a_{s} A_{\mu}(k^{(s)})
            \partial_{\sigma_{s}} \right)
       \ln \left| e^{i\sigma_{r}} - e^{i\sigma_{s}} \right|
      \right]
    \nonumber\\
&& 
~~\times
  \prod_{M+1\leq r<s \leq M+N} \exp \left[
   2 \alpha' G^{\mu\nu}
      \left(k_{\mu}^{(r)}-a_{r}A_{\mu}(k^{(r)})
            \partial_{\sigma_{r}}   \right)
     \left(k_{\mu}^{(s)}-a_{s} A_{\mu}(k^{(s)})
            \partial_{\sigma_{s}} \right)
       \ln \left| e^{i\sigma_{r}} - e^{i\sigma_{s}} \right|
      \right]
    \nonumber\\
&& 
~~\times
    F_{A}\left( q_{c}, \left\{ \sigma_{r}\right\},
                \left\{ k^{(r)} \right\} ; \left\{a_{r}\right\}
         \right)~. 
  \label{pre gluon amplitude by boundary state}
\end{eqnarray}
Here $F_{A}$ represents the sum of contributions 
from the massive modes of closed-string. 
It is given by the following infinite products :
\begin{eqnarray}
\lefteqn{F_{A}\left( q_{c}, \left\{ \sigma_{r}\right\},
                \left\{ k^{(r)} \right\} ; \left\{a_{r}\right\}
         \right)} \nonumber\\
&& = \prod_{n=1}^{\infty}
    \langle 0 | \exp \left[
       -\frac{q_{c}^{n}}{n} \alpha^{\mu}_{n} N_{\mu\nu}
              \tilde{\alpha}^{\nu}_{n} \right.
     \nonumber\\
&&   \hspace{6.5em} -\frac{\sqrt{2\alpha'} q_{c}^{\frac{n}{2}}}{n}
    \sum_{r=M+1}^{M+N} \left\{
      \left(k_{\mu}^{(r)}+ina_{r} A_{\mu}(k^{(r)})\right)
      {\left(\frac{1}{E^{T}}g\right)^{\mu}}_{\nu} \alpha^{\nu}_{n}
      e^{-in\sigma_{r}} \right. \nonumber\\
&& \hspace{14em} \left. \left.
      +\left(k_{\mu}^{(r)} -ina_{r} A_{\mu}(k^{(r)})\right)
      {\left(\frac{1}{E} g \right)^{\mu}}_{\nu}
      \tilde{\alpha}^{\nu}_{n} e^{in\sigma_{r}} \right\}\right]
    \nonumber\\
&& \hspace{3em} \times \exp
    \left[ -\frac{1}{n}\alpha^{\mu}_{-n}N_{\mu\nu}
                       \tilde{\alpha}^{\nu}_{-n} \right.
    \nonumber\\
&&  \hspace{6.5em} +\frac{\sqrt{2\alpha'}}{n}
      \sum_{r=1}^{M}\left\{ 
        \left( k_{\mu}^{(r)} -ina_{r} A_{\mu}(k^{(r)}) \right)
        {\left(\frac{1}{E^{T}}g\right)^{\mu}}_{\nu}
        \alpha^{\nu}_{-n} e^{in\sigma_{r}} \right.\nonumber\\
&&  \hspace{12.5em} \left.\left.
       +\left(k_{\mu}^{(r)}+ina_{r} A_{\mu}(k^{(r)})\right)
       {\left(\frac{1}{E}g\right)^{\mu}}_{\nu} \tilde{\alpha}^{\nu}_{-n}
       e^{-in\sigma_{r}}\right\}\right]
      |0\rangle~.
\nonumber \\
  \label{def of FA}
\end{eqnarray}

We need to evaluate the above infinite products. 
These are carried out in Appendix \ref{sec:formulae}. 
We just quote the result obtained there. 
These turn out to be as follows :  
\begin{eqnarray}
&&  F_{A}\left( q_{c}, \left\{ \sigma_{r}\right\},
                \left\{ k^{(r)} \right\} ; \left\{a_{r}\right\}
         \right)  \nonumber\\
&&= \prod_{n=1} \left(1-q_{c}^{n}\right)^{-p-1}\nonumber\\
&& \quad \times \prod_{1\leq r<s \leq M}
   \exp \Bigg[ 2\alpha' G^{\mu\nu}
     \left(k_{\mu}^{(r)}
          -a_{r}A_{\mu}(k^{(r)})\partial_{\sigma_{r}}\right)
     \left(k_{\nu}^{(s)}
          -a_{s} A_{\mu}(k^{(s)}) \partial_{\sigma_{s}}\right)
      \nonumber\\
  && \hspace{10em} \times
       \ln  \left\{
       \frac{ \prod_{n=1}^{\infty}
              \left(1-e^{i(\sigma_{r}-\sigma_{s})}q_{c}^{n}\right)
              \left(1-e^{-i(\sigma_{r}-\sigma_{s})} q_{c}^{n} \right)}
            {\prod_{n=1}^{\infty} \left(1-q_{c}^{n}\right)^{2}}
       \right\}  \Bigg] \nonumber\\
&& \quad \times \prod_{M+1\leq r<s \leq M+N}
   \exp \Bigg[ 2\alpha' G^{\mu\nu}
     \left(k_{\mu}^{(r)}
          -a_{r} A_{\mu}(k^{(r)})\partial_{\sigma_{r}}\right)
     \left(k_{\nu}^{(s)}
          -a_{s} A_{\mu}(k^{(s)}) \partial_{\sigma_{s}}\right)
      \nonumber\\
  && \hspace{12em} \times
       \ln  \left\{
       \frac{ \prod_{n=1}^{\infty}
              \left(1-e^{i(\sigma_{r}-\sigma_{s})}q_{c}^{n}\right)
              \left(1-e^{-i(\sigma_{r}-\sigma_{s})} q_{c}^{n} \right)}
            {\prod_{n=1}^{\infty} \left(1-q_{c}^{n}\right)^{2}}
       \right\}  \Bigg] \nonumber\\
 && \quad \times \prod_{r=1}^{M} \prod_{s=M+1}^{M+N}
      \exp \Bigg[ 2\alpha' G^{\mu\nu}
     \left(k_{\mu}^{(r)}
          -a_{r} A_{\mu}(k^{(r)})\partial_{\sigma_{r}}\right)
     \left(k_{\nu}^{(s)}
          -a_{s} A_{\mu}(k^{(s)}) \partial_{\sigma_{s}}\right)
      \nonumber\\
  && \hspace{12em} \times
       \ln  \left\{
       \frac{ \prod_{m=0}^{\infty}
              \left(1-e^{i(\sigma_{r}-\sigma_{s})}q_{c}^{m+\frac{1}{2}}
              \right)
              \left(1-e^{-i(\sigma_{r}-\sigma_{s})} q_{c}^{m+\frac{1}{2}}
               \right)}
            {\prod_{n=1}^{\infty} \left(1-q_{c}^{n}\right)^{2}}
       \right\}  \Bigg] \nonumber\\
  && \quad \times
     \exp \left[ 2\alpha'G^{\mu\nu}
          \sum_{r=1}^{M+N} \left(a_{r}\right)^{2}
                  A_{\mu}(k^{(r)}) A_{\nu}(k^{(r)})
                  \;
          \ln \left\{\prod_{n=1}^{\infty} \left(1-q_{c}^{n}\right)\right\}
          \right]~.
    \label{result on FA}
\end{eqnarray}
We have used the total momentum conservation, 
$\delta^{(p+1)} \left(\sum_{r=1}^{M+N}k^{(r)}\right)$ 
to obtain the above expression. 
The last exponential will be ignored since 
the exponent is proportional to $(a_{r})^{2}$ and 
this term brings about nothing when   
$\left.\prod_{r=1}^{M+N}
\frac{\partial}{\partial a_{r}}\right|_{a_r=0}$ 
operate on the amplitudes.

The amplitudes which are 
obtained by plugging Eq.(\ref{result on FA}) into
Eq.(\ref{pre gluon amplitude by boundary state}) 
may be written down 
by using the elliptic $\theta$-functions. 
With a similar manipulation which leads
Eq.(\ref{tachyon amplitude by boundary state}) 
we can recast the amplitudes into the following forms :
\begin{eqnarray}
  \lefteqn{
  \left\langle  \hat{B}_{N}[A];\{a_{r}\};
    (\sigma_{M+1},k^{(M+1)}), \cdots, (\sigma_{M+N},k^{(M+N)}) \right|
} \nonumber\\
&& \hspace{4em}
  \times q_{c}^{\frac{1}{2} \left(L_{0}+\tilde{L}_{0}-2 \right)}
 \left| \hat{B}_{N}[A];\{a_{r}\};
    (\sigma_{1},k^{(1)}), \cdots, (\sigma_{M},k^{(M)}) \right \rangle
 \nonumber\\
&&=\left(\frac{-\det G_{\mu\nu}}{(2\alpha')^{p+1}}
   \right)^{\frac{1}{2}}\;
    q_{c}^{\frac{p-25}{24}}
    \eta \left(\tau^{(c)}\right)^{-p+1}
    \;\delta^{(p+1)}
       \left(\sum_{r=1}^{M+N} k^{(r)} \right)
  \nonumber\\
&& \quad \times
   \prod_{1\leq r <s \leq M}
   e^{\frac{i}{2}k_{\mu}^{(r)}\theta^{\mu\nu} k_{\nu}^{(s)}
      \epsilon( \sigma_{r} - \sigma_{s}) }
   \prod_{M+1 \leq r<s \leq M+N}
   e^{-\frac{i}{2}k_{\mu}^{(r)}\theta^{\mu\nu} k_{\nu}^{(s)}
      \epsilon( \sigma_{r} - \sigma_{s}) }
   \nonumber\\
&& \quad \times
   q_{c}^{-\frac{1}{16\pi^{2} \alpha'} 
           K_{\mu}(\theta G \theta)^{\mu\nu}K_{\nu}}
   \;
   \prod_{r=1}^{M+N}
   e^{\frac{i}{2\pi} \left( a_{r}A_{\mu}(k^{(r)})
            -\sigma_{r} k_{\mu}^{(r)} \right)
      \theta^{\mu\nu}K_{\nu}} \nonumber\\
&&  \quad \times
    \prod_{1\leq r <s \leq M}
    \exp \Bigg[ 2\alpha' G^{\mu\nu}
       \left(k_{\mu}^{(r)}
          -a_{r} A_{\mu}(k^{(r)})\partial_{\sigma_{r}}
       \right)  \nonumber\\
&&    \hspace{10em} \times
       \left(k_{\mu}^{(s)}-a_{s} A_{\mu}(k^{(s)})\partial_{\sigma_{s}}
       \right)
       \ln \left\{\frac{\theta_{1}\left.\left(
                           \frac{|\sigma_{r}-\sigma_{s}|}{2\pi}
                           \right|   \tau^{(c)}\right)}
                        {\eta\left(\tau^{(c)}\right)^{3}}
           \right\} \Bigg] \nonumber\\
&&  \quad \times
      \prod_{M+1\leq r<s \leq M+N}
       \exp \Bigg[ 2\alpha' G^{\mu\nu}
       \left(k_{\mu}^{(r)}
           -a_{r}A_{\mu}(k^{(r)})\partial_{\sigma_{r}}
       \right)  \nonumber\\
&&    \hspace{12em} \times
       \left(k_{\mu}^{(s)}
          -a_{s} A_{\mu}(k^{(s)})\partial_{\sigma_{s}}
       \right)
       \ln \left\{\frac{\theta_{1}\left.\left(
                           \frac{|\sigma_{r}-\sigma_{s}|}{2\pi}
                           \right|   \tau^{(c)}\right)}
                        {\eta\left(\tau^{(c)}\right)^{3}}
           \right\} \Bigg] \nonumber\\
&& \quad \times
      \prod_{r=1}^{M} \prod_{s=M+1}^{M+N}
      \exp \Bigg[ 2\alpha' G^{\mu\nu}
       \left(k_{\mu}^{(r)}
          -a_{r} A_{\mu}(k^{(r)})\partial_{\sigma_{r}}
       \right)  \nonumber\\
&&    \hspace{12em} \times
       \left(k_{\mu}^{(s)}
          -a_{s} A_{\mu}(k^{(s)})\partial_{\sigma_{s}}
       \right)
       \ln \left\{\frac{\theta_{4}\left.\left(
                           \frac{\sigma_{r}-\sigma_{s}}{2\pi}
                           \right|   \tau^{(c)}\right)}
                        {\eta\left(\tau^{(c)}\right)^{3}}
           \right\} \Bigg]~,
\nonumber \\
\label{aux gluon amplitude by boundary state}
\end{eqnarray}
where we put 
$K_{\mu}\equiv \sum_{r=1}^{M}k_{\mu}^{(r)}$.
In the above expression 
we have included the contribution of the ghosts,  
$\prod_{n=1}^{\infty} (1-q_{c}^{n})^{2}$.

\subsection{Comparison with open-string one-loop calculation}

We compare the amplitudes 
(\ref{aux gluon amplitude by boundary state}), 
taking account of closed-string propagations 
along the Dirichlet directions, 
with open-string gluon one-loop amplitudes. 
The discussion goes parallel to the case of the tachyon. 
We first describe the open-string calculation. 
We use the same conventions as in subsection
\ref{sec:open-string tachyon one-loop}.

\subsubsection{Open-string one-loop calculation}
Open-string gluon vertex operator of momentum $k$ 
is given by 
\begin{eqnarray}
V_A^{open}(\xi,\sigma;k)
\equiv
\left. 
iA_{\mu}(k)
:
(\partial_{\rho}+\overline{\partial}_{\rho})
   X^{\mu}(\rho,\bar{\rho})
      e^{ik_{\nu}X^{\nu}(\rho,\bar{\rho})}
:
\right|_{\rho=\xi e^{i\sigma}},
\label{def of open-string gluon vertex}
\end{eqnarray}
where $\rho$ and $(\xi,\sigma)$ are 
respectively  the complex and the radial coordinates 
of the upper half-plane (open-string world sheet). 
$A_{\mu}(k)$ are the polarization vectors or 
the Fourier modes of the $U(1)$ gauge field 
${\cal A}_{\mu}(x)$. 
We consider the scattering process of $M+N$ gluons with 
momenta $k^{(r)}$. 
Diagram relevant to 
the one-loop scattering process can be drawn 
on the upper half-plane as depicted in Figure 
\ref{upper-half-plane}. 
The corresponding gluon amplitude, which we call  
$
I_{A}
\left( 
   (k^{(1)},\cdots,k^{(M)})
    ;
   (k^{(M+1)},\cdots,k^{(M+N)})
\right)
$, 
is given by  a sum of traces of 
their products arranged 
in cyclically distinct orders with keeping 
their partial orderings at the each end : 
\begin{eqnarray}
&&
I_{A}
\left( \left( k^{(1)},\cdots,k^{(M)} \right) ;
       \left( k^{(M+1)},\cdots,k^{(M+N)} \right)
\right)
\nonumber \\
&&
~~
\equiv
\mbox{Tr} 
\left\{ 
\Delta 
V_A^{open}(1,0;k^{(1)})
\cdots 
\Delta 
V_A^{open}(1,0;k^{(M)})
\right.
\nonumber \\
&&
~~~~~~~~~~~~~~~~~~~~~
\times
\left.
\Delta 
V_A^{open}(1,\pi;k^{(M+1)})
\cdots 
\Delta 
V_A^{open}(1,\pi;k^{(M+N)})
\right\}
\nonumber \\
&&
~~~~~~~~
+~~\cdots\cdots. 
\label{def of open-string gluon amplitude}
\end{eqnarray}

In order to obtain the above amplitude 
it becomes convenient to introduce 
the gluon vertex operator in an exponential form. 
Let $a \in \mathbb{R}$ be an auxiliary parameter. 
We put \cite{IM}
\begin{eqnarray}
\hat{V}_A^{open}(\xi,\sigma; k; a)
\equiv  
\left.
:
\exp
\left\{
   ik_{\mu}X^{\mu}(\rho,\bar{\rho})
   +
   iaA_{\mu}(k)
   (\partial_{\rho}+\overline{\partial}_{\rho})
   X^{\mu}(\rho,\bar{\rho})
\right\}
:
\right|_{\rho=\xi e^{i\sigma}}~.
\label{def of open-string aux gluon vertex}
\end{eqnarray}
This operator is related with the vertex operator 
(\ref{def of open-string gluon vertex}) by 
\begin{eqnarray}
V_A^{open}(\xi,\sigma; k)
=
\left.
\frac{\partial}{\partial a}
\hat{V}_A^{open}(\xi,\sigma;k;a) 
\right|_{a=0}.
\label{open-string gluon vertex by aux gluon vertex}
\end{eqnarray}
The corresponding amplitude constructed from  
$\hat{V}_A^{open}$ instead of $V_A^{open}$ will be called 
$\hat{I}_A$. 
It is given by 
\begin{eqnarray}
&&
\hat{I}_{A}
\left( \left( \left( k^{(1)},a_1 \right),
              \cdots ,
              \left( k^{(M)},a_M \right) 
       \right) ;
       \left( \left( k^{(M+1)}, a_{M+1} \right), 
              \cdots , 
              \left( k^{(M+N)}, a_{M+N} \right) 
       \right)
\right)
\nonumber \\
&&
~~
\equiv
\mbox{Tr} 
\left\{ 
\Delta 
\hat{V}_A^{open}
\left(
   1,0;k^{(1)};a_1
\right)
\cdots 
\Delta 
\hat{V}_A^{open}
\left(
   1,0;k^{(M)};a_M
\right)
\right.
\nonumber \\
&&
~~~~~~~~~~~~~~~~~~~~~
\times
\left.
\Delta 
\hat{V}_A^{open}
\left(
   1,\pi;k^{(M+1)};a_{M+1}
\right)
\cdots 
\Delta 
\hat{V}_A^{open}
\left(
    1,\pi;k^{(M+N)};a_{M+N}
\right)
\right\}
\nonumber \\
&&
~~~~~~~~
+~~\cdots\cdots~.
\label{def of open-string aux gluon amplitude}
\end{eqnarray}
The gluon scattering amplitude $I_A$ can be obtained from 
$\hat{I}_A$ by differentiating it with respect to $a_r$ 
and making them vanish:
\begin{eqnarray}
&&
I_{A}
\left( \left( k^{(1)},\cdots,k^{(M)} \right) ;
       \left( k^{(M+1)},\cdots,k^{(M+N)} \right)
\right)
\nonumber \\
&&
=
\left.
\prod_{r=1}^{M+N}
\frac{\partial}{\partial a_r}
\right|_{a_r=0}
\hat{I}_{A}
\left( \left( \left( k^{(1)},a_1 \right),
              \cdots ,
              \left( k^{(M)},a_M \right) 
       \right) ;
       \left( \left( k^{(M+1)}, a_{M+1} \right), 
              \cdots , 
              \left( k^{(M+N)}, a_{M+N} \right) 
       \right)
\right)~.
\nonumber \\
\label{IA by hatIA}
\end{eqnarray}

Due to the existence of the auxiliary parameter 
the Virasoro algebra acts on 
the auxiliary gluon vertex operator 
$\hat{V}_A^{open}$ 
in an unfamiliar form. 
The generators have been given in 
(\ref{open-string Virasoro generator}). 
Their action turns out to be as follows : For $n \in \mathbb{Z}$
\begin{eqnarray}
&&
[ L_n, \hat{V}_A^{open}(\xi,0;k;a) ]
\nonumber \\
&&
=
\left\{ 
 \xi^{n+1}\frac{\partial }{\partial \xi}
 +
 (n+1)\alpha'k_{\mu}G^{\mu \nu}k_{\nu}\xi^n
 +
 (n+1)\xi^n a \frac{\partial }{\partial a}
\right.
\nonumber \\
&&
~~~~~~~~~~~~
 +
 n(n+1)a \alpha'k_{\mu}G^{\mu \nu}A_{\nu}(k)\xi^{n-1}
\nonumber \\
&&
~~~~~~~~~~~~~~~~~~~~~~
\left.
 +
 \frac{n(n^2-1)}{6}a^2 \alpha' A_{\mu}(k)G^{\mu \nu}A_{\nu}(k)
 \xi^{n-2}
\right\}
\hat{V}_A^{open}(\xi,0;k;a),
\end{eqnarray}
and the same form on $\hat{V}^{open}_A(\xi,\pi;k;a)$ 
except replacing $\xi$ by $-\xi$ in the RHS. 
By applying the relation 
(\ref{open-string gluon vertex by aux gluon vertex}) 
to the above equation we obtain the standard Virasoro action 
on the gluon vertex operators. 
The auxiliary operators become the tachyon 
vertex operators simply by letting $a$ vanish in 
(\ref{def of open-string aux gluon vertex}), and  
the above action reduces to the standard action on 
the tachyon vertex operators.

We can evaluate the amplitude $\hat{I}_A$ 
in the standard manner.  
The open-string propagators in 
(\ref{def of open-string aux gluon amplitude}) 
may be replaced by the integral forms. 
The above Virasoro action implies :
\begin{eqnarray}
\xi^{L_0}\hat{V}_A^{open}(1,0;k;a)\xi^{-L_0}
&=&
\xi^{\alpha'k_{\mu}G^{\mu \nu}k_{\nu}}
\hat{V}_A^{open}(\xi,0;k;\xi a),
\nonumber \\
\xi^{L_0}\hat{V}_A^{open}(1,\pi;k)\xi^{-L_0}
&=&
\xi^{\alpha'k_{\mu}G^{\mu \nu}k_{\nu}}
\hat{V}_A^{open}(\xi,\pi;k;\xi a).
\label{Evolution of open-string aux gluon vertex}
\end{eqnarray}
Utilizing these properties we can 
write the amplitude in the following form :
\begin{eqnarray}
&&
\hat{I}_{A}
\left( \left( \left(k^{(1)}, a_1 \right),
               \cdots,
              \left(k^{(M)},a_M \right) 
       \right) ; 
       \left( \left(k^{(M+1)},a_{M+1} \right),
               \cdots, 
              \left(k^{(M+N)},a_{M+N} \right) 
       \right)
\right)
\nonumber \\
&&
=
\int_{0 \leq \xi_{M+N} \leq \xi_{M+N-1} 
        \leq \cdots \leq \xi_1 \leq 1}
\prod_{r=1}^{M+N}
\frac{d \xi_r}{\xi_r} 
\times 
\prod_{r=1}^{M+N}
\xi_r^{\alpha'k_{\mu}^{(r)}G_{\mu \nu}k_{\nu}^{(r)}}
\nonumber \\
&&
~~~~~~
\times 
\mbox{Tr}
\left\{
\hat{V}_A^{open}
\left( 
   \xi_1,0 ; k^{(1)} ; \xi_1a_1
\right)
\cdots
\hat{V}_A^{open}
\left( 
   \xi_M,0 ; k^{(M)} ; \xi_Ma_M
\right) 
\right.
\nonumber \\
&&
~~~~~~~~~~
\left.
\times 
\hat{V}_A^{open}
\left(
   \xi_{M+1},\pi ; k^{(M+1)} ; \xi_{M+1}a_{M+1}
\right)
\cdots
\hat{V}_A^{open}
\left(
   \xi_{M+N},\pi ; k^{(M+N)} ; \xi_{M+N}a_{M+N}
\right) 
\xi_{M+N}^{L_0-1}
\right\} 
\nonumber \\
&&
~~~~
+
~~
\cdots\cdots. 
\nonumber \\
\label{integral open-string aux gluon amplitude}
\end{eqnarray}
The coordinates $\xi_r$ in the RHS are 
insertion points of the auxiliary gluon vertices 
and provide a parametrization of the diagram. 
Another parametrization can be obtained 
by mapping the diagram to the cylinder 
with width $\pi$ 
as depicted in Figure \ref{open-string diagram}. 
Correspondingly $\xi_r$ are mapped to $\nu_r$ 
by $\nu_r=-\ln \xi_r$. 
We put $\tau^{(o)}\equiv -\ln \xi_{M+N}$. 
These are the open-string parameters.

To evaluate the RHS of 
Eq.(\ref{integral open-string aux gluon amplitude}),
we first normal-order 
the products of the auxiliary operators 
within the traces
by taking account of their OPEs.   
For instance we use   
\begin{eqnarray}
&&
\hat{V}_A^{open}\left(\xi_1,0;k^{(1)};a_1 \right)
\hat{V}_A^{open}\left(\xi_2,0;k^{(2)};a_2 \right)
\nonumber \\
&&
~~~~~~~~~
=
e^{-\frac{i}{2}k_{\mu}^{(1)}\theta^{\mu \nu}k_{\nu}^{(2)}}
\times 
e^{2\alpha'G^{\mu \nu}
     \left(
        k_{\mu}^{(1)}
          +
        a_1A_{\mu}(k^{(1)})\frac{\partial}{\partial \xi_1}
     \right)
     \left(
        k_{\nu}^{(2)}
           +
        a_2A_{\nu}(k^{(2)})\frac{\partial}{\partial \xi_2}
     \right)
    \ln(\xi_1-\xi_2)} 
\nonumber \\
&&
~~~~~~~~~~~~~~~~~~~~
\times 
:
\hat{V}_A^{open}\left(\xi_1,0;k^{(1)};a_1 \right)
\hat{V}_A^{open}\left(\xi_2,0;k^{(2)};a_2 \right)
:~~, 
\nonumber \\
&&
\hat{V}_A^{open}\left(\xi_1,0;k^{(1)};a_1 \right)
\hat{V}_A^{open}\left(\xi_2,\pi;k^{(2)};a_2 \right)
\nonumber \\
&&
~~~~~~~~~
=
e^{-\frac{i}{2}k_{\mu}^{(1)}\theta^{\mu \nu}k_{\nu}^{(2)}}
\times 
e^{ 2\alpha'G^{\mu \nu}
     \left(
        k_{\mu}^{(1)}
          +
        a_1A_{\mu}(k^{(1)})\frac{\partial}{\partial \xi_1}
     \right)
     \left(
        k_{\nu}^{(2)}
          -
        a_2A_{\nu}(k^{(2)})\frac{\partial}{\partial \xi_2}
     \right)
    \ln(\xi_1+\xi_2)} 
\nonumber \\
&&
~~~~~~~~~~~~~~~~~~~~~
\times 
:
\hat{V}_A^{open}\left(\xi_1,0;k^{(1)};a_1 \right)
\hat{V}_A^{open}\left(\xi_2,\pi;k^{(2)};a_2 \right)
:~~, 
\label{OPE of open-string aux gluon vertex} 
\end{eqnarray}
for $\xi_1 > \xi_2$. 
We still need to take traces of these normal-ordered operators.   
After straightforward but wearisome calculations  
the amplitude can be written down in terms of the open-string 
parameters. It turns out to be the following integral :  
\begin{eqnarray}
&&
\hat{I}_{A}
\left( 
\left( 
  \left(k^{(1)},a_1 \right),
   \cdots,
  \left(k^{(M)},a_M \right) 
\right) ; 
\left(
   \left(k^{(M+1)},a_{M+1} \right),
   \cdots,
   \left(k^{(M+N)},a_{M+N} \right)
\right)
\right)
\nonumber \\
&&
=
\sqrt{-\mbox{det}G_{\mu \nu}}
\delta^{(p+1)}
     \left(\sum_{r=1}^{M+N}k^{(r)}
     \right)
\nonumber \\
&&
~~~~
\times 
\prod_{1 \leq r < s \leq M}
e^{-\frac{i}{2}
          k_{\mu}^{(r)}\theta^{\mu \nu}k_{\nu}^{(s)}}
\prod_{M+1 \leq r < s \leq M+N}
e^{\frac{i}{2} 
         k_{\mu}^{(r)}\theta^{\mu \nu}k_{\nu}^{(s)}}
\nonumber \\
&&
~~~~
\times 
\int 
   d\tau^{(o)} 
\prod_{r=1}^{M+N-1}
   d\nu_r
\left( 
    \frac{\pi}{\alpha' \tau^{(o)}} 
\right)^{\frac{p+1}{2}} 
e^{\frac{26-D}{24}\tau^{(o)}}
\eta 
\left(
     -\frac{\tau^{(o)}}{2\pi i} 
\right)^{-D+2} 
\times 
e^{\frac{ K_{\mu}(\theta G \theta)^{\mu \nu}K_{\nu}}
        {4\alpha' \tau^{(o)}} }
\nonumber \\
&&
~~~~~~~
\times 
\prod_{r=1}^M
\prod_{s=M+1}^{M+N}
\exp 
\left[ 
  2\alpha' G^{\mu \nu}
  \left( 
     k_{\mu}^{(r)}
      -
     a_rA_{\mu}(k^{(r)})\frac{\partial}{\partial \nu_r}
  \right)
  \left( 
     k_{\nu}^{(s)}
      +
     a_sA_{\nu}(k^{(s)})\frac{\partial}{\partial \nu_s}
  \right)
\right. 
\nonumber \\
&&
~~~~~~~~~~~~~~~~~~~~~~~~~~~   
\left.
\times   
\ln 
\left\{
  e^{-\frac{(\nu_s-\nu_r)^2}{2\tau^{(0)}}}
       \eta 
           \left( \frac{-\tau^{(o)}}{2\pi i} \right)^{-3}
       \theta_1 
           \left( \frac{(\nu_s+i\pi) -\nu_r}{2\pi i} \right| 
           \left. \frac{-\tau^{(o)}}{2\pi i} \right) 
\right\}
\right]
\nonumber \\
&&
~~~~~~~
\times 
\prod_{r=1}^M
\exp 
\left\{
 -a_r\alpha'k_{\mu}^{(r)}G^{\mu \nu}A_{\nu}(k^{(r)})
\right\}
\prod_{r=1}^M
\exp 
\left\{
    \frac{i}{\tau^{(o)}}
      K_{\mu}\theta^{\mu \nu}
             \left( 
                 k_{\nu}^{(r)}\nu_r-a_rA_{\nu}(k^{(r)})
             \right)
\right\}
\nonumber \\
&&
~~~~~~~
\times
\prod_{1 \leq r < s \leq M}
\exp 
\left[ 
    2\alpha' G^{\mu \nu}
    \left( 
      k_{\mu}^{(r)}
       -
      a_rA_{\mu}(k^{(r)})\frac{\partial}{\partial \nu_r}
    \right)
    \left( 
      k_{\nu}^{(s)}
       -
      a_sA_{\nu}(k^{(s)})\frac{\partial}{\partial \nu_s}
    \right)
\right.
\nonumber \\
&&
~~~~~~~~~~~~~~~~~~~~~~~~~~~  
\left.
\times 
\ln 
\left\{
     ie^{-\frac{(\nu_s-\nu_r)^2}{2\tau^{(0)}}}
       \eta 
           \left( \frac{-\tau^{(o)}}{2\pi i} \right)^{-3}
       \theta_1 
           \left( \frac{\nu_s-\nu_r}{2\pi i} \right| 
           \left. \frac{-\tau^{(o)}}{2\pi i} \right) 
\right\}
\right]
\nonumber \\
&&
~~~~~~~
\times 
\prod_{r=M+1}^{M+N}
\exp 
\left\{
   a_r\alpha'k_{\mu}^{(r)}G^{\mu \nu}A_{\nu}(k^{(r)})
\right\}
\prod_{r=M+1}^{M+N}
\exp 
\left\{
  \frac{i}{\tau^{(o)}}
     K_{\mu}\theta^{\mu \nu}
         \left(
            k_{\nu}^{(r)}\nu_r+a_rA_{\nu}(k^{(r)})
         \right)
\right\}
\nonumber \\
&&
~~~~~~~
\times
\prod_{M+1 \leq r < s \leq M+N}
\exp 
\left[ 
   2\alpha' G^{\mu \nu}
   \left( 
      k_{\mu}^{(r)}
        +
      a_rA_{\mu}(k^{(r)})\frac{\partial}{\partial \nu_r}
   \right)
   \left( 
      k_{\nu}^{(s)}
        +
      a_sA_{\nu}(k^{(s)})\frac{\partial}{\partial \nu_s}
   \right)
\right.
\nonumber \\
&&
~~~~~~~~~~~~~~~~~~~~~~~~~~~~~~~~  
\left. 
\times  
\ln 
\left\{
     ie^{-\frac{(\nu_s-\nu_r)^2}{2\tau^{(0)}}}
       \eta 
           \left( \frac{-\tau^{(o)}}{2\pi i} \right)^{-3}
       \theta_1 
           \left( \frac{\nu_s-\nu_r}{2\pi i} \right| 
           \left. \frac{-\tau^{(o)}}{2\pi i} \right) 
\right\}
\right]~, 
\nonumber \\
\label{aux gluon one-loop by open-string parameters}  
\end{eqnarray}
where the integration of the open-string parameters 
are performed over the region 
(\ref{moduli by open-string parameters}),  
and $K_{\mu}$ is given by 
$K_{\mu}=\sum_{r=1}^Mk_{\mu}^{(r)}$. 
Contribution of the world-sheet reparametrization ghosts 
has been included in the above.

In order to compare the amplitude
(\ref{aux gluon one-loop by open-string parameters}) 
with Eq.(\ref{aux gluon amplitude by boundary state}) 
we map the cylinder drawn on the $u$-plane 
(Figure \ref{open-string diagram}) 
to the cylinder on the $v$-plane 
(Figure \ref{closed-string diagram}) 
by conformal transformation 
$v=2\pi u/\tau^{(o)}$.  
$\nu_r$ are mapped to $\sigma_r$ by 
$\sigma_r=2\pi \nu_r/\tau^{(o)}$. We put 
$\tau^{(c)}\equiv 2\pi i/\tau^{(o)}$. 
These are the closed-string parameters. 
We note that an insertion point of the $(M+N)$-th auxiliary 
gluon vertex operator is fixed at $\sigma_{M+N}=2\pi$.  
By making use of the modular transformations 
(\ref{modular trans of tachyon correlations}) 
the above one-loop amplitude  
can be described as an integral with respect to 
$\tau^{(c)}$ and $\sigma_r$. 
Subsequently we use the relation (\ref{IA by hatIA}). 
An integral form of the $M+N$ gluon scattering amplitude 
is obtained in terms of the closed-string parameters.  
This turns out to be as follows~:
\begin{eqnarray}
&&
I_{A}
\left( 
    \left( k^{(1)},\cdots,k^{(M)} \right) ; 
    \left( k^{(M+1)},\cdots,k^{(M+N)}\right)
\right)
\nonumber \\
&&
=
\left(
\frac{-\mbox{det}G_{\mu \nu}}{(2\alpha')^{p+1}}
\right)^{1/2}
\delta^{(p+1)}
     \left(\sum_{r=1}^{M+N}k^{(r)}
     \right)
\nonumber \\
&&
~~
\times 
\prod_{1 \leq r < s \leq M}
e^{-\frac{i}{2}
          k_{\mu}^{(r)}\theta^{\mu \nu}k_{\nu}^{(s)}}
\prod_{M+1 \leq r < s \leq M+N}
e^{\frac{i}{2} 
         k_{\mu}^{(r)}\theta^{\mu \nu}k_{\nu}^{(s)}}
\nonumber \\
&&
~~
\times 
(-2\pi i)
\left.
\prod_{r=1}^{M+N}
\frac{\partial}{\partial a_r}
\right|_{a_r=0}
\int_{i0}^{i \infty} 
d\tau^{(c)} 
\int 
\prod_{r=1}^{M+N-1}
d\sigma_r~
  e^{\frac{26-D}{24}\frac{2\pi i}{\tau^{(c)}}}
  \left( -i\tau^{(c)} \right)^{-\frac{d}{2}}
  \eta (\tau^{(c)})^{-D+2} 
\nonumber \\
&&
~~~~
\times 
e^{\frac{\tau^{(c)}}{8\pi i\alpha'}
     K_{\mu}(\theta G \theta)^{\mu \nu}K_{\nu}} 
\times  
\prod_{r=1}^{M+N}
  (-i\tau^{(c)})^{\alpha'k_{\mu}^{(r)}G^{\mu \nu}k_{\nu}^{(r)}}
\nonumber \\
&&
~~~~
\times 
\prod_{r=1}^M
\prod_{s=M+1}^{M+N}
\exp 
\left[ 
  2\alpha' G^{\mu \nu}
  \left( 
     k_{\mu}^{(r)}
      -
     a_rA_{\mu}(k^{(r)})\frac{\partial}{\partial \sigma_r}
  \right)
  \left( 
     k_{\nu}^{(s)}
      +
     a_sA_{\nu}(k^{(s)})\frac{\partial}{\partial \sigma_s}
  \right)
\right. 
\nonumber \\
&&
~~~~~~~~~~~~~~~~~~~~~~~~~~~~~~~~   
\left.
\times   
\ln 
\left\{
       \eta 
           \left( \tau^{(c)} \right)^{-3}
       \theta_4
           \left.\left( \frac{\sigma_s -\sigma_r}{2\pi i} \right| 
                  \tau^{(c)} \right)
\right\}
\right]
\nonumber \\
&&
~~~~
\times 
\prod_{r=1}^M
\exp 
\left\{
   -\frac{a_r}{(-i\tau^{(c)})}
   \alpha'k_{\mu}^{(r)}G^{\mu \nu}A_{\nu}(k^{(r)})
\right\}
\prod_{r=1}^M
\exp 
\left\{ 
 \frac{i}{2\pi}K_{\mu}\theta^{\mu \nu}
   \left(
      k_{\nu}^{(r)}\sigma_r-a_rA_{\nu}(k^{(r)})
   \right)
\right\}
\nonumber \\
&&
~~~~
\times
\prod_{1 \leq r < s \leq M}
\exp 
\left[ 
    2\alpha' G^{\mu \nu}
    \left( 
      k_{\mu}^{(r)}
       -
      a_rA_{\mu}(k^{(r)})\frac{\partial}{\partial \sigma_r}
    \right)
    \left( 
      k_{\nu}^{(s)}
       -
      a_sA_{\nu}(k^{(s)})\frac{\partial}{\partial \sigma_s}
    \right)
\right.
\nonumber \\
&&
~~~~~~~~~~~~~~~~~~~~~~~~~~~~~~~  
\left.
\times 
\ln 
\left\{
       \eta 
           \left( \tau^{(c)} \right)^{-3}
       \theta_1 
           \left.\left( \frac{\sigma_s-\sigma_r}{2\pi} \right| 
            \tau^{(c)} \right) 
\right\}
\right]
\nonumber \\
&&
~~~~
\times 
\prod_{r=M+1}^{M+N}
\exp 
\left\{
  \frac{a_r}{(-i\tau^{(c)})}
  \alpha'k_{\mu}^{(r)}G^{\mu \nu}A_{\nu}(k^{(r)})
\right\}
\prod_{r=M+1}^{M+N}
\exp 
\left\{ 
 \frac{i}{2\pi}K_{\mu}\theta^{\mu \nu} 
   \left(
       k_{\nu}^{(r)}\sigma_r+a_rA_{\nu}(k^{(r)})
   \right)
\right\}
\nonumber \\
&&
~~~~
\times
\prod_{M+1 \leq r < s \leq M+N}
\exp 
\left[ 
    2\alpha' G^{\mu \nu}
    \left( 
      k_{\mu}^{(r)}
        +
      a_rA_{\mu}(k^{(r)})\frac{\partial}{\partial \sigma_r}
    \right)
    \left( 
      k_{\nu}^{(s)}
        +
      a_sA_{\nu}(k^{(s)})\frac{\partial}{\partial \sigma_s}
    \right)
\right.
\nonumber \\
&&
~~~~~~~~~~~~~~~~~~~~~~~~~~~~~~~~~~~~  
\left.
\times 
\ln 
\left\{
       \eta 
           \left( \tau^{(c)} \right)^{-3}
       \theta_1 
           \left.\left( \frac{\sigma_s-\sigma_r}{2\pi} \right| 
            \tau^{(c)} \right) 
\right\}
\right], 
\nonumber \\
\label{aux gluon one-loop by closed-string parameters}  
\end{eqnarray}
where the integration of the closed-string parameters are 
performed over the region 
(\ref{moduli by closed-string parameters}). 
And we have rescaled the auxiliary parameters $a_r$ to 
$(-i\tau^{(c)})a_r$.

The above integral form allows us to express the one-loop 
amplitude of $M+N$ gluons by using the closed-string amplitudes  
given in (\ref{aux gluon amplitude by boundary state}). 
Including the contribution of closed-string propagations  
along the Dirichlet directions we finally obtain the following 
equality :
\begin{eqnarray}
&&
I_{A}
\left( \left( k^{(1)},\cdots,k^{(M)} \right) ; 
       \left( k^{(M+1)},\cdots,k^{(M+N)} \right)
\right)
\nonumber \\
&&
=
2\pi i
(-)^{N+1}
\prod_{r=1}^{M+N}
\left.
\frac{\partial}{\partial a_r}
\right|_{a_r=0}
\int_{i0}^{i \infty} 
d\tau^{(c)}  
~e^{
     2\pi i
        \left( 
         \tau^{(c)}+\frac{1}{\tau^{(c)}}
        \right) 
        \frac{26-D}{24}
    }
\prod_{r=1}^{M+N}
(-i \tau^{(c)})
^{\alpha'k_{\mu}^{(r)}G^{\mu \nu}k_{\nu}^{(r)}}
\nonumber \\
&&
~~~~~~~~~~
\times 
\prod_{r=1}^M
\exp 
\left\{
 - \frac{a_r}{(-i\tau^{(c)})}
   \alpha'k_{\mu}^{(r)}G^{\mu \nu}A_{\nu}(k^{(r)})
\right\}
\prod_{r=M+1}^{M+N}
\exp 
\left\{
   \frac{a_r}{(-i\tau^{(c)})}
   \alpha'k_{\mu}^{(r)}G^{\mu \nu}A_{\nu}(k^{(r)})
\right\}
\nonumber \\
&&
~~~~~~~~~~
\times 
\left[ 
\int \prod_{r=M+1}^{M+N-1}d\sigma_r
\left \langle 
   B_D 
\right | 
\otimes 
\Bigl \langle  
   \hat{B}_N[A] ; \{a_r\}; (\sigma_{M+1},k^{(M+1)}), \cdots, 
                     (\sigma_{M+N},k^{(M+N)})
\Bigr |
\right] 
\nonumber \\
&&
~~~~~~~~~~~~~~~~~~~~~~
\times q_c^{\frac{L_0+\tilde{L}_0-2}{2}} 
\left[ 
\int \prod_{r=1}^{M} d\sigma_r  
\Bigl | 
 \hat{B}_N[A] ; \{a_r\}; (\sigma_{1},k^{(1)}), \cdots, 
                     (\sigma_{M},k^{(M)})     
\Bigr \rangle  
\otimes 
\left |
     B_D 
\right \rangle   
\right]~, 
\nonumber \\
\label{equivalence with boundary state formalism for gluon}
\end{eqnarray}
where the integration is performed over the region  
(\ref{moduli by closed-string parameters}).  
All the additional terms appearing in the RHS
become completely vanishing when the on-shell conditions
(\ref{physical condition of gluon}) are imposed 
on the gluon vertices at the critical dimensions.


\section{UV limit of Non-commutative Gauge Theory}
\label{sec:UV NC gauge}

Low-energy world-volume theory of open-string gluons 
in the presence of a constant $B$ field is 
a $U(1)$ gauge theory on the non-commutative world-volume. 
In this section, 
taking the same route as the previous study  
of the non-commutative scalar field theory, 
we investigate 
the UV behavior of the non-commutative gauge theory. 
Our study in this section is restricted to the case of 
$25$-brane in the critical dimensions. 
We examine two zero-slope limits of the one-loop amplitudes 
of gluons. One is based on the open-string parameters  
and the other is on the closed-string parameters. 
These two limits, 
as we explained in the study of the non-commutative 
scalar field theory, 
are complementary.

We first take a zero-slope limit based on  
the open-string parameters $(\tau^{(o)},\nu_r)$. 
Eq.(\ref{aux gluon one-loop by open-string parameters}) 
may be used as an integral form 
of the amplitude. Strictly speaking, the amplitude is 
obtained from this integral by using the 
relation (\ref{IA by hatIA}). 
The zero-slope limit will be a gauge theory one-loop 
amplitude, particularly written in terms of the Schwinger 
parameters, $s^{(o)}$ and $T_r$. 
These parameters are related 
with $\tau^{(o)}$ and $\nu_r$ by the relations 
(\ref{Schwinger parameters of scalar field theory}).
The auxiliary parameters $a_r$ are also 
used in the amplitude 
(\ref{aux gluon one-loop by open-string parameters}). 
They are introduced in order to describe 
the gluon vertex operators 
(\ref{def of open-string gluon vertex})
in the auxiliary forms  
(\ref{def of open-string aux gluon vertex}) 
and to make the loop calculation tractable. 
At the zero-slope limit the gluon vertex operator  become   
$iA_{\mu}(k)\frac{dx^{\mu}(s)}{ds}e^{ik_{\nu}x^{\nu}(s)}$,  
where $x(s)$ is a world-line 
parametrized by the Schwinger parameter $s$. 
The auxiliary vertex operator is expected to be 
$
e^{i \left(k_{\mu}x^{\mu}(s)
   +
   \kappa A_{\mu}(k)
     \frac{dx^{\mu}(s)}{ds} \right)}
$ 
in the world-line description.  
Here $\kappa$ is an auxiliary parameter 
(a counterpart of $a$ in the field theory). 
A simple dimensional analysis shows that $\kappa$ is dimensionful 
and proportional to $\alpha' a$. 
Therefore the zero-slope limit must be taken by 
fixing the following field theory parameters in the amplitude :
\begin{eqnarray}
s^{(o)}=\alpha'\tau^{(o)},~~~
T_r=\alpha' \nu_r,~~~
\kappa_r=\alpha' a_r. 
\label{Schwinger parameters of gauge theory}
\end{eqnarray}
Simultaneously we also need to 
fix open-string tensors $G_{\mu \nu}$ and $\theta^{\mu \nu}$. 
We rewrite the amplitude 
(\ref{aux gluon one-loop by open-string parameters})  
in terms of the above parameters 
and then pick up the dominant contribution 
of the $\alpha'$-expansion. 
These are parallel 
to what we did in the previous section 
to obtain the zero-slope limit 
(\ref{open-string FTL of one-loop amplitude}) 
of the tachyon amplitude. 
By using the relation (\ref{IA by hatIA}), we find that 
the following integral turns out to be the zero-slope 
limit :
\begin{eqnarray}
&&
I_{A}
\left( \left( k^{(1)},\cdots,k^{(M)} \right) ; 
       \left( k^{(M+1)},\cdots,k^{(M+N)} \right)
\right)
\nonumber \\
&&
\approx 
\pi^{13}
\sqrt{-\mbox{det} G_{\mu \nu}}
\delta^{26} \left( \sum_{r=1}^{M+N}k^{(r)} \right)
\prod_{1 \leq r < s \leq M}
e^{-\frac{i}{2}
          k_{\mu}^{(r)}\theta^{\mu \nu}k_{\nu}^{(s)}}
\prod_{M+1 \leq r < s \leq M+N}
e^{\frac{i}{2} 
         k_{\mu}^{(r)}\theta^{\mu \nu}k_{\nu}^{(s)}}
\nonumber \\
&&
~~
\times 
\prod_{r=1}^{M+N}
\left.
\frac{\partial}{\partial \kappa_r}
\right|_{\kappa_r=0}
\int ds^{(o)} 
(s^{(o)})^{-13}
\exp 
\left \{ 
     \frac{s^{(o)}}{\alpha'}
      +
     \frac{K_{\mu}(\theta G \theta)^{\mu \nu} 
             K_{\nu}}{4 s^{(o)}}
\right \}   
\int 
\prod_{r=1}^{M+N-1} dT_r 
\nonumber \\
&&
~~~~~~~~
\times 
\prod_{r=1}^{M+N}
\exp 
\left\{
   -\epsilon^{(r)}\kappa_r
    k_{\mu}^{(r)}G^{\mu \nu}A_{\nu}(k^{(r)})
\right\}
\nonumber \\
&&
~~~~~~~~
\times 
\prod_{r=1}^{M+N}
\exp
\left\{
    -\frac{i}{s^{(0)}}
         \left( 
              T_rk_{\mu}^{(r)}
              -\epsilon^{(r)}\kappa_rA_{\mu}(k^{(r)})
         \right)
    \theta^{\mu \nu}
     K_{\nu}
\right\}
\nonumber \\
&&
~~~~~~~~
\times 
\prod_{r < s}^{M+N}
\exp 
\left[
  G^{\mu \nu}
  \left( 
    k_{\mu}^{(r)}-\epsilon^{(r)}\kappa_rA_{\mu}(k^{(r)})
    \frac{\partial}{\partial T_r}
  \right)
  \left( 
    k_{\nu}^{(s)}-\epsilon^{(s)}\kappa_sA_{\nu}(k^{(s)})
    \frac{\partial}{\partial T_s}
  \right)
\right.
\nonumber \\
&&
~~~~~~~~~~~~~~~~~~~~~~~~~~~~
\left.
\times
  \left\{
    -\frac{(T_r-T_s)^2}{s^{(o)}}
    +|T_s-T_r|
  \right\}
\right],
\label{open-string FTL of gluon one-loop amplitude}
\end{eqnarray}
where $\epsilon^{(r)}$ take values $\pm 1$ such that 
$+1$ for $1 \leq r \leq M$
and $-1$ for $M+1 \leq r \leq M+N$, 
and the integral is performed over the region 
(\ref{moduli in open-string FTL}).
If we neglect $\kappa_r$ besides their differentiations 
the above zero-slope limit reduces to 
Eq.(\ref{open-string FTL of one-loop amplitude}),  
modulo the factor $\alpha^{'-M-N}$. This power has a simple origin 
in the scaling relations, $\kappa_r = \alpha' a_r$. 
It was discussed previously that 
$\exp \left\{ \frac{s^{(o)}}{\alpha'} \right\}$ 
in the integral 
comes from the Schwinger representation of 
the open-string tachyon propagator,  
$\exp \left\{ -s^{(o)}(p^2-\frac{1}{\alpha'}) \right\}$. 
Presently the same term 
may be interpreted as an IR regularization of the amplitude 
by an analytic continuation of 
$-\frac{1}{\alpha^{'2}}$ 
to a small positive $m^2$.

The above zero-slope limit can be identified 
with the corresponding one-loop amplitude of the 
non-commutative $U(1)$ gauge theory. 
Some related calculation in the gauge theory 
may be found in
\cite{Armoni-Lopez}\cite{one-loop of NCG}
\footnote{
Eq.(\ref{open-string FTL of gluon one-loop amplitude})
may be compared with Eq.(4.9) of \cite{one-loop of NCG}.
The integration measure used in that paper is different
from ours. 
}. 
Similarly to the scalar field theory,  
$
\exp 
\left \{ 
  \frac{K_{\mu}(\theta G \theta)^{\mu \nu}
   K_{\nu}}{4 s^{(o)}}
\right \} 
$ 
in the above integral is understood 
as a UV regularization \cite{UV-IR mixing}
 which depends on the 
external momentum $K$. 
In the gauge theory as well,
this causes the problem of UV-IR mixing 
and makes the field theory 
description at high energy scale difficult.

As explained in Section 
\ref{sec:UV NC scalar}, 
the open- and closed-string parameters 
used in the descriptions  of the amplitude 
are two kinds of coordinates of the moduli 
space of conformal classes of cylinder 
with $M+N$ punctures at the boundaries.
Two ends of the moduli space which are located at 
$\tau^{(o)}=0$ and $\tau^{(o)}=+\infty$ play important 
roles in the zero-slope limits. 
The field theory amplitude 
(\ref{open-string FTL of gluon one-loop amplitude}) 
is obtained by a suitable magnification of the integral 
(\ref{aux gluon one-loop by open-string parameters}) 
on an infinitesimal neighbourhood around 
$\tau^{(o)}=+\infty$. 
We need to focus on the region $s^{(o)} \approx 0$ 
to know the UV behavior of the gauge theory. 
As is the case of the scalar field theory, 
a possible resolution is to take a zero-slope limit 
such that an infinitesimal neighbourhood 
around $\tau^{(o)}=0$ is magnified. 
It becomes effective to use the closed-string parameters
near this end.
The zero-slope limit which we examine is 
essentially the same as that was examined for the open-string 
tachyons. Explicitly we take the limit by  
fixing the following parameters :
\begin{eqnarray}
s^{(c)}=\alpha'|\tau^{(c)}|,~~~
\sigma_r,~~~
a_r.   
\label{gauge theory parameters of closed-string}
\end{eqnarray}
In the limiting process we fix 
open-string tensors $G_{\mu \nu}$ and $\theta^{\mu \nu}$
as well 
to capture the world-volume theory. 
The auxiliary parameters in the above 
are the rescaled ones used in the integral 
(\ref{aux gluon one-loop by closed-string parameters}).  
In contrast with the previous scaling 
we keep them intact. 
This is because the present limit 
is not expected to allow a naive world-line description.
As can be observed in the expression 
(\ref{equivalence with boundary state formalism for gluon}) 
these parameters are nothing but 
the auxiliary parameters used for the description of 
gluon vertex operators in the boundary state 
formalism.

Relevant integral form of the one-loop amplitude is given by 
Eq.(\ref{aux gluon one-loop by closed-string parameters}) 
or equivalently 
Eq.(\ref{equivalence with boundary state formalism for gluon}) 
using the boundary states. 
We slightly generalize the amplitude by dropping out the factors 
$
\prod_{r=1}^{M+N}
$
$
(-i\tau^{(c)})^{\alpha'k_{\mu}^{(r)}G^{\mu \nu}k_{\nu}^{(r)}}
e^{
-\frac{a_r\epsilon^{(r)}}{(-i\tau^{(c)})}
\alpha'k_{\mu}^{(r)}G^{\mu \nu}A_{\nu}(k^{(r)})
}
$
and introduce the following quantity :
\begin{eqnarray}
&&
J_{A}
\left( \left( k^{(1)},\cdots,k^{(M)} \right) ; 
       \left( k^{(M+1)},\cdots,k^{(M+N)} \right)
\right)
\nonumber \\
&&
=
2\pi i (-)^{N+1}
\int_{i0}^{i \infty} 
d\tau^{(c)}  
\left. 
\prod_{r=1}^{M+N}
\frac{\partial}{\partial a_r}
\right|_{a_r=0}
\nonumber \\
&&
~~~~~~~~~~~~~~
\times
\left[ 
\int \prod_{r=M+1}^{M+N-1}d\sigma_r
\left \langle 
   B_D 
\right | 
\otimes 
\Bigl \langle  
   \hat{B}_N[A] ; \{a_r\}; 
       (\sigma_{M+1},k^{(M+1)}), 
             \cdots, 
       (\sigma_{M+N},k^{(M+N)})        
\Bigr |
\right] 
\nonumber \\
&&
~~~~~~~~~~~~~~~
\times q_c^{\frac{L_0+\tilde{L}_0-2}{2}} 
\left[ 
\int \prod_{r=1}^{M} d\sigma_r  
\Bigl | 
     \hat{B}_N[A] ; \{a_r\}; (\sigma_{1},k^{(1)}), 
                    \cdots, 
                     (\sigma_{M},k^{(M)})     
\Bigr \rangle  
\otimes 
\left |
     B_D 
\right \rangle   
\right]~, 
\nonumber \\
\label{def of JA}
\end{eqnarray}
where the integration is performed on 
the region (\ref{moduli by closed-string parameters}). 
The $U(1)$ symmetry of the one-loop diagram is 
fixed by putting $\sigma_{M+N}=2\pi$. 
When the gluon vertices satisfy the on-shell conditions 
the above $J_A$ coincides with the original 
amplitude. It is an off-shell generalization 
slightly different from the original one.

The zero-slope limit can be obtained by a calculation 
parallel to that of open-string tachyons. 
The operations 
$
\left. \prod_{r=1}^{M+N}
\frac{\partial}{\partial a_r}
\right|_{a_r=0}
$
are carried out 
on the dominant term of the $\alpha'$-expansion
without difficulty. 
The following integral turns out to be the zero-slope limit :
\begin{eqnarray}
&&
J_{A}
\left( \left( k^{(1)},\cdots,k^{(M)} \right) ; 
       \left( k^{(M+1)},\cdots,k^{(M+N)} \right)
\right)
\nonumber \\
&&
\approx 
\frac{2\pi}{\alpha'}
\delta^{26} 
  \left(
       \sum_{r=1}^{M+N}k^{(r)}
  \right) 
\int ds^{(c)}
\exp \left \{ -\frac{\pi s^{(c)}}{2}
        \left( 
        K_{\mu}g^{\mu \nu}K_{\nu}-\frac{4}{\alpha'}
        \right) 
     \right \}
\nonumber \\
&&
~~
\times 
\left[ 
    \left( 
       \frac{-\mbox{det}G_{\mu \nu}}{(2\alpha')^{26}}
    \right)^{1/4}
    \int \prod_{r=1}^{M} d\sigma_r
        \prod_{1 \leq r < s \leq M}
            e^{-\frac{i}{2}k_{\mu}^{(r)}
                 \theta^{\mu \nu}k_{\nu}^{(s)}}
\right.
\nonumber \\
&&
~~~~~~~~~~~~~~~~~~~~~~~~~~~~~~~
\times 
\left.
        \prod_{r=1}^M
           \left\{
                  -\frac{i}{2\pi}K_{\mu}\theta^{\mu \nu}
                  A_{\nu}(k^{(r)})
                e^{-ik_{\mu}^{(r)}\theta^{\mu \nu}K_{\nu}
                   \frac{\sigma_r}{2\pi}}
           \right\}
\right]
\nonumber \\
&&
~~
\times 
\left[
    \left( 
       \frac{-\mbox{det}G_{\mu \nu}}{(2\alpha')^{26}}
    \right)^{1/4}
    \int \prod_{r=M+1}^{M+N-1} d\sigma_r
          \prod_{M+1 \leq r < s \leq M+N}
              e^{\frac{i}{2}k_{\mu}^{(r)}
                    \theta^{\mu \nu}k_{\nu}^{(s)}}
\right.
\nonumber \\
&&
~~~~~~~~~~~~~~~~~~~~~~~~~~~~~~~
\times
\left.
          \prod_{r=M+1}^{M+N}
            \left\{
               \frac{i}{2\pi}K_{\mu}\theta^{\mu \nu}
                   A_{\nu}(k^{(r)})
                e^{-ik_{\mu}^{(r)}\theta^{\mu \nu}K_{\nu}
                    \frac{\sigma_r}{2\pi}}
            \right\}
\right]~,
\label{closed-string FTL of one-loop amplitude for gluon}
\end{eqnarray}
where the integral is performed over the region 
(\ref{moduli in closed-string FTL}) and the closed-string 
metric is understood as 
$
g^{\mu \nu}
\approx 
-\frac{(\theta G \theta)^{\mu \nu}}{(2\pi \alpha')^2}
$.

As was pointed out in the previous study of the scalar field theory,  
the zero-slope limit based on the closed-string parameters 
describes physics at the trans-string scale of the world-volume theory. 
All the perturbative stringy states of open-string 
contribute to the limit. 
They bring about a striking contrast between the two limits.
We can find the propagator of closed-string tachyon of momentum 
$K$ in the limit 
(\ref{closed-string FTL of one-loop amplitude for gluon})
while the counterpart in  
(\ref{open-string FTL of gluon one-loop amplitude})  
is the curious regularization factor. 
In that limit there appear terms 
describing correlations between 
two gluons inserted at $T_r$ and $T_s$. 
These correlations are caused by the kinetic energies 
of the propagating gluons. 
But they are lost in 
(\ref{closed-string FTL of one-loop amplitude for gluon})  
and gluons become topological in this limit.
The disappearance originates in the modular 
transforms (\ref{modular trans of tachyon correlations}).

It is also possible to rewrite the limit 
(\ref{closed-string FTL of one-loop amplitude for gluon}) 
in terms of the Moyal products. 
The translation can be made in the same manner 
as we translated the tachyon amplitude 
(\ref{closed-string FTL of one-loop amplitude}) into 
(\ref{closed-string FTL by Moyal products}). 
We obtain
\begin{eqnarray} 
&&
\int \prod_{r=1}^{M+N}
\frac{dk^{(r)}}{(2\pi)^{13}}
(-)^{M+N}
J_A
\left(
\left( k^{(1)},\cdots,k^{(M)}
\right)
;
\left(-k^{(M+1)},\cdots,-k^{(M+N)}
\right)
\right)
\nonumber \\
&&
\approx 
\frac{2\pi}{\alpha'}
\int dp_0 
\left[
 \int_{0}^{\infty}ds^{(c)}
    \exp 
     \left\{ -\frac{\pi s^{(c)}}{2}
        \left( 
           p_{0 \mu}g^{\mu \nu}p_{0 \nu}-\frac{4}{\alpha'}
        \right)
     \right\}
\right]
\nonumber \\
&&
~~
\times 
\left( 
     \frac{-\mbox{det} G_{\mu \nu}}
          {(2\alpha')^{26}}
\right)^{\frac{1}{4}}
\int 
\frac{dx}{(2\pi)^{26}}
    e^{ip_{0 \mu}x^{\mu}} \star
\left[
  \int 
    \prod_{r=1}^{M}d\sigma_r 
        \left\{ 
               i {\cal A}_{\mu}(x_0(\sigma_1))
               \frac{dx_0^{\mu}(\sigma_1)}{d\sigma_1}
        \right\}
             \star \cdots 
\right.
\nonumber \\
&&
~~~~~~~~~~~~~~~~~~~~~~~~~~~~~~~~~~~~~~~~~~~~~~~~~~~~~~~~~
\left.
\cdots \star 
        \left\{
               i {\cal A}_{\mu}(x_0(\sigma_M))
               \frac{dx_0^{\mu}(\sigma_M)}{d\sigma_M}
        \right\}
\right]
\nonumber \\
&&
~~
\times  
  \left( 
     \frac{-\mbox{det} G_{\mu \nu}}
          {(2\alpha')^{26}}
  \right)^{\frac{1}{4}}
\overline{
\int 
\frac{dx}{(2\pi)^{26}}
    e^{ip_{0 \mu}x^{\mu}} \star
\left[
  \int 
      \prod_{r=M+1}^{M+N-1}d\sigma_r 
           \left\{ 
               i {\cal A}_{\mu}(x_0(\sigma_{M+1}))
               \frac{dx_0^{\mu}(\sigma_{M+1})}{d\sigma_{M+1}}
            \right\}
                  \star  \cdots 
\right.}
\nonumber \\
&&
~~~~~~~~~~~~~~~~~~~~~~~~~~~~~~~~~~~~~~~~~~~~~~~~~~~~~~~~~
\overline{
\left.
\cdots \star 
           \left\{
               i {\cal A}_{\mu}(x_0(\sigma_{M+N}))
               \frac{dx_0^{\mu}(\sigma_{M+N})}{d\sigma_{M+N}}
            \right\}
\right]
}, 
\nonumber\\
\label{closed-string FTL for gluon by Moyal products}
\end{eqnarray}
where the integrations on $\sigma_r$ are performed over the 
region (\ref{moduli in closed-string FTL}). 
And $x_0(\sigma)$ is the straight line 
given by  
$
x^{\mu}_0(\sigma)=
x^{\mu}+\theta^{\mu \nu}p_{0 \nu}\frac{\sigma}{2\pi}
$. 
This allows us to write down 
the generating function of the amplitudes 
in terms of open Wilson lines. 
By summing up the above equation with respect to $M$ and $N$ 
the zero-slope limit becomes as follows :
\begin{eqnarray}
&&
\sum_{M=0}^{\infty}\sum_{N=1}^{\infty}
\int \prod_{r=1}^{M+N}
\frac{dk^{(r)}}{(2\pi)^{13}}
(-)^{M+N}
J_A
\left(
\left( k^{(1)},\cdots,k^{(M)}
\right)
;
\left(-k^{(M+1)},\cdots,-k^{(M+N)}
\right)
\right)
\nonumber \\
&&
\approx 
\frac{2\pi}{\alpha'}
\int dp_0 
\left[
 \int_{0}^{\infty}ds^{(c)}
  \exp 
     \left\{ -\frac{\pi s^{(c)}}{2}
        \left( 
           p_{0 \mu}g^{\mu \nu}p_{0 \nu}-\frac{4}{\alpha'}
        \right)
     \right\}
\right]
\nonumber \\
&&
~~
\times 
  \left( 
     \frac{-\mbox{det} G_{\mu \nu}}
          {(2\alpha')^{26}}
  \right)^{\frac{1}{4}}
\int 
\frac{dx}{(2\pi)^{26}}
    e^{ip_{0 \mu}x^{\mu}}\star
\left[
{\cal P}_{\star}
     e^{ i
         \int_{0}^{2\pi}d\sigma 
         {\cal A}_{\mu}(x_0(\sigma))
          \frac{dx_0^{\mu}}{d\sigma}(\sigma)}
\right]   
\nonumber \\
&&
~~
\times 
  \left( 
     \frac{-\mbox{det} G_{\mu \nu}}
          {(2\alpha')^{26}}
  \right)^{\frac{1}{4}}
\overline{
\int 
\frac{dx}{(2\pi)^{26}}
    e^{ip_{0 \mu}x^{\mu}} \star
\left[
{\cal P}_{\star}
   e^{ i
       \int_{0}^{2\pi}d\sigma 
       {\cal A}_{\mu}(x_0(\sigma))
       \frac{dx_0^{\mu}}{d \sigma}(\sigma)}
\right]
\star 
\left\{
i 
{\cal A}_{\mu}(x_0(2\pi))
\frac{dx_0}{d\sigma}(2\pi)  
\right\}
}. 
\nonumber\\
\label{closed-string FTL for gluon by straight open Wilson lines}
\end{eqnarray}
The last term, which makes the equation asymmetric, 
appears owing to fixing the $U(1)$ symmetry of 
the open-string diagrams. Modulo this asymmetry 
the generating function is factorized at the zero-slope limit 
into a sum of products of straight open Wilson lines. 
Open Wilson lines with the same velocity 
$\frac{\theta^{\mu \nu}p_{0 \nu}}{2\pi}$ 
interact with each other 
by exchanging closed-string tachyon with momentum $p_{0 \mu}$.

\section{Open Wilson Lines in Closed-String Theory (II)}
\label{sec:open Wilson lines (II)}

\subsection{Straight open Wilson line and closed-string tachyon}
\label{sec:straight OWL}

It was shown in \cite{Okawa-Ooguri}
disk amplitudes of a closed-string tachyon
scattering with arbitrary number of gluons 
turn out to generate 
a {\it straight\/} open Wilson line
in Seiberg-Witten's zero-slope limit.
The path is a straight line
connecting $x^{\mu}$ and $x^{\mu}+\theta^{\mu\nu}p_{0\nu}$, 
where $p_0$ is the momentum of the tachyon. 
The displacement $\theta^{\mu \nu}p_{0 \nu}$ is 
required by the gauge invariance 
\cite{IIKK}. 
In this section we first reproduce the above result 
by using the boundary states constructed in
Section \ref{sec:gluon boundary state}. 
Computations provided below 
become also helpful for our subsequent 
investigations of generically curved open Wilson lines.

Closed-string tachyon state with momentum $p_{0\mu}$
and its BPZ dual state are given by 
\begin{equation}
|p_{0N}\rangle
=\lim_{\tau \rightarrow -\infty}V_{T}(\sigma,\tau;p_{0})
 |\mathbf{0}\rangle =e^{ip_{0\mu}\hat{x}_{0}^{\mu}}
  |\mathbf{0}\rangle~,
\quad
\langle -p_{0N}| =\langle \mathbf{0}| e^{ip_{0\mu}\hat{x}^{\mu}_{0}}~.
\label{eq:closed tachyon state}
\end{equation}
Disk amplitude of the closed-string tachyon scattering with
$M$ gluons is obtained in the boundary state formalism 
by integrating the overlap 
$\left\langle -p_{0N} \left|
  B_{N}[A]; (\sigma_{1},k^{(1)}),\cdots, (\sigma_{M},k^{(M)})
  \right\rangle\right.$
on its moduli space.
To evaluate this overlap, 
we start by computing the overlap 
with the auxiliary boundary states. 
This becomes
\begin{eqnarray}
\lefteqn{
\left\langle -p_{0N} \left|
  \hat{B}_{N}[A];\{a_r\};
   (\sigma_{1},k^{(1)}),\cdots, (\sigma_{M},k^{(M)})
  \right\rangle\right.
}\nonumber\\
&& = \left(\frac{-\det G_{\mu\nu}}
{(2\alpha')^{p+1}}\right)^{\frac{1}{4}}
\,
\prod_{r<s}^{M}
e^{\frac{i}{2}k_{\mu}^{(r)}\theta^{\mu\nu}k_{\nu}^{(s)}
                  \epsilon (\sigma_{r}-\sigma_{s})}
\delta^{(p+1)}
    \left(p_{0}+\sum_{r=1}^{M}k^{(r)} \right) \nonumber\\
&& 
\quad \times
\prod_{r<s}^{M} 
\exp 
\left[ 2\alpha' G^{\mu\nu}
     \left(k_{\mu}^{(r)}-a_r A_{\mu}(k^{(r)})
            \partial_{\sigma_{r}}  \right)
     \left(k_{\nu}^{(s)}-a_s A_{\nu}(k^{(s)})
            \partial_{\sigma_{s}}
     \right) \ln \left|e^{i\sigma_{r}}-e^{i\sigma_{s}}
                 \right| \right]\nonumber\\
&& \quad \times
\exp \left[-\frac{i}{2\pi}\sum_{r=1}^{M}
        \left(a_rA_{\mu}(k^{(r)})
        -\sigma_{r} k_{\mu}^{(r)}\right)
        \theta^{\mu\nu}p_{0\nu}\right]~.
 \label{eq:overlap with tachyon 1}
\end{eqnarray}
The zero-slope limit can be read from the RHS as follows :
\begin{eqnarray}
\lefteqn{
  \left\langle -p_{0N}\left|
  \hat{B}_{N}[A];\{a_r\}; (\sigma_{1},k^{(1)}),
   \cdots,(\sigma_{M},k^{(M)})
  \right\rangle  \right.
}\nonumber\\
&& \approx 
\left(\frac{-\det G_{\mu\nu}}
    {(2\alpha')^{p+1}}\right)^{\frac{1}{4}}
\,
\prod_{r<s}^{M}
e^{\frac{i}{2}k_{\mu}^{(r)}\theta^{\mu\nu}k_{\nu}^{(s)}
                  \epsilon (\sigma_{r}-\sigma_{s})}
\delta^{(p+1)}
    \left(p_{0}+\sum_{r=1}^{M}k^{(r)} \right) \nonumber\\
&& \quad \times
\exp \left[-\frac{i}{2\pi}
         \sum_{r=1}^{M}\left(a_rA_{\mu}(k^{(r)})
              -\sigma_{r} k_{\mu}^{(r)}\right)
        \theta^{\mu\nu}p_{0\nu}\right]~.
 \label{eq:overlap with tachyon 2}
\end{eqnarray}
This gives 
\begin{eqnarray}
&&
  \left\langle -p_{0N}\left|
  B_{N}[A];(\sigma_{1},k^{(1)}),\cdots,(\sigma_{M},k^{(M)})
  \right\rangle \right.\nonumber\\
&& \quad
= \left.
\prod_{r=1}^{M} 
    \left( i \frac{\partial}{\partial a_r}\right)
\right|_{a_r=0}
  \left\langle -p_{0N}\left|
   \hat{B}_{N}[A];\{ a_r\};(\sigma_{1},k^{(1)}),
         \cdots,(\sigma_{M},k^{(M)})
   \right\rangle  \right. 
\nonumber\\
&&\quad 
\approx
\left(\frac{-\det G_{\mu\nu}}{(2\alpha')^{p+1}}
        \right)^{\frac{1}{4}}
\,
\prod_{r<s}^{M}
e^{\frac{i}{2}k_{\mu}^{(r)}\theta^{\mu\nu}k_{\nu}^{(s)}
                  \epsilon (\sigma_{r}-\sigma_{s})}
\delta^{(p+1)}
    \left(p_{0}+\sum_{r=1}^{M}k^{(r)} \right)
\nonumber\\
&& \qquad \quad \times
  \prod_{r=1}^{M} \left(
   \frac{dy_{0}^{\mu} (\sigma_{r})}{d\sigma_{r}}
   A_{\mu}(k^{(r)}) e^{ik_{\nu}^{(r)} y^{\nu}_{0} (\sigma_{r})}
   \right) \nonumber\\
&&\quad
 = \left(\frac{-\det G_{\mu\nu}}{(2\alpha')^{p+1}}\right)^{\frac{1}{4}}
\, \int \frac{d^{p+1}x}{(2\pi)^{p+1}}
 e^{ip_{0\mu}x^{\mu}}
 e^{-\frac{i}{2}(k_{\mu}^{(1)}+\cdots +k^{(M)}_{\mu})
        \theta^{\mu\nu}p_{0\nu}}
 \prod_{r<s}^{M} e^{\frac{i}{2}k_{\mu}^{(r)}\theta^{\mu\nu}k_{\nu}^{(s)}
                  \epsilon (\sigma_{r}-\sigma_{s})} \nonumber\\
&& \qquad \quad \times
  \prod_{r=1}^{M} \left\{
   \frac{dy_{0}^{\mu} (\sigma_{r})}{d\sigma_{r}}
   A_{\mu}(k^{(r)})
   e^{ik_{\nu}^{(r)}\left(x^{\nu}+ y^{\nu}_{0} (\sigma_{r})\right)}
   \right\}~,
\label{eq:overlap with tachyon 3}
\end{eqnarray}
where $y_{0}^{\mu} (\sigma)$ $(0\leq \sigma \leq 2\pi)$
is a straight line defined as
\begin{equation}
y_{0}^{\mu}(\sigma) 
= \theta^{\mu\nu}p_{0\nu} \frac{\sigma}{2\pi}~. 
\label{eq:straight-line}
\end{equation}
In the above 
we have used the following equations to obtain 
the last equality :
\begin{eqnarray}
\delta^{(p+1)}
    \left(p_{0}+\sum_{r=1}^{M}k^{(r)} \right)
  &=& \delta^{(p+1)}
    \left(p_{0}+\sum_{r=1}^{M}k^{(r)} \right)
    \, e^{-\frac{i}{2}(k_{\mu}^{(1)}+\cdots+k_{\mu}^{(M)})
             \theta^{\mu\nu} p_{0\nu} }\nonumber\\
&=& \int \frac{d^{p+1}x}{(2\pi)^{p+1}}
    e^{i(k_{\mu}^{(1)}+\cdots+k_{\mu}^{(M)})x^{\mu}}
    e^{ip_{0\mu}x^{\mu}}
    e^{-\frac{i}{2}(k_{\mu}^{(1)}+\cdots+k_{\mu}^{(M)})
             \theta^{\mu\nu} p_{0\nu} }~.
  \label{eq:rearrangement of delta}
\end{eqnarray}

The relevant moduli parameters are $\sigma_r$ with 
$0 \leq \sigma_1$ 
$\leq \cdots \leq$ 
$\sigma_M \leq 2\pi$. 
Their integration will give us the amplitude. 
Eq.(\ref{eq:overlap with tachyon 3}) yields
\begin{eqnarray}
\lefteqn{
\int_{0\leq \sigma_{1}\leq \sigma_{2} 
         \leq \cdots \leq \sigma_{M}
       \leq 2\pi} \prod_{r=1}^{M} d\sigma_{r} \;
   \left\langle -p_{0N}\left|
    B_{N}[A];(\sigma_{1},k^{(1)}),\cdots,(\sigma_{M},k^{(M)})
    \right\rangle
   \right. }\nonumber\\
&& \approx \left(\frac{-\det G_{\mu\nu}}{(2\alpha')^{p+1}}
                 \right)^{\frac{1}{4}}
   \int \frac{d^{p+1}x}{(2\pi)^{p+1}}
   \int_{0\leq \sigma_{1}\leq \sigma_{2} 
              \leq \cdots \leq \sigma_{M}
       \leq 2\pi} \prod_{r=1}^{M} d\sigma_{r}
  \nonumber\\
&& \qquad
  \left\{
   \frac{dy_{0}^{\mu} (\sigma_{1})}{d\sigma_{r}}
   A_{\mu}(k^{(1)})
   e^{ik_{\nu}^{(1)}
      \left(x^{\nu}+ y^{\nu}_{0} (\sigma_{1})\right)}
   \right\} \star
  \left\{
   \frac{dy_{0}^{\mu} (\sigma_{2})}{d\sigma_{r}}
   A_{\mu}(k^{(2)})
   e^{ik_{\nu}^{(2)}
         \left(x^{\nu}+ y^{\nu}_{0} (\sigma_{2})\right)}
   \right\} \star \cdots\nonumber\\
 && \qquad \qquad \qquad
    \cdots \star
    \left\{
   \frac{dy_{0}^{\mu} (\sigma_{M})}{d\sigma_{r}}
   A_{\mu}(k^{(M)})
   e^{ik_{\nu}^{(M)}
      \left(x^{\nu}+ y^{\nu}_{0} (\sigma_{M})\right)}
   \right\}
   \star e^{ip_{0\mu}x^{\mu}}~,
   \label{eq:overlap with tachyon 4}
\end{eqnarray}
where the Moyal products are taken with respect to $x$.
The straight open Wilson line is obtained 
from the above equation by summing up with respect to $M$ 
as follows :
\begin{eqnarray}
&&\sum_{M=0}^{\infty} i^{M}
  \int \prod_{r=1}^{M} \frac{d^{p+1}k^{(r)}}{(2\pi)^{\frac{p+1}{2}}}
  \int_{0\leq \sigma_{1}\leq \sigma_{2} \leq \cdots \leq \sigma_{M}
       \leq 2\pi} \prod_{r=1}^{M} d\sigma_{r} \;
   \left\langle -p_{0N}\left| 
   B_{N}[A];(\sigma_{1},k^{(1)}),\cdots,(\sigma_{M},k^{(M)})
   \right\rangle \right. \nonumber\\
 && \approx \left(\frac{-\det G_{\mu\nu}}{(2\alpha')^{p+1}}
            \right)^{\frac{1}{4}}
     \int \frac{d^{p+1}x}{(2\pi)^{p+1}}
     \left[
     \mathcal{P}_{\star}
     \exp \left( i \int_{0}^{2\pi} d\sigma 
     \frac{dy^{\mu}_{0}(\sigma)}{d\sigma} \mathcal{A}_{\mu}
     \left(x+y_{0}(\sigma)\right) \right) \right]
     \star e^{ip_{0\mu}\hat{x}^{\mu}}~,
   \label{eq:straight open wilson line}
\end{eqnarray}
where the path is taken along the straight line 
$x_{0}^{\mu}(\sigma)\equiv x^{\mu}+y^{\mu}_{0}(\sigma)$.

\subsection{Curved open Wilson lines and closed-string states}
\label{sec:curved OWL}

Taking account of the fact that 
closed-string tachyons give rise to straight open Wilson 
lines, Dhar and Kitazawa suggested in 
\cite{Dhar-Kitazawa} that 
fluctuations of open Wilson lines  
should originate in the massive states of closed-string. 
They conjectured 
a possible correspondence between the perturbative massive 
states of closed-string and the gauge theory operators 
obtained as the coefficients in a perturbative expansion 
of open Wilson line 
(the harmonic expansion at the straight line). 
In this subsection we prove their conjecture.
We show, in a self-contained fashion, 
how one can obtain curved open Wilson lines and 
present an explicit correspondence between
their fluctuations and the closed-string states.

Let $P_{\mu}(\sigma)$ be a loop in the momentum space. 
The harmonic expansion is given by 
$P_{\mu}(\sigma)=\frac{1}{2\pi}$ 
$\left[
  p_{0\mu}+\frac{1}{\sqrt{2}} \sum_{n=1}^{\infty}
   \left(\psi_{n\mu}e^{-in\sigma}+\bar{\psi}_{n\mu}
          e^{in\sigma}\right)\right]$.
We first introduce the following 
{\it out\/}-state of closed-string : 
\begin{equation}
\left\langle \Omega (\psi_{n},\bar{\psi}_{n};p_{0})\right|
\equiv \langle \mathbf{0} |
: \exp \left(i\int^{2\pi}_{0} d\sigma P_{\mu}(\sigma)
       \hat{X}^{\mu}(\sigma) \right):~.
   \label{eq:momentum loop state}
\end{equation}
One may think of this state 
as a stringy extension, 
(or a generalization to include the massive modes),
of the closed-string tachyon state.
The oscillator representation can be read as 
\begin{equation}
\left\langle \Omega (\psi_{n},\bar{\psi}_{n};p_{0})\right|
= \left\langle -p_{0N}\right| \prod_{n=1}^{\infty}
\exp \left[ -\frac{\sqrt{\alpha'}}{2}
    \frac{1}{n}
    \left(\bar{\psi}_{n\mu}\alpha^{\mu}_{n}
          +\psi_{n\mu}\tilde{\alpha}^{\mu}_{n}
     \right)\right]~.  
     \label{eq:oscillator expression of momentum loop state}
\end{equation}
This tells us that we can  write the state in terms of 
the coherent state given in Appendix \ref{sec:formulae} 
as follows : 
\begin{equation}
\left\langle \Omega (\psi_{n},\bar{\psi}_{n};p_{0})\right|
=\left\langle -p_{0N}\right|
 \otimes \prod_{n=1}^{\infty}
 \left( -\frac{\sqrt{\alpha'}}{2} \psi_{n},
        -\frac{\sqrt{\alpha'}}{2} \bar{\psi}_{n}
  \right|~,
  \label{eq:momentum loop and coherent state}
\end{equation}
where 
$
\left( -\frac{\sqrt{\alpha'}}{2} \psi_{n},
        -\frac{\sqrt{\alpha'}}{2} \bar{\psi}_{n}
\right|
$ 
are the coherent states 
$(\lambda^{+}_{n},\lambda_{n}^{-}|$ 
of the $n$-th levels defined in
Eq.(\ref{eq:dual of coherent state}) 
with setting the complex variables 
$
\lambda^{+}_{n}=
 -\frac{\sqrt{\alpha'}}{2} \psi_{n}$ 
and 
$
\lambda_{n}^{-}=
  -\frac{\sqrt{\alpha'}}{2} \bar{\psi}_{n}
$.  
One can readily find that for $n\geq 1$
\begin{eqnarray}
\left\langle \Omega (\psi_{n},\bar{\psi}_{n};p_{0})\right|
 \alpha^{\mu}_{-n}
 &=&-\frac{\sqrt{\alpha'}}{2}g^{\mu\nu}\bar{\psi}_{n\nu}
 \left\langle \Omega (\psi_{n},\bar{\psi}_{n};p_{0})\right|~,
\nonumber\\
\left\langle \Omega (\psi_{n},\bar{\psi}_{n};p_{0})\right|
  \tilde{\alpha}^{\mu}_{-n}
  &=&-\frac{\sqrt{\alpha'}}{2}g^{\mu\nu}\psi_{n\nu}
 \left\langle \Omega (\psi_{n},\bar{\psi}_{n};p_{0})\right|~.
 \label{eq:momentum loop and coherent state 2}
\end{eqnarray}
Recalling $\overline{\psi}_n$ is the complex conjugate 
of $\psi_n$, one can think of the state
$\left\langle \Omega (\psi_{n},\bar{\psi}_{n};p_{0})\right|$
as a real section of the coherent state of closed-string.
Eq.(\ref{eq:oscillator expression of momentum loop state}) 
also yields the following equalities :
\begin{eqnarray}
&& \left.
\prod_{n=1}^{\infty} \prod_{\mu=0}^{p}
 \left( -\frac{2n}{\sqrt{\alpha'}}
        \frac{\partial}{\partial \bar{\psi}_{n\mu}}
 \right)^{m(n,\mu)}
 \left( -\frac{2n}{\sqrt{\alpha'}}
        \frac{\partial}{\partial \psi_{n\mu}}
 \right)^{m'(n,\mu)}
\left\langle \Omega (\psi_{n},\bar{\psi}_{n};p_{0})\right|
\right|_{\psi,\bar{\psi}=0}\nonumber\\
&& \hspace{5em}
  = \langle -p_{0N}|
  \prod_{n=1}^{\infty} \prod_{\mu=0}^{p}
  \left(\alpha_{n}^{\mu}\right)^{m(n,\mu)}
  \left(\tilde{\alpha}_{n}^{\mu}\right)^{m'(n,\mu)}~,
\label{eq:generating function}
\end{eqnarray}
where $m(n,\mu)$ and $m'(n,\mu)$ are integers
greater than or equal to zero. 
This implies that the state
$\left\langle \Omega (\psi_{n},\bar{\psi}_{n};p_{0})\right|$
is a generating function of the closed-string states
which are off-shell in general.

Now 
we wish to compute overlaps 
of the above state,
instead of the tachyon, 
with the boundary states. 
Since it is a generating function 
of the closed-string states,
these overlaps give us 
a generating function of amplitudes 
between closed-string states and gluons.  
We will show in the below that 
the zero-slope limit of this generating 
function is nothing but an open 
Wilson line taken along a curve  
parametrized by
$\psi_{n}$ and $\bar{\psi}_{n}$.

By using Eq.(\ref{eq:momentum loop and coherent state 2}),
we obtain
\begin{eqnarray}
\lefteqn{
\left\langle \Omega (\psi_{n},\bar{\psi}_{n};p_{0})\left|
  \hat{B}_{N}[A];\{a_r\};(\sigma_{1},k^{(1)}),\cdots,(\sigma_{M},k^{(M)})
  \right\rangle\right.}
\nonumber\\
&&= \left(\frac{-\det G_{\mu\nu}}{(2\alpha')^{p+1}}\right)^{\frac{1}{4}}
  \prod_{r<s}^{M}
  e^{\frac{i}{2}k_{\mu}^{(r)}\theta^{\mu\nu}k_{\nu}^{(s)}
     \epsilon (\sigma_{r}-\sigma_{s})}
  \delta^{(p+1)} 
    \left(p_{0} +\sum_{r=1}^{M}k^{(r)}\right)
  \nonumber\\
&& \quad \times \prod_{r<s}^{M}
   \exp \left[2\alpha' G^{\mu\nu}
    \left( k_{\mu}^{(r)}-a_r A_{\mu}(k^{(r)})
          \partial_{\sigma_{r}} \right)
    \left( k_{\nu}^{(s)}-a_s A_{\nu}(k^{(s)})
          \partial_{\sigma_{s}} \right)
    \ln \left|e^{i\sigma_{r}}-e^{i\sigma_{s}}\right|
  \right] \nonumber\\
&& \quad \times \prod_{n=1}^{\infty}
  \exp \left[ -\frac{\alpha'}{4n}\bar{\psi}_{n\mu}
      \left(g^{-1}Ng^{-1}\right)^{\mu\nu}\psi_{n\nu}
      \right] \nonumber\\
&& \quad \times
\exp \left[-i \sum_{r=1}^{M} a_rA_{\mu}(k^{(r)})
   \right.\nonumber\\
&& \hspace{8em} \left. \times
   \left\{
     \frac{\theta^{\mu\nu}}{2\pi}p_{0\nu}+
      \frac{\alpha'}{\sqrt{2}}
       \sum_{n=1}^{\infty}\left(
        \left(\frac{1}{E}\right)^{\mu\nu}\psi_{n\nu}e^{-in\sigma_{r}}
         -\left(\frac{1}{E^{T}}\right)^{\mu\nu} \bar{\psi}_{n\nu}
             e^{in\sigma_{r}}\right)\right\}\right]
    \nonumber\\
&&  \quad \times
 \exp\left[ i\sum_{r=1}^{M}k_{\mu}^{(r)}
   \left\{\frac{\theta^{\mu\nu}}{2\pi}p_{0\nu}\sigma_{r}+
     \frac{\alpha'}{\sqrt{2}} \sum_{n=1}^{\infty} \frac{i}{n}
     \left( \left(\frac{1}{E}\right)^{\mu\nu}\psi_{n\nu} e^{-in\sigma_{r}}
     +\left(\frac{1}{E^{T}}\right)^{\mu\nu}\bar{\psi}_{n\nu}
     e^{in\sigma_{r}}\right) \right\}\right]~.\nonumber\\
   \label{eq:overlap with momentum loop 1}
\end{eqnarray}
The zero-slope limit can be read from the RHS. 
In the present limiting procedure, we have
\begin{equation}
\alpha'\left(\frac{1}{E}\right)^{\mu\nu} \approx
  \frac{\theta^{\mu\nu}}{2\pi}~,\quad
\alpha'\left(\frac{1}{E^{T}}\right)^{\mu\nu} \approx
  -\frac{\theta^{\mu\nu}}{2\pi}~,\quad
\alpha'\left(g^{-1}Ng^{-1}\right)^{\mu\nu}
    \approx \frac{1}{4\pi^{2}\alpha'}
    \left(\theta G \theta\right)^{\mu\nu}~.
\end{equation}
This enables us to find out that
\begin{eqnarray}
&&
\left\langle \Omega (\psi_{n},\bar{\psi}_{n};p_{0})\left|
  \hat{B}_{N}[A];\{a_r\};(\sigma_{1},k^{(1)}),\cdots,(\sigma_{M},k^{(M)})
  \right\rangle\right.
\nonumber\\
&& \approx
  \left(\frac{-\det G_{\mu\nu}}{(2\alpha')^{p+1}}\right)^{\frac{1}{4}}
  \prod_{r<s}^{M}
  e^{\frac{i}{2}k_{\mu}^{(r)}\theta^{\mu\nu}k_{\nu}^{(s)}
     \epsilon (\sigma_{r}-\sigma_{s})}
  \delta^{(p+1)} \left(p_{0} +\sum_{r=1}^{M}k^{(r)}\right)
  \nonumber\\
&& \quad \times\prod_{n=1}^{\infty}\exp \left[
    -\frac{\bar{\psi}_{n\mu}\left(\theta G \theta\right)^{\mu\nu}
           \psi_{n\nu}}
          {16\pi^{2}n\alpha'} \right] 
   \exp \left[ -i\sum_{r=1}^{M}\left\{
       a_r \frac{dy^{\mu}(\sigma_{r})}{d\sigma_{r}}
           A_{\mu}(k^{(r)})
      -k^{(r)}_{\mu} y^{\mu}(\sigma_{r})\right\} \right]~.
\nonumber \\
  \label{eq:overlap with momentum loop 2}
\end{eqnarray}
Here we introduce $y^{\mu}(\sigma)$ with 
\begin{eqnarray}
y^{\mu}(\sigma) = y_{0}^{\mu}(\sigma) + \tilde{y}^{\mu}(\sigma)~,
\label{eq:curved path}
\end{eqnarray}
where 
\begin{eqnarray}
\tilde{y}^{\mu}(\sigma) \equiv 
  \theta^{\mu\nu} \frac{1}{2\pi}\frac{1}{\sqrt{2}}
    \sum_{n=1}^{\infty}\frac{i}{n}
     \left(\psi_{n\nu}e^{-in\sigma}
           -\bar{\psi}_{n\nu} e^{in\sigma}\right)~.
\end{eqnarray}
It is a curve 
deviating from the straight line $y_{0}^{\mu}(\sigma)$ 
and the deviation is denoted by $\tilde{y}^{\mu}(\sigma)$.

It follows from
Eq.(\ref{eq:overlap with momentum loop 2}) that
\begin{eqnarray}
\lefteqn{
\left\langle \Omega (\psi_{n},\bar{\psi}_{n};p_{0})\left|
  B_{N}[A];(\sigma_{1},k^{(1)}),\cdots,(\sigma_{M},k^{(M)})
  \right\rangle\right.
}\nonumber\\
&&= \left.\prod_{r=1}^{M}
\left(i \frac{\partial}{\partial a_r}\right)
\right |_{a_r=0}
\left\langle \Omega (\psi_{n},\bar{\psi}_{n};p_{0})\left|
  \hat{B}_{N}[A];\{a_r\}(\sigma_{1},k^{(1)}),\cdots,(\sigma_{M},k^{(M)})
  \right\rangle\right.
\nonumber\\
&&\approx
  \left(\frac{-\det G_{\mu\nu}}{(2\alpha')^{p+1}}
         \right)^{\frac{1}{4}}
  \prod_{r<s}^{M}
  e^{\frac{i}{2}k_{\mu}^{(r)}\theta^{\mu\nu}k_{\nu}^{(s)}
     \epsilon (\sigma_{r}-\sigma_{s})}
  \delta^{(p+1)} 
        \left(p_{0} +\sum_{r=1}^{M}k^{(r)}\right)
  \nonumber\\
&& \quad \times\prod_{n=1}^{\infty}\exp \left[
    -\frac{\bar{\psi}_{n\mu}\left(\theta G \theta\right)^{\mu\nu}
           \psi_{n\nu}}
          {16\pi^{2}n\alpha'} \right] 
   \prod_{r=1}^{M}
   \left\{
      \frac{dy^{\mu}(\sigma_{r})}{d\sigma_{r}}
           A_{\mu}(k^{(r)})
      e^{ik^{(r)}_{\nu} y^{\nu}(\sigma_{r})} \right\}~.
  \label{eq:overlap with momentum loop 3}
\end{eqnarray}
This takes essentially the same form as
Eq.(\ref{eq:overlap with tachyon 3})
with the straight line $y_{0}^{\mu}(\sigma)$ replaced by
the curved one $y^{\mu}(\sigma)$.
Therefore,
performing the same rearrangement as carried out
in the last subsection,
we obtain
\begin{eqnarray}
&&
\sum_{M=0}^{\infty} i^{M}
\int \prod_{r=1}^{M} \frac{d^{p+1}k^{(r)}}{(2\pi)^{\frac{p+1}{2}}}
\int_{0\leq\sigma_{1} \leq \cdots \leq \sigma_{M} \leq 2\pi}
\prod_{r=1}^{M} d\sigma_{r}
 \nonumber\\
&& \hspace{5em} \times
\left\langle \Omega (\psi_{n},\bar{\psi}_{n};p_{0})\left|
  B_{N}[A];(\sigma_{1},k^{(1)}),\cdots,(\sigma_{M},k^{(M)})
  \right\rangle\right.
  \nonumber\\
&& \approx
  \left(\frac{-\det G_{\mu\nu}}{(2\alpha')^{p+1}}\right)^{\frac{1}{4}}
  \prod_{n=1}^{\infty}\exp \left[
    -\frac{\bar{\psi}_{n\mu}\left(\theta G \theta\right)^{\mu\nu}
           \psi_{n\nu}}
          {16\pi^{2}n\alpha'} \right] \nonumber\\
&& \qquad \quad \times
   \int \frac{d^{p+1}x}{(2\pi)^{p+1}}
   \left[\mathcal{P}_{\star}
    \exp \left( i\int_{0}^{2\pi} d\sigma
    \frac{dy^{\mu}(\sigma)}{d\sigma} \mathcal{A}_{\mu}(x+y(\sigma)
    \right)\right] \star
    e^{ip_{0\mu}x^{\mu}}~.
  \label{eq:overlap with momentum loop 4}
\end{eqnarray}
This is the open Wilson line taken along the curve 
$x^{\mu}(\sigma)\equiv x^{\mu}+y^{\mu}(\sigma)$.

Thus we have shown that the zero-slope limit
of the generating function of the amplitudes 
between closed-string states and gluons becomes 
the open Wilson line multiplied by a Gaussian weight.
The path is curved by $\psi_n$ and $\overline{\psi}_n$. 
As can be seen 
in Eq.(\ref{eq:generating function}),
these variables originally measure condensations 
of the $n$-th massive modes of closed-string.

This completes the proof of the conjecture.

\subsubsection*{Overlap with graviton}
As has been mentioned, the overlap
$\left\langle \Omega (\psi_{n},\bar{\psi}_{n};p_{0})\left|
  B_{N}[A];(\sigma_{1},k^{(1)}),\cdots,(\sigma_{M},k^{(M)})
  \right\rangle\right.$
serves as a generating function of the amplitudes between 
closed-string states and gluons. 
As an illustration, let us consider the case of gravitons.

The graviton states can be written as follows :  
\begin{equation}
\left\langle -p_{0N}\right|h_{\mu\nu}(p_{0})
   \alpha_{1}^{\mu}\tilde{\alpha}^{\nu}_{1}
= \left.  \frac{4}{\alpha'}h_{\mu\nu}(p_{0})
  \frac{\partial^{2}}{\partial \bar{\psi}_{1\mu}\partial \psi_{1\nu}}
  \left\langle \Omega \left(\psi_{n},\bar{\psi}_{n};p_{0}\right)
  \right| \right|_{\psi,\bar{\psi}=0}~,
  \label{eq:graviton out-state}
\end{equation}
where $h_{\mu\nu}(p_{0})$ denotes
the polarization tensor of graviton
which is symmetric and traceless.
Using Eq.(\ref{eq:overlap with momentum loop 3}),
we obtain
\begin{eqnarray}
\lefteqn{
\langle -p_{0N}|\alpha^{\mu}_{1}\tilde{\alpha}_{1}^{\nu}
  \left| B_{N}[A];(\sigma_{1},k^{(1)}),\cdots,(\sigma_{M},k^{(M)})
   \right\rangle}
  \nonumber \\
&&= \left. \frac{4}{\alpha'}
  \frac{\partial^{2}}{\partial \bar{\psi}_{1\mu}\partial \psi_{1\nu}}
  \left\langle \Omega \left(\psi_{n},\bar{\psi}_{n};p_{0}\right)
  \left| B_{N};(\sigma_{1},k^{(1)}),\cdots,(\sigma_{M},k^{(M)})
   \right\rangle\right.
   \right|_{\psi,\bar{\psi}=0} \nonumber\\
&& \approx
   \left(\frac{-\det G_{\mu\nu}}{(2\alpha')^{p+1}}\right)^{\frac{1}{4}}
   \prod_{r=1}^{M} e^{ik_{\mu}^{(r)}\theta^{\mu\nu}p_{0\nu}
                       \frac{\sigma_{r}}{2\pi}}
   \prod_{r<s}^{M}e^{\frac{i}{2}k_{\mu}^{(r)}\theta^{\mu\nu}k_{\nu}^{(s)}
     \epsilon (\sigma_{r}-\sigma_{s})}
   \delta^{(p+1)} \left(
     p_{0\mu}+\sum_{r=1}^{M}k_{\mu}^{(r)}\right) \nonumber\\
&& \quad \times \frac{2}{\alpha'} \left(-\frac{1}{2\pi}\right)^{M}
   \left[-\frac{\left(\theta G\theta \right)^{\mu\nu}}{8\pi^{2}\alpha'}
         \prod_{u=1}^{M}\left(p_{0}\theta A(k^{(u)}) \right)
    \right.\nonumber\\
&& \hspace{2em} + \sum_{r=1}^{M} \left\{
       \left(\theta A (k^{(r)}) \right)^{\mu}
       \frac{\left(\theta k^{(r)} \right)^{\nu}}{2\pi}
       -\frac{\left(\theta k^{(r)}\right)^{\mu}}{2\pi}
         \left(\theta A(k^{(r)})\right)^{\nu}
      \right.\nonumber\\
   && \hspace{8em} \left.
         +i\frac{\left(\theta k^{(r)}\right)^{\mu}}{2\pi}
          i\frac{\left(\theta k^{(r)}\right)^{\nu}}{2\pi}
            \left(p_{0}\theta A(k^{(r)})\right)
        \right\} \prod_{u\neq r}\left(p_{0}\theta A(k^{(u)})
                                \right)\nonumber\\
&& \hspace{2em} +\sum_{r\neq s}
   \left\{ \left(i\frac{\left(\theta k^{(r)}\right)^{\mu}
                        \left(p_{0}\theta A(k^{(r)})\right)}
                       {2\pi}
                 -i\left(\theta A(k^{(r)})\right)^{\mu}\right)
            e^{i\sigma_{r}}
   \right. \nonumber\\
&& \hspace{4.5em} \left. \left. \times
    \left(i\frac{\left(\theta k^{(s)}\right)^{\nu}
                 \left(p_{0}\theta A(k^{(s)})\right)}
                {2\pi}
          + i\left(\theta A(k^{(s)})\right)^{\nu}\right)
          e^{-i\sigma_{s}}\right\}
     \prod_{u\neq r,s}\left(p_{0}\theta 
          A(k^{(u)})\right) \right]~. 
\nonumber \\
  \label{eq:overlap with graviton 2}
\end{eqnarray}

This reproduces the result obtained in
Appendix A of \cite{Okawa-Ooguri}. In fact, 
combining the above equation 
with Eq.(\ref{eq:graviton out-state}),
we obtain
\begin{eqnarray}
\lefteqn{
\langle -p_{0N}|h_{\mu\nu}(p_{0})
    \alpha^{\mu}_{1}\tilde{\alpha}_{1}^{\nu}
  \left| B_{N}[A];(\sigma_{1},k^{(1)}),\cdots,(\sigma_{M},k^{(M)})
   \right\rangle}
  \nonumber \\
&& \approx
   \left(\frac{-\det G_{\mu\nu}}{(2\alpha')^{p+1}}\right)^{\frac{1}{4}}
   \prod_{r=1}^{M} e^{ik_{\mu}^{(r)}\theta^{\mu\nu}p_{0\nu}
                       \frac{\sigma_{r}}{2\pi}}
   \prod_{r<s}^{M}e^{\frac{i}{2}k_{\mu}^{(r)}\theta^{\mu\nu}k_{\nu}^{(s)}
     \epsilon (\sigma_{r}-\sigma_{s})}
   \prod_{\mu=0}^{p} \delta \left(
     p_{0\mu}+\sum_{r=1}^{M}k_{\mu}^{(r)}\right) \nonumber\\
&& \quad \times \frac{2}{\alpha'} \left(-\frac{1}{2\pi}\right)^{M}
   \left[ 
     \sum_{r=1}^{M} \frac{1}{(2\pi)^{2}}
           k_{\mu}^{(r)}\left(\theta h(p_{0})\theta\right)^{\mu\nu}
                  k_{\nu}^{(r)}
          \prod_{u=1}^{M}\left(p_{0}\theta A(k^{(u)}) \right)
    \right.\nonumber\\
&& \hspace{3.5em} + \sum_{r\neq s}\left\{
       \left( \frac{\left(p_{0}\theta A(k^{(r)})\right)}{2\pi}
              k_{\mu}^{(r)}
              -A_{\mu}(k^{(r)}) \right)
       \left(\theta h(p_{0})\theta\right)^{\mu\nu}
       \left(\frac{\left(p_{0}\theta A(k^{(s)})\right)}{2\pi}
             k_{\nu}^{(s)}+A_{\nu}(k^{(s)})\right)
       \right. \nonumber\\
&& \hspace{8em} \left.\left.\times e^{i(\sigma_{r}-\sigma_{s})}
       \prod_{u\neq r,s}\left(p_{0}\theta A(k^{(u)})\right)
      \right\}\right]~,
\label{overlap with graviton 3}
\end{eqnarray}
where the properties of $h_{\mu\nu}(p_{0})$ are used 
in the following manner :
\begin{equation}
0=\alpha'g^{\mu\nu} h_{\mu\nu}(p_{0})
  =\left(\alpha'G^{\mu\nu}
  -\frac{\left(\theta G\theta\right)^{\mu\nu}}{(2\pi)^{2}\alpha'}
   \right) h_{\mu\nu}(p_{0})
   \approx
   -\frac{\left(\theta G\theta\right)^{\mu\nu}}{(2\pi)^{2}\alpha'}
   h_{\mu\nu}(p_{0})~.
\end{equation}
It is worth emphasizing that Eq.(\ref{overlap with graviton 3}) 
is derived without using the on-shell conditions of graviton, 
i.e.\ neither $g^{\mu\nu}p_{0\mu}h_{\nu\lambda}(p_{0})=0$ or 
      $\alpha' g^{\mu\nu} p_{0\mu}p_{0\nu}=0$.

\subsection{Reparametrization invariance of open Wilson lines}

Wilson lines in gauge theories 
are invariant under reparametrizations of the paths. 
Transformations analogous to the reparametrizations 
are generated by $L_n-\tilde{L}_{-n}$ 
in closed-string theory. 
In fact, $L_n-\tilde{L}_{-n}$ can be identified with 
vector fields 
$z^{n+1}\partial_z-\bar{z}^{-n+1}\partial_{\bar{z}}$ 
on the world-sheet 
\footnote{$z=e^{\tau+i\sigma}$.}  
and 
at $\tau=0$, 
where the boundary states reside, 
these vector fields have the forms of 
$e^{in\sigma}\partial_{\sigma}$.  
We have observed that the action of diff$S^1$ on the boundary 
states is identified with the action of the closed-string 
BRST charge. The boundary states are in general not BRST-closed. 
Hence the reparametrization invariance of open Wilson lines 
indicates that the action of diff$S^1$ or the BRST charge 
$Q_c$ becomes null at the zero-slope limit.

Let us verify the above observation. 
By making use of Eqs.(\ref{Ishibashi condition}) 
and (\ref{eq:Ishibashi-Virasoro}),
we obtain for $\forall n \in \mathbb{Z}$
\begin{eqnarray}
&&\left(L_{n}-\tilde{L}_{-n}\right)
 \left|B_{N}[A];(\sigma_{1},k^{(1)}),\cdots,(\sigma_{M},k^{(M)})
 \right\rangle \nonumber\\
&& \quad = \sum_{r=1}^{M} e^{in\sigma_{r}}
  \left\{-i\frac{\partial}{\partial\sigma_{r}}
         +n\left(\alpha' k^{(r)}_{\mu}G^{\mu\nu}k_{\nu}^{(r)}
                  +1\right)\right\}
  \left|B_{N}[A];(\sigma_{1},k^{(1)}),\cdots,(\sigma_{M},k^{(M)})
 \right\rangle \nonumber\\
&& \qquad +\sum_{r=1}^{M} e^{in\sigma_{r}}\alpha'
    k_{\mu}^{(r)}G^{\mu\nu} A_{\nu}(k^{(r)})
    \left|B_{N}[A]_{\check{r}};
       (\sigma_{1},k^{(1)}),\cdots,(\sigma_{M},k^{(M)})
    \right\rangle~,
 \label{eq:Ishibashi-Virasoro 2}
\end{eqnarray}
where 
$\left|B_{N}[A]_{\check{r}};
       (\sigma_{1},k^{(1)}),\cdots,(\sigma_{M},k^{(M)})
    \right\rangle$ denotes
\begin{equation}
\lim_{\forall \tau_{r} \rightarrow 0+}
V^{ren}_{A}(\sigma_{1},\tau_{1};k^{(1)})
\cdots V^{ren}_{T}(\sigma_{r},\tau_{r};k^{(r)})
\cdots V^{ren}_{A}(\sigma_{M},\tau_{M};k^{(M)})
\left| B_{N}\right\rangle~.
\end{equation}
Combining Eq.(\ref{eq:Ishibashi-Virasoro 2}) with
Eq.(\ref{eq:overlap with momentum loop 4}),
we find that in the zero-slope limit
\begin{eqnarray}
&& 
\int \prod_{r=1}^{M} \frac{d^{p+1}k^{(r)}}{(2\pi)^{\frac{p+1}{2}}}
\int_{0\leq\sigma_{1} \leq \cdots \leq \sigma_{M} \leq 2\pi}
\prod_{r=1}^{M} d\sigma_{r}
 \nonumber\\
&& \hspace{3em} \times
\left\langle \Omega (\psi_{n},\bar{\psi}_{n};p_{0})\right|
 i\epsilon_{n}\left(L_{n}-\tilde{L}_{-n}\right)
  \left|  B_{N}[A];(\sigma_{1},k^{(1)}),\cdots,(\sigma_{M},k^{(M)})
  \right\rangle
  \nonumber\\
&& \approx
   \int \prod_{r=1}^{M} \frac{d^{p+1}k^{(r)}}{(2\pi)^{\frac{p+1}{2}}}
\int_{0\leq\sigma_{1} \leq \cdots \leq \sigma_{M} \leq 2\pi}
\prod_{s=1}^{M} d\sigma_{s}
 \nonumber\\
&& \hspace{3em} \times
\sum_{r=1}^{M}
\frac{\partial}{\partial \sigma_{r}} \epsilon_{n}e^{in\sigma_{r}}
    \left\langle \Omega (\psi_{n},\bar{\psi}_{n};p_{0})\left|
       B_{N}[A];(\sigma_{1},k^{(1)}),\cdots,(\sigma_{M},k^{(M)})
  \right\rangle\right. \nonumber\\
&& \approx
  \left(\frac{-\det G_{\mu\nu}}{(2\alpha')^{p+1}}\right)^{\frac{1}{4}}
  \prod_{n=1}^{\infty} \exp \left[
     -\frac{\bar{\psi}_{n\mu} \left(\theta G\theta\right)^{\mu\nu}
            \psi_{n\nu}}{16\pi^{2}n\alpha'}  \right]
   \int \frac{d^{p+1}x}{(2\pi)^{p+1}}
   \int_{0\leq\sigma_{1} \leq \cdots \leq \sigma_{M} \leq 2\pi}
\prod_{s=1}^{M} d\sigma_{s} \nonumber\\
&&\hspace{1.3em}\sum_{r=1}^{M}
    \frac{\partial}{\partial\sigma_{r}}
    \epsilon_{n}e^{in\sigma_{r}}
     \left\{ \frac{dy^{\mu}(\sigma_{1})}{d\sigma_{1}}
           \mathcal{A}_{\mu}(x+y(\sigma_{1})) \right\}\star
         \cdots\star
   \left\{ \frac{dy^{\mu}(\sigma_{M})}{d\sigma_{M}}
           \mathcal{A}_{\mu}(x+y(\sigma_{M})) \right\}
     \star e^{ip_{0\mu}x^{\mu}}~,\nonumber\\
    \label{eq:reparametrization 1}
\end{eqnarray}
where $\epsilon_{n}$ denotes an infinitesimal parameter.
This shows that at the zero-slope limit $L_n-\tilde{L}_{-n}$ 
give rise to infinitesimal reparametrizations 
$\sigma$ 
$\mapsto$ 
$\sigma'=\sigma+\delta_{\epsilon}^{(n)}\sigma$, 
where 
$\delta_{\epsilon}^{(n)}\sigma 
\equiv -\epsilon_ne^{in \sigma}$.
This can be checked by recalling that the pull-back 
$ \tilde{\mathcal{A}}(\sigma)= 
      \frac{dy^{\mu}(\sigma)}{d\sigma}
     \mathcal{A}_{\mu}(x+y(\sigma))$ 
on the path transforms 
under the infinitesimal reparametrizations as  
$
\tilde{\mathcal{A}}(\sigma)
  \mapsto \tilde{\mathcal{A}}(\sigma)+
     \delta^{(n)}_{\epsilon}
         \tilde{\mathcal{A}}(\sigma) 
$,
where    
\begin{eqnarray}
\delta^{(n)}_{\epsilon}
  \tilde{\mathcal{A}}(\sigma)
  =\frac{\partial}{\partial \sigma}
    \left( \epsilon_{n}e^{in\sigma}
    \frac{dy^{\mu}(\sigma)}{d\sigma}
     \mathcal{A}_{\mu}(x+y(\sigma)) \right)~.
\end{eqnarray}
{}From this, we can write the integrand 
in the RHS of Eq.(\ref{eq:reparametrization 1}) 
as follows :
\begin{eqnarray}
&&
\sum_{r=1}^{M}
   \frac{\partial}{\partial\sigma_{r}} \epsilon_{n} e^{in\sigma_{r}}
  \left\{\frac{dy^{\mu}(\sigma_{1})}{d\sigma_{1}}
  \mathcal{A}_{\mu}(x+y(\sigma_{1}))\right\}\star
  \cdots \star
  \left\{  \frac{dy^{\mu}(\sigma_{M})}{d\sigma_{M}}
  \mathcal{A}_{\mu}(x+y(\sigma_{M}))\right\}
  \nonumber\\
&& \quad =
  \delta^{(n)}_{\epsilon}
  \left(
     \left\{\frac{dy^{\mu}(\sigma_{1})}{d\sigma_{1}}
     \mathcal{A}_{\mu}(x+y(\sigma_{1}))\right\}\star
     \cdots \star
     \left\{  \frac{dy^{\mu}(\sigma_{M})}{d\sigma_{M}}
      \mathcal{A}_{\mu}(x+y(\sigma_{M}))\right\}
   \right)~.
\end{eqnarray}

The above integrand is a total derivative 
with respect to each $\sigma_{r}$. 
The $\sigma_{r}$ integrations in 
(\ref{eq:reparametrization 1}) 
give rise to surface terms. 
Let us show that these surface terms actually cancel out. 
Taking account of the integration region of $\sigma_{r}$
being $[\sigma_{r-1},\sigma_{r+1}]$ for $2 \leq r \leq M-1$, 
the following equality holds for arbitrary 
$\mathcal{F}$ : 
\begin{eqnarray}
&&\int_{0\leq\sigma_{1} \leq \cdots \leq \sigma_{M} \leq 2\pi}
\prod_{s=1}^{M} d\sigma_{s} 
\left(\frac{\partial}{\partial \sigma_{r}} e^{in\sigma_{r}}
     + \frac{\partial}{\partial \sigma_{r+1}} e^{in\sigma_{r+1}}
\right) \mathcal{F} (\sigma_{1},\cdots,\sigma_{M})
 \nonumber\\
&&= \int_{0\leq \tilde{\sigma}_{1}\leq \cdots \leq \tilde{\sigma}_{M-1}}
     \prod_{s=1}^{M-1} d\tilde{\sigma}_{s}
     \Big\{e^{in\tilde{\sigma}_{r+1}}
           \mathcal{F}
           \left(\tilde{\sigma}_{1},\cdots,\tilde{\sigma}_{r},
           \tilde{\sigma}_{r+1}\tilde{\sigma}_{r+1},\tilde{\sigma}_{r+2},
           \cdots,\tilde{\sigma}_{M-1}\right)
       \nonumber\\
&& \hspace{11em}
       -e^{in\tilde{\sigma}_{r-1}}
         \mathcal{F}\left(\tilde{\sigma}_{1},\cdots,
            \tilde{\sigma}_{r-2},\tilde{\sigma}_{r-1},\tilde{\sigma}_{r-1},
            \tilde{\sigma}_{r},\cdots,\tilde{\sigma}_{M-1}\right)
       \Big\}~.
\end{eqnarray}
This leads us to find
\begin{eqnarray}
&&\int \frac{d^{p+1}x}{(2\pi)^{p+1}}
   \int_{0\leq\sigma_{1} \leq \cdots \leq \sigma_{M} \leq 2\pi}
\prod_{s=1}^{M} d\sigma_{s}
   \nonumber\\
&& \sum_{r=1}^{M}
   \frac{\partial}{\partial\sigma_{r}}\epsilon_{n}e^{in\sigma_{r}}
   \left\{ \frac{dy^{\mu}(\sigma_{1})}{d\sigma_{1}}
           \mathcal{A}_{\mu}(x+y(\sigma_{1})) \right\}\star
   \cdots\star
   \left\{ \frac{dy^{\mu}(\sigma_{M})}{d\sigma_{M}}
           \mathcal{A}_{\mu}(x+y(\sigma_{M})) \right\}
    \star e^{ip_{0\mu}x^{\mu}}
    \nonumber\\
&&=\int \frac{d^{p+1}x}{(2\pi)^{p+1}}
    \int_{0\leq\sigma_{1} \leq \cdots \leq \sigma_{M-1} \leq 2\pi}
     \prod_{s=1}^{M-1} d\sigma_{s}
     \nonumber\\
&& \qquad
     \left\{ \frac{dy^{\mu}(\sigma_{1})}{d\sigma_{1}}
           \mathcal{A}_{\mu}(x+y(\sigma_{1})) \right\}\star
      \cdots
      \star
      \left\{ \frac{dy^{\mu}(\sigma_{M-1})}{d\sigma_{M-1}}
           \mathcal{A}_{\mu}(x+y(\sigma_{M-1})) \right\}
    \nonumber\\
&& \qquad\;
     \star \epsilon_{n}
     \left[ \left\{ \frac{dy^{\mu}(2\pi)}{d\sigma}
           \mathcal{A}_{\mu}(x+y(2\pi)) \right\}\star
           e^{ip_{0\mu}x^{\mu}}
           -e^{ip_{0\mu}x^{\mu}}
           \star
           \left\{ \frac{dy^{\mu}(0)}{d\sigma}
           \mathcal{A}_{\mu}(x+y(0)) \right\}
    \right]~.
\nonumber \\
\label{eq:moduli-sekibun}
\end{eqnarray}
Here we have used the cyclic property of the Moyal product 
inside the integration:
$\int d^{p+1}x\, f(x)\star g(x) =\int d^{p+1}x \, g(x)\star f(x)$.
Let us recall that $e^{ip_{0\mu}x^{\mu}}$ plays a role
of a translation generator on the non-commutative space-time:
\begin{equation}
f(x)\star e^{ip_{0\mu}x^{\mu}}
  =e^{ip_{0\mu}x^{\mu}} \star f(x-\theta p_{0})~.
\end{equation}
Combined with the relations
$\frac{dy^{\mu}}{d\sigma}(2\pi)=\frac{dy^{\mu}}{d\sigma}(0)$
and $y^{\mu}(2\pi)-y^{\mu}(0)=\theta^{\mu\nu}p_{0\nu}$,
this tells us that Eq.(\ref{eq:moduli-sekibun}) is vanishing
and so is Eq.(\ref{eq:reparametrization 1}).

Therefore we have shown that 
the action of diff$S^1$ on the boundary states 
reduces to infinitesimal reparametrizations 
of paths of open Wilson lines  
and becomes null at the zero-slope limit. 
The situation is summarized as follows : 
\begin{eqnarray}
\lefteqn{\sum_{M=0}^{\infty} i^{M}
\int \prod_{r=1}^{M} \frac{d^{p+1}k^{(r)}}{(2\pi)^{\frac{p+1}{2}}}
\int_{0\leq\sigma_{1} \leq \cdots \leq \sigma_{M} \leq 2\pi}
\prod_{r=1}^{M} d\sigma_{r}
 }
 \nonumber\\
&& \hspace{3em} \times
\left\langle \Omega (\psi_{n},\bar{\psi}_{n};p_{0})\right|
 i\epsilon_{n}\left(L_{n}-\tilde{L}_{-n}\right)
  \left|  B_{N}[A];(\sigma_{1},k^{(1)}),\cdots,(\sigma_{M},k^{(M)})
  \right\rangle
  \nonumber\\
&& \approx 
   \left(\frac{-\det G_{\mu\nu}}{(2\alpha')^{p+1}}\right)^{\frac{1}{4}}
   \prod_{n=1}^{\infty} \exp \left[
      -\frac{\bar{\psi}_{n\mu} \left(\theta G \theta\right)^{\mu\nu}
               \psi_{n\nu}}
            {16 \pi^{2}n\alpha'} \right] \nonumber\\
&& \qquad
   \delta^{(n)}_{\epsilon} \int \frac{d^{p+1}x}{(2\pi)^{p+1}}
     \mathcal{P}_{\star} \left[ \exp \left( i\int^{2\pi}_{0} d\sigma
      \frac{y^{\mu}(\sigma)}{d\sigma}
      \mathcal{A}_{\mu}(x+y(\sigma))\right)\right]
      \star e^{ip_{0\mu}x^{\mu}}\nonumber\\
&& =0~.
\end{eqnarray}

\subsection{Factorization by closed-string momentum
eigenstates}

We have seen in Section \ref{sec:open Wilson lines (I)} 
that analogues of open Wilson line, 
$
\mathcal{P}_{\star}
\left[ 
  \exp 
    \left\{\int_0^{2\pi}d\sigma \phi(x(\sigma))\right\}
\right]
$,
factorize the generating function of 
one-loop amplitudes of open-string tachyons 
at the zero-slope limit. 
Decompositions of the string amplitudes made 
by insertions of the unity,  
$1=\int [dP_{N}] |P_{N}\rangle \langle P_{N}|$,
play an important role 
to gain the factorization 
(\ref{closed-string FTL by open Wilson lines}).  
With the same manipulation one can expect 
that a similar factorization is also obtainable 
for the amplitudes of gluons. 
Prior to the actual computations, 
let us explain briefly why the decompositions  
via the momentum eigenstates give us 
open Wilson lines or their analogues.

To make the discussion transparent we start with the 
momentum eigenstate $\langle -P_N |$. 
As described in Eq.(\ref{eq:PN-shift}),
we can write the state as 
\begin{equation}
\left\langle -P_{N} \right|
  =\prod_{n=1}^{\infty} \exp \left[-\frac{\alpha'}{4n}
     \bar{\psi}_{n\mu}g^{\mu\nu}\psi_{n\nu} \right]
     \times
  {}_{B=0}\!\left\langle B_{N}\right|
     :\exp\left(i\int_{0}^{2\pi} d\sigma P_{\mu}(\sigma)
               \hat{X}^{\mu}(\sigma)\right):~.
\end{equation}
Taking account of its use in the decompositions 
of the amplitudes  
we should consider the state 
$\langle -P_N |q_c^{\frac{1}{4}(L_0+\tilde{L}_0-2)}$ 
rather than 
$\langle -P_N |$. 
Let us also recall that 
the above $\psi_n$ and $\overline{\psi}_n$ 
are rescaled appropriately in order  
to obtain the factorization 
(\ref{closed-string FTL by open Wilson lines}) 
and that 
the rescaled variables are kept intact 
under taking the limit. 
To be explicit, we denote the rescaled
$\psi_{n}$ and $\bar{\psi}_{n}$ by
the following 
$\psi_n'$ and $\overline{\psi}_n'$ :
\begin{equation}
(\psi_{n\mu},\bar{\psi}_{n\mu}) \mapsto 
  (\psi'_{n\mu},\bar{\psi}'_{n\mu})
    =(2q_{c}^{\frac{n}{4}} \psi_{n\mu} ,
      2q_{c}^{\frac{n}{4}}\bar{\psi}_{n\mu})~.
   \label{eq:rescaled psi}
\end{equation}
Oscillator representation of the state 
$\langle -P_N |q_c^{\frac{1}{4}(L_0+\tilde{L}_0-2)}$ 
can be read as follows by using these rescaled variables : 
\begin{eqnarray}
&&\left\langle -P_{N}\right|
  q_{c}^{\frac{1}{4}\left(L_{0}+\tilde{L}_{0}-2\right)}
  = q_{c}^{\frac{\alpha'}{8}g^{\mu\nu}
                         p_{0\mu}p_{0\nu}-\frac{1}{2}}
   \times \nonumber\\
&&\hspace{2em}\times
      \langle -p_{0N}|
      \prod_{n=1}^{\infty}
      \exp\left[ -\frac{q_{c}^{\frac{n}{2}}}{n}
                     \alpha_{n}^{\mu}g_{\mu\nu}\tilde{\alpha}_{n}^{\nu}
              -\frac{\sqrt{\alpha'}}{2n}
                    \left(\bar{\psi}'_{n\mu}\alpha^{\mu}_{n}
                    +\psi'_{n\mu}\tilde{\alpha}^{\mu}_{n}\right)
             -\frac{\alpha'}{8nq_{c}^{\frac{n}{2}}}
                     \bar{\psi}'_{n\mu}g^{\mu\nu}\psi'_{n\nu}
            \right]~.
\end{eqnarray}
Now it is clear that this state reduces to the state 
$\left \langle \Omega(\psi_n',\bar{\psi}_n';p_0)\right|$ 
at the zero-slop limit :
\begin{eqnarray}
&&\left\langle -P_{N}\right|
  q_{c}^{\frac{1}{4}\left(L_{0}+\tilde{L}_{0}-2\right)}
 \nonumber\\
 &&\approx e^{-\frac{\pi s^{(c)}}{4}
              \left(p_{0\mu} g^{\mu\nu}p_{0\nu}-\frac{4}{\alpha'}\right)}
  \langle -p_{0N}| \prod_{n=1}^{\infty}
    \exp\left[ -\frac{\sqrt{\alpha'}}{2n}
                    \left(\bar{\psi}'_{n\mu}\alpha^{\mu}_{n}
                    +\psi'_{n\mu}\tilde{\alpha}^{\mu}_{n}\right)
             +\frac{\psi'_{n\mu}\left(\theta G\theta\right)^{\mu\nu}
                    \bar{\psi}'_{n\nu}}
                    {32\pi^{2}nq_{c}^{\frac{n}{2}}\alpha'}
          \right] \nonumber\\
 &&=  e^{-\frac{\pi s^{(c)}}{4}
              \left(p_{0\mu} g^{\mu\nu}p_{0\nu}-\frac{4}{\alpha'}\right)}
 \prod_{n=1}^{\infty}
     \exp\left\{\frac{\psi'_{n\mu}\left(\theta G\theta\right)^{\mu\nu}
                    \bar{\psi}'_{n\nu}}
                    {32\pi^{2}nq_{c}^{\frac{n}{2}}\alpha'}
          \right\}
     \left\langle \Omega (\psi'_{n},\bar{\psi}'_{n};p_{0})\right|~.
\end{eqnarray}
As we have seen in subsection \ref{sec:curved OWL} 
the above state is a generating function of the closed-string states 
and its overlaps with the boundary states lead the open Wilson line 
taken along the corresponding path.

We will compute the factorization of the generating 
function of one-loop amplitudes of gluons. 
We will restrict ourselves to $25$-brane in the critical 
dimensions. 
Computations 
become parallel to those given 
in Section \ref{sec:open Wilson lines (I)} 
but much complicated.  
The factorization at the zero-slope limit 
by open Wilson lines are given 
in Eq.(\ref{eq:factorization of gluons 6}).

Let us study factorizations of the following 
amplitudes :
\begin{eqnarray}
&&
\left\langle B_{N}[A]; (\sigma_{M+1},k^{(M+1)}),\cdots,
   (\sigma_{M+N},k^{(M+N)}) \right|
  q_c^{\frac{1}{2}(L_0+\tilde{L}_0-2)}
\nonumber\\
&& \hspace{4em} \times
 \left| B_{N}[A]; (\sigma_{1},k^{(1)}),\cdots,(\sigma_{M},k^{(M)})
  \right\rangle~.
\label{amplitude 1}
\end{eqnarray}
Oscillator representations of the boundary states of gluons 
are unsuitable for computations of the amplitudes. 
We instead consider the corresponding amplitudes 
in auxiliary forms.
Let us decompose the string amplitudes  
(\ref{def of aux gluon amplitude by bs})
by using the momentum eigenstates as follows :
\begin{eqnarray}
&&\left\langle \hat{B}_{N}[A];\{a_{r}\};
    (\sigma_{M+1},k^{(M+1)}),\cdots,(\sigma_{M+N},k^{(M+N)}) \right|
   \nonumber\\
 && \hspace{5em}\times
q_{c}^{\frac{1}{2}\left(L_{0}+\tilde{L}_{0}-2\right)}
 \left| \hat{B}_{N}[A];\{a_{r}\};
 (\sigma_{1},k^{(1)}),\cdots,(\sigma_{M},k^{(M)})
  \right\rangle
 \nonumber\\
&& =\int \left[dP_{N}\right]
\Bigl\langle \hat{B}_{N}[A];\{a_{r}\};
    (\sigma_{M+1},k^{(M+1)}),\cdots,(\sigma_{M+N},k^{(M+N)}) \Bigr|
q_{c}^{\frac{1}{4}\left(L_{0}+\tilde{L}_{0}-2\right)}
\Bigl|-P_{N}\Bigr\rangle \nonumber\\
&& \hspace{4em} \times
 \Bigl\langle -P_{N} \Bigr|
 q_{c}^{\frac{1}{4}\left(L_{0}+\tilde{L}_{0}-2\right)}
 \Bigl| \hat{B}_{N}[A];\{a_{r}\};
 (\sigma_{1},k^{(1)}),\cdots,(\sigma_{M},k^{(M)})
  \Bigr\rangle~.
 \label{eq:factorization of gluons 1}
\end{eqnarray}
Here we have changed the integration variables from
$P_{\mu}(\sigma)$ to $-P_{\mu}(\sigma)$ for
the later convenience. 
The amplitudes (\ref{amplitude 1}) are 
obtained from the above by the operation 
$\left.
\prod_{r=1}^Mi\frac{\partial}{\partial a_r} 
\right|_{a_r=0}$.

Each factor in the above decomposition can be evaluated  
by using the oscillator representations. 
These are given in 
Eq.(\ref{aux M gluon bs}) 
(and Eq.(\ref{dual aux M gluon bs})) for the 
boundary states, and Eq.(\ref{state PN}) 
(and its hermitian conjugate) for the momentum eigenstates. 
We then need to compute matrix 
elements similar to 
Eq.(\ref{difficult matrix element}). 
Eq.(\ref{eq:vevee}) enables us to calculate them.   
After the rescaling (\ref{eq:rescaled psi}) of 
$\psi_n$ and $\overline{\psi}_n$ we obtain 
the following expressions : 
\begin{eqnarray}
&&\int \left[dP_{N}\right]
\Bigl\langle \hat{B}_{N}[A];\{a_{r}\};
    (\sigma_{M+1},k^{(M+1)}),\cdots,(\sigma_{M+N},k^{(M+N)}) \Bigr|
q_{c}^{\frac{1}{4}\left(L_{0}+\tilde{L}_{0}-2\right)}
\Bigl|-P_{N} \Bigr\rangle\ \nonumber\\
&& \hspace{4em} \times
 \left\langle -P_{N} \left|
 q_{c}^{\frac{1}{4}\left(L_{0}+\tilde{L}_{0}-2\right)}
 \left| \hat{B}_{N}[A];\{a_{r}\};
 (\sigma_{1},k^{(1)}),\cdots,(\sigma_{M},k^{(M)})
  \right\rangle \right.\right.
 \nonumber\\
&& = \left(\frac{-\det G_{\mu\nu}}{(2\alpha')^{26}}\right)^{\frac{1}{2}}
 \int d^{26}p_{0}\
   \delta^{26} \left(p_{0}+\sum_{r=1}^{M}k^{(r)}\right)
   \delta^{26} \left( p_{0}-\sum_{r=M+1}^{M+N}k^{(r)}\right)
   \, q_{c}^{\frac{\alpha'}{4}g^{\mu\nu}p_{0\mu}p_{0\nu} -1}
 \nonumber\\
&& \quad \times
   \prod_{1\leq r<s \leq M}
   e^{\frac{i}{2}k_{\mu}^{(r)}\theta^{\mu\nu} k_{\nu}^{(s)}
          \epsilon (\sigma_{r}-\sigma_{s})}
   \prod_{M+1 \leq r<s \leq M+M}
   e^{-\frac{i}{2}k_{\mu}^{(r)}\theta^{\mu\nu} k_{\nu}^{(s)}
          \epsilon (\sigma_{r}-\sigma_{s})}
 \nonumber\\
&& \quad \times
   \prod_{1\leq r<s \leq M} \exp \left[ \alpha' G^{\mu\nu}
     \left(k_{\mu}^{(r)}-a_{r} A_{\mu}(k^{(r)})\partial_{\sigma_{r}}
     \right)
     \left(k_{\nu}^{(s)}-a_{s} A_{\nu}(k^{(s)})\partial_{\sigma_{s}}
     \right)
     \ln \left|e^{i\sigma_{r}} -e^{i\sigma_{s}}\right|^{2}
     \right] \nonumber\\
&& \quad \times
   \prod_{M+1\leq r<s \leq M+N} \exp \left[ \alpha' G^{\mu\nu}
     \left(k_{\mu}^{(r)}-a_{r} A_{\mu}(k^{(r)})\partial_{\sigma_{r}}
     \right)
     \left(k_{\nu}^{(s)}-a_{s} A_{\nu}(k^{(s)})\partial_{\sigma_{s}}
     \right)
     \ln \left|e^{i\sigma_{r}} -e^{i\sigma_{s}}\right|^{2}
     \right] \nonumber\\
&& \quad \times  \prod_{n=1}^{\infty} \exp \left[
   -\frac{2\alpha'q_{c}^{\frac{n}{2}}}{n}
    \left(\frac{1}{E^{T}-q_{c}^{\frac{n}{2}}E} g \frac{1}{E^{T}}
    \right)^{\mu\nu} \right.\nonumber\\
&& \hspace{6.5em} \times
    \left\{ \sum_{r,s=1}^{M}
    \left(k_{\mu}^{(r)}-ina_{r} A_{\mu}(k^{(r)})\right)
    \left(k_{\nu}+ina_{s} A_{\nu}(k^{(s)}) \right)
    e^{in(\sigma_{r}-\sigma_{s})}   \right.\nonumber\\
&&  \hspace{8em} + \left.\left. \sum_{r,s=M+1}^{M+N}
    \left(k_{\mu}^{(r)}+ina_{r} A_{\mu}(k^{(r)})\right)
    \left(k_{\nu}-ina_{s} A_{\nu}(k^{(s)}) \right)
    e^{-in(\sigma_{r}-\sigma_{s})}  \right\} \right]\nonumber\\
&& \quad \times
 \int \prod_{n=1}^{\infty}
  \left[ \frac{d^{26}\bar{\psi}_{n} d^{26}\psi_{n}}{(2i)^{26}}
         \left(\frac{\alpha'}{4n\pi q_{c}^{\frac{n}{2}}}
         \right)^{26}
         \frac{-\det g_{\mu\nu}}
              {\det^{2} \left(g-q_{c}^{\frac{n}{2}}N\right)_{\mu\nu}}
  \right] \nonumber\\
&& \quad \times
 \exp \left[ -\sum_{n=1}^{\infty} \frac{\alpha'}{4n}
     \bar{\psi}_{n\mu}\left(
      \frac{1}{q_{c}^{\frac{n}{2}}g}
      +\frac{1}{g-q_{c}^{\frac{n}{2}}N^{T}} N^{T} \frac{1}{g}
      +\frac{1}{g}N \frac{1}{g-q_{c}^{\frac{n}{2}}N}
      \right)^{\mu\nu}\psi_{n\nu}\right]
 \nonumber\\
&& \quad \times \exp \left[
   -i \sum_{r=1}^{M} a_{r} A_{\mu}(k^{(r)})\left[
     \frac{\theta^{\mu\nu}}{2\pi}p_{0\nu}
      +\frac{\alpha'}{\sqrt{2}} \sum_{n=1}^{\infty}
       \left\{ \left(\frac{1}{E-q_{c}^{\frac{n}{4}}E^{T}}\right)^{\mu\nu}
       \psi_{n\nu}e^{-in\sigma_{r}}
    \right.\right.\right.\nonumber\\
&& \hspace{21em} \left.\left.\left.
       -\left(\frac{1}{E^{T}-q_{c}^{\frac{n}{4}}E}\right)^{\mu\nu}
         \bar{\psi}_{n\nu} e^{in\sigma_{r}}\right\}
         \right]\right]\nonumber\\
&& \quad \times
   \exp \left[i\sum_{r=1}^{M}k_{\mu}^{(r)}\left[
       \frac{\theta^{\mu\nu}}{2\pi}p_{0\nu}\sigma_{r}
       +\frac{\alpha'}{\sqrt{2}}\sum_{n=1}^{\infty}\frac{i}{n}
       \left\{ \left(\frac{1}{E-q_{c}^{\frac{n}{4}}E^{T}}\right)^{\mu\nu}
       \psi_{n\nu}e^{-in\sigma_{r}}
    \right.\right.\right.\nonumber\\
&& \hspace{20em} \left.\left.\left.
      +\left(\frac{1}{E^{T}-q_{c}^{\frac{n}{4}}E}\right)^{\mu\nu}
         \bar{\psi}_{n\nu} e^{in\sigma_{r}}\right\}
         \right]\right]\nonumber\\
&& \quad \times \exp \left[
   -i \sum_{r=M+1}^{M+N} a_{r} A_{\mu}(k^{(r)})\left[
     \frac{\theta^{\mu\nu}}{2\pi}p_{0\nu}
      -\frac{\alpha'}{\sqrt{2}} \sum_{n=1}^{\infty}
       \left\{ \left(\frac{1}{E^{T}-q_{c}^{\frac{n}{4}}E}\right)^{\mu\nu}
       \psi_{n\nu}e^{-in\sigma_{r}}
    \right.\right.\right.\nonumber\\
&& \hspace{22em} \left.\left.\left.
       -\left(\frac{1}{E-q_{c}^{\frac{n}{4}}E^{T}}\right)^{\mu\nu}
         \bar{\psi}_{n\nu} e^{in\sigma_{r}}\right\}
         \right]\right]\nonumber\\
&& \quad \times
   \exp \left[i\sum_{r=M+1}^{M+N}k_{\mu}^{(r)}\left[
       \frac{\theta^{\mu\nu}}{2\pi}p_{0\nu}\sigma_{r}
       -\frac{\alpha'}{\sqrt{2}}\sum_{n=1}^{\infty}\frac{i}{n}
       \left\{ \left(\frac{1}{E^{T}-q_{c}^{\frac{n}{4}}E}\right)^{\mu\nu}
       \psi_{n\nu}e^{-in\sigma_{r}}
    \right.\right.\right.\nonumber\\
&& \hspace{20em} \left.\left.\left.
      +\left(\frac{1}{E-q_{c}^{\frac{n}{4}}E^{T}}\right)^{\mu\nu}
         \bar{\psi}_{n\nu} e^{in\sigma_{r}}\right\}
         \right]\right]~,
   \label{eq:factorization of gluons 2}
\end{eqnarray}
where we have newly written $(\psi'_{n\mu},\bar{\psi}'_{n\mu})$
as $(\psi_{n\mu},\bar{\psi}_{n\mu})$.

We examine the zero-slope limit of the above expression.  
The limiting procedure we consider is the same that was investigated
in  Section \ref{sec:UV NC gauge} to capture the UV behavior of 
the non-commutative gauge theory. 
It is taken by fixing parameters 
$s^{(c)}=\alpha'|\tau^{(c)}|$, $\sigma_r$ and $a_r$ 
besides the open-string tensors. 
$\psi_n$ and $\overline{\psi}_n$ 
in Eq.(\ref{eq:factorization of gluons 2}) are also 
left intact. 
This yields 
\begin{eqnarray}
&&\int \left[dP_{N}\right]
\Bigl\langle \hat{B}_{N}[A];\{a_{r}\};
    (\sigma_{M+1},k^{(M+1)}),\cdots,(\sigma_{M+N},k^{(M+N)}) \Bigr|
q_{c}^{\frac{1}{4}\left(L_{0}+\tilde{L}_{0}-2\right)}
\Bigl|-P_{N} \Bigr\rangle \nonumber\\
&& \hspace{4em} \times
 \Bigl\langle -P_{N} \Bigr|
 q_{c}^{\frac{1}{4}\left(L_{0}+\tilde{L}_{0}-2\right)}
 \Bigl| \hat{B}_{N}[A];\{a_{r}\};
 (\sigma_{1},k^{(1)}),\cdots,(\sigma_{M},k^{(M)})
  \Bigr\rangle
  \nonumber\\
&& \approx 
   \left(\frac{-\det G_{\mu\nu}}{(2\alpha')^{26}}
           \right)^{\frac{1}{2}} 
   \int d^{26}p_{0} \exp\left\{  -\frac{\pi s^{(c)}}{2}
     \left(p_{0\mu}g^{\mu\nu}p_{0\nu}-\frac{4}{\alpha'}\right)\right\}
 \nonumber\\
&& \quad \times
   \int \prod_{n=1}^{\infty}
  \left[ \frac{d^{26}\bar{\psi}_{n} d^{26}\psi_{n}}{(2i)^{26}}
         \left(\frac{1}{\pi} \right)^{26}
         \left\{ -\det
            \left( - \frac{\left(\theta G\theta\right)^{\mu\nu}}
                       {16 \pi^{2}nq_{c}^{\frac{n}{2}}\alpha'}
            \right)  \right\}
   \exp\left(\frac{\bar{\psi}_{n\mu}\left(\theta G\theta\right)^{\mu\nu}
                    \psi_{n\nu}}
                   {16 \pi^{2}nq_{c}^{\frac{n}{2}}\alpha'}
       \right) \right] \nonumber\\
&& \qquad \times 
      \delta^{26}\left(p_{0}+\sum_{r=1}^{M}k^{(r)} \right)
      \prod_{1\leq r<s \leq M}
     e^{\frac{i}{2}k_{\mu}^{(r)}\theta^{\mu\nu}k_{\nu}^{(s)}
         \epsilon (\sigma_{r}-\sigma_{s})}  
     \nonumber\\
&& \hspace{6em} \times 
   \exp\left[-i\sum_{r=1}^{M}\left\{ 
       a_{r}
       \frac{dy^{\mu}(\sigma_{r})}{d\sigma_{r}} A_{\mu}(k^{(r)})
         -k_{\mu}^{(r)} y^{\mu}(\sigma_{r}) \right\}\right]
  \nonumber\\
&&\qquad \times 
      \delta^{26}\left(p_{0}-\sum_{r=M+1}^{M+N}k^{(r)} \right)
           \prod_{M+1\leq r<s \leq M+N}
           e^{-\frac{i}{2}k_{\mu}^{(r)}\theta^{\mu\nu}k_{\nu}^{(s)}
         \epsilon (\sigma_{r}-\sigma_{s})}
   \nonumber\\
&& \hspace{6em} \times
   \exp\left[-i\sum_{r=M+1}^{M+N}\left\{ 
       a_{r}
       \frac{dy^{\mu}(\sigma_{r})}{d\sigma_{r}} A_{\mu}(k^{(r)})
         -k_{\mu}^{(r)} y^{\mu}(\sigma_{r}) \right\}\right]~,
   \label{eq:factorization of gluons 3}
\end{eqnarray}
where $y^{\mu}(\sigma)$ 
is the curve given by Eq.(\ref{eq:curved path}).

The zero-slope limit of the amplitudes 
(\ref{amplitude 1}) can be obtained 
from the above by the operations 
$
\left.
\prod_{r=1}^{M+N}
i \frac{\partial}{\partial a_r}
\right|_{a_r=0}
$, 
which are carried out without difficulty. 
The zero-slope limits turn out to be as follows :
\begin{eqnarray}
&&\int \left[dP_{N}\right]
\left\langle B_{N}[A]; (\sigma_{M+1},k^{(M+1)}),\cdots,
   (\sigma_{M+N},k^{(M+N)}) \right|
 q_{c}^{\frac{1}{4}\left(L_{0}+\tilde{L}_{0}-2\right)}
 \left|-P_{N}\right\rangle \nonumber\\
&& \hspace{4em} \times
 \left\langle -P_{N}\right|
 q_{c}^{\frac{1}{4}\left(L_{0}+\tilde{L}_{0}-2\right)}
 \left| B_{N}[A]; (\sigma_{1},k^{(1)}),\cdots,(\sigma_{M},k^{(M)})
  \right\rangle
 \nonumber\\
&& \approx \left(\frac{-\det G_{\mu\nu}}{(2\alpha')^{26}}
           \right)^{\frac{1}{2}} 
   \int d^{26}p_{0} \exp \left\{
    -\frac{\pi s^{(c)}}{2}
     \left(p_{0\mu}g^{\mu\nu}p_{0\nu}-\frac{4}{\alpha'}\right)
     \right\} \nonumber\\
&& \quad \times
   \int \prod_{n=1}^{\infty}
  \left[ \frac{d^{26}\bar{\psi}_{n} d^{26}\psi_{n}}{(2i)^{26}}
         \left(\frac{1}{\pi} \right)^{26}
         \left\{ -\det
            \left(-\frac{\left(\theta G\theta\right)^{\mu\nu}}
                       {16 \pi^{2}nq_{c}^{\frac{n}{2}}\alpha'}
            \right)  \right\}
   \exp\left(\frac{\bar{\psi}_{n\mu}\left(\theta G\theta\right)^{\mu\nu}
                    \psi_{n\nu}}
                   {16 \pi^{2}nq_{c}^{\frac{n}{2}}\alpha'}
       \right) \right] \nonumber\\
&& \quad\qquad \times 
      \delta^{26}\left(p_{0}+\sum_{r=1}^{M}k^{(r)} \right)
   \prod_{1\leq r<s \leq M}
     e^{\frac{i}{2}k_{\mu}^{(r)}\theta^{\mu\nu}k_{\nu}^{(s)}
         \epsilon (\sigma_{r}-\sigma_{s})}
    \nonumber\\
 && \hspace{7em} \times
    \prod_{r=1}^{M} \left\{
    \frac{dy^{\mu}(\sigma_{r})}{d\sigma_{r}} A_{\mu}(k^{(r)})
    e^{ik_{\nu}^{(r)}y^{\nu}(\sigma_{r})}\right\}
      \nonumber\\
&& \quad\qquad \times 
      \delta^{26}\left(p_{0}-\sum_{r=M+1}^{M+N}k^{(r)} \right)
   \prod_{M+1\leq r<s \leq M+N}
     e^{-\frac{i}{2}k_{\mu}^{(r)}\theta^{\mu\nu}k_{\nu}^{(s)}
         \epsilon (\sigma_{r}-\sigma_{s})}
    \nonumber\\
&& \hspace{7em} \times
    \prod_{r=M+1}^{M+N} \left\{
    \frac{dy^{\mu}(\sigma_{r})}{d\sigma_{r}} A_{\mu}(k^{(r)})
    e^{ik_{\nu}^{(r)}y^{\nu}(\sigma_{r})}\right\}~.
  \label{eq:factorization of gluons 4}
\end{eqnarray}

Factorized form of the $M+N$ gluon amplitude 
is obtained by integrating the above amplitude  
over the moduli 
(\ref{moduli in closed-string FTL}), where we set 
$\sigma_{M+N}=2\pi$ in order to fix the $U(1)$ 
symmetry. This gives rise to the asymmetric term 
of the factorization 
(\ref{closed-string FTL for gluon by straight open Wilson lines}) 
at the zero-slope limit. 
In order to avoid complexity of expressions 
we ignore this gauge fixing in the below and 
integrate $\sigma_{M+N}$ over 
$\sigma_{M+N-1} \leq \sigma_{M+N} \leq 2\pi$.  
The integrations over $\sigma_r$ turn out to be 
written by using the Moyal products : 
\begin{eqnarray}
&& \!\!\!
  \int_{0\leq\sigma_{1}\leq\cdots\leq\sigma_{M}\leq 2\pi}
    \prod_{r=1}^{M} d\sigma_{r}
   \int_{0\leq\sigma_{M+1}\leq\cdots\leq\sigma_{M+N}\leq 2\pi}
    \prod_{s=M+1}^{M+N} d\sigma_{s}
  \int \left[ dP_{N}\right]
\nonumber\\
&& \quad
   \left\langle B_{N}[A];(\sigma_{M+1},k^{(M+1)}), \cdots,
                   (\sigma_{M+N},k^{(M+N)})\right|
   q_{c}^{\frac{1}{4}\left(L_{0}+\tilde{L}_{0}-2\right)}
   \left|-P_{N}\right\rangle\nonumber\\
&& \hspace{3em}\times
   \left\langle -P_{N}\right|
   q_{c}^{\frac{1}{4}\left(L_{0}+\tilde{L}_{0}-2\right)}
   \left| B_{N}[A];(\sigma_{1},k^{(1)}),\cdots,
                    (\sigma_{M},k^{(M)})\right\rangle
   \nonumber\\
&&\approx \left(\frac{-\det G_{\mu\nu}}{(2\alpha')^{26}}
           \right)^{\frac{1}{2}} 
   \int d^{26}p_{0} \exp \left\{
    -\frac{\pi s^{(c)}}{2}
     \left(p_{0\mu}g^{\mu\nu}p_{0\nu}-\frac{4}{\alpha'}\right)
     \right\} \nonumber\\
&& \quad \times
   \int \prod_{n=1}^{\infty}
  \left[ \frac{d^{26}\bar{\psi}_{n} d^{26}\psi_{n}}{(2i)^{26}}
         \left(\frac{1}{\pi} \right)^{26}
         \left\{ -\det
            \left(-\frac{\left(\theta G\theta\right)^{\mu\nu}}
                       {16 \pi^{2}nq_{c}^{\frac{n}{2}}\alpha'}
            \right)  \right\}
   \exp\left(\frac{\bar{\psi}_{n\mu}\left(\theta G\theta\right)^{\mu\nu}
                    \psi_{n\nu}}
                   {16 \pi^{2}nq_{c}^{\frac{n}{2}}\alpha'}
       \right) \right] \nonumber\\
&& \hspace{2em} \times\int \frac{d^{26}x}{(2\pi)^{26}}
     \left\{\frac{dy^{\mu}(\sigma_{1})}{d\sigma_{1}} A_{\mu}(k^{(1)})
            e^{ik^{(1)}_{\nu}\left(x^{\nu}+y^{\nu}(\sigma_{1})\right)}
     \right\} 
      \nonumber\\
&& \hspace{8em} \star \cdots \star
     \left\{\frac{dy^{\mu}(\sigma_{M})}{d\sigma_{M}} A_{\mu}(k^{(M)})
            e^{ik^{(M)}_{\nu}\left(x^{\nu}+y^{\nu}(\sigma_{M})\right)}
     \right\} \star
     e^{ip_{0\mu}x^{\mu}} \nonumber\\
&& \hspace{2em} \times \int \frac{d^{26}\tilde{x}}{(2\pi)^{26}}
     \; e^{-ip_{0\mu}\tilde{x}^{\mu}}\star
     \left\{\frac{dy^{\mu}(\sigma_{M+N})}{d\sigma_{1}}
      A_{\mu}(k^{(M+N)})
            e^{ik^{(M+N)}_{\nu}
               \left(\tilde{x}^{\nu}+y^{\nu}(\sigma_{M+N})\right)}
     \right\}
    \nonumber\\
&& \hspace{8em} \star \cdots \star
     \left\{\frac{dy^{\mu}(\sigma_{M+1})}{d\sigma_{M+1}}
        A_{\mu}(k^{(M+1)})
            e^{ik^{(M+1)}_{\nu}\left(\tilde{x}^{\nu}+y^{\nu}
                (\sigma_{M+1})\right)}
     \right\}~.
     \label{eq:factorization of gluons 5}
\end{eqnarray}

Factorization of the generating function of 
the amplitudes (\ref{amplitude 1}) 
at the zero-slope limit can be obtained from 
the above by integrating out the gluon momenta $k_r$ 
and then summing up with respect to $M$ and $N$. 
It turns out to have the following form : 
\begin{eqnarray}
&& 
2\pi 
\int_{0}^{+\infty}
d|\tau^{(c)}|
\sum_{M=0}^{\infty}
         \sum_{N=0}^{\infty}i^{M}(-i)^{N}
  \int_{0\leq\sigma_{1}\leq\cdots\leq\sigma_{M}\leq 2\pi}
    \prod_{r=1}^{M} d\sigma_{r}
   \int_{0\leq\sigma_{M+1}\leq\cdots\leq\sigma_{M+N}\leq 2\pi}
    \prod_{s=M+1}^{M+N} d\sigma_{s}
\nonumber\\
&& \hspace{2em}
\int \prod_{r=1}^{M}\frac{d^{26}k^{(r)}}{(2\pi)^{13}}
     \int \prod_{s=M+1}^{M+N} \frac{d^{26}k^{(s)}}{(2\pi)^{13}}
   \Bigl\langle B_{N}[A];(\sigma_{M+1},-k^{(M+1)}), \cdots,
                   (\sigma_{M+N},-k^{(M+N)})\Bigr|
   \nonumber\\
&& \hspace{15em} \times
    q_{c}^{\frac{1}{2}\left(L_{0}+\tilde{L}_{0}-2\right)}
     \Bigl| B_{N}[A];(\sigma_{1},k^{(1)}),\cdots,
                    (\sigma_{M},k^{(M)})\Bigr\rangle
   \nonumber\\
&&\approx 
\int d^{26}p_0 
\left[
\frac{2\pi}{\alpha'}
\int_0^{+\infty}ds^{(c)}       
\exp \left\{
    -\frac{\pi s^{(c)}}{2}
     \left(p_{0\mu}g^{\mu\nu}p_{0\nu}-\frac{4}{\alpha'}\right)
     \right\} 
\right]
\nonumber\\
&& \quad \times
\int  
\prod_{n=1}^{\infty}
\left[
\frac{d^{26}\bar{\psi}_{n} d^{26}\psi_{n}}{(2\pi i)^{26}}
\left\{ 
-\det \left(
           -\frac{\theta G\theta }
                       {16 \pi^{2}nq_{c}^{\frac{n}{2}}\alpha'}
       \right)  
\right\}
\exp
\left(
\frac{\bar{\psi}_{n\mu}\left(\theta G\theta\right)^{\mu\nu}
        \psi_{n\nu}}
                   {16 \pi^{2}nq_{c}^{\frac{n}{2}}\alpha'}
\right) 
\right] \nonumber\\
&& \hspace{2em} \times \left(\frac{-\det G_{\mu\nu}}{(2\alpha')^{26}}
           \right)^{\frac{1}{4}}
         \int \frac{d^{26}x}{(2\pi)^{26}}
     \left[ \mathcal{P}_{\star} \exp
        \left(i\int^{2\pi}_{0} d\sigma \frac{dy^{\mu}(\sigma)}{d\sigma}
          \mathcal{A}_{\mu}\left(x+y(\sigma)\right)
        \right) \right]\star e^{ip_{0\mu}x^{\mu}} 
   \nonumber\\
&&\hspace{2em} \times \left(\frac{-\det G_{\mu\nu}}{(2\alpha')^{26}}
           \right)^{\frac{1}{4}}
     \overline{    \int \frac{d^{26}x}{(2\pi)^{26}}
     \left[ \mathcal{P}_{\star} \exp
        \left(i\int^{2\pi}_{0} d\sigma \frac{dy^{\mu}(\sigma)}{d\sigma}
          \mathcal{A}_{\mu}\left(x+y(\sigma)\right)
        \right) \right]\star e^{ip_{0\mu}x^{\mu}} 
     }~.
\nonumber \\ 
\label{eq:factorization of gluons 6}
\end{eqnarray}

\section{Duality between Open and Closed Strings}
\label{sec:discussions}

So far, 
momentum eigenstates of closed-string play 
a crucial role in our study. 
Their overlaps with the boundary states 
provide open Wilson lines at the zero-slope limit. 
The eigenvalues, that is, loops in the momentum space 
become paths of open Wilson lines after suitable rescalings. 
In this section we wish to deliver the other side of the story. 
The following discussions include much speculation and 
therefore they are incomplete.

Let us first observe that 
the auxiliary boundary states of gluons 
are eigenstates of the momentum operator 
$\hat{P}^{(B)}_{\mu}(\sigma)$.
The boundary state 
$
\Bigl|
\hat{B}_{N}[A];\{a_{r}\};(\sigma_{1},k^{(1)}),
\Bigr. 
$
$ 
\cdots,
$
$
\Bigl.
(\sigma_{M},k^{(M)})
\Bigr \rangle
$ 
has the eigenvalue equal to\footnote{
The eigenvalue becomes complex 
but this causes no trouble in the 
subsequent arguments at least formally. 
Such subtlety  does not occur 
in the tachyon case.}
$
P^{(B)}_{\mu}(\sigma)
=
$
$
\sum_{r=1}^{M}
\left(
k_{\mu}^{(r)}-a_{r}A_{\mu}(k^{(r)})\partial_{\sigma_{r}}
\right)
$
$
\delta \left(\sigma -\sigma_{r}\right)
$.   
Therefore this state is proportional to the eigenstate  
$\left| P_N^{(B)} \right \rangle$ of 
the corresponding eigenvalues. 
This can be seen by a comparison between 
oscillator representations 
(\ref{aux M gluon bs}) and (\ref{eq:PN-state}) 
of the boundary state and the momentum eigenstate. 
In terms of the parametrization (\ref{eq:PBN-eigenvalue}) 
the above eigenvalue corresponds to  
\begin{eqnarray}
&&
   \varrho_{n\mu}=\sqrt{2} \sum_{r=1}^{M}
      \left(k^{(r)}_{\mu}-ina_{r}A_{\mu}(k^{(r)})\right)
      e^{in\sigma_{r}}~,
\nonumber\\
&&
 \bar{\varrho}_{n\mu}=\sqrt{2} \sum_{r=1}^{M}
      \left(k^{(r)}+ina_{r}A_{\mu}(k^{(r)})\right)
      e^{-in\sigma_{r}}~,
  \quad
p_{0\mu} = \sum_{r=1}^{M}k^{(r)}_{\mu}~.
\label{eq:p-varrho-barvarrho}
\end{eqnarray}
By using these values of 
$\varrho_n$ and $\overline{\varrho}_n$,  
the precise relation between the two states 
can be written as follows : 
\begin{eqnarray}
\lefteqn{
\Bigl|\hat{B}_{N}[A];\{a_{r}\};
  (\sigma_{1},k^{(1)}),\cdots,(\sigma_{M},k^{(M)})
  \Bigr\rangle} \nonumber\\
&&=\prod_{r=1}^{M} \exp
  \left[\alpha' G^{\mu\nu} \sum_{n=1}^{\infty} \left(
       \frac{1}{n} k^{(r)}_{\mu}k^{(r)}_{\nu}
     +(a_{r})^{2} n A_{\mu}(k^{(r)})A_{\nu}(k^{(r)}) \right)\right]
 \nonumber\\
&& \quad \times
      e^{\frac{i}{2\pi}\sum_{r=1}^{M}
          \left( a_{r}A_{\mu}(k^{(r)})-\sigma_{r}k_{\mu}^{(r)}\right)
          \theta^{\mu\nu}\sum_{s=1}^{M}k^{(s)}_{\nu}} 
      \prod_{r<s}^{M}
           e^{\frac{i}{2}k_{\mu}^{(r)}\theta^{\mu\nu}k_{\nu}^{(s)}
                \epsilon (\sigma_{r}-\sigma_{s})}
     \nonumber\\
&& \quad \times
    \left| P^{(B)}_{\mu}(\sigma)=\sum_{r=1}^{M}
               \left(k_{\mu}^{(r)}
               -a_{r} A_{\mu}(k^{(r)}) \partial_{\sigma_{r}}\right)
           \delta (\sigma -\sigma_{r}) \right\rangle~,
     \label{eq:aux state as eigenstate}
\end{eqnarray}
where we have used 
$\mathcal{C}_{P_N^{(B)}}$ in (\ref{CPB}) 
as the normalization constant
of the momentum eigenstate and represented it in terms
of open-string tensors.

Let us comment on the multiplicative factors 
appearing in the above relation. 
As can be seen from 
Eq.(\ref{boundary limit of disk Green function}),  
the second line on the RHS comes from 
the terms proportional to 
$\theta$ of disk Green's functions at the world-sheet 
boundary (the boundary circle). 
As regard the exponential in the first line
on the RHS, we might think of it as being
related to short distance singularities
between the gluons on the boundary circle. 
In fact, by recasting the exponent into the following form,
\begin{eqnarray}
\lefteqn{\alpha' G^{\mu\nu} \sum_{n=1}^{\infty} \left(
       \frac{1}{n} k^{(r)}_{\mu}k^{(r)}_{\nu}
     +(a_{r})^{2} n A_{\mu}(k^{(r)})A_{\nu}(k^{(r)}) \right)}
 \nonumber\\
&&=-\alpha' G^{\mu\nu} \lim_{s\rightarrow r}
    \left(k_{\mu}^{(r)}-a_{r}A_{\mu}(k^{(r)})\partial_{\sigma_{r}}
    \right)
    \left(k_{\nu}^{(s)}-a_{s}A_{\nu}(k^{(s)})\partial_{\sigma_{s}}
    \right)
    \ln \left|e^{i\sigma_{r}}-e^{i\sigma_{s}}\right|~,
\end{eqnarray}
we find a singularity similar 
to that appearing in the OPE 
between auxiliary gluon vertex operators.

In the same way,
concerning the dual boundary states, 
we obtain the following relation as well:
\begin{eqnarray}
\lefteqn{\Bigl\langle \hat{B}_{N};\{a_{r}\};
  (\sigma_{1},k^{(1)}),\cdots, (\sigma_{M},k^{(M)})
   \Bigr| }\nonumber\\
&&= \prod_{r=1}^{M}\left[\alpha' G^{\mu\nu}
       \sum_{n=1}^{\infty}\left(\frac{1}{n}k_{\mu}^{(r)}k_{\nu}^{(r)}
               +(a_{r})^{2}nA_{\mu}(k^{(r)})A_{\nu}(k^{(r)})\right)
       \right]
\nonumber\\
&& \quad \times
e^{-\frac{i}{2}\sum_{r=1}
       \left(a_{r}A_{\mu}(k^{(r)})-\sigma_{r}k_{\mu}^{(r)}\right)
       \theta^{\mu\nu}\sum_{s=1}^{M}k_{\nu}^{(s)}}
    \prod_{r<s}^{M} e^{-\frac{i}{2}k_{\mu}^{(r)} \theta^{\mu\nu}
                    k_{\nu}^{(s)} \epsilon (\sigma_{r}-\sigma_{s})}
    \nonumber\\
&& \quad \times
\left\langle P^{(B)}_{\mu}(\sigma)= - \sum_{r=1}^{M}
               \left(k_{\mu}^{(r)}
               -a_{r} A_{\mu}(k^{(r)}) \partial_{\sigma_{r}}\right)
           \delta (\sigma -\sigma_{r}) \right|~,
    \label{eq:dual gluon bs as eigenstate}
\end{eqnarray}
where $\langle P_{N}^{(B)} |$
denotes the hermitian conjugate of $|P_{N}^{(B)}\rangle$.

Boundary states of open-string tachyons are obtained from 
auxiliary boundary states of gluons by letting 
their auxiliary parameters vanish. 
Hence these boundary states are also eigenstates of 
$\hat{P}_{\mu}^{(B)}(\sigma)$. 
The boundary state 
$\Bigl| B_{N}; (\sigma_{1},k^{(1)}), \cdots, (\sigma_{M},k^{(M)})
 \Bigr \rangle$
has the eigenvalue 
$P^{(B)}_{\mu}(\sigma)
  = \sum_{r=1}^{M}k_{\mu}^{(r)}\delta (\sigma-\sigma_{r})$  
and we can write the state as follows :
\begin{eqnarray}
\lefteqn{
  \Bigl| B_{N}; (\sigma_{1},k^{(1)}),\cdots,
                (\sigma_{M},k^{(M)})\Bigr\rangle }
  \nonumber\\
 && = \prod_{r=1}^{M} \exp \left( \alpha'
             k_{\mu}^{(r)}G^{\mu\nu}k_{\nu}^{(r)}
             \sum_{n=1}^{\infty} \frac{1}{n}  \right)
       \times e^{-\frac{i}{2\pi}\sum_{r=1}^{M}\sigma_{r} k_{\mu}^{(r)}
                  \theta^{\mu\nu}
                  \sum_{s=1}^{M} k_{\nu}^{(s)} }
         \prod_{r<s}^{M}
              e^{\frac{i}{2}k_{\mu}^{(r)}\theta^{\mu\nu}k_{\nu}^{(s)}
                \epsilon (\sigma_{r}-\sigma_{s})}
  \nonumber\\
 && \quad \times
    \left| P^{(B)}_{\mu}(\sigma)
       = \sum_{r=1}^{M}k_{\mu}^{(r)}\delta (\sigma-\sigma_{r})
    \right\rangle~.
    \label{eq:tachyon bs as eigenstate}
\end{eqnarray}

We would like to discuss physical meanings of 
the momentum eigenstates appearing 
in Eqs.(\ref{eq:aux state as eigenstate}) and
(\ref{eq:tachyon bs as eigenstate}).
We henceforth concentrate on gluons.
It is because
analyses for the tachyon 
can be carried out in a parallel way
to the gluon case 
and the formulae for gluons reduce
to those of tachyons 
by setting the auxiliary parameters 
$a_{r}=0$.
Let us regard 
the momentum eigenstate
$
\Bigl \langle P^{(B)}_{\mu}(\sigma)= \Bigr.
$
$
\Bigl. 
-\sum_{r}\left(k_{\mu}^{(r)}
        -a_{r}A_{\mu}(k^{(r)})\partial_{\sigma_{r}}\right)           
 \delta(\sigma -\sigma_{r}) 
\Bigr |$
as a boundary state. 
The associated boundary action $S_{b} \left[X;P^{(B)}\right]$
can be obtained by following the prescription in \cite{CLNY3}.
It is given by
\begin{equation}
e^{-S_{b} \left[X;P^{(B)}\right]}
 =\left.\left\langle
  P^{(B)}_{\mu}(\sigma)
     =-\sum_{r=1}^{M}\left(k_{\mu}^{(r)}
               -a_{r}A_{\mu}(k^{(r)})\partial_{\sigma_{r}}\right)
             \delta(\sigma -\sigma_{r}) \right|
   X_{N}\right\rangle~,
 \label{eq:boundary action1}
\end{equation}
where $\left|X_{N} \right\rangle$ is
the coordinate eigenstate in (\ref{coordinates 2}).
Overlaps 
$\left.\left \langle P_N^{(B)} \right| X_N \right \rangle$ 
can be computed 
for arbitrary eigenvalues 
by using the formulae (\ref{eq:vevee}) 
in the oscillator representations of the eigenstates.
The overlaps take the following forms :
\begin{eqnarray}
\left.\left\langle P^{(B)}_{N}    \right|X_{N}\right\rangle\
&=& \frac{1}{(8\pi)^{\frac{p+1}{2}}}
   \mathcal{C}_{X}^{(N)}\mathcal{C}_{P^{(B)}_{N}}
   \frac{\det E_{\mu\nu}}{\det g_{\mu\nu}}
   \prod_{n=1}^{\infty}
    e^{-\frac{1}{4\pi n}\bar{\varrho}_{n\mu} \theta^{\mu\nu}
                        \varrho_{n\nu}}\nonumber\\
&&\times \exp \left[ i\int^{2\pi}_{0} d\sigma
    \left( \frac{1}{2}B_{\mu\nu}X^{\mu}(\sigma)
            \partial_{\sigma}X^{\nu}(\sigma)
         -X^{\mu}(\sigma)P^{(B)}_{\mu}(\sigma)
    \right)\right]~.
\end{eqnarray}
We apply the above formula to 
Eq.(\ref{eq:boundary action1}) 
with putting
$$P^{(B)}_{\mu}(\sigma)=-\sum_{r=1}^{M}\left(k_{\mu}^{(r)}-
  a_{r}A_{\mu}(k^{(r)})\partial_{\sigma_{r}}\right)
  \delta(\sigma-\sigma_{r})~.$$
This enables us to find that the boundary action
$S_{b} \left[X;P^{(B)}\right]$ becomes,
modulo constant terms,
\begin{eqnarray}
&&
S_{b} \left[X;P^{(B)}\right] 
\nonumber \\
&&
=
-i \int^{2\pi}_{0} d\sigma
     \left(\frac{1}{2} B_{\mu\nu}
      X^{\mu}(\sigma)\partial_{\sigma}X^{\nu}(\sigma)
      +
     X^{\mu}(\sigma)
   \sum_{r=1}^{M}\left(k_{\mu}^{(r)}-
  a_{r}A_{\mu}(k^{(r)})\partial_{\sigma_{r}}\right)
  \delta(\sigma-\sigma_{r})  \right)~.
\nonumber \\
  \label{eq:boundary action2}
\end{eqnarray}

The above boundary action appears naturally in the 
path-integral formalism of world-sheet theory of string. 
Let $\hat{\mathcal{G}}_{A}\left( (\sigma_{1},k^{(1)}),\cdots,
(\sigma_{M},k^{(M)}) \right)$
be the correlation function
of $M$ gluons in the auxiliary forms
with momenta $k_{\mu}^{(r)}$ and
polarization vectors $A_{\mu}(k^{(r)})$
on the world-sheet disk.
In the path-integral approach 
the correlation function can be expressed as 
\begin{eqnarray}
\lefteqn{
 \hat{\mathcal{G}}_{A}\left( (\sigma_{1},k^{(1)}),\cdots,
   (\sigma_{M},k^{(M)}) \right)
}\nonumber\\
 &&= \int [dX^{\mu}]\left[
  e^{i\left(k_{\mu}^{(1)}
      -a_{1}A_{\mu}(k^{(1)})\partial_{\sigma_{1}}\right)
      X^{\mu}(\sigma_{1})}
  \cdots e^{i\left( k^{(M)}_{\mu}
       -a_{M}A_{\mu}(k^{(M)})\partial_{\sigma_{M}}\right)
       X^{\mu}(\sigma_{M})}\right]
  e^{-S[X]}~,
\end{eqnarray}
where $S[X]$ is the world-sheet action (\ref{eq:action2}).
We can recast the RHS of this equation into
\begin{equation}
\hat{\mathcal{G}}_{A}\left( (\sigma_{1},k^{(1)}),\cdots,
(\sigma_{M},k^{(M)}) \right)
= \int [dX^{\mu}] e^{-S_{\mathrm{eff}}[X]}~,
\end{equation}
where $S_{\mathrm{eff}}[X]$ is 
the sum of the bulk free action $S_{0}[X]$
and the boundary action $S_{b}[X]$:
$S_{\mathrm{eff}} [X] =S_{0}[X] + S_{b}[X]$
with
\begin{eqnarray}
&&S_{0}[X]= \frac{1}{4\pi\alpha'} \int_{\Sigma} d\tau d\sigma
           \partial_{a}X^{M}(\sigma,\tau)
           \partial^{a}X^{N}(\sigma,\tau) g_{MN}~, \\
&& S_{b}[X]= -i \int_{\partial \Sigma} d\sigma
    \left[ \frac{1}{2}
     B_{\mu\nu}X^{\mu}(\sigma)\partial_{\sigma}X^{\nu}(\sigma)
     +X^{\mu}(\sigma)  \sum_{r=1}^{M}
        \left(k_{\mu}^{(r)}-a_{r}A_{\mu}(k^{(r)})\partial_{\sigma_{r}}
        \right) \delta (\sigma-\sigma_{r})\right]~.\nonumber
\end{eqnarray}
Here we have used
$\partial_{\sigma}X^{\mu}(\sigma)\cdot \delta (\sigma-\sigma_{r})
  =-X^{\mu}(\sigma) \partial_{\sigma}\delta(\sigma-\sigma_{r})
  =X^{\mu}(\sigma) \partial_{\sigma_{r}}\delta(\sigma-\sigma_{r})$.
We find that the above boundary action $S_{b}[X]$ is
identical with $S_{b}\left[X;P^{(B)}\right]$ 
given in Eq.(\ref{eq:boundary action2}).
Boundary conditions of $X^{\mu}$ can be read 
from variation of $S_{\mathrm{eff}}[X]$. 
It turns out to be
$P^{(B)}_{\mu}(\sigma,0)
  =-\sum_{r=1}^{M}\left(k_{\mu}^{(r)}-
  a_{r}A_{\mu}(k^{(r)})\partial_{\sigma_{r}}\right)
  \delta(\sigma-\sigma_{r})$, 
as expected.

These observations on the boundary states and the momentum 
eigenstates seem to indicate a chance to interpret boundary 
states of open-string legs as momentum eigenstates 
of closed-string with eigenvalues 
being delta functions on the boundary circle. 
Our expectation is, in fact, beyond this. 
Let us recall that 
the momentum eigenstates are expressed 
in Eqs.(\ref{eq:PNB-shift}) and (\ref{eq:PN-shift})
by using operators of the form 
$$
:
\exp 
\left(
i \int_{\partial \Sigma}
d\sigma 
P_{\mu}(\sigma)\hat{X}^{\mu}(\sigma)
\right)
:
$$
where $\hat{X}(\sigma)$ are the closed-string coordinate 
operators and $:~:$ denotes the standard normal-ordering 
of closed-string. 
If we forget about $\hat{X}(\sigma)$ being quantum operators, 
we could write down the exponential, without any hesitation,  
in the path-ordered form along the boundary circle, 
$
\mathcal{P}
\left[
\exp 
\left(
i \int_{\partial \Sigma}
d\sigma 
P_{\mu}^{}(\sigma)\hat{X}^{\mu}(\sigma)
\right)
\right] 
$.
Since $\hat{X}(\sigma)$ are quantum operators 
the path-ordered exponential becomes vague without 
a prescription. Namely we need to regularize 
the path-ordered integral. 
Relevant regularization 
in place of the above normal-ordering 
will be discretization of the boundary circle.  
In such a regularization scheme the eigenvalues 
$P(\sigma)$ are regarded as sums of 
delta functions on the original boundary circle. 
This leads us to the conjecture : 
{\it 
Momentum eigenstates of closed-string have expansions 
by means of boundary states with open-string legs
}.

Finally we would like to mention some related issues. 
As we carried out to a certain extent in this paper, 
our boundary states enable us to perform off-shell 
calculations. This suggests that these states 
can be used in a covariant formulation of a field theory 
of interacting open- and closed-strings. 
String field theory that has prediction power for the 
open-closed mixed systems has been required also in our 
understanding of unstable $D$-branes and tachyon condensations 
associated with them \cite{Sen}.
It seems probable that boundary states with open-string legs 
constructed here are generalized to vertex functions between 
open- and closed-strings in such a field theory.  
We plan to discuss these issues elsewhere.

\section*{\sc Acknowledgements}
T.N. would like to thank members of particle theory group of KEK 
for their hospitality during his stay in summer, 2002.  
The final part of this work was done there. 
K.M. would like to thank Y.~Kitazawa, N.~Ishibashi,
E.~Sezgin,
T.~Suyama, T.~Masuda and P.~Sundell for discussions and comments.
The work of K.M. is
supported in part by NSF Grant PHY-0070964.

\newpage
\appendix
\section{Open-String and Closed-String Tensors}
\label{sec:open-string tensors}

Let $g_{\mu \nu}$ and $B_{\mu \nu}$ be a flat space-time 
metric and a constant two-form gauge field of closed-string. 
These are called closed-string tensors in the text. 
$E_{\mu \nu}$ is given by their combination as 
$E_{\mu\nu}\equiv g_{\mu\nu}+2\pi\alpha'B_{\mu\nu}$.
In \cite{Seiberg-Witten}, 
open-string metric $G_{\mu\nu}$ 
and non-commutativity parameter $\theta^{\mu\nu}$ are obtained 
respectively from symmetric and anti-symmetric
parts of $\left(\frac{1}{E}\right)^{\mu\nu}$ :
\begin{equation}
\left(\frac{1}{E}\right)^{\mu\nu}
 =G^{\mu\nu}+\frac{\theta^{\mu\nu}}{2\pi\alpha'}~,
\end{equation}
where 
$G^{\mu \nu}=(G^{-1})^{\mu \nu}$.  
Tensors $G_{\mu \nu}$ and $\theta^{\mu \nu}$ are called 
open-string tensors in the text. 
The above relation implies :
\begin{eqnarray}
G^{\mu\nu} &=& \frac{1}{2}\left(\frac{1}{E}+\frac{1}{E^{T}}\right)^{\mu\nu}
   =\left(\frac{1}{E}g\frac{1}{E^{T}}\right)^{\mu\nu}
   =\left(\frac{1}{E^{T}}g\frac{1}{E}\right)^{\mu\nu}~,
  \nonumber\\
\frac{\theta^{\mu\nu}}{2\pi\alpha'}
   &=& \frac{1}{2}\left( \frac{1}{E}-\frac{1}{E^{T}} \right)^{\mu\nu}
   = -2\pi\alpha' \left(\frac{1}{E} B \frac{1}{E^{T}}\right)^{\mu\nu}
   = -2\pi\alpha' \left(\frac{1}{E^{T}}B\frac{1}{E}\right)^{\mu\nu}~.
   \label{eq:ot}
\end{eqnarray}
It is also possible to express 
the closed-string tensors 
by means of the open-string ones : 
\begin{eqnarray}
g^{\mu\nu}&=&\left(g^{-1}\right)^{\mu\nu}
  = G^{\mu\nu}
  -\frac{1}{(2\pi\alpha')^{2}} \left( \theta G\theta\right)^{\mu\nu}~,
\nonumber\\
B_{\mu\nu} &=& -\frac{1}{(2\pi\alpha')^{2}}
   \left(\frac{1}{G^{-1}+\frac{\theta}{2\pi\alpha'}}
         \theta
         \frac{1}{G^{-1}-\frac{\theta}{2\pi\alpha'}}
   \right)_{\mu\nu}~.
 \label{eq:ct}
\end{eqnarray}

The tensor 
$N_{\mu\nu} \equiv \left(g\frac{1}{E}E^{T} \right)_{\mu\nu}$ 
is used frequently in the text. 
This tensor enjoys the following relations :
\begin{eqnarray}
&&\left( \frac{1}{E^{T}}N\frac{1}{E^{T}}\right)^{\mu\nu}
 =G^{\mu\nu}~,
\label{eq:ANA-G} \\
&& \left(g^{-1} N g^{-1} \right)^{\mu\nu}
  =G^{\mu\nu}+\frac{1}{\left(2\pi\alpha'\right)^{2}}
               \left(\theta G\theta\right)^{\mu\nu}
   +\frac{\theta^{\mu\nu}}{\pi\alpha'}~,
\label{eq:N-ot} \\
&& \left(N g^{-1} N^{T} \right)_{\mu\nu}
   =\left(N^{T}g^{-1}N\right)_{\mu\nu}=g_{\mu\nu}~.
\label{eq:NgN-g}
\end{eqnarray}
Eq.(\ref{eq:N-ot}) implies 
\begin{equation}
\left(g^{-1}\left(N-N^{T}\right)g^{-1}\right)^{\mu\nu}
 =\frac{2}{\pi\alpha'} \theta^{\mu\nu}~.
 \label{eq:N-theta}
\end{equation}
Combination of Eq.(\ref{eq:N-ot}) 
with Eq.(\ref{eq:ot}) gives the following equalities : 
\begin{eqnarray}
g^{\mu\nu} + \left(g^{-1}Ng^{-1}\right)^{\mu\nu}
  &=& 2\left(\frac{1}{E} \right)^{\mu\nu}
  =2G^{\mu\nu}+\frac{\theta^{\mu\nu}}{\pi\alpha'}~,
  \nonumber\\
g^{\mu\nu}+\left(g^{-1}N^{T}g^{-1}\right)^{\mu\nu}
  &=&2\left(\frac{1}{E^{T}}\right)^{\mu\nu}
  =2G^{\mu\nu}-\frac{\theta^{\mu\nu}}{\pi\alpha'}~.
 \label{eq:N-A}
\end{eqnarray}


\section{Some Formulae of Creation and Annihilation Modes}
\label{sec:formulae}

We present formulae which become very useful 
for computations of string amplitudes in the text. 
As an illustration we derive  
$F_A$ in Eq.(\ref{result on FA}) 
(and $F$ in Eq.(\ref{result on F})).

\subsection{The formulae}

\begin{formula}
Let 
$u_{\mu}$, $v_{\mu}$, $w_{\mu}$ and $y_{\mu}$ 
be arbitrary complex $(p+1)$-vectors. 
Let 
$\Omega^{(1)}_{\mu\nu}$ and $\Omega^{(2)}_{\mu\nu}$ 
be any complex $(p+1)\times (p+1)$-matrices. 
The following equality holds : 
\begin{eqnarray}
\lefteqn{ \exp \left[\frac{1}{n}\alpha_{n}^{\mu}\Omega^{(1)}_{\mu\nu}
              \tilde{\alpha}^{\nu}_{n}
              +u_{\mu}\alpha^{\mu}_{n}
              +v_{\mu}\tilde{\alpha}^{\mu}_{n}\right]
     \exp \left[\frac{1}{n}\alpha_{-n}^{\mu}\Omega^{(2)}_{\mu\nu}
               \tilde{\alpha}_{-n}^{\nu}
               +w_{\mu}\alpha^{\mu}_{-n}
               +y_{\mu}\tilde{\alpha}^{\mu}_{-n}\right]
     |\mathbf{0}\rangle}  \nonumber\\
&&= \frac{\det g_{\mu\nu}}
         {\det \left(g-\Omega^{(2)T}g^{-1}\Omega^{(1)}\right)_{\mu\nu}}
             \nonumber\\
&& \quad \times \exp \left[
    n u_{\mu}\left(\frac{1}{g-\Omega^{(2)}g^{-1}\Omega^{(1)T}}
             \right)^{\mu\nu}w_{\nu}
    +n v_{\mu}\left(\frac{1}{g-\Omega^{(2)T}g^{-1}\Omega^{(1)}}
             \right)^{\mu\nu}y_{\nu} \right.\nonumber\\
&& \hspace{5em} +n u_{\mu}\left(g^{-1}\Omega^{(2)}
        \frac{1}{g-\Omega^{(1)T}g^{-1}\Omega^{(2)}}\right)^{\mu\nu}
        v_{\nu}  \nonumber\\
&& \hspace{5em} \left.
    +nw_{\mu}\left(g^{-1}\Omega^{(1)}
        \frac{1}{g-\Omega^{(2)T}g^{-1}\Omega^{(1)}}\right)^{\mu\nu}
        y_{\nu} \right]     \nonumber\\
&&  \quad \times \exp \left[
       \frac{1}{n}\alpha^{\mu}_{-n}
      \left(\Omega^{(2)}\frac{1}{g-\Omega^{(1)T}g^{-1}\Omega^{(2)}}
           g\right)_{\mu\nu} \tilde{\alpha}^{\nu}_{-n}
     \right.
    \nonumber\\
&& \hspace{5em}
+\left\{v_{\lambda}
            {\left(g^{-1}\Omega^{(2)T}\right)^{\lambda}}_{\mu}
            +w_{\mu} \right\}
     {\left(\frac{1}{g-\Omega^{(1)}g^{-1}\Omega^{(2)T}} g
      \right)^{\mu}}_{\nu} \alpha^{\nu}_{-n} 
\nonumber\\
&& \hspace{5em} \left.
    +\left\{ u_{\lambda}{\left(g^{-1}\Omega^{(2)}\right)^{\lambda}}_{\mu}
            + y_{\mu}\right\}
     {\left(\frac{1}{g-\Omega^{(1)T}g^{-1}\Omega^{(2)}} g
       \right)^{\mu}}_{\nu} \tilde{\alpha}^{\nu}_{-n}
      \right] |\mathbf{0}\rangle~.\label{eq:ee}
\end{eqnarray}
\label{f-1}
\end{formula}

Use of coherent states becomes convenient 
to see the above equality. 
Let $\lambda_{n \mu}^{\pm}$ be complex $(p+1)$-vectors   
$(n=1,2,\ldots)$. 
Coherent state 
$\left| \lambda^{+}_{n},\lambda^{-}_{n} \right)$
of the $n$-th level oscillators 
is defined as 
\begin{equation}
\left|\lambda^{+}_{n},\lambda^{-}_{n}\right)
 =\exp\left[\frac{1}{n}\left(\lambda^{+}_{n\mu}\alpha^{\mu}_{-n}
           +\lambda^{-}_{n\mu}\tilde{\alpha}^{\mu}_{-n}
      \right)\right]
  |0\rangle~.
\label{eq:coherent state}
\end{equation}
$\lambda_{n \mu}^{\pm}$ becomes 
eigenvalues of the annihilation operators 
$\alpha^{\mu}_{n}$ and $\tilde{\alpha}^{\mu}_{n}$ 
respectively :
\begin{equation}
\alpha^{\mu}_{n} \left|\lambda^{+}_{n},\lambda^{-}_{n} \right)
  = g^{\mu\nu}\lambda^{+}_{n\nu}
      \left|\lambda^{+}_{n}, \lambda^{-}_{n}\right)~,
      \quad
\tilde{\alpha}^{\mu}_{n}
   \left|\lambda^{+}_{n},\lambda^{-}_{n} \right)
  = g^{\mu\nu}\lambda^{-}_{n\nu}
      \left|\lambda^{+}_{n},\lambda^{-}_{n}\right)~.
\end{equation}
Foe each $n$, 
the coherent states 
$\left|\lambda^{+}_{n},\lambda^{-}_{n} \right)$
constitute a (over)complete basis 
of the Fock space built by 
$\alpha_{-n}^{\mu}$ and $\tilde{\alpha}_{-n}^{\mu}$. 
The completeness relation reads
\begin{eqnarray}
1 &=& \int 
     \frac{1}{(\det g_{\mu\nu})^{2}} \prod_{\mu=0}^{p}\left(
      \frac{d\bar{\lambda}^{+}_{n\mu} d\lambda^{+}_{n\mu}}{2\pi in}
      \frac{d\bar{\lambda}^{-}_{n\mu}
            d\lambda^{-}_{n\mu}}{2\pi in}\right)
     \nonumber\\
  && \quad \times
      \left|\lambda^{+}_{n},\lambda^{-}_{n}\right)
     \exp \left[-\frac{1}{n}\left(
         \bar{\lambda}^{+}_{n\mu}g^{\mu\nu}\lambda^{+}_{n\nu}+
         \bar{\lambda}^{-}_{n\mu}g^{\mu\nu}
         \lambda^{-}_{n\nu}\right)\right]
     \left(\lambda^{+}_{n},\lambda^{-}_{n} \right|~.
 \label{eq:complete-cohe}
\end{eqnarray}
Here
$\left(\lambda^{+}_{n},\lambda^{-}_{n}\right|$
denotes the hermitian conjugate of the state
$\left|\lambda^{+}_{n}, \lambda^{-}_{n} \right)$.  
It takes the form of
\begin{equation}
\left(\lambda^{+}_{n},\lambda^{-}_{n}\right|
= \langle \mathbf{0}|
  \exp\left[\frac{1}{n}\left(\bar{\lambda}^{+}_{n\mu}\alpha^{\mu}_{n}
            +\bar{\lambda}^{-}_{n\mu}\tilde{\alpha}^{\mu}_{n}
     \right)\right]~, \label{eq:dual of coherent state}
\end{equation}
where $\bar{\lambda}^{\pm}_{n\mu}$ 
are complex conjugate to $\lambda^{\pm}_{n\mu}$ 
and become eigenvalues of the creation operators 
$\alpha^{\mu}_{-n}$ and $\tilde{\alpha}^{\mu}_{-n}$.

Formula \ref{f-1} can be shown by making use of the above 
partition of unity. 
We only describe an outline of the proof. 
We first 
insert the unity given in Eq.(\ref{eq:complete-cohe})
between the two exponentials
on the LHS of Eq.(\ref{eq:ee}).
This makes the LHS into  Gaussian
integrals with respect to $(\lambda^{+}_{n},\bar{\lambda}_{n}^{+})$
and $(\lambda^{-}_{n},\bar{\lambda}^{-}_{n})$.
These Gaussian integrals are performed successively
by using the relation
\begin{equation}
\int \prod_{\mu=0}^{p}\left(\frac{d\bar{\lambda}_{\mu} d\lambda_{\mu}}
                                  {2i}   \right)
     \exp \left[ -\left(\bar{\lambda}_{\mu}+\alpha_{\mu}\right)
                   M^{\mu\nu}
                   \left(\lambda_{\nu}+\beta_{\nu}\right)\right]
      = \frac{\pi^{p+1}}{\det M^{\mu\nu}}~,
\end{equation}
for $\forall \alpha_{\mu},\beta_{\mu} \in \mathbb{C}^{p+1}$
and any $(p+1)\times (p+1)$ matrix $M^{\mu\nu}$.
Then we obtain the RHS of Eq.(\ref{eq:ee}).

The following formula is a corollary of Formula \ref{f-1}:
\begin{formula}
\begin{eqnarray}
\lefteqn{
  \langle \mathbf{0}|
  \exp \left[\frac{1}{n}\alpha_{n}^{\mu}\Omega^{(1)}_{\mu\nu}
              \tilde{\alpha}^{\nu}_{n}
              +u_{\mu}\alpha^{\mu}_{n}
              +v_{\mu}\tilde{\alpha}^{\mu}_{n}\right]
     \exp \left[\frac{1}{n}\alpha_{-n}^{\mu}\Omega^{(2)}_{\mu\nu}
               \tilde{\alpha}_{-n}^{\nu}
               +w_{\mu}\alpha^{\mu}_{-n}
               +y_{\mu}\tilde{\alpha}^{\mu}_{-n}\right]
     |\mathbf{0} \rangle } \nonumber\\
&&= \frac{\det g_{\mu\nu}}
    {\det \left(g-\Omega^{(2)T}g^{-1}\Omega^{(1)}\right)_{\mu\nu}}
    \nonumber\\
&& \quad \times \exp \left[
    n u_{\mu}\left(\frac{1}{g-\Omega^{(2)}g^{-1}\Omega^{(1)T}}
             \right)^{\mu\nu}w_{\nu}
    +n v_{\mu}\left(\frac{1}{g-\Omega^{(2)T}g^{-1}\Omega^{(1)}}
             \right)^{\mu\nu}y_{\nu} \right.\nonumber\\
&& \hspace{4em} +n u_{\mu}\left(g^{-1}\Omega^{(2)}
        \frac{1}{g-\Omega^{(1)T}g^{-1}\Omega^{(2)}}\right)^{\mu\nu}
        v_{\nu}  \nonumber\\
&& \hspace{4em} \left.
    +nw_{\mu}\left(g^{-1}\Omega^{(1)}
        \frac{1}{g-\Omega^{(2)T}g^{-1}\Omega^{(1)}}\right)^{\mu\nu}
        y_{\nu} \right]~. 
\label{eq:vevee} 
\end{eqnarray}
\label{f-2}
\end{formula}

It is worth noting that a similar formula to
Formula \ref{f-1} is used in
open string field theory
(see e.g.\ Eq.(B.2) of \cite{Kishi}). 
In particular, 
when we restrict $\Omega^{(1)}_{\mu\nu}$ and
$\Omega^{(2)}_{\mu\nu}$
to symmetric matrices and change the variables from
$\alpha_{\pm n}^{\mu}$ and $\tilde{\alpha}_{\pm n}^{\mu}$
to $\hat{a}^{(i)\mu}_{n}$ and $\hat{a}^{\dagger (i)\mu}_{n}$
($i=\mathrm{I}, \mathrm{II}$) defined in Eq.(\ref{eq:2harmonics}),
Formula~\ref{f-1} reduces to
the formula (B.2) of \cite{Kishi}.
Formula~\ref{f-1} can be regarded as a
closed-string extension  of it.

\subsection{Applications}

In this appendix,
we derive Eqs.(\ref{result on F}) and (\ref{result on FA}).
This also serves as an illustration of usage 
of the formulae.

Contributions of the massive states of closed-string propagating
between boundary states of tachyons
are denoted by $F \left(q_{c},\{\sigma_{r}\},\{k^{(r)}\}\right)$
in Eq.(\ref{pre tachyon amplitude by boundary state}).
Those between boundary states of gluons
are denoted by $F_{A}\left(q_{c},\{\sigma_{r}\},\{k^{(r)}\};\{a_{r}\}
                     \right)$
in Eq.(\ref{pre gluon amplitude by boundary state}).
Their oscillator representations 
(\ref{def of F}) and (\ref{def of FA}) 
imply that $F_{A}$ 
reduces to $F$ 
by setting $\forall a_{r}=0$. 
We therefore focus on Eq.(\ref{result on FA}).

The representation (\ref{def of FA}) 
allows us to evaluate 
$F_A$ by applying Formula \ref{f-2} 
with the following substitution : 
\begin{eqnarray}
\Omega^{(1)}_{\mu\nu}&=&-q_{c}^{n}N_{\mu\nu}~,
\qquad
\Omega^{(2)}_{\mu\nu}=-N_{\mu\nu}~,\nonumber\\
u_{\mu}&=&
  -\frac{\sqrt{2\alpha'}}{n} q_{c}^{\frac{n}{2}}
   \sum_{r=M+1}^{M+N}
   \left(k_{\mu}^{(r)}+ina_{r} A_{\nu}(k^{(r)})\right)
     {\left(\frac{1}{E^{T}}g\right)^{\nu}}_{\mu}
     e^{-in\sigma_{r}}~,\nonumber\\
v_{\mu}&=&-\frac{\sqrt{2\alpha'}}{n} q_{c}^{\frac{n}{2}}
   \sum_{r=M+1}^{M+N}
    \left( k_{\nu}^{(r)}-ina_{r} A_{\nu}(k^{(r)})\right)
    {\left(\frac{1}{E}g\right)^{\nu}}_{\mu}
    e^{in\sigma_{r}}~,\nonumber\\
w_{\mu}&=& \frac{\sqrt{2\alpha'}}{n} \sum_{r=1}^{M}
   \left(k_{\nu}^{(r)}-ina_{r} A_{\mu}(k^{(r)})\right)
   {\left(\frac{1}{E^{T}}g\right)^{\nu}}_{\mu}
   e^{in\sigma_{r}}~,\nonumber\\
y_{\mu} &=& \frac{\sqrt{2\alpha'}}{n} \sum_{r=1}^{M}
   \left(k_{\nu}^{(r)}+ina_{r} A_{\nu}(k^{(r)})\right)
   {\left(\frac{1}{E}g\right)^{\nu}}_{\mu}
   e^{-in\sigma_{r}}~.
 \label{eq:substitution}
\end{eqnarray}

In what follows, we will calculate each term
in Eq.(\ref{eq:vevee}) with the above substitution.
In this course the following relations will be used 
implicitly :
\begin{eqnarray} 
\left(g-\Omega^{(2)}g^{-1}\Omega^{(1)T}\right)_{\mu\nu}
   =\left(g-\Omega^{(2)T} g^{-1} \Omega^{(1)}\right)_{\mu\nu}
    =\left(1-q_{c}^{n}\right)g_{\mu\nu}~.
\end{eqnarray}
First, 
the determinants in the RHS of Eq.(\ref{eq:vevee}) 
become as follows :
\begin{equation}
\frac{\det g_{\mu\nu}}
     {\det \left(g-\Omega^{(2)T} g^{-1} \Omega^{(1)}\right)_{\mu\nu}}
 = \left(\frac{1}{1-q_{c}^{n}}\right)^{p+1}~.
\label{eq:tensors-and-det}
\end{equation}
As regards the exponential in the RHS of Eq.(\ref{eq:vevee}), 
the first two terms of the exponent are translated to 
\begin{eqnarray}
\lefteqn{
  nu_{\mu}\left(\frac{1}{g-\Omega^{(2)}g^{-1}\Omega^{(1)T}}
          \right)^{\mu\nu} w_{\nu}
  +nv_{\mu}\left(\frac{1}{g-\Omega^{(2)T}g^{-1}\Omega^{(1)}}
          \right)^{\mu\nu} y_{\nu}
}\nonumber\\
&&= -2\alpha' G^{\mu\nu} \sum_{r=1}^{M} \sum_{s=M+1}^{M+N}
  \left(k_{\mu}^{(r)}-a_{r} A_{\mu}(k^{(r)})
        \partial_{\sigma_{r}}\right)
  \left(k_{\mu}^{(s)}-a_{s} A_{\mu}(k^{(s)})
        \partial_{\sigma_{s}}\right)
\nonumber\\
&& \hspace{12em} \times
     \frac{q^{\frac{n}{2}}_{c}}{n\left(1-q_{c}^{n}\right)}
     \left\{ e^{in(\sigma_{r}-\sigma_{s})} +
           e^{-in(\sigma_{r}-\sigma_{s})} \right\}~.
   \label{eq:uw+vy}
\end{eqnarray}
We can recast the third term as follows : 
\begin{eqnarray}
\lefteqn{
  nu_{\mu} \left( g^{-1}\Omega^{(2)}
   \frac{1}{g-\Omega^{(1)T}g^{-1}\Omega^{(2)}}
   \right)^{\mu\nu} v_{\nu}} \nonumber\\
&&=-2\alpha' G^{\mu\nu}
   \sum_{r,s=M+1}^{M+N}
     \left(k_{\mu}^{(r)}+ina_{r} A_{\mu}(k^{(r)})\right)
     \left(k_{\nu}^{(s)}-ina_{s} A_{\nu}(k^{(s)})\right)
     \frac{q_{c}^{n} e^{-in(\sigma_{r}-\sigma_{s})} }
          {n\left(1-q_{c}^{n}\right)}
\nonumber\\
&& = -2\alpha' G^{\mu\nu} \sum_{M+1\leq r<s \leq M+N}
     \left(k_{\mu}^{(r)}-a_{r} A_{\mu}(k^{(r)})
        \partial_{\sigma_{r}}\right)
  \left(k_{\nu}^{(s)}-a_{s} A_{\nu}(k^{(s)})
        \partial_{\sigma_{s}}\right) \nonumber\\
&& \hspace{14em} \times
     \frac{q^{n}_{c}}{n\left(1-q_{c}^{n}\right)}
     \left\{ e^{in(\sigma_{r}-\sigma_{s})} +
           e^{-in(\sigma_{r}-\sigma_{s})} \right\}
      \nonumber\\
&& \hspace{2em} -2\alpha' G^{\mu\nu}
   \sum_{r=M+1}^{M+N} \left\{k_{\mu}^{(r)}k_{\nu}^{(r)}
      +n^{2} \left(a_{r}\right)^{2} A_{\mu}(k^{(r)})A_{\nu}(k^{(r)})
      \right\}
      \frac{q^{n}_{c}}{n\left(1-q_{c}^{n}\right)}~.
   \label{eq:uv}
\end{eqnarray}
The last term of the exponent 
turns out to be the same as (\ref{eq:uv}) 
with shifting the indices $r$ and $s$ 
to $1\leq r,s \leq M$.

Contribution of the $n$-th level oscillators 
is obtained by gathering all the above results. 
Contributions of the massive states are given 
by the infinite products taken over all the levels. 
These turn out to be written as follows :  
\begin{eqnarray}
\lefteqn{
F_{A}\left(q_{c},\left\{\sigma_{r}\right\}
    \left\{k^{(r)}\right\};\left\{ a_{r}\right\}
    \right)
}\nonumber\\
&&=\left( \frac{1}{1-q_{c}^{n}}\right)^{p+1} \nonumber\\
&& \quad \times
  \prod_{1\leq r<s \leq M} \exp \Bigg[ -2\alpha'G^{\mu\nu}
    \left(k_{\mu}^{(r)}-a_{r} A_{\mu}(k^{(r)})
        \partial_{\sigma_{r}}\right)
  \left(k_{\nu}^{(s)}-a_{s} A{\nu}(k^{(s)})
        \partial_{\sigma_{s}}\right) \nonumber\\
&& \hspace{13em} \times \sum_{n=1}^{\infty}
   \frac{ \left(q_{c}e^{i(\sigma_{r}-\sigma_{s})}\right)^{n}
          +\left(q_{c} e^{-i(\sigma_{r}-\sigma_{s})}\right)^{n}
          -2q_{c}^{n}}
         {n\left(1-q_{c}^{n}\right)}  \Bigg] \nonumber\\
&& \quad \times
  \prod_{M+1\leq r<s \leq M+N}
  \exp \Bigg[-2\alpha'G^{\mu\nu}
    \left(k_{\mu}^{(r)}-a_{r} A_{\mu}(k^{(r)})
        \partial_{\sigma_{r}}\right)
  \left(k_{\nu}^{(s)}-a_{s} A_{\nu}(k^{(s)})
        \partial_{\sigma_{s}}\right) \nonumber\\    
&& \hspace{16em} \times \sum_{n=1}^{\infty}
   \frac{ \left(q_{c}e^{i(\sigma_{r}-\sigma_{s})}\right)^{n}
          +\left(q_{c} e^{-i(\sigma_{r}-\sigma_{s})}\right)^{n}
          -2q_{c}^{n}}
         {n\left(1-q_{c}^{n}\right)}  \Bigg] \nonumber\\
&& \quad \times
    \prod_{r=1}^{M} \prod_{s=M+1}^{M+N}
      \exp \Bigg[
    -2\alpha'G^{\mu\nu}
    \left(k_{\mu}^{(r)}-a_{r} A_{\mu}(k^{(r)})
        \partial_{\sigma_{r}}\right)
  \left(k_{\nu}^{(s)}-a_{s} A_{\nu}(k^{(s)})
        \partial_{\sigma_{s}}\right) \nonumber\\
&& \hspace{13em} \times \sum_{n=1}^{\infty}
   \frac{ \left(q_{c}^{\frac{1}{2}}e^{i(\sigma_{r}-\sigma_{s})}
          \right)^{n}
         +\left( q_{c}^{\frac{1}{2}} e^{-i(\sigma_{r}-\sigma_{s})}
          \right)^{n}
         -2q_{c}^{n}}
        {n\left(1-q_{c}^{n}\right)}\Bigg] \nonumber\\
&& \quad \times \exp
    \left[ -2\alpha' G^{\mu\nu} \left\{
         \left(\sum_{r=1}^{M+N}k_{\mu}^{(r)}\right)
         \left(\sum_{s=1}^{M+N}k_{\nu}^{(s)}\right)
         +\sum_{r=1}^{M+N}\left(a_{r}\right)^{2}
            A_{\mu}(k^{(r)}) A_{\nu}(k^{(r)}) \right\}\right.
    \nonumber\\
&& \hspace{5em} \left.  \times
    \sum_{n=1}^{\infty} \frac{q_{c}^{n}}{n\left(1-q_{c}^{n}\right)}
    \right]~,
\label{eq:preFAresult}
\end{eqnarray}
where we have used the following rearrangement : 
\begin{eqnarray}
\sum_{r=1}^{M+N} k_{\mu}^{(r)}G^{\mu\nu} k_{\nu}^{(r)}
 &=&\left(\sum_{r=1}^{M+N}k_{\mu}^{(r)}\right)G^{\mu\nu} 
         \left(\sum_{s=1}^{M+N}k_{\nu}^{(s)}\right)
     - 2\sum_{1\leq r<s \leq M}k_{\mu}^{(r)}G^{\mu\nu}k_{\nu}^{(s)}
  \nonumber\\
 && \quad
       - 2\sum_{r=1}^{M}\sum_{s=M+1}^{M+N} G^{\mu\nu}
            k_{\mu}^{(r)}k_{\nu}^{(s)}
    - 2\sum_{M+1\leq r<s \leq M+N}
            k_{\mu}^{(r)} G^{\mu\nu}k_{\nu}^{(s)}~.
\end{eqnarray}

The infinite sums in Eq.(\ref{eq:preFAresult}) 
can be translated into infinite products by using 
the following relation :
\begin{equation}
\sum_{n=1}^{\infty} \frac{x^{n}}{n\left(1-y^{n}\right)}
 = - \ln \prod_{m=0}^{\infty}\left(1-xy^{m}\right)~.
\end{equation}
After these translations we obtain Eq.(\ref{result on FA}).

\section{Eigenstates}\label{sec:eigen}

We provide oscillator realizations of eigenstates 
of the closed-string operators 
$\hat{X}^{\mu}(\sigma)$, $\hat{P}_{\mu}(\sigma)$
and $\hat{P}^{(B)}_{\mu}(\sigma)$. 
These are used in the text.

To start with, 
it is useful to recall 
coordinate and momentum eigenstates 
of a harmonic oscillator in quantum mechanics. 
Description of this system is made by 
an annihilation and a creation operators 
$\hat{a}$ and $\hat{a}^{\dagger}$ satisfying 
$[\hat{a},\hat{a}^{\dagger}]=1$. 
Let $\hat{q}$ and $\hat{p}$ be the coordinate 
and the momentum operators satisfying 
$[\hat{q}, \hat{p}]=1$. 
The operators $(\hat{a},\hat{a}^{\dagger})$ 
are related with the canonical pair 
$(\hat{q},\hat{p})$  as 
\begin{eqnarray}
&& \hat{a} = \frac{1}{\sqrt{2}} \left( \frac{1}{\Gamma}\hat{q}
    +i\bar{\Gamma}\hat{p} \right)~,
\quad
\hat{a}^{\dagger}
=\frac{1}{\sqrt{2}} \left(\frac{1}{\bar{\Gamma}}\hat{q}
      -i\Gamma \hat{p}\right)~,\nonumber\\
&& \qquad \Longleftrightarrow \quad 
    \hat{q}=\frac{1}{\sqrt{2}} \left(\Gamma \hat{a}
             +\bar{\Gamma}\hat{a} \right)~,
    \quad
    \hat{p} =\frac{-i}{\sqrt{2}} \left(
             \frac{1}{\bar{\Gamma}}\hat{a}-\frac{1}{\Gamma}\hat{a}^{\dagger}
             \right)~,
   \label{eq:def of a}
\end{eqnarray}
where $\Gamma \in \mathbb{C}$ is chosen so that 
$1/\omega |\Gamma|^2$ becomes the mass of harmonic oscillator. 
Here $\omega$ is the frequency. 
Let $|q \rangle$ and $|p \rangle$ be the eigenstates of 
$\hat{q}$ and $\hat{p}$. 
They are normalized by   
$\langle q'|q\rangle = \delta (q'-q)$
and $\langle p'|p\rangle = \delta(p'-p)$.
It is possible to realize these states 
on the Fock vacuum $|0 \rangle$ by using $\hat{a}^{\dagger}$. 
They are given by :
\begin{eqnarray}
| q\rangle &=& \left(\frac{1}{\pi |\Gamma|^{2}}\right)^{\frac{1}{4}}
 \exp \left[ 
  -\frac{\bar{\Gamma}}{2\Gamma} \hat{a}^{\dagger} \hat{a}^{\dagger}
  +\frac{\sqrt{2}}{\Gamma} q \hat{a}^{\dagger}
  -\frac{q^{2}}{2|\Gamma|^{2}} 
  \right] |0\rangle~,\nonumber\\
 |p\rangle &=& \left(\frac{|\Gamma|^{2}}{\pi}\right)^{\frac{1}{4}}
   \exp \left[ \frac{\bar{\Gamma}}{2\Gamma} \hat{a}^{\dagger}
               \hat{a}^{\dagger}
               +i\sqrt{2}\bar{\Gamma}p\hat{a}^{\dagger}
               -\frac{|\Gamma|^{2}}{2}p^{2}
         \right]|0\rangle~. 
   \label{eq:ho-eigenstates}
\end{eqnarray}

The above realizations of coordinate and momentum eigenstates 
are generalized to the case of string. 
We first expand the coordinate and the momentum operators 
of closed-string by suitable canonical pairs, 
typically denoted by $(\hat{\phi}^{(i)}_n,\hat{\pi}_n^{(i)})$ 
($i=\mathrm{I,II}$; $n=1,2,..$). 
For the each pair, 
we introduce creation and annihilation operators, 
typically $(\hat{a}_n^{(i)},\hat{a}_n^{(i)\dagger})$.  
We then realize eigenstates of the canonical operators 
$\hat{\phi}^{(i)}_n$ and $\hat{\pi}_n^{(i)}$
by using the creation operators 
$\hat{a}_n^{(i)\dagger}$. 
Eigenstates of the coordinate and the momentum operators 
of closed-string are given by their infinite products.

In order to obtain canonical pairs,
we expand $\hat{X}^{\mu}(\sigma)$, $\hat{P}_{\mu}(\sigma)$
and $\hat{P}^{(B)}_{\mu}(\sigma)$ by a real basis
of the periodic functions
on a circle:
$\left\{ 1, \cos n\sigma, \sin n\sigma \right\}_{n=1}^{\infty}$,
\begin{eqnarray}
\hat{X}^{\mu}(\sigma) &=&
  \hat{x}_{0}^{\mu} + \sqrt{2} \sum_{n=1}^{\infty}
    \left( \hat{\phi}^{(\mathrm{I})\mu}_{n} \cos n\sigma
           +\hat{\phi}^{(\mathrm{II})\mu}_{n} \sin n\sigma
    \right)~, \nonumber\\
\hat{P}_{\mu}(\sigma) &=& \frac{1}{2\pi}
   \left[ \hat{p}_{0\mu}+\sqrt{2}
         \sum_{n=1}^{\infty} \left(
           \hat{\pi}^{(\mathrm{I})}_{n\mu} \cos n\sigma
           +\hat{\pi}^{(\mathrm{II})}_{n\mu} \sin n\sigma
       \right)\right]~, \nonumber\\
\hat{P}^{(B)}_{\mu} (\sigma) &=& 
       \frac{1}{2\pi}
   \left[ \hat{p}_{0\mu}+\sqrt{2}
         \sum_{n=1}^{\infty} \left(
           \hat{\varpi}^{(\mathrm{I})}_{n\mu} \cos n\sigma
           +\hat{\varpi}^{(\mathrm{II})}_{n\mu} \sin n\sigma
       \right)\right]~.
   \label{eq:mode-expansion}
\end{eqnarray}
The canonical commutation relations
between $\hat{X}^{\mu}(\sigma)$ and $\hat{P}_{\mu}(\sigma)$
($\hat{P}^{(B)}_{\mu}(\sigma)$) are converted
into the following relations
among the hermitian operators
$\hat{\phi}^{(i)\mu}_{n}$, $\hat{\pi}^{(i)}_{n\mu}$
and $\hat{\varpi}^{(i)}_{n\mu}$
($i=\mathrm{I},\mathrm{II}$; $n=1,2,\ldots$):
\begin{equation}
[\hat{x}_{0}^{\mu},\hat{p}_{0\nu}] = i\delta^{\mu}_{\nu}~,
\quad
[\hat{\phi}^{(i)\mu}_{m},\hat{\pi}^{(j)}_{n\nu}]
  =i\delta^{\mu\nu}\delta^{i,j}\delta_{m,n}~,
\quad
[\hat{\phi}^{(i)\mu}_{m},\hat{\varpi}^{(j)}_{n\nu}]
  =i\delta^{\mu}_{\nu}~,
\end{equation}
and the others are vanishing.
These can be derived from the following
relations as well:
\begin{eqnarray}
&&\left\{
  \begin{array}{l}
  \displaystyle
  \hat{\phi}^{(\mathrm{I})\mu}_{n}
  =\frac{\sqrt{\alpha'}}{2n}i
     \left(\alpha^{\mu}_{n}+\tilde{\alpha}^{\mu}_{n}
     - \alpha^{\mu}_{-n}  -  \tilde{\alpha}^{\mu}_{-n}\right)\\[1.5ex]
  \displaystyle
   \hat{\phi}^{(\mathrm{II})\mu}_{n}
   = \frac{\sqrt{\alpha'}}{2n} \left(
    \alpha^{\mu}_{n}-\tilde{\alpha}^{\mu}_{n}
    +\alpha^{\mu}_{-n}-\tilde{\alpha}^{\mu}_{-n} \right)
  \end{array}  \right., ~
\left\{
   \begin{array}{l}
   \displaystyle
    \hat{\pi}^{(\mathrm{I})}_{n\mu}=\frac{1}{2\sqrt{\alpha'}}g_{\mu\nu}
    \left(\alpha^{\nu}_{n}+\tilde{\alpha}^{\nu}_{n}
         +\alpha^{\nu}_{-n}+\tilde{\alpha}^{\nu}_{-n}\right)\\[1.5ex]
   \displaystyle
     \hat{\pi}^{(\mathrm{II})}_{n\mu}=\frac{-i}{2\sqrt{\alpha'}}g_{\mu\nu}
     \left(\alpha^{\nu}_{n}-\tilde{\alpha}^{\nu}_{n}
       -\alpha^{\nu}_{-n}+\tilde{\alpha}^{\nu}_{-n} \right)\\
   \end{array} \right., \nonumber\\
&&
\left\{
  \begin{array}{l}
    \displaystyle
      \hat{\varpi}^{(\mathrm{I})}_{n\mu} =
       \frac{1}{2\sqrt{\alpha'}}\left(
       E_{\mu\nu}\alpha^{\nu}_{n}
       +E^{T}_{\mu\nu}\tilde{\alpha}^{\nu}_{n}
       +E_{\mu\nu}\alpha^{\nu}_{-n}
       +E^{T}_{\mu\nu}\tilde{\alpha}^{\nu}_{-n}\right)\\[1.5ex]
   \displaystyle
      \hat{\varpi}^{(\mathrm{II})}_{n\mu}
      =\frac{-i}{2\sqrt{\alpha'}}
      \left( E_{\mu\nu}\alpha^{\nu}_{n}
            -E^{T}_{\mu\nu}\tilde{\alpha}^{\nu}_{n}
            -E_{\mu\nu}\alpha^{\nu}_{-n}
            +E^{T}_{\mu\nu} \tilde{\alpha}^{\nu}_{-n} \right)
  \end{array} \right..
\label{eq:among-modes}
\end{eqnarray}

Let us begin with the canonical pairs
$\{\hat{\phi}^{(i)\mu}_{n},\hat{\pi}^{(i)}_{n\mu}\}$.
For these pairs, we introduce annihilation
and creation modes
$(\hat{a}^{(i)\mu}_{n},\hat{a}^{\dagger (i)\mu}_{n})$ by
\begin{equation}
\hat{\phi}^{(i)\mu}_{n}
  =\sqrt{ \frac{\alpha'}{2n}}i
   \left(\hat{a}^{(i)\mu}_{n}-\hat{a}^{\dagger (i)\mu}_{n}
   \right)~, \quad
\hat{\pi}^{(i)}_{n\mu} = \sqrt{\frac{n}{2\alpha'}} g_{\mu\nu}
   \left(\hat{a}^{(i)\nu}_{n}+\hat{a}^{\dagger (i)\nu}_{n}\right)~.
  \label{eq:def of a2}
\end{equation}
They satisfy
\begin{equation}
[\hat{a}^{(i)\mu}_{m},\hat{a}^{\dagger (j) \nu}_{n}]
  =g^{\mu\nu} \delta^{i,j}\delta_{m,n}~.
  \label{eq:commutator of a}
\end{equation}
It follows from Eqs.(\ref{eq:among-modes}) that 
these modes are expressed as
\begin{eqnarray}
\hat{a}^{(\mathrm{I}) \mu}_{n}
  = \frac{1}{\sqrt{2n}}
     \left(\alpha_{n}^{\mu}+\tilde{\alpha}_{n}^{\mu} \right)
 &,& \quad
 \hat{a}^{\dagger (\mathrm{I}) \mu}
   =\frac{1}{\sqrt{2n}}
      \left( \alpha^{\mu}_{-n}+\tilde{\alpha}_{-n}^{\mu}\right)~,
 \nonumber\\
\hat{a}^{(\mathrm{II})\mu}_{n}
   =\frac{-i}{\sqrt{2n}}
      \left(\alpha^{\mu}_{n}-\tilde{\alpha}^{\mu}_{n} \right)
  &,& \quad
  \hat{a}^{\dagger (\mathrm{II})\mu}_{n}
   = \frac{i}{\sqrt{2n}}
     \left(\alpha^{\mu}_{-n} - \tilde{\alpha}^{\mu}_{-n}\right)~.  
   \label{eq:2harmonics}
\end{eqnarray}

Eqs.(\ref{eq:def of a2})  
take the same forms as Eqs.(\ref{eq:def of a}) 
with
$\Gamma=i \sqrt{\frac{\alpha'}{n}}$.
Therefore, 
Eqs.(\ref{eq:ho-eigenstates}) 
enable us to write 
eigenstates
$\left|\phi^{(\mathrm{I})}_{n},\phi^{(\mathrm{II})}_{n}\right\rangle$
of $\hat{\phi}^{(\mathrm{I})\mu}_{n}$
and $\hat{\phi}^{(\mathrm{II})\mu}_{n}$ with
eigenvalues
$(\phi^{(\mathrm{I})\mu}_{n},\phi^{(\mathrm{II})\mu}_{n})
  \in \mathbb{R}^{p+1}\times\mathbb{R}^{p+1}$
as follows : 
\begin{eqnarray}
\lefteqn{
\left|\phi^{(\mathrm{I})}_{n},\phi^{(\mathrm{II})}_{n}\right\rangle
 = \left(\frac{n}{\pi\alpha'} \right)^{\frac{p+1}{2}}
     \sqrt{-\det g_{\mu\nu}}
} \nonumber\\
 && \hspace{2em}\times \prod_{i=\mathrm{I},\mathrm{II}}
   \exp \left[  \frac{1}{2}\hat{a}^{\dagger (i)\mu}_{n}
                g_{\mu\nu} \hat{a}^{\dagger (i)\nu}_{n}
      -i\sqrt{\frac{2n}{\alpha'}} 
          \phi^{(i)\mu}_{n}g_{\mu\nu}\hat{a}_{n}^{\dagger(i)\nu}
      -\frac{n}{2\alpha'} \phi^{(i)\mu}_{n}g_{\mu\nu}
                          \phi^{(i)\nu}_{n} \right]
    |\mathbf{0}\rangle~.
    \label{eq:phiphi-a}
\end{eqnarray}
In the same way, 
eigenstates
$\left|\pi^{(\mathrm{I})}_{n},\pi^{(\mathrm{II})}_{n} \right\rangle$
of $\hat{\pi}^{(\mathrm{I})\mu}_{n}$
and $\hat{\pi}^{(\mathrm{II})\mu}_{n}$ with eigenvalues
$(\pi^{(\mathrm{I})\mu}_{n},\pi^{(\mathrm{II})\mu}_{n})
  \in \mathbb{R}^{p+1}\times \mathbb{R}^{p+1}$
become : 
\begin{eqnarray}
\lefteqn{
  \left|\pi^{(\mathrm{I})}_{n},\pi^{(\mathrm{II})}_{n}\right\rangle
 =\left(\frac{\alpha'}{\pi n}\right)^{\frac{p+1}{2}}
  \frac{1}{\sqrt{-\det g_{\mu\nu}}}
}\nonumber\\
&&\hspace{2em} \times \prod_{i=\mathrm{I},\mathrm{II}}
   \exp \left[ -\frac{1}{2} \hat{a}^{\dagger (i)\mu}_{n} g_{\mu\nu}
                \hat{a}^{\dagger (i)\nu}_{n}
               +\sqrt{\frac{\alpha'}{2n}}
                 \pi^{(i)}_{n\mu}\hat{a}^{\dagger (i)\mu}_{n}
               -\frac{\alpha'}{2n} \pi^{(i)}_{n\mu} g^{\mu\nu}
                 \pi^{(i)}_{n\nu} \right]
   |\mathbf{0}\rangle~.
  \label{eq:pipi-a}
\end{eqnarray}
By the construction, these states are normalized as
follows :
\begin{eqnarray}
\left\langle \phi^{\prime (\mathrm{I})}_{n},
             \phi^{\prime (\mathrm{II}}_{n} \left. \right|
             \phi^{(\mathrm{I})}_{n},
             \phi^{(\mathrm{II}}_{n}  \right\rangle
  &=& \prod_{\mu=0}^{p} \delta \left(
      \phi^{\prime (\mathrm{I})\mu}_{n} - \phi^{(\mathrm{I})\mu}_{n}
       \right)
       \delta \left(
         \phi^{\prime (\mathrm{II})\mu}_{n} - \phi^{(\mathrm{II})\mu}_{n}
        \right)~,\nonumber\\
\left\langle \pi^{\prime (\mathrm{I})}_{n},
             \pi^{\prime (\mathrm{II}}_{n} \left. \right|
             \pi^{(\mathrm{I})}_{n},
             \pi^{(\mathrm{II}}_{n}  \right\rangle
  &=& \prod_{\mu=0}^{p} \delta \left(
      \pi^{\prime (\mathrm{I})}_{n\mu} - \pi^{(\mathrm{I})}_{n\mu}
       \right)
       \delta \left(
         \pi^{\prime (\mathrm{II})}_{n\mu} - \pi^{(\mathrm{II})}_{n\mu}
        \right)~.
\end{eqnarray}
Eqs.(\ref{eq:2harmonics}) make it possible to write these 
eigenstates in the forms of 
Eqs.(\ref{eq:chi-chi N}) and (\ref{eq:psi-psi}).
In the text we use complex variables
$(\chi^{\mu}_{n},\bar{\chi}^{\mu}_{n})
=(\phi^{(\mathrm{I})\mu}_{n}+i \phi^{(\mathrm{II})\mu}_{n},
  \phi^{(\mathrm{I})\mu}_{n}-i \phi^{(\mathrm{II})\mu}_{n})$
and write the coordinate eigenstates as
$\left|\chi_{n},\bar{\chi}_{n}\right\rangle
   =\left|\phi^{(\mathrm{I})}_{n},\phi^{(\mathrm{II})}_{n}\right\rangle$.
As for the momentum eigenstates,
we use complex variables
$(\psi_{n\mu},\bar{\psi}_{n\mu})=
  (\pi^{(\mathrm{I})}_{n\mu}+i\pi^{(\mathrm{II})}_{n\mu},
  \pi^{(\mathrm{I})}_{n\mu}-i\pi^{(\mathrm{II})}_{n\mu})$
and write them as
$\left| \psi_{n},\bar{\psi}_{n}\right\rangle
  =\left|\pi^{(\mathrm{I})}_{n},\pi^{(\mathrm{II})}_{n}\right\rangle$.

Next we consider eigenstates of 
$\hat{P}^{(B)}_{\mu} (\sigma)$. 
Taking account of Eqs.(\ref{eq:among-modes}), 
let us write the expansion modes $\hat{\varpi}_{n\mu}^{(i)}$
($i=\mathrm{I},\mathrm{II}$; $n=1,2,\ldots$)  as
\begin{equation}
\hat{\varpi}^{(i)}_{n\mu} = \sqrt{\frac{n}{2\alpha'}}
  G_{\mu\nu} \left(\hat{\eta}^{(i)\mu}_{n}
      +\hat{\eta}^{\dagger (i) \mu}_{n} \right)~,
    \label{eq:def of eta}
\end{equation}
where $\hat{\eta}^{(i)\mu}_{n}$ and
$\hat{\eta}^{\dagger (i)\mu}_{n}$
denote
\begin{eqnarray}
 \hat{\eta}^{(\mathrm{I})\mu}_{n} &=& \frac{1}{\sqrt{2n}}
      \left\{ {\left(\frac{1}{E^{T}} g\right)^{\mu}}_{\nu} 
               \alpha^{\nu}_{n}
      +{\left( \frac{1}{E} g\right)^{\mu}}_{\nu}
              \tilde{\alpha}^{\nu}_{n} \right\}~, \nonumber\\
\hat{\eta}^{\dagger (\mathrm{I}) \mu}_{n} &=&
        \frac{1}{\sqrt{2n}} \left\{
        { \left(\frac{1}{E^{T}}g\right)^{\mu}}_{\nu} \alpha^{\nu}_{-n}
        +{\left(\frac{1}{E} g\right)^{\mu}}_{\nu} \tilde{\alpha}^{\nu}_{-n}
      \right\}~,
    \nonumber\\
\hat{\eta}^{(\mathrm{II})\mu}_{n}&=&
      \frac{-i}{\sqrt{2n}} \left\{
       {\left( \frac{1}{E^{T}}g\right)^{\mu}}_{\nu}\alpha^{\nu}_{n}
       -{\left(\frac{1}{E} g\right)^{\mu}}_{\nu} \tilde{\alpha}^{\nu}_{n}
       \right\}~,\nonumber\\
\hat{\eta}^{\dagger (\mathrm{II})\mu}_{n} &=&
     \frac{i}{\sqrt{2n}} \left\{
      {\left(\frac{1}{E^{T}}g\right)^{\mu}}_{\nu}\alpha^{\nu}_{-n}
        -{\left(\frac{1}{E} g\right)^{\mu}}_{\nu}
           \tilde{\alpha}^{\nu}_{-n} \right\}~.
     \label{eq:def of eta2}
\end{eqnarray}
They turn out to satisfy
\begin{equation}
 [\hat{\eta}^{(i)\mu}_{m}, \hat{\eta}^{\dagger (j) \nu}_{n}]
  =G^{\mu\nu}\delta^{i,j}\delta_{m,n}~.
  \label{eq:commutator of eta}
\end{equation}

Comparing Eqs.(\ref{eq:def of eta}) and (\ref{eq:commutator of eta})
with Eqs.(\ref{eq:def of a2}) and (\ref{eq:commutator of a}), 
we can find that 
eigenstates
$\left| \varpi^{(\mathrm{I})}_{n},
       \varpi^{(\mathrm{II})}_{n}\right\rangle$
of $\hat{\varpi}^{(\mathrm{I})}_{n\mu}$
and $\hat{\varpi}^{(\mathrm{II})}_{n\mu}$ 
are obtained from 
$\left| \pi^{(\mathrm{I})}_{n}, \pi^{(\mathrm{II})}_{n}
  \right\rangle$ 
by the following replacements :  
\begin{equation}
g_{\mu\nu} \rightarrow G_{\mu\nu}~;
\quad
\left(\hat{a}^{(i)\mu}_{n},\hat{a}^{\dagger (i)\mu}_{n}\right)
 \rightarrow
 \left(\hat{\eta}^{(i)\mu}_{n},\hat{\eta}^{\dagger (i)\mu}_{n}\right)~.
 \label{eq:replacement}
\end{equation}
Thus we have : 
\begin{eqnarray}
\lefteqn{
\left| \varpi^{(\mathrm{I})}_{n},
       \varpi^{(\mathrm{II})}_{n}\right\rangle
  =\left(\frac{\alpha'}{\pi n}\right)^{\frac{p+1}{2}}
   \frac{1}{\sqrt{-\det G_{\mu\nu}}} }\nonumber\\
&& \times \prod_{i=\mathrm{I},\mathrm{II}}
   \exp \left[ -\frac{1}{2}\hat{\eta}^{\dagger (i)\mu}G_{\mu\nu}
                \hat{\eta}^{\dagger (i)\nu}_{n}
    +\sqrt{\frac{2\alpha'}{n}} 
       \varpi^{(i)}_{n\mu}\hat{\eta}^{\dagger (i)\mu}_{n}
   -\frac{\alpha'}{2n}
    \varpi^{(i)}_{n\mu}G^{\mu\nu}\varpi^{(i)}_{n\nu}\right]
    |\mathbf{0}\rangle~.             
\end{eqnarray}
Substitutions of Eqs.(\ref{eq:def of eta2}) 
into the above realizations make the states in the forms 
(\ref{eq:varrho-varrho}).
We use complex variables 
$(\varrho_{n\mu},\bar{\varrho}_{n\mu})
  =(\varpi^{(\mathrm{I})}_{n\mu}+i\varpi^{(\mathrm{II})}_{n\mu},
    \bar{\varpi}^{(\mathrm{I})}_{n\mu}-i\varpi^{(\mathrm{II})}_{n\mu})$
in the text and write
$\left|\varrho_{n},\bar{\varrho}_{n}\right\rangle =
  \left|\varpi^{(\mathrm{I})}_{n},\varpi^{(\mathrm{II})}_{n}
  \right\rangle$.


\end{document}